\documentclass[12pt,twoside,english]{extarticle}
\usepackage{lmodern}
\usepackage{helvet}
\usepackage[T1]{fontenc}
\usepackage[latin9]{inputenc}
\usepackage{geometry}
\usepackage{color}
\usepackage{babel}
\usepackage{array}
\usepackage{float}
\usepackage{mathrsfs}
\usepackage{multirow}
\usepackage{amsmath}
\usepackage{amsthm}
\usepackage{amssymb}
\usepackage{graphicx}
\usepackage{setspace}
\usepackage[unicode=true,pdfusetitle,
 bookmarks=true,bookmarksnumbered=false,bookmarksopen=false,
 breaklinks=true,pdfborder={0 0 0},pdfborderstyle={},backref=false,colorlinks=true]
 {hyperref}

\makeatletter

\providecommand{\tabularnewline}{\\}

\usepackage[natbibapa]{apacite}
  \theoremstyle{definition}
  \newtheorem{defn}{\protect\definitionname}[section]
  \theoremstyle{remark}
  \newtheorem{rem}{\protect\remarkname}[section]
  \theoremstyle{plain}
  \newtheorem{assumption}{\protect\assumptionname}
  \theoremstyle{definition}
  \newtheorem*{example*}{\protect\examplename}
  \theoremstyle{definition}
  \newtheorem{condition}{\protect\conditionname}
  \theoremstyle{plain}
  \newtheorem{thm}{\protect\theoremname}[section]
  \theoremstyle{plain}
  \newtheorem{cor}{\protect\corollaryname}[section]
  \theoremstyle{plain}
  \newtheorem{lem}{\protect\lemmaname}[section]

\usepackage{graphicx}
\usepackage{amsmath}
\usepackage{dcolumn}
\usepackage{listings}
\usepackage{color}
\usepackage{setspace}
\usepackage[normalem]{ulem}  % For using \uline
\definecolor{hellgelb}{rgb}{1,1,0.8}
\definecolor{colKeys}{rgb}{0,0,1}
\definecolor{colIdentifier}{rgb}{0,0,0}
\definecolor{colComments}{rgb}{1,0,0}
\definecolor{colString}{rgb}{0,0.5,0}

\usepackage{lmodern}
\usepackage[T1]{fontenc}
\usepackage{geometry}
\usepackage{fancyhdr}
\usepackage{amsthm}
\usepackage{thmtools}
\usepackage{amsmath}
\usepackage{amssymb}
\usepackage{esint}
\usepackage{multirow}
\usepackage{mathrsfs} 
\usepackage{textgreek}
\usepackage{amsfonts}
\usepackage{amssymb}
\usepackage{mathrsfs} 

\usepackage[USenglish]{isodate}
\usepackage{chngcntr}

\numberwithin{equation}{section}
\numberwithin{table}{section}
\numberwithin{assumption}{section}
\counterwithout{figure}{section}
\counterwithout{table}{section}

\makeatother

  \providecommand{\assumptionname}{Assumption}

  \providecommand{\corollaryname}{Corollary}
  
  \providecommand{\definitionname}{Definition}
  \providecommand{\examplename}{Example}

  \providecommand{\lemmaname}{Lemma}

  \providecommand{\remarkname}{Remark}
  
  \providecommand{\theoremname}{Theorem}
 \providecommand{\corollaryname}{Corollary}
 \providecommand{\theoremname}{Theorem}
 
\usepackage{thmtools}

\newtheoremstyle{MyTheoremstyle}
  {\topsep} % Space above
  {\topsep} % Space below
  {} % Body font
  {} % Indent amount
  {\bfseries} % Theorem head font
  {.} % Punctuation after theorem head
  {.90em} % Space after theorem head
  {} % Theorem head spec (can be left empty, meaning `normal')
\theoremstyle{MyTheoremstyle} 
\theoremstyle{MyTheoremstyle} 
\theoremstyle{MyTheoremstyle} 
\theoremstyle{MyTheoremstyle} 
\theoremstyle{MyTheoremstyle}

\declaretheoremstyle[
    headfont=\bfseries,
    notefont=\normalfont,
    bodyfont=\itshape,
    headpunct=\newline,
    headformat={%
        \makebox{\NAME\ \NUMBER\ }{\NOTE}%
    },
]{theorem}

\newlength{\spacelength}
\declaretheoremstyle[
    headfont=\bfseries,
    notefont=\normalfont,
    bodyfont=\itshape,
    headpunct=\newline,
    headformat={%
        \makebox[0pt][l]{\NAME\ \NUMBER\ }\hskip-\spacelength{\NOTE}%
    },
]{theore}

 \usepackage{fancyhdr}
\usepackage{booktabs}
\usepackage{float}
\usepackage{graphicx}
\usepackage{epstopdf}
\usepackage{morefloats}
\usepackage[referable]{threeparttablex}
\usepackage{footnote}

\usepackage{setspace} 
\usepackage{graphicx,color}

\usepackage{listings}
\RequirePackage[T1]{fontenc}
\RequirePackage{ae,fancyvrb}
\DefineVerbatimEnvironment{Sinput}{Verbatim}{fontshape=sl}
\DefineVerbatimEnvironment{Soutput}{Verbatim}{}
\DefineVerbatimEnvironment{Scode}{Verbatim}{fontshape=sl}

\title{\bf Tests for Forecast Instability and Forecast Failure under a Continuous Record Asymptotic Framework}
\author{%\hfill
\textsc{\textcolor{MyBlue}{Alessandro Casini}}\thanks{Department of Economics, Boston University, 270 Bay State Rodad, Boston, MA 02215, US. %Tel: + . Fax: +. 
Email: %\email{acasini@bu.edu} 
\texttt{\textcolor{MyBlue}{{acasini@bu.edu}}}. Web Homepage: \url{http://alessandro-casini.com}.} 
\\
\small{Boston University}
}

 \makeatletter

\numberwithin{equation}{section}
\makeatother

\usepackage{babel}
\addto\captionsenglish{%
}

\usepackage{color}
\usepackage{xcolor}

\renewcommand*{\thesection}{\arabic{section}}

\definecolor{MyRed}{rgb}{0.8,0,0}
\definecolor{MyBlue}{rgb}{0,0,0.7}
\definecolor{Green}{rgb}{0,0.5,0}
\definecolor{hellgelb}{rgb}{1,1,0.8}
\definecolor{colKeys}{rgb}{0,0,1}
\definecolor{colIdentifier}{rgb}{0,0,0}
\definecolor{colComments}{rgb}{1,0,0}
\definecolor{colString}{rgb}{0,0.5,0}

\definecolor{MyLightRed}{rgb}{2.2,0.2,0.4} % changed this
\definecolor{MyLightRed2}{rgb}{0.6,0.2,0.3} %amaranto
\definecolor{MyLightRed2temp}{rgb}{0.6,0.2,0.3}
\definecolor{MyLightRed3}{rgb}{0.8,0.1,0.1} %red
\definecolor{MyRed}{rgb}{0.7,0.0,0}
\definecolor{MyLigthBlue13}{rgb}{0,0.2,0.7}
 \definecolor{MyLigthBlack}{rgb}{0.2,0.25,0.3} % Light Black for font

\hypersetup{%
  bookmarks=true
  pdftitle = {Title Here},
  pdfsubject = {},
  pdfkeywords = {Keywords},
  pdfauthor = {Alessandro Casini},}
\hypersetup{ 
  colorlinks = {true}, % false if surround link by color fram. True if color the text
  citecolor = {MyBlue},
  linkcolor = {MyRed},
  filecolor= {Green},	
  urlcolor = {MyBlue},
  hyperindex = {true},
  linktocpage = {true},
  linkbordercolor ={1 0 0},
 citebordercolor ={0 1 1},
 urlbordercolor ={0 1 1}
}

\usepackage{bookmark}

\usepackage[bottom]{footmisc}
\makeatother

  \providecommand{\assumptionname}{Assumption}
  \providecommand{\conditionname}{Condition}
  \providecommand{\definitionname}{Definition}
  \providecommand{\examplename}{Example}
  \providecommand{\lemmaname}{Lemma}
  \providecommand{\remarkname}{Remark}
\providecommand{\corollaryname}{Corollary}
\providecommand{\theoremname}{Theorem}

\begin{document}
\begin{titlepage}
\newcommand{\HRule}{\rule{\linewidth}{0.5mm}} % Defines a new command for the horizontal lines, change thickness here
\center % Center everything on the page   %---------------------------------------------------------------------------------------- %   HEADING SECTIONS %----------------------------------------------------------------------------------------
%\textsc{\LARGE DR. B R AMBEDKAR NATIONAL }\\[0.3cm] 
% Name of your university/college 
%\textsc{\LARGE INSTITUTE OF TECHNOLOGY  }\\[0.3cm] 
% \textsc{\Large JALANDHAR-144011, PUNJAB(INDIA) }\\[0.3cm]
\textsc{\Large \textcolor{MyLigthBlack}{research paper and supplemental material}}\\[0.5cm] 
%\textsc{\Large Research Working Paper}\\[0.5cm] 
% Major heading such as course name  
% Minor heading such as course title
%---------------------------------------------------------------------------------------- %   TITLE SECTION %----------------------------------------------------------------------------------------
\rule{\textwidth}{1.6pt}\vspace*{-\baselineskip}\vspace*{2pt} \rule{\textwidth}{0.4pt}\\[\baselineskip] {\huge \bfseries \textcolor{MyLightRed2!75!white}{TESTS FOR FORECAST INSTABILITY AND FORECAST FAILURE UNDER A CONTINUOUS RECORD ASYMPTOTIC FRAMEWORK}\\[0.2\baselineskip] \rule{\textwidth}{0.4pt}\vspace*{-\baselineskip}\vspace{3.2pt} \rule{\textwidth}{1.6pt}\\[\baselineskip]}
\bigskip
\bigskip
\bigskip

%\HRule \\[0.4cm] { \huge \bfseries CONTINUOUS RECORD ASYMPTOTIC FRAMEWORK FOR INFERENCE %IN STRUCTURAL CHANGE MODELS\\}%[0.03cm] % Title of your document 
%\HRule \\[1.5cm]

%---------------------------------------------------------------------------------------- %   AUTHOR SECTION %----------------------------------------------------------------------------------------
 \centering \large \textbf{\textcolor{MyLigthBlue13!70!white}{ALESSANDRO CASINI}}
\\\textcolor{MyLigthBlack}{Department of Economics}
\\\textcolor{MyLigthBlack}{Boston University} % Your name 
% If you don't want a supervisor, uncomment the two lines below and remove the section above %\Large \emph{Author:}\\ %John \textsc{Smith}\\[3cm] % Your name

\bigskip
\bigskip
\bigskip
%---------------------------------------------------------------------------------------- %   DATE SECTION %----------------------------------------------------------------------------------------
\textcolor{MyLigthBlack}{\normalfont{\today} 
\\
%\normalfont{First Vesrion: \printdate{28.10.2015}}
\bigskip
\bigskip
\bigskip
%\uline{\texttt{\small{Preliminary and Incomplete}}}\\
\smallskip
%\uline{\texttt{\small{Please do not cite or circulate}}}
}
%{\large June-December, 2015 \\Lab Practicals Record}\\[1cm] % Date, change the \today to a set date if you want to be precise
%---------------------------------------------------------------------------------------- %   LOGO SECTION %----------------------------------------------------------------------------------------
%\includegraphics{logo.png}\\[1cm] % Include a department/university logo - this will require the graphicx package   %----------------------------------------------------------------------------------------
\vfill % Fill the rest of the page with whitespace 
\end{titlepage}
\noindent
\textbf{Correspondence}:
\medskip\\
\begin{minipage}{0.8\textwidth}
\begin{flushleft} \normalfont \textbf{\textcolor{MyLigthBlue13!70!white}{Alessandro Casini}}
\\Department of Economics
\\Boston University 
\\270 Bay State Road
\\Boston, MA 02215, US
%\\Tel: +1 (857) 919-9787
\\Email: %\email{acasini@bu.edu}  
\texttt{\textcolor{MyLigthBlue13!70!white}{{acasini@bu.edu}}}
\\Webpage: 
\textcolor{MyLigthBlue13!70!white}{{\url{http://alessandro-casini.com}}}
\end{flushleft}  
\end{minipage}
% If you don't want a supervisor, uncomment the two lines below and remove the section above %\Large \emph{Author:}\\ %John \textsc{Smith}\\[3cm] % Your name
\medskip
\bigskip
\noindent\\ 
\textbf{Ducument Information}:
This electronic document includes the working paper and supplemental materials. The Supplement starts on pdf page 48. Monte Carlo simulations were conducted in \textsc{Matlab}. The most recent working paper version can be downloaded from the author's webpage. The associated computer package, if available, should be downloaded from the same webpage. %This is a preliminaty version. %\uline{\texttt{please do not cite or circulate}}. 
%The Supplement contains the Mathematical Appendix and additional materials (examples, illustrations, empirical applications, etc...).
\smallskip
\begin{singlespace}
\noindent\\
\footnotesize{}{
\textbf{Ducument History}}\\
%First Vesrion: \printdate{28.10.2015}\\
First Version: \today
\medskip
\noindent\\
\textbf{Abstract}:
We develop a novel continuous-time asymptotic framework for inference on whether the predictive ability of a given forecast model remains stable over time. We formally define forecast instability from the economic forecaster's perspective and highlight that the time duration of the instability bears no relationship with stable period. Our approach is applicable in forecasting environment involving low-frequency as well as high-frequency macroeconomic and financial variables. As the sampling interval between observations shrinks to zero the sequence of forecast losses is approximated by a continuous-time stochastic process (i.e., an It� semimartingale) possessing certain pathwise properties. We build an hypotheses testing problem based on the local properties of the continuous-time limit counterpart of the sequence of losses. The null distribution follows an extreme value distribution. While controlling the statistical size well, our class of test statistics feature uniform power over the location of the forecast failure in the sample. The test statistics are designed to have power against general form of insatiability and are robust to common forms of nonstationarity such as heteroschedasticity and serial correlation. The gains in power are substantial relative to extant methods, especially when the instability is short-lasting and when occurs toward the tail of the sample.
\medskip
\noindent\\
\textbf{Acknowledgements}: I am indebted to Pierre Perron for introducing me to the issue of structural changes in economics and for sharing insightful discussions on the topic. I am grateful to Raffaella Giacomini for helpful disccusions and to Viktor Todorov for comments on the continous-time approach. I thank Simon Gilchrist for useful remarks on the GZ credit spread index from \citet{gilchrist/zakrajsek:12}. I thank Raffaella Giacomini and Barbara Rossi for making their programs available online. Partial financial support from the Bank of Italy is gratefully acknowledged.
\end{singlespace}
\thispagestyle{empty}

\pagebreak{}

\setcounter{page}{1}

\raggedbottom
% This eliminates unwanted space between paragraphs.

\title{\textbf{Tests for Forecast Instability and Forecast Failure under
a Continuous Record Asymptotic Framework}\thanks{I am indebted to Pierre Perron for introducing me to the issue of
structural changes in economics and for sharing insightful discussions
on the topic. I am grateful to Raffaella Giacomini for helpful disccusions
and to Viktor Todorov for comments on the continous-time approach.
I thank Simon Gilchrist for useful remarks on the GZ credit spread
index from \citet{gilchrist/zakrajsek:12}. I thank Raffaella Giacomini
and Barbara Rossi for making their programs available online. Partial
financial support from the Bank of Italy is gratefully acknowledged.}}
\maketitle
\begin{abstract}
We develop a novel continuous-time asymptotic framework for inference
on whether the predictive ability of a given forecast model remains
stable over time. We formally define forecast instability from the
economic forecaster's perspective and highlight that the time duration
of the instability bears no relationship with stable period. Our approach
is applicable in forecasting environment involving low-frequency as
well as high-frequency macroeconomic and financial variables. As the
sampling interval between observations shrinks to zero the sequence
of forecast losses is approximated by a continuous-time stochastic
process (i.e., an It� semimartingale) possessing certain pathwise
properties. We build an hypotheses testing problem based on the local
properties of the continuous-time limit counterpart of the sequence
of losses. The null distribution follows an extreme value distribution.
While controlling the statistical size well, our class of test statistics
feature uniform power over the location of the forecast failure in
the sample. The test statistics are designed to have power against
general form of insatiability and are robust to common forms of non-stationarity
such as heteroskedasticty and serial correlation. The gains in power
are substantial relative to extant methods, especially when the instability
is short-lasting and when occurs toward the tail of the sample.
\end{abstract}

\onehalfspacing

\pagebreak{}

\section{Introduction }

Since the seminal contribution of Klein \citeyearpar{klein:69,klein:71},
economic forecasts had been built upon the presumption that the relationships
between economic variables remain stable over time. However, the last
decades have been subject to many social-economic episodes and technological
advancements that have led economists to reconsider the assumption
of model stability. The resonant empirical evidences documented in,
among others, \citet{perron:89} and \citet{stock/watson:96} {[}see
also the recent survey by \citet{ng/wright:13}{]} have motivated
the development of econometric methods that detect such instabilities\textemdash most
work directed toward structural changes\textemdash and estimate the
actual dates at which economic relationships change. Yet, the issue
of parameter insatiability is not limited to model estimation. In
the forecasting literature, there has been a widespread concordance
that the major issue that prevents good forecasts for economic variables
is parameter instability\textemdash and structural changes as a special
case\textemdash {[}cf. \citet{banerjee/marcellino/masten:08}, \citeauthor{clements/hendry:98}
(1998, 2006)\nocite{clements/hendry:06}, \citet{elliott/timmermann:16},
\citet{giacomini:15}, \citet{giacomini/rossi:15}, \citet{inoue/rossi:11},
\citet{clark/mccracken:05}, \citet{pesaran/pettenuzzo/timmermann:06}
and \citet{rossi:13a}{]}.

This paper develops a statistical setting under infill asymptotics
to address the issue of testing whether the predictive ability of
a given forecast model remains stable over time. \citet{ng/wright:13}
and \citet{stock/watson:03} explain that there has been abundant
evidence for which a predictor that has performed well over a certain
time period may not perform as well during other subsequent periods.
For example, \citet{gilchrist/zakrajsek:12} proposed a new credit
spread index and showed that a residual component labeled as the excess
bond premium\textemdash the credit spread adjusted for expected default
risk\textemdash has considerable predictive content for future economic
activity. They documented that this forecasting ability is stronger
over the subsample 1985-2010 rather than over the full sample starting
from 1973.\footnote{They reported that structural change tests provide some statistical
evidence for a break in a coefficient associated with financial indicators\textemdash more
specifically the coefficient on the federal funds rate. Given the
latter evidence and the well-documented change in the conduct of monetary
policy in the late 1970s and the early 1980s, it seems plausible to
split the sample in 1985 (see p. 1709 and footnote 11 in their paper).} The latter finding can be attributed to a more developed bond market
in the 1985-2010 subsample. Relatedly, \citet{giacomini/rossi:10}
and \citet{ng/wright:13} further examined this finding and found
that indeed the predictive ability of commonly used term and credit
spreads is unstable and somehow episodic. The latter authors suggested
that credit spreads may be more useful predictors of economic activity
in a more highly leveraged economy and that recent developments in
financial markets translate into credit spreads containing more information
than they had previously. We refer to such temporal instability for
a given forecasting method as \textit{forecast instability} or more
specifically, as \textit{forecast failure}. These terminologies are
not new to professional forecasters as they were informally introduced
by \citet{clements/hendry:98} and generalized in econometric terms
by \citet{giacomini/rossi:09} who interpreted \textit{forecast breakdown}
(or \textit{forecast failure}) as a situation in which the out-of-sample
performance of a forecast model significantly deteriorates relative
to its in-sample performance. Our approach is to formally define forecast
instability from the economic forecaster's perspective.\footnote{We use the terminology ``instability'' because not only the deterioration
but also the improvement of the performance of a given forecast model
over time can provide useful information to the forecaster.} We emphasize that a forecast failure may well result from a short
period of instability within the out-of-sample and not necessarily
require that the instability be systematic in the sense of persisting
throughout the whole out-of-sample period. That is, consistency of
a forecast model's performance with expected performance given the
past should hold not only throughout the out-of-sample but also in
any sub-sample of the latter. Indeed, many documented episodes of
forecast failure seemed to arise from parameters nonconstancy data-generating
processes over relatively short time periods compared to the total
sample size. Hence, the desire of focusing on statistical tests being
able to detect short-lasting instabilities is intuitive: if a test
for forecast failure needs the deterioration of the forecasting ability
to last for, say, at least half of the total sample in order to have
sufficiently high power to reject the null hypotheses, then this test
would not perform very well in practice because instability can be
short-lasting. Furthermore, the occurrence of recurrent structural
instabilities or multiple breaks that compensate each other in the
out-of-sample might lead a forecast model to perform, on average,
in a similar fashion as in the in-sample period. However, should a
forecaster know about those recurrent changes she would conceivably
revise its forecast model to adapt to the unstable environment. Hence,
we introduce the following definition.
\begin{defn}
(Forecast Instability)

Forecast Instability refers to a situation of either sustained deterioration
or improvement of the predictive ability of a given forecast model
relative to the historical performance that would had led a forecaster
to revise or reconsider its forecast model if she had known the occurrence
of such instability. The time lengths of these two distinct periods
need not bear any relationship.\footnote{\textit{Forecast Failure} constitutes a special case of the definition\textemdash namely,
a sustained deterioration of predictive ability.} 
\end{defn}
Th definition poses at the center the economic forecaster and consequently
it is not merely a statistical definition; rather, it is based on
an equilibrium concept. Since forecasting constitutes a decision theoretic
problem, it should be from the forecaster perspective that a given
forecast model is deemed to have failed. It is implicit from the definition
to distinguish between forecasting method and model. Two forecasters
may share the same forecast model\textemdash the relationship between
the variable of interest and the predictor\textemdash but use different
methods (e.g., recursive scheme versus rolling scheme). Thus, instability
refers to a given method-model pair. The object of the definition
is predictive ability. Since the latter can be measured differently
by different loss functions, then the definition applies to a given
choice of the loss function. A notable aspect of the definition is
the reference to the time span of the historical performance and of
the putative period of instability. They need not be related. Consider
a given forecasting strategy which has performed well during, say,
the Great Moderation (i.e., from mid-1980s up to prior the beginning
of the Great Recession in 2007). Assume that during the years 2007-2012
this method endures a time of poor performance and returns to perform
well thereafter. According to our definition, this episode constitutes
an example of forecast instability. However, if one designs the forecasting
exercise in such a way that half of the sample is used for estimation
and the remaining half for prediction, then this relatively short
period of instability gets ``averaged-out'' from tests which simply
compare the in-sample and out-of-sample averages. Conceivably, such
tests would not reject the null hypotheses of no forecast failure
while it seems that a forecaster would had revised its strategy during
the crisis if she had known about such occurring under-performance
in the present and immediate future period. Finally, detection of
forecast instability does not necessarily mean that a forecast model
should be abandoned. In fact, its performance may have improved over
time. Yet, even if forecast instability is induced by performance
deterioration, a forecaster might not end up switching to a new predictor.
For example, entering a state of high variability might lead to poor
performance even if the forecast model is still correct. Hence, our
definition uses the term \textit{reconsider}. Continuing with the
above example, a forecaster may \textit{reconsider} the choice of
the forecasting window since a longer window may now produce better
forecasts while keeping the \textit{same} forecast model. In other
words, knowledge of forecast instability is important because indicates
that care must be exercised to assess the source of the changes.\footnote{Economists have documented episodes of forecast failure in many areas
of macroeconomics. In the empirical literature on exchange rates a
prominent forecast failure is associated with the Meese and Rogoff's
puzzle {[}cf. \citet{meese/rogoff:83}, \citet{cheung/chinn/garcia:05},
and \citet{rossi:13b} for an up-to-date account{]}. In the context
of inflation forecasting, forecast failures have been reported by
\citet{atkeson/ohanian:01} and \citet{stock/watson:09}. For forecast
instability concerning other macroeconomic variables see the surveys
of \citet{stock/watson:03} and \citet{ng/wright:13}.}

The theoretical implication is that in this paper our tests for forecast
instability shall be based on the local behavior of the sequence of
realized forecast losses. This is opposite to existing tests for forecast
instability\textemdash and classical structural change tests more
generally\textemdash which instead rely on a global and retrospective
methodology merely comparing the average of in-sample losses with
the average of out-of-sample losses. While maintaining approximately
correct nominal size, our class of test statistics achieves substantial
gains in statistical power relative to previous methods. Furthermore,
as the initial timing of the instability moves away from middle sample
toward the tail of the out-of-sample, the gains in power become considerable.

In this paper, we set out a continuous record asymptotic framework
for a forecasting environment where $T$ observations at equidistant
time intervals $h$ are made over a fixed time span $\left[0,\,N\right],$
with $N=Th.$ These observations are realizations from a continuous-time
model for the variable to be forecast and for the predictor. From
these discretely observed realizations we compute a sequence of forecasts
using either a fixed, recursive or rolling scheme. To this sequence
of forecasts there corresponds a continuous-time process which satisfies
mild regularity conditions and that under the null hypotheses possesses
a continuous sample-path. We exploit this pathwise property to base
an hypothesis testing problem on the relative performance of a given
forecast model over time. Under the hypotheses we expect the sequence
of losses to display a smooth and stable path. Any discontinuous or
jump behavior followed by a (possibly short) period of substantial
discrepancy from the same path over the in-sample period provides
evidence against the hypotheses. Our asymptotic theory involves a
continuous record of observations where we let the sample size $T$
grow to infinity by shrinking the sampling interval $h$ to zero with
the time span kept fixed at $N$, thereby approaching the continuous-time
limit. 

Our underlying probabilistic model is specified in terms of continuous
It� semimartingales which are standard building blocks for analysis
of macro and financial high-frequency data {[}cf. \citet*{andersen/bollerslev/diebold/labys:01},
\citet{andersen/fusari/todorov:16}, \citet{bandi/reno:16} and \citet{barndorff/shephard:04}{]};
the theoretical methodology is thus related to that of \citet{casini/perron_CR_Single_Break},
\citet{li/todorov/tauchen:17}, \citet{li/xiu:16} and \citet{mykland/zhang:09}.\footnote{Recent work by \citet{li/patton:17} extends standard methods for
testing predictive accuracy of forecasts to a high-frequency financial
setting.} The framework is not only useful for high-frequency data; in particular,
recent work of \citeauthor{casini/perron_CR_Single_Break} (2017a,
2017b)\nocite{casini/perron_Lap_CR_Single_Inf} has adopted this continuous-time
approach for modeling time series regression models with structural
changes fitted to low-frequency data (e.g., macroeconomic data that
are sampled at weekly, monthly, quarterly, annual frequency, etc.).
They have showed that this continuous-time approach delivers a 
better approximation to the finite-sample distributions of estimators
in structural change models and inference is more reliable than previous
methods based on classical long-span asymptotics. 

The classical approach to economic forecasting for macroeconomic variables
is to formulate models in discrete-time and then base inference on
long-span asymptotics where the sample size increases without bound
and the sampling interval remains fixed {[}cf. \citet{diebold/mariano:95},
\citet{giacomini/white:06} and \citet{west:96}{]}. There are crucial
distinctions between this classical approach and the setting introduced
in this paper. Under long-span asymptotics, identification of parameters
hinges on assumptions on the distributions or moments of the studied
processes {[}cf. the specification of the null hypotheses in \citet{giacomini/rossi:09}{]},
whereas within a continuous-time framework, unknown structural parameters
are identified from the sample paths of the studied processes. Hence,
we only need to assume rather mild pathwise regularity conditions
for the underlying continuous-time model and avoid any ergodic or
weak-dependence assumption.  As in \citet{casini/perron_CR_Single_Break},
our framework encompasses any time series regression model allowing
for general forms of non-stationarity such as heteroskedasticty and
serial correlation.

Given a null hypotheses stated in terms of the path properties of
the sequence of losses, we propose a test statistic which compares
the local behavior of the sequence of surprise losses defined as the
difference between the out-of-sample and in-sample losses. More specifically,
our maximum-type statistic examines the smoothness of the sequence
of surprise losses as the continuous-time limit is approached. Under
the hypotheses, the continuous-time analogue of the sequence of losses
follows a continuous motion and any deviation from such smooth path
is interpreted as evidence against the hypotheses. The null distribution
of the test statistic is non-standard and follows an extreme value
distribution. Therefore, our limit theory exploits results from extreme
value theory as elaborated by \citet{bickel/rosenblatt:73} and \citet{galambos:87}.\footnote{In nonparametric change-point testing, related works are \citet{wu/zhao:07}
and \citet{bibinger/jirak/vetter:16}.} 

We propose two versions of the test statistic: one that is self-normalized
and one that uses an appropriate estimator of the asymptotic variance.
The test statistic is defined as the maximal deviation between the
average surprise losses over asymptotically vanishing time blocks.
Further, we consider extensions of each of these statistics which
use overlapping rather than non-overlapping blocks. Although they
should be asymptotically equivalent, the statistics based on overlapping
blocks are more powerful in finite-samples. In a framework where one
allows for model misspecification, the problem of non-stationarity
such as heteroskedastcity and serial correlation in the forecast losses
should be taken seriously. Given the block-based form our test statistics
we derive an alternative estimator of the long-run variance of the
forecast losses. This estimator differs from the popular estimators
of \citet{andrews:91} and \citet{newey/west:87} {[}see \citet{muller:07}
for a review{]} and it is of independent interest. Finally, we extend
results to settings that allow for stochastic volatility, and we conduct
a local power analysis and highlight a few differences of our testing
framework from the structural change test of \citet{andrews:93}.
Related aspects, such as estimating the timing of the instability
and covering high-frequency setting with jumps, are being considered
in a companion paper. 

The rest of the paper is organized as follows. Section \ref{Section: Statistical Enviromnent}
introduces the statistical setting, the hypotheses of interest and
the test statistics. Section \ref{Section CR Distribution Theory Test Stats}
derives the asymptotic null distribution under a continuous record.
We discuss the estimation of the asymptotic variance in Section \ref{Section Estimation-of-Asymptotic Variance}.
Some extensions and a local power analysis are presented in Section
\ref{Section Ito Vol}. Additional elements that are covered in our
companion paper are briefly described in Section \ref{Section Extensions}.
A simulation study is contained in Section \ref{Section Simulation Study}.
Section \ref{Section Conclusions} concludes the paper. The supplemental
material to this paper contains all mathematical proofs and additional
simulation experiments.

\section{The Statistical Environment\label{Section: Statistical Enviromnent}}

Section \ref{Subsection The-Forecasting-Problem} introduces the statistical
setting with a description of the forecasting problem and the sampling
scheme considered throughout. The underlying continuous-time model
and its assumptions are introduced in Section \ref{Subsection The Continuous-Time Model}.
In Section \ref{Subsection The-Hypotheses-of} we set out the testing
problem and state the relevant null and alternative hypotheses. The
test statistics are presented in Section \ref{Subsection The-Test-Statistics}.
Throughout we adopt the following notational conventions. All limits
are taken as $T\rightarrow\infty$, or equivalently as $h\downarrow0$,
where $T$ is the sample size and $h$ is the sampling interval.
All vectors are column vectors and for two vectors $a$ and $b$,
we write $a\leq b$ if the inequality holds component-wise. For a
sequence of matrices $\left\{ A_{T}\right\} ,$ we write $A_{T}=o_{\mathbb{P}}\left(1\right)$
if each of its elements is $o_{\mathbb{P}}\left(1\right)$ and likewise
for $O_{\mathbb{P}}\left(1\right).$ If $x$ is a non-stochastic vector,
$\left\Vert x\right\Vert $ denotes the its Euclidean norm, whereas
if $x$ is a stochastic vector, the same notation is used for the
$L^{2}$ norm. We use $\left\lfloor \cdot\right\rfloor $ to denote
the largest smaller integer function and for a set $A,$ the indicator
function of $A$ is denoted by $\mathbf{1}_{A}$. A sequence $\left\{ u_{kh}\right\} _{k=1}^{T}$
is $\textrm{i.i.d.}$ if the $u_{kh}$ are independent and identically
 distributed. We use $\overset{\mathbb{P}}{\rightarrow},\,\Rightarrow$
to denote convergence in probability and weak convergence, respectively.
$\mathcal{M}_{p}^{\textrm{c�dl�g}}$ is used for the space of $p\times p$
positive define real-valued matrices whose elements are c�dl�g. The
symbol ``$\triangleq$'' is definitional equivalence. 

\subsection{\label{Subsection The-Forecasting-Problem}The Forecasting Problem}

The continuous-time stochastic process $Z\triangleq\left(Y,\,X'\right)$
is defined on a filtered probability space $\left(\Omega,\,\mathscr{F},\,\left\{ \mathscr{F}_{t}\right\} _{t\geq0},\,\mathbb{P}\right)$
and takes value in $\mathbf{Z}\subseteq\mathbb{R}^{q+1}$ where $\left\{ Y_{t}\right\} _{t\geq0}$
is the variable to be forecast and $\left\{ X_{t}\right\} _{t\geq0}$
are the predictor variables. The index $t$ is defined as the continuous-time
index and we have $t\in\left[0,\,N\right]$, where $N$ is referred
to as the time span. In this paper, $N$ will remain fixed. That is,
the unobserved process $Z_{t}$ evolves within the fixed time horizon
$\left[0,\,N\right]$ and the econometrician records $T$ of its realizations,
with a sampling interval $h$, at discrete-time points $h,\,2h\ldots,\,Th$,
where accordingly $Th=N.$ A continuous record asymptotic framework
involves letting the sample size $T$ grow to infinity by shrinking
the time interval $h$ to zero at the same rate so that $N$ remains
fixed. The index $k$ is used for the observation (or tick) times
$k=1,\ldots,\,T$.

The objective is to generate a series $\left\{ Y_{\left(k+\tau\right)h}\right\} $
of $\tau$-step ahead forecasts. We shall adopt an out-of-sample
precedure whereby splitting the time span $\left[0,\,N\right]$ into
an \textit{in-sample} and \textit{out-of-sample} window, $\left[0,\,N_{\mathrm{in}}\right]$
and $\left[N_{\mathrm{in}}+h,\,N\right]$, respectively.\footnote{Indeed, $\left[0,\,N_{\mathrm{in}}\right]$ corresponds to the in-sample
window only for the fixed forecasting scheme to be introduced later\textemdash e.g.,
the rolling scheme only uses the most recent span of data of length
$N_{\mathrm{in}}$. A minor and straightforward modification to this
notation should be applied when the recursive and rolling schemes
are considered. However, for all methods $N_{\mathrm{in}}$ indicates
the artificial separation such that $N_{\mathrm{in}}+h$ is the beginning
of the out-of-sample period.} The latter two time horizons are supposed to be fixed and therefore
within the in-sample (or prediction) window a sample of size $T_{m}$
is observed whereas within the out-of-sample (or estimation) window
the sample is of size $T_{n}=T-T_{m}-\tau+1$. We consider a general
framework that allows for the three traditional forecasting schemes:
(1) a \textit{fixed} forecasting scheme with discrete-time observations
$h,\,2h,\ldots,\,\left(T_{m}-1\right)h,\,T_{m}h=N_{\mathrm{in}}$;
(2) a \textit{recursive} forecasting scheme where at time $kh$ the
prediction sample includes observations $h,\ldots,\,\left(k-1\right)h,\,kh$;
(3) a \textit{rolling} forecasting scheme where the time span of the
rolling window is fixed and of the same length as $N_{\mathrm{in}}$
(i.e., at time $kh$ the in-sample window includes observations $kh-T_{m}h+h,\ldots,\,\left(k-1\right)h,\,kh$.\footnote{Equivalently, the observation times within the rolling widow at the
$k$th's observation are $k-T_{m}+1,\ldots,\,k$.}

The forecasts may be based on a parametric model whose time-$kh$
parameter estimates are then collected into the $q\times1$ random
vector $\widehat{\beta}_{k}$. If no parametric assumption is made,
then $\widehat{\beta}_{k}$ represents whatever semiparametric or
nonparametric estimator used for generating the forecasts. The time-$kh$
forecast is denoted by $\widehat{f}_{k}\left(\widehat{\beta}_{k}\right)\triangleq f\left(Z_{kh},\,Z_{\left(k-1\right)h},\ldots,\,Z_{\left(k-m_{f}+1\right)h};\,\widehat{\beta}_{k}\right)$,
where $f$ is some measurable function. The notation indicates that
the $kh$-time forecast is generated from information contained in
a sample of size $m_{f}$.\footnote{$m_{f}$ varies with the forecastis scheme; e.g., for the rolling
scheme we have $m_{f}=T_{m}$ while for the recursive scheme we have
$m_{f}=k$.}

Next, we introduce a loss function $L\left(\cdot\right)$ which serves
for evaluating the performance of a given forecast model. More specifically,
each out-of-sample loss $L_{\left(k+\tau\right)h}\left(\widehat{\beta}_{k}\right)\triangleq L\left(Y_{\left(k+\tau\right)h},\,\widehat{f}_{k}\left(\widehat{\beta}_{k}\right)\right)$
constitutes a statistical measure of accuracy of the $\tau$-step
forecast made at time $kh$. However, given the objective of detecting
potential instability of a certain forecasting method over time, we
need additionally to introduce the in-sample losses $L_{jh}\left(\widehat{\beta}_{k}\right)\triangleq L\left(Y_{jh},\,\widehat{y}_{j}\left(\widehat{\beta}_{k}\right)\right),$
where $\widehat{y}_{j}\left(\widehat{\beta}_{k}\right)$ is an in-sample
fitted value with $j$ varying over the specific in-sample window.
That is, for each time-$kh$ forecast there corresponds a sequence
(indexed by $j$) of in-sample fitted values $\widehat{y}_{j}\left(\widehat{\beta}_{k}\right)$.\footnote{We have $j=\tau+1,\ldots,\,T_{m}$ for the fixed scheme, $j=\tau+1,\ldots,\,k$
for the recursive scheme and $j=k-T_{m}+\tau+1,\ldots,\,k$ for the
rolling scheme. } Then, the testing problem turns into the detection of any ``systematic
difference'' between the sequence of out-of-sample and in-sample
losses; the formal measure of such difference under our context is
provided below. 

\subsection{\label{Subsection The Continuous-Time Model}The Underlying Continuous-Time
Model}

The process $Z$ is a $\mathbb{R}^{q+1}$-valued semimartingale on
$\left(\Omega,\,\mathscr{F},\,\left\{ \mathscr{F}_{t}\right\} _{t\geq0},\,\mathbb{P}\right)$
and we further assume that all processes considered in this paper
are c�dl�g adapted and possess a $\mathbb{P}$-a.s. continuous path
on $\left[0,\,N\right]$.\footnote{For accessible treatments of the probabilistic elements used in this
section we refer to \citet{sahalia/jacod:14}, \citet{jacod/shiryaev:03},
\citet{jacod/protter:12}, \citet{karatzas/shreve:96} and \citet{protter:05}.} The continuity property represents a key assumption in our setting
and implies that $Z$ is a continuous It\^o semimartignale. The integral
form for $X_{t}$ is given by, 
\begin{align}
X_{t} & =x_{0}+\int_{0}^{t}\mu_{X,s}ds+\int_{0}^{t}\sigma_{X,s}dW_{X,s},\label{Mode for X}
\end{align}
 where $\left\{ W_{X,t}\right\} _{t\geq0}$ is a $q\times1$ Wiener
process, $\mu_{X,s}\in\mathbb{R}^{q}$ and $\sigma_{X,s}\in\mathcal{M}_{q}^{\textrm{c�dl�g}}$
are the drift and spot covariance process, respectively, and $x_{0}$
is $\mathscr{F}_{0}$-measurable. We incorporate model misspeficication
into our framework by allowing for a large non-zero drift which adds
to the residual process: 
\begin{align}
Y_{t} & \triangleq y_{0}+\left(\beta^{*}\right)'X_{t-}+\int_{0}^{t}\mu_{e,s}h^{-\vartheta}ds+e_{t},\qquad\qquad e_{t}\triangleq\int_{0}^{t}\sigma_{e,s}dW_{e,s}\label{Mode for Y}
\end{align}
where $\beta^{*}\in\mathbb{R}^{q}$, $\left\{ W_{e,t}\right\} _{t\geq0}$
is a standard Wiener process, $\sigma_{e,s}\in\mathbb{R}_{+}$ is
its associated volatility, $\mu_{e,s}\in\mathbb{R}$ and $y_{0}$
is $\mathscr{F}_{0}$-measurable. In \eqref{Mode for Y}, the last
two terms on the right-hand side account for the residual part of
$Y_{t}$ which is not explained by $X_{t-},$ where $X_{t-}=\lim_{s\uparrow t}X_{s}$.
We assume $\vartheta\in[0,\,1/8)$ so that the factor $h^{-\vartheta}$
inflates the infinitesimal mean of the residual component thereby
approximating a setting with arbitrary misspecification. 
\begin{rem}
In \eqref{Mode for Y}, misspecification manifests itself in the
form of (time-varying) non-zero conditional mean of the residual process,
and in giving rise to serial dependence in the disturbances which
in turn leads to dependence in the sequence of forecast losses.\footnote{Asymptotically, these features can be dealt with basic arguments used
in the high-frequency financial statistics literature; however, when
$h$ is not too small one needs methods that are robust in finite-samples
to such misspecification-induced properties. More precisely, we will
propose an appropriate estimator of the long-run variance of the sequence
of forecast losses in Section \ref{Section Estimation-of-Asymptotic Variance}.
} Hence, this specification is similar in spirit to the near-diffusion
assumption of \citet{foster/nelson:96} who studied the impact of
misspecification in ARCH models. On the other hand, \citet{casini/perron_CR_Single_Break}
introduced a ``large-drift'' asymptotics with $h^{-1/2}$ to deal
with non-identification of the drift in their context. Technically,
the latter specification implies that as $h$ becomes small the drift
features larger oscillations that add to the local Gaussianity of
the stochastic part. \citet{casini/perron_CR_Single_Break} referred
to this specification as \textit{small-dispersion} assumption. Finally,
note that the presence of $h^{-\vartheta}$ can also be related to
the signal plus small Gaussian noise of \citet{ibragimov/has:81}
if one sets $\varepsilon_{h}=h^{\vartheta}$ in their model in Section
VII.2.
\end{rem}
\begin{assumption}
\label{Assumption 1, CT}We have the following assumptions: (i) The
processes $\left\{ X_{t}\right\} _{t\geq0}$ and $\Sigma^{0}\triangleq\left\{ \sigma_{X,t},\,\sigma_{e,t}\right\} _{t\geq0}$
have $\mathbb{P}$-a.s. continuous sample paths; (ii) The processes
$\left\{ \mu_{X,t}\right\} _{t\geq0},\,\left\{ \mu_{e,t}\right\} _{t\geq0},\,\left\{ \sigma_{X,t}\right\} _{t\geq0}$
and $\left\{ \sigma_{e,t}\right\} _{t\geq0}$ are locally bounded;
(iii) There exists $0<\sigma_{-}<\sigma_{+}<\infty$ such that $\mathbb{P}$-a.s.
$\inf_{t\in\left[0,\,N\right]}\sigma_{V,t}^{2}\geq\sigma_{-}^{2}$
and $\sigma_{+}^{2}\geq\sup_{t\in\left[0,\,N\right]}\sigma_{V,t}^{2}$
with $V=X,\,e$; (iv) $\sigma_{X,t}\in\mathcal{M}_{q}^{\mathrm{c\grave{a}dl\grave{a}g}}$
and $\sigma_{e,t}\in\mathcal{M}_{1}^{\mathrm{c\grave{a}dl\grave{a}g}}$
and the conditional variance (or spot covariance) is defined as $\Sigma_{X,t}=\sigma_{X,t}\sigma'_{X,t}$,
which for all $t<\infty$ satisfies $\int_{0}^{t}\Sigma_{X,s}^{\left(j,j\right)}ds<\infty,\,\left(j=1,\ldots,\,q\right)$
where $\Sigma_{X,t}^{\left(j,r\right)}$ denotes the $\left(j,r\right)$-th
element of $\Sigma_{X,t}$. Furthermore, for every $j=1,\ldots,\,q,$
and $k=1,\,2,\ldots,\,T$, the quantity $h^{-1}\int_{\left(k-1\right)h}^{kh}\Sigma_{X,s}^{\left(j,j\right)}ds$
is bounded away from zero and infinity, uniformly in $k$ and $h$;
(v) The disturbance process $e_{t}$ is orthogonal (in martingale
sense) to $X_{t}:$ $\left\langle e,\,X\right\rangle _{t}=0$ identically
for all $t\geq0$.\footnote{The angle brackets notation $\left\langle \cdot,\cdot\right\rangle $
is used for the predictable quadratic variation process.}
\end{assumption}
Part (i) rules out jump processes from our setting. We relax this
restriction in our companion paper; see Section \ref{Section Extensions}.
Part (ii) restricts those processes to be locally bounded. These should
be viewed as regularity conditions rather than assumptions and are
standard in the financial econometrics literature {[}see \citet{barndorff/shephard:04},
\citet{li/xiu:16} and \citet*{li/todorov/tauchen:17}{]}; recently,
they have been used by \citet{casini/perron_CR_Single_Break} in the
context of structural change models.

The continuous-time model in \eqref{Mode for X}-\eqref{Mode for Y}
is not observable. The econometrician only has access to $T$ realizations
of $Y_{t}$ and $X_{t}$ with a sampling interval $h>0$ over the
horizon $\left[0,\,N\right]$. For each $h>0,$ $Z_{kh}\in\mathbb{R}^{q+1}$
is a random vector step function that jumps only at time $0,\,h,\,2h,\ldots$,
and so on. The discretized processes $Y_{kh}$ and $X_{kh}$ are assumed
to be adapted to the increasing and right-continuous filtration $\left\{ \mathscr{F}_{t}\right\} _{t\geq0}$.
The increments of a process $U$ are denoted by $\Delta_{h}U_{k}=U_{kh}-U_{\left(k-1\right)h}$.
A seminal result known as Doob-Meyer Decomposition {[}cf. the original
sources are \citet{doob:53} and \citet{meyer:67}; see also Section
III.3 in \citet{protter:05}{]} allows us to decompose the semimartingale
process $X_{t}$ into a predictable part and a local martingale part.
Hence, it follows that we can write for $k=1,\ldots,T$, $\Delta_{h}X_{k}\triangleq\mu_{X,kh}\cdot h+\Delta_{h}M_{X,k}$
where the drift $\mu_{X,t}\in\mathbb{R}^{q}$ is $\mathscr{F}_{t-h}$
measurable, and $M_{X,kh}\in\mathbb{R}^{q}$ is a continuous local
martingale with finite conditional covariance matrix $\mathbb{P}$-a.s.
$\mathbb{E}\left(\Delta_{h}M_{X,k}\Delta_{h}M'_{X,k}|\,\mathscr{F}_{\left(k-1\right)h}\right)=\Sigma_{X,\left(k-1\right)h}\cdot h$.
Turning to equation \eqref{Mode for Y}, the error process $\left\{ \Delta_{h}e_{k}^{*},\,\mathscr{F}_{t}\right\} $,
with $\Delta_{h}e_{k}^{*}\triangleq\sigma_{e,\left(k-1\right)h}\Delta_{h}W_{e,k}$,
is then a continuous local martingale difference sequence taking its
values in $\mathbb{R}$ with finite conditional variance $\mathbb{E}\left[\left(\Delta_{h}e_{k}^{*}\right)^{2}|\,\mathscr{F}_{\left(k-1\right)h}\right]=\sigma_{e,\left(k-1\right)h}^{2}\cdot h,\,\mathbb{P}$-a.s.
Therefore, we express the discretized analogue of \eqref{Mode for Y}
as 
\begin{align}
\Delta_{h}Y_{k} & =\left(\beta^{*}\right)'\Delta_{h}X_{k-\tau}+\mu_{e,kh}\cdot h^{1-\vartheta}+\Delta_{h}e_{k},\qquad\qquad k=\tau+1,\ldots,T.\label{Discretized Model Y}
\end{align}
\begin{rem}
As explained above, we accommodate possible model misspecification
by adding the component $\mu_{e,k}\cdot h^{1-\vartheta}$. In the
forecasting literature, often one directly imposes restrictions on
the sequence of losses, say, $L\left(e_{k}\right)$ where $e_{k}=Y_{k}-\widehat{f}_{k}\left(\widehat{\beta}_{k}\right)$
is a forecast error. There are two main differences from our approach.
First, in order to facilitate illustrating our novel framework to
the reader, we have chosen, without loss of generality, to express
directly the relationship between $\Delta_{h}Y_{k+\tau}$ and $\Delta_{h}X_{k}$
while at the same time, allowing for misspecification by including
$\mu_{e,kh}\cdot h^{1-\vartheta}$.  A second distinction from the
classical approach is that the latter imposes restrictions on the
sequences of losses such as mixing and ergodicity conditions, covariance
stationary and so on. In contrast, our infill asymptotics does not
require us to impose any ergodic or mixing condition {[}cf. \citet{casini/perron_CR_Single_Break}{]}.
\end{rem}
Finally, we have an additional assumption on the path of the volatility
process $\left\{ \sigma_{e,t}^{2}\right\} _{t\geq0}$. This turns
out be important because it partly affects the local behavior of the
forecast losses.
\begin{assumption}
\label{Assumption Lipchtitz cont of Sigma}For small $\eta>0$, define
the modulus of continuity of $\left\{ \sigma_{e,t}\right\} _{t\geq0}$
on the time horizon $\left[0,\,N\right]$ by $\phi_{\sigma,\eta,N}=\sup_{s,t\in\left[0,\,N\right]}\left\{ \left|\sigma_{t}-\sigma_{s}\right|:\,\left|t-s\right|<\eta\right\} .$
We assume that $\phi_{\sigma,\eta,\tau_{h}\wedge N}\leq K_{h}\eta$
for some sequence of stopping times $\tau_{h}\rightarrow\infty$ and
some $\mathbb{P}$-a.s. finite random variable $K_{h}$. 
\end{assumption}
The assumption essentially states that $\phi_{\sigma,\eta,N}$ is
locally bounded and $\left\{ \sigma_{e,t}\right\} _{t\geq0}$ is Lipschitz
continuous. Lipschitz volatility is a more than reasonable specification
for the macroeconomic and financial data to which our analysis is
primarily directed. Indeed, the basic case of constant variance $\sigma^{2}$
is easily accommodated by the assumption. Time-varying volatility
is also covered provided $\sigma_{e,t}^{2}$ is sufficiently smooth.
However, the assumption rules out some standard stochastic volatility
models often used in finance. We relax that assumption in Section
\ref{Section Ito Vol}, so that we can extend our results to, for
example, stochastic volatility models driven by a Wiener process.

\subsection{\label{Subsection The-Hypotheses-of}The Hypotheses of Interest}

As time evolves, a forecast model can suffer instability for multiple
reasons. However, incorporating model misspecification into our framework
necessarily implies that the exact form of the instability is unknown
and thus one has to leave it unspecified. This differs from the classical
setting for estimation of structural change models {[}cf. \citet{bai/perron:98}
and \citet{casini/perron_CR_Single_Break}{]} where (i) the break
date is well-defined as it is part of the definition of the econometric
problem, and (ii) the form of the instability is explicitly specified
through a discrete shift in a regression parameter. In contrast, under
our context we remain agnostic regarding both (i) and (ii). There
may be multiple dates at which the forecast model suffers instability
and they might be interrelated in a complicated way. Forecast instability
may manifest itself in several forms, including gradual, smooth or
recurrent changes in the predictive relationship between $Y_{\left(k+\tau\right)h}$
and $X_{kh}$; certainly, there could also be discrete shifts in $\beta^{*}$\textemdash arguably
the most common case in practice\textemdash but this is only a possibility
in our setting and not an assumption as in structural change models.
A forecast failure then reflects the forecaster's failure to recognize
the shift in the predictive power of $X_{kh}$ on $Y_{\left(k+\tau\right)h}$.
On the other hand, even if one can rule out shits in $\beta^{*}$,
a forecast instability may be induced by an increase/decrease in the
uncertainty in the data which might result, for example, from changes
in the unconditional variance of the target variable. In this case,
the predictive ability of $X_{kh}$ on $Y_{\left(k+\tau\right)h}$,
as described for instance by a parameter $\beta,$ remains stable
while due to an increase in the unconditional variance of $Y_{\left(k+\tau\right)h}$
it might become weak and in turn the forecasting power might breakdown.
Tests for forecast failure such as those proposed in this paper and
the ones proposed in \citet{giacomini/rossi:09} are designed to have
power against both of the above hypotheses.\footnote{Recently, \citet{perron/yamamoto:18} proposed to apply modified versions
of classical structural break tests to the forecast failure setting.
However, their testing framework and hence their null hypotheses are
different from ours because they do not fix a model-method pair but
only fix the forecast model under the null.}

\subsubsection{The Null and Alternative Hypotheses on Forecast Instability}

Define at time $\left(k+\tau\right)h$ a \textit{surprise loss} given
by the deviation between the time-$\left(k+\tau\right)h$ out-of-sample
loss and the average in-sample loss: $SL_{\left(k+\tau\right)h}\left(\widehat{\beta}_{k}\right)\triangleq L_{\left(k+\tau\right)h}\left(\widehat{\beta}_{k}\right)-\overline{L}_{kh}\left(\widehat{\beta}_{k}\right),$
for $k=T_{m},\ldots,\,T-\tau$, where $\overline{L}_{kh}\left(\widehat{\beta}_{k}\right)$
is the average in-sample loss computed according to the specific
forecasting scheme. One can then define the average of the out-of-sample
surprise losses 
\begin{align}
\overline{SL}{}_{N_{0}}\left(\widehat{\beta}_{k}\right) & \triangleq N_{0}^{-1}\sum_{k=T_{m}}^{T-\tau}SL_{\left(k+\tau\right)h}\left(\widehat{\beta}_{k}\right),\label{eq. SL_bar}
\end{align}
 where $N_{0}\triangleq N-N_{\mathrm{in}}-h$ denotes the time span
of the out-of-sample window.\footnote{By definition $N_{0}$ is fixed and should not be confused with $T_{n},$
which indicates the number of observations in the out-of-sample window.
Indeed, $N_{0}=T_{n}h.$ } In the classical discrete-time setting, under the hypotheses of no
forecast instability one would naturally test whether $\overline{SL}{}_{N_{0}}\left(\beta^{*}\right)$
has zero mean, where $\beta^{*}$ is the pseudo-true value of $\beta$.
If the forecasting perfomance remains stable throughout the whole
sample then there should be no systematic surprise losses in the out-of-sample
window and thus $\mathbb{E}\left[N_{0}^{-1}\sum_{k=T_{m}}^{T-\tau}SL_{\left(k+\tau\right)h}\left(\beta^{*}\right)\right]=0.$
This reasoning motivated the forecast breakdown test of \citet{giacomini/rossi:09}.
Therefore, under the classical asymptotic setting one exploits time
series properties of the process $SL_{\left(k+\tau\right)h}\left(\beta^{*}\right)$
such as ergodicity and mixing together with the representation of
the hypotheses by a global moment restriction.\footnote{Global refers to the property that the zero-mean restriction involves
the entire sequence of forecast losses.} By letting the span $N\rightarrow\infty$, this method underlies
the classical approach to statistical inference but does not directly
extend to an infill asymptotic setting. Under continuous-time asymptotics,
identification of parameters is achieved by properties of the paths
of the involved processes and not by moment conditions. This constitutes
the key difference and requires one to recast the above hypotheses
into an infill setting thereby making use of assumptions on an underlying
continuous-time data-generating mechanism which is assumed to govern
the observed data.

We begin with observing that the sequence of losses $\left\{ L_{kh}\left(\cdot\right)\right\} $
can be viewed as realizations from an underlying continuous-time process
$\left\{ \widetilde{L}_{t}\right\} _{t\geq0}$, with $\widetilde{L}_{t}\triangleq\int_{0}^{t}L_{s}\left(Y_{s},\,X_{s-};\,\beta^{*}\right)ds$.
That is, $\widetilde{L}_{t}$ consists of temporally integrated forecast
losses where $L_{t}$ is the loss at time $t$ and is defined by some
transformation of the target variable $Y_{t}$ and of the predictor
$X_{t-}$.\footnote{The definition of $\widetilde{L}_{t}$ uses that so long as the forecast
step $\tau$ is small and finite one can approximate $X_{s-\tau h}$
by $X_{s-}$ for sufficiently small $h>0.$} In order to provide a general theory, we focus on families of loss
functions that depend only on the forecast error.\footnote{The most popular loss functions used in economic forecasting are within
this category {[}see \citet{elliott/timmermann:16} for a recent incisive
account of the literature{]}. Extension to \textit{ad hoc} loss functions
requires specific treatment that might vary from case to case.} We denote this class by $\boldsymbol{L}_{e}$ and we say that the
loss function $L_{\cdot}\left(\cdot,\,\cdot;\,\cdot\right)\in\boldsymbol{L}_{e}$
if $L_{t}\left(Y_{t},\,X_{t-};\,\beta\right)=L_{t}\left(e_{t};\,\beta\right)$
for all $t\in\left[0,\,N\right]$, where $e_{t}=Y_{t}-\widehat{f}_{t}\left(\beta\right)$.
The class $\boldsymbol{L}_{e}$ comprises the vast majority of loss
functions employed in empirical work, including among others the popular
Quadratic loss, Absolute error loss and Linex loss. The following
examples illustrate how these loss functions are constructed under
our setting. For the rest of this section, assume for simplicity $y_{0}=0,\,\mu_{e,\cdot}=0$
and that $X_{t}$ is one-dimensional in \eqref{Mode for Y}.
\begin{example*}
($\mathbf{QL}$: Quadratic Loss)\\
The \textit{Mean Squared Error} or \textit{Quadratic} loss function
is symmetric and is by far the most commonly used by practitioners.
Given \eqref{Mode for Y}, we have $e_{t}=Y_{t}-\beta^{*}X_{t-}$.
Then $L\left(e\right)=ae^{2}$ or $L_{t}\left(Y_{t},\,X_{t-};\,\beta^{*}\right)=ae_{t}^{2}$
with $a>0$.
\end{example*}

\begin{example*}
($\mathbf{LL}$: Linex Loss)\\
The \textit{Linear-exponential} or \textit{Linex }loss was introduced
by \citet{varian:75} and it is an example of asymmetric loss function.
By the same reasoning as in the Quadratic loss case, we have $L\left(e\right)=a_{1}\left(\exp\left(a_{2}e\right)-a_{2}e-1\right)$
or $L_{t}\left(Y_{t},\,X_{t-};\,\beta^{*}\right)=a_{1}\left(\exp\left(a_{2}e_{t}\right)-a_{2}e_{t}-1\right)$
with $a_{1}>0,\,a_{2}\neq0$.
\end{example*}
Below we make very mild pathwise assumptions on the process $Z$ which
imply restrictions on $\left\{ \widetilde{L}_{t}\right\} _{t\geq0}$.
We derive asymptotic results under Lipschitz continuity (in $t$)
of the coefficients of the system of stochastic differential equations
driving the data $\left\{ Z_{t}\right\} _{t\geq0}$. We apply the
techniques of stochastic calculus to formulate our testing problem.
To avoid clutter, we introduce the notation $g\left(Y_{t},\,X_{t-};\,\beta^{*}\right)=L_{t}\left(Y_{t},\,X_{t-};\,\beta^{*}\right)$
and its shorthand $g\left(e_{t};\,\beta^{*}\right)=L_{t}\left(e_{t};\,\beta^{*}\right)$.\footnote{The notation implicitly assumes that the same loss function is used
for estimation and prediction which in turn implies that the subscript
$t$ in $L_{t}\left(e_{t};\,\beta^{*}\right)$ can be omitted since
it can be understood from that of the argument $e_{t}$.} By It\^o Lemma, {[}cf. Section II.7 in \citet{protter:05}{]}, under
smoothness of $g\left(e_{t};\,\beta^{*}\right)$, 
\begin{align*}
dL_{t}\left(e_{t};\,\beta^{*}\right) & =\frac{\sigma_{e,t}^{2}}{2}\frac{\partial^{2}g\left(e_{t};\,\beta^{*}\right)}{\partial e^{2}}dt+\sigma_{e,t}\frac{\partial g\left(e_{t};\,\beta^{*}\right)}{\partial e}dW_{e,t}.
\end{align*}
Let $\mathbb{E}_{\sigma}$ denote the expectation conditional on the
path $\left\{ \sigma_{e,t}\right\} _{t\geq0}$. The instantaneous
mean of $dL\left(e_{t};\,\beta^{*}\right)$ is $\mathbb{E}_{\sigma}\left[dL\left(e_{t};\,\beta^{*}\right)/dt\right]=2^{-1}\sigma_{e,t}^{2}\mathbb{E}_{\sigma}\left(\partial^{2}g\left(e_{t};\,\beta^{*}\right)/\partial e^{2}\right)$.
Note that the latter is a symbolic abbreviation for 
\begin{align*}
\mathbb{E}_{\sigma}\left[L_{t}\left(e_{t};\,\beta^{*}\right)-L_{s}\left(e_{s};\,\beta^{*}\right)\right] & =\frac{\sigma_{e,t}^{2}}{2}\mathbb{E}_{\sigma}\left[\frac{\partial^{2}g\left(e_{t};\,\beta^{*}\right)}{\partial e^{2}}\right]\left(t-s\right)+o\left(t-s\right),\qquad\qquad\mathrm{as\,}s\uparrow t.
\end{align*}
Since the coefficients of the original system of stochastic equations
are Lipschitz continuous in $t$, one can verify that $\mathbb{E}_{\sigma}\left[dL\left(e_{t};\,\beta^{*}\right)/dt\right]$
is also Lipschitz upon regularity conditions on $g\left(\cdot,\,\beta^{*}\right)$
and time-$t$ information.

We denote by $\boldsymbol{Lip}\left(\left[0,\,N\right]\right)$ the
class of Lipschitz continuous functions on $\left[0,\,N\right]$.
Let $\left\{ c_{t}\right\} _{t\geq0}$ denote a continuous-time stochastic
process that is $\mathbb{P}$-a.s. locally bounded and adapted. 
\begin{defn}
\label{def: Lipschitz}The process $\left\{ c_{t}\right\} _{t\geq0}$
belongs to $\boldsymbol{Lip}\left(\left[0,\,N\right]\right)$ if $\sup_{s,t\in\left[0,\,\tau_{h}\wedge N\right],t\neq s}\left|c_{t}-c_{s}\right|<K_{h}\left|t-s\right|$
for some sequence of stopping times $\tau_{h}\rightarrow\infty$ and
some $\mathbb{P}$-a.s. finite random variable $K_{h}$.
\end{defn}
We are in a position to formulate the testing problem in terms of
the pathwise property of $L_{t}\left(e_{t};\,\beta^{*}\right).$ This
implies that the hypotheses are specified in terms of random events
which differs from classical hypotheses testing but it is typical
under continuous-time asymptotics; see \citet{sahalia/jacod:12} (for
many references), \citet{li/todorov/tauchen:16} and \citet{reiss/todorov/tauchen:15}.
We consider the following hypotheses: for any $L\left(\cdot;\,\cdot\right)\in\boldsymbol{L}_{e}$,\footnote{Precise assumptions will be stated below.}
\begin{align}
H_{0}: & \quad\left\{ \lim_{s\uparrow t}\mathbb{E}_{\sigma}\left[L_{t}\left(e_{t};\,\beta^{*}\right)-L_{s}\left(e_{s};\,\beta^{*}\right)\right]\right\} \in\boldsymbol{Lip}\left(\left[N_{\mathrm{in}}+h,\,N\right]\right),\label{eq H0}
\end{align}
which means that we wish to discriminate between the following two
events that divide $\Omega:$ 
\begin{align*}
\Omega_{0}\triangleq\left\{ \omega\in\Omega:\,\left\{ \lim_{s\uparrow t}\mathbb{E}_{\sigma}\left[L_{t}\left(e_{t}\left(\omega\right);\,\beta^{*}\right)-L_{s}\left(e_{s}\left(\omega\right);\,\beta^{*}\right)\right]\right\} \in\boldsymbol{Lip}\left(\left[N_{\mathrm{in}}+h,\,N\right]\right)\right\} , & \qquad\Omega_{1}\triangleq\Omega\backslash\Omega_{0}.
\end{align*}
The dependence of the hypotheses on $\omega$ is appropriate because
each event $\omega$ generates a certain path of $L\left(e_{t}\left(\cdot\right);\,\beta^{*}\right)$
on $\left[0,\,N\right]$, where $de_{t}\left(\omega\right)=\sigma_{e,t}\left(\omega\right)dW_{e,t}\left(\omega\right)$.
 The hypotheses requires a Lipschitz condition to hold on $\left[N_{\mathrm{in}}+h,\,N\right]$,
where $N_{\mathrm{in}}$ is the usual artificial separation date after
which the first forecast is made. $N_{\mathrm{in}}$ is taken as given
here because the testing problem applies to a specific method-model
pair and $N_{\mathrm{in}}$ is part of the chosen forecasting method.
From a practical standpoint, it would be helpful if this separation
date is such that the forecast model is stable on $\left[0,\,N_{\mathrm{in}}\right]$
{[}see \citet{casini/perron_Oxford_Survey} for more details{]}. The
latter property is, however, unknown a priori by the practitioner.
We cover this case in Section \ref{Section Extensions}.
\begin{example*}
($\mathbf{QL}$; cont'd)\\
For the Quadratic loss $L\left(e\right)=ae^{2},$ It\^o Lemma yields
$\mathbb{E}_{\sigma}\left[dL_{t}\left(e_{t};\,\beta^{*}\right)/dt\right]=a\sigma_{e,t}^{2}$.
If $\sigma_{e,t}$ is Lipschitz continuous, then the hypothesis $H_{0}$
holds. 
\end{example*}

\begin{example*}
($\mathbf{LL}$; cont'd)\\
From It\^o Lemma, $dL_{t}\left(e_{t};\,\beta^{*}\right)=a_{1}\left\{ a_{2}\left[2^{-1}a_{2}\sigma_{e,t}^{2}\exp\left(a_{2}e_{t}\right)dt+\left(\exp\left(a_{2}e_{t}\right)-1\right)\sigma_{e,t}dW_{e,t}\right]-1\right\} $.
Consequently, by It\^o Isometry {[}cf. Section 3.3.2 in \citet{karatzas/shreve:96}
or Lemma 3.1.5 in \citet{oksendal:00}{]} $\mathbb{E}_{\sigma}\left[dL\left(e_{t};\,\beta^{*}\right)/dt\right]=a_{1}\left(a_{2}^{2}\left(\sigma_{e,t}^{2}/2\right)\right)\exp\left(a_{2}^{2}\left(\int_{0}^{t}\sigma_{e,s}^{2}ds\right)/2\right)$
and hypotheses $H_{0}$ is seen to hold under Lipschitz continuity
of $\sigma_{e,t}$.\footnote{Recall that composition of Lipschitz functions is Lipschitz and that
under our context $\exp\left(a_{2}\left(\int_{0}^{t}\sigma_{e,s}^{2}ds\right)/2\right)$
is Lipschitz because (i) $\sigma_{e,s}^{2}$ is locally bounded and
Lipschitz, and (ii) $t\leq N$ and $N$ remains fixed. } 
\end{example*}
We have reduced the forecast instability problem to examination of
the local properties of the path of $L_{t}$. However, we still have
to face the question on how to use the data to test $H_{0}$ in practice.
Even if we could observe $\widetilde{L}_{t}$, it would not be clear
how to formulate a testing problem on the stability of $L_{t}$ by
using path properties of $\widetilde{L}_{t}$. The reason is that
$\widetilde{L}_{t}$ is always absolutely continuous by definition,
and thus it would provide little information on the large deviations
of the forecast error $e_{t}$. In order to study the local behavior
of $L_{t}$ one needs to consider the small increments of $L_{t}$
close to time $t$. Leaving the definition of $\widetilde{L}_{t}$
aside for a moment, observe that $\mathbb{P}$-a.s. continuity of
$Z_{t}$ is equivalent to having the relationship between $Y_{t}$
and $X_{t}$ holding for any infinitesimal interval of time. For the
basic parametric linear model: $dY_{t}=\beta^{*}dX_{t}+de_{t}$. Then,
the forecast loss is $L\left(de_{t}\right)$, which is difficult to
interpret in rigorous probabilistic terms. However, we can consider
its discrete-time analogue. We normalize the forecast error by the
factor $\psi_{h}=h^{1/2}$ and redefine $L_{\psi,kh}\left(\Delta_{h}e_{k};\,\beta^{*}\right)\triangleq L_{kh}\left(\psi_{h}^{-1}\Delta_{h}e_{k};\,\beta^{*}\right)$.\footnote{Alternatively, $L_{\psi,kh}\left(\Delta_{h}Y_{k},\,\Delta_{h}X_{k};\,\beta^{*}\right)=L_{kh}\left(\psi_{h}^{-1}\left(\Delta_{h}Y_{k}-\beta^{*}\Delta_{h}X_{k}\right)\right)$.}
Then, for all $k$, the mean of $L_{\psi,kh}\left(\Delta_{h}e_{k};\,\beta^{*}\right)$\textemdash conditional
on $\sigma_{e,kh}$\textemdash depends on the parameters of the model
and its local behavior can be used as a proxy for the local behavior
of the infinitesimal mean of $dL_{t}\left(e_{t};\,\beta^{*}\right)$.
If the corresponding structural parameters of the continuous-time
data-generating process satisfy a Lipschitz continuity in $t$, then\textemdash knowing
$\sigma_{e,kh}$\textemdash also $\mathbb{E}_{\sigma}\left[L_{\psi,kh}\left(\Delta_{h}e_{k};\,\beta^{*}\right)\right]$
should be Lipschitz in the continuous-time limit. Under the hypotheses
$H_{0}$ there should be no break in $\mathbb{E}_{\sigma}\left[L_{\psi,kh}\left(\Delta_{h}e_{k};\,\beta^{*}\right)\right]$
and an appropriately defined right local average of $L_{\psi,kh}\left(\Delta_{h}e_{k};\,\beta^{*}\right)$
should not differ too much from its left local average. That is, one
can test for forecast instability by using a two-sample t-test over
asymptotically vanishing time blocks.
\begin{example*}
($\mathbf{QL}$; cont'd)\\
Conditional on $\left\{ \sigma_{t}\right\} _{t\geq0}$, $\Delta_{h}e_{k}\sim\mathscr{N}\left(0,\,\sigma_{e,\left(k-1\right)h}^{2}\cdot h\right)$.
Thus, $\mathbb{E}_{\sigma}\left[L_{\psi,kh}\left(\Delta_{h}e_{k};\,\beta^{*}\right)\right]=a\sigma_{e,\left(k-1\right)h}^{2}$.
If $\sigma_{e,t}$ is Lipschitz continuous, then the hypothesis $H_{0}$
holds. 
\end{example*}

\begin{example*}
($\mathbf{LL}$; cont'd)\\
Similar to the Quadratic loss case, we have $\mathbb{E}_{\sigma}\left[L_{\psi,kh}\left(\Delta_{h}e_{k};\,\beta^{*}\right)\right]=a_{1}\left(\exp\left(a_{2}^{2}\sigma_{e,\left(k-1\right)h}^{2}/2\right)-1\right)$.
Again, the hypotheses $H_{0}$ is satisfied if $\sigma_{e,t}$ is
Lipschitz. 
\end{example*}
Both examples demonstrate that pathwise assumptions on the data-generating
process implies restrictions on the properties of the sequence of
loss functions. For the QL example, if there is a structural break
at the observation $k=T_{b},$ then this would result in the mean
of $L_{\psi,kh}\left(\Delta_{h}e_{k};\,\beta^{*}\right)$ shifting
to a new level after time $T_{b}h$. Given that the same reasoning
extends to the sequence of surprise losses, one may consider to construct
a test statistic on the basis of the local behavior of the surprise
losses over time. If there is no instability in the predictive ability
of a certain model, then the sequence of out-of-sample surprise losses
should display a certain stability. Under the framework of \citet{giacomini/rossi:09},
this stability is interpreted in a retrospective and global sense
as a zero-mean restriction on the sequence over the entire out-of-sample.
In contrast, under our continuous-time setting, this stability manifests
itself as a continuity property of the path of the continuous-time
counterpart of the sequence.

\subsection{\label{Subsection The-Test-Statistics}The Test Statistics }

By inspection of the null hypotheses in \eqref{eq H0}, it is evident
that a considerable number of forms of instabilities are allowed.
These may result from discrete shifts in a model's structural parameter
and/or in structural properties of the processes considered such as
conditional and unconditional moments and so on. This first set of
non-stationarities relates to the popular case of structural changes
which are designed to be detected with high probability by the structural
break tests of, among others, \citet{andrews:93} and \citet{andrews/ploberger:94},
\citet{bai/perron:98} and \citet{elliott/mueller:06} in univariate
settings and of \citet{qu/perron:07} in multivariate settings. However,
a forecast instability may be generated by many other forms of non-stationarities
against which such classical tests for structural breaks are not designed
for and consequently they might have little power against. For example,
consider the case of smooth changes in model parameters, or in the
unconditional variance of $Y_{kh}$. Even more serious would be the
presence of recurrent smooth changes in the marginal distribution
of the predictor since in this case the above-mentioned tests are
likely to falsely reject $H_{0}$ too often {[}cf. \citet{hansen:00}{]}.
Thus, the null hypotheses of no forecast instability calls for a new
statistical hypotheses testing framework. Ideally, in this context
one needs a test statistic that retains power against any discontinuity,
jump and recurrent switch at any point in the out-of-sample and for
any magnitude of the shift. We propose a test statistic which aims
asymptotically at distinguishing any discontinuity from a regular
Lipschitz continuous motion. We introduce a sequence of two-sample
t-tests over asymptotically vanishing adjacent time blocks. This should
lead to significant gains in power whenever on fixed time intervals
the out-of-sample losses exhibit instabilities of any form such as
breaks, jumps and relatively large deviations. Such gains are likely
to occur especially when instabilities take place within a small portion
of the sample relative to the whole time span\textemdash a common
case in practice that has characterized many episodes of forecast
failure in economics.

Interestingly, for the Quadratic loss function we can exploit properties
of the local quadratic variation and propose a self-normalized test
statistic. Thus, we separate the discussion on the Quadratic loss
from that on general loss functions. Let $SL_{\psi,\left(k+\tau\right)h}\left(\widehat{\beta}_{k}\right)\triangleq L_{\psi,\left(k+\tau\right)h}\left(\widehat{\beta}_{k}\right)-\overline{L}_{\psi,kh}\left(\widehat{\beta}_{k}\right)$,
$k=T_{m},\ldots,\,T-\tau$. Next, we partition the out-of-sample
into $m_{T}\triangleq\left\lfloor T_{n}/n_{T}\right\rfloor $ blocks
each containing $n_{T}$ observations. Let $B_{h,b}\triangleq n_{T}^{-1}\sum_{j=1}^{n_{T}}SL_{\psi,T_{m}+\tau+bn_{T}+j-1}\left(\widehat{\beta}_{T_{m}+bn_{T}+j-1}\right)$
and $\overline{B}_{h,b}\triangleq n_{T}^{-1}\sum_{j=1}^{n_{T}}L_{\psi,\left(T_{m}+\tau+bn_{T}+j-1\right)h}\left(\widehat{\beta}_{T_{m}+bn_{T}+j-1}\right)$
for $b=0,\ldots,\,,\,\left\lfloor T_{n}/n_{T}\right\rfloor -1$.

\subsubsection{Test Statistics under Quadratic Loss}

We propose the following statistic
\begin{align*}
\mathrm{B}_{\mathrm{max},h}\left(T_{n},\,\tau\right) & \triangleq\max_{b=0,\ldots,\,\left\lfloor T_{n}/n_{T}\right\rfloor -2}\left|\frac{B_{h,b+1}-B_{h,b}}{\overline{B}_{h,b+1}}\right|.
\end{align*}
The quantity $B_{h,b}$ is a local average of the surprise losses
within the block $b$. We have partitioned the out-of-sample window
into $m_{T}$ blocks of asymptotically vanishing length $\left[bn_{T}h,\,\left(b+1\right)n_{T}h\right]$.
We consider an asymptotic experiment in which the number of blocks
$m_{T}$ increases at a controlled rate to infinity while the per-block
sample size $n_{T}$ grows without bound at a slower rate than the
out-of-sample size $T_{n}$. The appeal of the $\mathrm{B}_{\mathrm{max},h}\left(T_{n},\,\tau\right)$
statistic is that a large deviation $B_{h,b+1}-B_{h,b}$ suggests
the existence of either a discontinuity or non-smooth shift in the
surprise losses close to time $bn_{T}h$ and thus it provides evidence
against $H_{0}$. We comment on the nature of the normalization $\overline{B}_{h,b+1}$
in the denominator of $\mathrm{B}_{\mathrm{max},h}$ below, after
we introduce a version of $\mathrm{B}_{\mathrm{max},h}$ statistic
which uses all admissible overlapping blocks of length $n_{T}h$:
\begin{align*}
\mathrm{MB}{}_{\mathrm{max},h}\left(T_{n},\,\tau\right) & \triangleq\max_{i=n_{T},\ldots,\,T_{n}-n_{T}}\left|\left(n_{T}^{-1}\sum_{j=i-n_{T}+1}^{i}SL_{\psi,T_{m}+\tau+j-1}\left(\widehat{\beta}_{T_{m}+j-1}\right)\right.\right.\\
 & \left.\left.\quad-n_{T}^{-1}\sum_{j=i+1}^{i+n_{T}}SL_{\psi,T_{m}+\tau+j-1}\left(\widehat{\beta}_{T_{m}+j-1}\right)\right)/\overline{B}_{h,i}\right|,
\end{align*}
 where $\overline{B}_{h,i}=n_{T}^{-1}\sum_{j=i+1}^{i+n_{T}}L_{\psi,T_{m}+\tau+j-1}\left(\widehat{\beta}_{T_{m}+j-1}\right)$.
Since under the alternative hypotheses the exact location of the change-point\textendash -or
possibly the locations of the multiple change-points\textemdash within
the block might actually affect the power of the $\mathrm{B}_{\mathrm{max},h}$-based
test in small samples, we indeed find in our simulation study that
the test statistic $\mathrm{MB}{}_{\mathrm{max},h}$ which uses overlapping
blocks is more powerful especially when the instability arises in
forms other than the simple one-time structural change. Thus, the
power of the $\mathrm{B}_{\mathrm{max},h}$ test is slightly sensible
to the actual location of the change-point within the block, with
higher power achieved when the change-point is close to either the
beginning or the end of the block. In contrast, the statistical power
of $\mathrm{MB}_{\mathrm{max},h}$ is uniform over the location of
the change-point in the sample. The latter property is not shared
by the exiting test of \citet{giacomini/rossi:09} given that its
power tends to be substantially lower if the instability is not located
at about middle sample.

An important characteristic of both $\mathrm{B}_{\mathrm{max},h}$
and $\mathrm{MB}_{\mathrm{max},h}$ is that they are self-normalized;
no asymptotic variance appears in their definition. The reason for
why $\overline{B}_{h,b+1}$ appears in the denominator of, for example,
$\mathrm{B}_{\mathrm{max},h}$ is that even though $B_{h,b+1}$ constitutes
a more logical self-normalizing term, it might be close to zero in
some cases. This would occur under Quadratic loss if, for example,
$\sigma_{e,t}=\sigma_{e}$ for all $t\geq0.$ This is not true for
the factor $\overline{B}_{h,b+1}$.

In addition, observe that allowing for misspecification naturally
leads one to deal carefully with artificial serial dependence in the
forecast losses in small samples. Thus, we consider a version of the
statistics $\mathrm{B}_{\mathrm{max},h}$ and $\mathrm{MB}_{\mathrm{max},h}$
that are normalized by their asymptotic variance: $\mathrm{Q}_{\mathrm{max},h}\left(T_{n},\,\tau\right)\triangleq\nu_{L}^{-1}\max_{b=0,\ldots,\,\left\lfloor T_{n}/n_{T}\right\rfloor -2}\left|B_{h,b+1}-B_{h,b}\right|,$
and similarly, 
\begin{align*}
\mathrm{MQ}{}_{\mathrm{max},h} & \left(T_{n},\,\tau\right)\\
 & \triangleq\nu_{L}^{-1}\max_{i=n_{T},\ldots,\,T_{n}-n_{T}}\left|n_{T}^{-1}\sum_{j=i+1}^{i+n_{T}}SL_{\psi,T_{m}+\tau+j-1}\left(\widehat{\beta}_{T_{m}+j-1}\right)-n_{T}^{-1}\sum_{j=i-n_{T}+1}^{i}SL_{\psi,T_{m}+\tau+j-1}\left(\widehat{\beta}_{T_{m}+j-1}\right)\right|.
\end{align*}
 The quantity $\nu_{L}$ standardizes the test statistic so that under
the null hypotheses we obtain a distribution-free limit. This can
be useful because given the fully non-stationary setting together
with the possible consequences of misspecification in finite-samples,
standardization by the square-root of the asymptotic variance $\nu_{L}^{2}$
might lead to a more precise empirical size in small samples. We relegate
theoretical details on $\nu_{L}$ as well as on its estimation to
Section \ref{Section Estimation-of-Asymptotic Variance} where we
also present a discussion about its relation with the choice of the
number of blocks. 

\subsubsection{Test Statistics under General Loss Function}

For general loss $L\in\boldsymbol{L}_{e}$, we propose the following
statistic,

\begin{align*}
\mathrm{G}_{\mathrm{max},h}\left(T_{n},\,\tau\right) & \triangleq\max_{b=0,\ldots,\,\left\lfloor T_{n}/n_{T}\right\rfloor -2}\left|\frac{B_{h,b+1}-B_{h,b}}{\sqrt{D_{h,b+1}}}\right|,
\end{align*}
 where $B_{h,b},\,B_{h,b+1}$ are defined as in the quadratic case
and 
\begin{align*}
D_{h,b+1} & \triangleq n_{T}^{-1}\sum_{j=1}^{n_{T}}\left(L_{\psi,\left(T_{m}+\tau+\left(b+1\right)n_{T}+j-1\right)h}\left(\widehat{\beta}_{T_{m}+\left(b+1\right)n_{T}+j-1}\right)-\overline{L}_{\psi,b+1}\left(\widehat{\beta}\right)\right)^{2},
\end{align*}
 with $\overline{L}_{\psi,b}\left(\widehat{\beta}\right)\triangleq n_{T}^{-1}\sum_{j=1}^{n_{T}}L_{\psi,\left(T_{m}+\tau+bn_{T}+j-1\right)h}\left(\widehat{\beta}_{T_{m}+bn_{T}+j-1}\right)$.
The interpretation of $\mathrm{G}_{\mathrm{max},h}$ is essentially
the same as of $\mathrm{B}_{\mathrm{max},h}$, the only difference
arising from the denominator $\sqrt{D_{h,b+1}}$ that estimates the
within-block variance. A version that uses all overlapping blocks
is 
\begin{align*}
\mathrm{MG}{}_{\mathrm{max},h} & \left(T_{n},\,\tau\right)\\
 & \triangleq\max_{i=n_{T},\ldots,\,T_{n}-n_{T}}\left|\frac{n_{T}^{-1}\sum_{j=i+1}^{i+n_{T}}SL_{\psi,T_{m}+\tau+j-1}\left(\widehat{\beta}_{T_{m}+j-1}\right)-n_{T}^{-1}\sum_{j=i-n_{T}+1}^{i}SL_{\psi,T_{m}+\tau+j-1}\left(\widehat{\beta}_{T_{m}+j-1}\right)}{\sqrt{D_{h,i}}}\right|,
\end{align*}
 where $D_{h,i}\triangleq n_{T}^{-1}\sum_{j=i+1}^{i+n_{T}}\left(L_{\psi,\left(T_{m}+\tau+j-1\right)h}\left(\widehat{\beta}_{T_{m}+j-1}\right)-\overline{L}_{\psi,i}\left(\widehat{\beta}\right)\right)^{2}$,
with
\begin{align*}
\overline{L}_{\psi,i}\left(\widehat{\beta}\right) & \triangleq n_{T}^{-1}\sum_{j=i+1}^{i+n_{T}}L_{\psi,\left(T_{m}+\tau+j-1\right)h}\left(\widehat{\beta}_{T_{m}+j-1}\right).
\end{align*}
As argued above, it is useful to consider versions of the statistic
$\mathrm{B}_{\mathrm{max},h}$ and $\mathrm{MB}_{\mathrm{max},h}$
that are normalized by their asymptotic variance: 
\begin{align*}
\mathrm{Q}_{\mathrm{max},h}^{\mathrm{G}}\left(T_{n},\,\tau\right) & \triangleq\nu_{L}^{-1}\max_{b=0,\ldots,\,\left\lfloor T_{n}/n_{T}\right\rfloor -2}\left|B_{h,b+1}-B_{h,b}\right|,
\end{align*}
 and similarly, 
\begin{align*}
\mathrm{MQ}_{\mathrm{max},h}^{\mathrm{G}} & \left(T_{n},\,\tau\right)\\
 & \triangleq\nu_{L}^{-1}\max_{i=n_{T},\ldots,\,T_{n}-n_{T}}\left|n_{T}^{-1}\sum_{j=i+1}^{i+n_{T}}SL_{\psi,T_{m}+\tau+j-1}\left(\widehat{\beta}_{T_{m}+j-1}\right)-n_{T}^{-1}\sum_{j=i-n_{T}+1}^{i}SL_{\psi,T_{m}+\tau+j-1}\left(\widehat{\beta}_{T_{m}+j-1}\right)\right|.
\end{align*}

\section{Continuous Record Distribution Theory for the Test Statistics\label{Section CR Distribution Theory Test Stats}}

\subsection{Asymptotic Distribution under the Null Hypotheses}

We begin with a set of assumptions. Assumption \ref{Assumption Moments of Losses}
below is a finite-moment condition on the sequence of rescaled forecast
losses and and on its first-order derivative. It has a similar scope
to A4 in \citet{giacomini/rossi:09}. Assumption \ref{Assumption A5 in GR: Finite E L'}
is similar to A5 in \Citet{giacomini/rossi:09} and it imposes the
first-order derivative of the forecast losses to be constant over
time. It trivially holds when one employs the same loss function for
estimation and evaluation. Assumption \ref{Assumption Roo-T Consistent beta }
demands the existence of a consistent estimator for $\beta^{*}$ at
the parametric rate $\sqrt{T}$ and it encompasses many estimation
procedures. In model \eqref{Discretized Model Y}, the popular least-squares
method will satisfy the condition {[}cf. \citet{barndorff/shephard:04}
and \citet*{li/todorov/tauchen:17}{]}.
\begin{assumption}
\label{Ass the process}The process $\left\{ Y_{t}-\left(\beta^{*}\right)'X_{t-}\right\} _{t\in\left[0,\,N\right]}$
takes value in an open set $\mathcal{E}\subseteq\mathbb{R}$, and
$\beta^{*}$ takes value in a compact parameter space $\Theta\subset\mathbb{R}^{\mathrm{dim}\left(\beta\right)}$.
\end{assumption}

\begin{assumption}
\label{Assumption Differentiability Loss}For any $L\in\boldsymbol{L}_{e}$
we assume $L:\,\mathcal{E}\times\Theta\mapsto\mathbb{R}$ is a measurable
function and $L\in\boldsymbol{C}^{2,2}$ (i.e., twice continuously
differentiable in both arguments). For every open set $B$ that contains
$\beta^{*}$ there exists $C<\infty$ such that for all $k\geq1$,
$\sup_{\beta\in B}\left\Vert \partial^{2}L_{\psi,kh}\left(\Delta_{h}e_{k};\,\beta\right)/\partial\beta\partial\beta'\right\Vert <C$. 
\end{assumption}

\begin{assumption}
\label{Assumption Moments of Losses}We have $\sup_{k=1,\ldots,T}\mathbb{E}_{\sigma}\left\Vert \left(L_{\psi,kh}\left(\Delta_{h}e_{k};\,\beta^{*}\right),\,\partial L_{\psi,kh}\left(\Delta_{h}e_{k};\,\beta^{*}\right)/\partial\beta\right)'\right\Vert ^{4+\varpi}<\infty$,
for $\varpi>0$.
\end{assumption}

\begin{assumption}
\label{Assumption A5 in GR: Finite E L'}For all $k\geq1,$ $\mathbb{E}_{\sigma}\left[\partial L_{\psi,kh}\left(\Delta_{h}e_{k};\,\beta^{*}\right)/\partial\beta\right]=\overline{K}$,
for some $\overline{K}<\infty$. 
\end{assumption}

\begin{assumption}
\label{Assumption Bounded on Bounded Sets LX (2-iv)}For all $k\geq1$,
$\left|\partial L_{\psi,kh}\left(e;\,\cdot\right)/\partial e\right|$
is bounded on bounded sets.
\end{assumption}

\begin{assumption}
\label{Assumption Roo-T Consistent beta }There exists a sequence
$\left\{ \widehat{\beta}_{k}\right\} _{k=T_{m}}^{T-\tau}$ such that
$\left\Vert \widehat{\beta}_{k}-\beta^{*}\right\Vert =O_{\mathbb{P}}\left(1/\sqrt{T}\right)$
uniformly over $k=T_{m},\ldots,\,T-\tau$.
\end{assumption}
Our asymptotic results are valid under the following conditions on
the auxiliary sequence $n_{T}$. 
\begin{condition}
\label{Cond The-auxiliary-sequence}The sequence $\left\{ n_{T}\right\} $
satisfies for some $\epsilon>0,$ 
\begin{align}
n_{T}\rightarrow\infty\quad\mathrm{as\quad}T\rightarrow\infty & \qquad\mathrm{and}\qquad T^{\epsilon}n_{T}^{-1}+n_{T}^{3/2}h\sqrt{\log\left(T\right)}\rightarrow0.\label{eq. Condition auxiliary nT}
\end{align}
\end{condition}
Condition \ref{Cond The-auxiliary-sequence} imposes a lower bound
and an upper bound on the growth condition of the sequence $\left\{ n_{T}\right\} $.
The first part of \eqref{eq. Condition auxiliary nT} requires $n_{T}$
to grow to infinity at any faster rate than $T^{\epsilon}$ with $\epsilon>0$,
which we interpret as saying that the number of observations $n_{T}$
in each block cannot be too small. The second part of \eqref{eq. Condition auxiliary nT}
provides an upper bound on the growth of $n_{T}$ and relates to Assumption
\ref{Assumption Lipchtitz cont of Sigma} concerning the smoothness
of $\left\{ \sigma_{e,t}\right\} _{t\geq0}$ thereby ensuring that,
for example, the random oscillations of $\mathrm{B}_{\mathrm{max},h}\left(T_{n},\,\tau\right)$
can be controlled. As we shall explain in the simulation study
of Section \ref{Section Simulation Study}, we recommend to set $n_{T}\propto T_{n}^{2/3-\epsilon}$
for small $\epsilon>0$.

\subsubsection{Asymptotic Distribution Under Quadratic Loss Function}
\begin{thm}
\label{Theoem Asymptotic H0 Distrbution Bmax and Qmax}Let $\gamma_{m_{T}}=\left[4\log\left(m_{T}\right)-2\log\left(\log\left(m_{T}\right)\right)\right]^{1/2}$
and recall $m_{T}=\left\lfloor T_{n}/n_{T}\right\rfloor $. Assume
Assumption \ref{Assumption 1, CT}-\ref{Assumption Lipchtitz cont of Sigma},
\ref{Ass the process}-\ref{Assumption Roo-T Consistent beta }, and
Condition \ref{Cond The-auxiliary-sequence} hold. Let $\mathscr{V}$
denote a random variable defined by $\mathbb{P}\left(\mathscr{V}\leq v\right)=\exp\left(-\pi^{-1/2}\exp\left(-v\right)\right).$
Under $H_{0}$, we have\\
(i) $\sqrt{\log\left(m_{T}\right)}\left(2^{-1/2}n_{T}^{1/2}\mathrm{B}_{\mathrm{max},h}\left(T_{n},\,\tau\right)-\gamma_{m_{T}}\right)\overset{}{\Rightarrow}\mathscr{V};$
\\
(ii) $2^{-1/2}\sqrt{\log\left(m_{T}\right)}n_{T}^{1/2}\mathrm{MB}_{\mathrm{max},h}\left(T_{n},\,\tau\right)-2\log\left(m_{T}\right)-\frac{1}{2}\log\log\left(m_{T}\right)-\log3\overset{}{\Rightarrow}\mathscr{V}.$
\end{thm}
\begin{cor}
\label{Corollary CR Null Distrb Quadratic}Under the same assumptions
of the previous theorem, we have under $H_{0}$, $\sqrt{\log\left(m_{T}\right)}$
~ $\left(n_{T}^{1/2}\nu_{L}^{-1}\mathrm{Q}_{\mathrm{max},h}\left(T_{n},\,\tau\right)-\gamma_{m_{T}}\right)\overset{}{\Rightarrow}\mathscr{V}$
and 
\begin{align*}
\sqrt{\log\left(m_{T}\right)}\left(n_{T}^{1/2}\nu_{L}^{-1}\right)\mathrm{MQ}_{\mathrm{max},h}\left(T_{n},\,\tau\right)-2\log\left(m_{T}\right)-\frac{1}{2}\log\log\left(m_{T}\right)-\log3\overset{}{\Rightarrow}\mathscr{V} & ,
\end{align*}
 where $\mathscr{V},\,m_{T}$ and $\gamma_{m_{T}}$ are defined as
in the previous theorem.
\end{cor}
Theorem \ref{Theoem Asymptotic H0 Distrbution Bmax and Qmax} shows
that the asymptotic null distribution of our test statistics follows
an extreme vale distribution whose critical values can be computed
directly. In nonparametric change-point analysis, \citet{wu/zhao:07}
and \citet*{bibinger/jirak/vetter:16} have derived an extreme value
null distribution for tests statistics which share a similar form
to ours. As it is stated, the tests statistics are not yet feasible
because the asymptotic variances $\nu_{L}^{2}$ is unknown. However,
we can find statistical consistent estimators which can be used in
place of $\nu_{L}^{2}$ to make the test feasible. We relegate the
treatment of its consistent estimation to Section \ref{Section Estimation-of-Asymptotic Variance}.

\subsubsection{Asymptotic Distribution Under General Loss Function}
\begin{thm}
\label{Theoem Asymptotic H0 Distrbution Gmax QGmax}Under the same
assumptions of the previous theorem and with $\mathscr{V},\,m_{T}$
and $\gamma_{m_{T}}$ defined analogously, we have under $H_{0}$,
\\
(i) $2^{-1/2}\sqrt{\log\left(m_{T}\right)}\left(n_{T}^{1/2}\mathrm{G}_{\mathrm{max},h}\left(T_{n},\,\tau\right)-\gamma_{m_{T}}\right)\overset{}{\Rightarrow}\mathscr{V};$
\\
(ii) $2^{-1/2}\sqrt{\log\left(m_{T}\right)}n_{T}^{1/2}\mathrm{MG}_{\mathrm{max},h}\left(T_{n},\,\tau\right)-2\log\left(m_{T}\right)-\frac{1}{2}\log\log\left(m_{T}\right)-\log3\overset{}{\Rightarrow}\mathscr{V}.$ 
\end{thm}
\begin{cor}
\label{Corollary CR Null Distrb General}Under the same assumptions
of the previous theorem, we have under $H_{0}$, $\sqrt{\log\left(m_{T}\right)}$
~ $\left(n_{T}^{1/2}\nu_{L}^{-1}\mathrm{Q}_{\mathrm{max},h}^{\mathrm{G}}\left(T_{n},\,\tau\right)-\gamma_{m_{T}}\right)\overset{}{\Rightarrow}\mathscr{V}$
and $\sqrt{\log\left(m_{T}\right)}n_{T}^{1/2}\nu_{L}^{-1}\mathrm{MQ}_{\mathrm{max},h}^{\mathrm{G}}\left(T_{n},\,\tau\right)-2\log\left(m_{T}\right)-\frac{1}{2}\log\log\left(m_{T}\right)-\log3\overset{}{\Rightarrow}\mathscr{V}.$
\end{cor}

\section{\label{Section Estimation-of-Asymptotic Variance}Estimation of Asymptotic
Variance}

The purpose of this section is to show how to construct an asymptotically
valid estimator of the variance $\nu_{L}^{2}$ that enters the definition
of our test statistics. This is an important aspect that together
with the selection of the block length might affect statistical inferences
based on the proposed tests in finite-samples. Allowing for misspecification
is customary in the forecasting literature, and as a consequence this
may result in forecast losses that artificially exhibit heteroskedasticity
and serial dependence in small samples.

\subsection{\label{subsec:Asymptotic-Variance-Estimator}Estimation of the Asymptotic
Variance}

We begin with the case of stationary forecast losses, including constant
$\nu_{L}$ as a special case.

\subsubsection{Stationary Forecast Losses}

Recall that our test statistics are related to a maximum over blocks
of data. Thus, for i.i.d. forecast losses one can use the following
estimator for $\nu_{L}$ in $\mathrm{Q}_{\mathrm{max},h}\left(T_{n},\,\tau\right)$:
\begin{align*}
\widehat{\nu}_{\mathrm{Q}1,b}^{2} & \triangleq\frac{2}{n_{T}}\sum_{j=1}^{n_{T}}\left(SL_{\psi,T_{m}+\tau+bn_{T}+j-1}\left(\widehat{\beta}_{T_{m}+bn_{T}+j-1}\right)-\overline{SL}{}_{\psi,b}\right)^{2},
\end{align*}
 where $\overline{SL}{}_{\psi,b}\triangleq n_{T}^{-1}\sum_{j=1}^{n_{T}}SL_{\psi,T_{m}+\tau+bn_{T}+j-1}\left(\widehat{\beta}_{T_{m}+bn_{T}+j-1}\right)$.
The estimator $\widehat{\nu}_{\mathrm{Q}1,b}^{2}$ normalizes the
difference in the out-of-sample forecast losses between the $b+1$
and $b$ blocks. The statistic $\mathrm{Q}_{\mathrm{max},h}\left(T_{n},\,\tau\right)$
then results in $\mathrm{Q}_{\mathrm{max},h}\left(T_{n},\,\tau\right)=\max_{b=0,\ldots,\,\left\lfloor T_{n}/n_{T}\right\rfloor -2}\left|\left(B_{h,b+1}-B_{h,b}\right)/\widehat{\nu}_{\mathrm{Q}1,b+1}\right|$.
For the overlapping blocks case, the estimator is $\widehat{\nu}_{\mathrm{MQ}1,i}^{2}\triangleq2n_{T}^{-1}\sum_{j=i+1}^{i+n_{T}}\left(SL_{\psi,T_{m}+\tau+j-1}\left(\widehat{\beta}_{T_{m}+j-1}\right)-\overline{SL}{}_{\psi,i}\right)^{2},$
where $\overline{SL}{}_{\psi,i}\triangleq n_{T}^{-1}\sum_{j=i+1}^{i+n_{T}}SL_{\psi,T_{m}+\tau+j-1}\left(\widehat{\beta}_{T_{m}+j-1}\right)$
so that we can write 
\begin{align*}
\mathrm{M} & \mathrm{Q}{}_{\mathrm{max},h}\left(T_{n},\,\tau\right)\\
 & \triangleq\max_{i=n_{T},\ldots,\,T_{n}-n_{T}}\left|\widehat{\nu}_{\mathrm{MQ}1,i}^{-1}n_{T}^{-1}\left(\sum_{j=i+1}^{i+n_{T}}SL_{\psi,T_{m}+\tau+j-1}\left(\widehat{\beta}_{T_{m}+j-1}\right)-\sum_{j=i-n_{T}+1}^{i}SL_{\psi,T_{m}+\tau+j-1}\left(\widehat{\beta}_{T_{m}+j-1}\right)\right)\right|.
\end{align*}
 Both $\widehat{\nu}_{\mathrm{Q}1,b+1}^{2}$ and $\widehat{\nu}_{\mathrm{MQ}1,i}^{2}$
apply a natural block-wise normalization in order to guarantee a distribution-free
limit under $H_{0}$. However, it is useful to consider estimators
that use all of the observations in the out-of-sample period. Thus,
one exploits covariance stationarity of the sequence of forecast losses.
Let $\Phi_{0.75}=0.647...$ denote the third quartile of the standard
normal distribution and define
\begin{align*}
\widehat{\nu}_{2,h}\triangleq\frac{\sqrt{\pi n_{T}}}{2\left(m_{T}-1\right)}\sum_{b=1}^{m_{T}-1}\left|B_{h,b}-B_{h,b-1}\right|, & \qquad\qquad\widehat{\nu}_{3,h}\triangleq\frac{\sqrt{n_{T}}}{\sqrt{2\left(m_{T}-1\right)}}\left(\sum_{b=1}^{m_{T}-1}\left|B_{h,b}-B_{h,b-1}\right|^{2}\right)^{1/2},\\
\widehat{\nu}_{4,h}\triangleq\frac{\sqrt{n_{T}}}{\sqrt{2\Phi_{0.75}}}\mathrm{median} & \left(\left|B_{h,b}-B_{h,b-1}\right|\right),\qquad1\leq b\leq m_{T}-1.
\end{align*}
Note that $\widehat{\nu}_{2,h},\,\widehat{\nu}_{3,h}$ and $\widehat{\nu}_{4,h}$
can be used to implement both $\mathrm{Q}{}_{\mathrm{max},h}\left(T_{n},\,\tau\right)$
and $\mathrm{MQ}{}_{\mathrm{max},h}\left(T_{n},\,\tau\right)$.\footnote{They can be also applied to the test statistics $\mathrm{Q}{}_{\mathrm{max},h}^{\mathrm{G}}\left(T_{n},\,\tau\right)$
and $\mathrm{MQ}{}_{\mathrm{max},h}^{\mathrm{G}}\left(T_{n},\,\tau\right)$
with $G_{h,b}$ in place of $B_{h,b}$.} $\widehat{\nu}_{3,h}$ is related to \citeauthor{carlstein:86}'s
(1986) subseries variance estimate in the context of strong mixing
processes and it was also used by \citet{wu/zhao:07}. Each of the
estimators $\widehat{\nu}_{2,h},\,\widehat{\nu}_{3,h}$ and $\widehat{\nu}_{4,h}$
allows for dependence but requires stationarity. The simulation study
in \citet{wu/zhao:07} suggests that $\widehat{\nu}_{4,h}$ is more
robust whereas $\widehat{\nu}_{2,h}$ and $\widehat{\nu}_{3,h}$ are
less precise when there are large instabilities or jumps. For two
sequences $\left\{ a_{k}\right\} $ and $\left\{ b_{k}\right\} $,
we write $a_{k}\asymp b_{k}$ if for some $c\geq1$, $b_{k}/c\leq a_{k}\leq cb_{k}$
for all $T.$ The following theorem is similar to Theorem 3 in \citet{wu/zhao:07}
and in particular, part (ii) states that if $n_{T}\asymp T_{n}^{1/3}$
then $\widehat{\nu}_{3,h}^{2}$ achieves the optimal MSE $O\left(n_{T}^{-2/3}\right)$.
\begin{condition}
\label{Cond The-auxiliary-sequence LR VAR}The sequence $\left\{ n_{T}\right\} $
satisfies
\begin{align}
n_{T}\rightarrow\infty\quad\mathrm{as\quad}T\rightarrow\infty & \qquad\mathrm{and}\qquad\sqrt{T_{n}}n_{T}^{-1}\log\left(T_{n}\right)^{3}+n_{T}T_{n}^{-2/3}\left(\log\left(T\right)\right)^{1/3}\rightarrow0.\label{eq. Condition auxiliary nT-1-1}
\end{align}
\end{condition}
\begin{thm}
\label{Theorem: Var Est 3 in Wu and Zhao}In addition to the assumptions
of Theorem \ref{Theoem Asymptotic H0 Distrbution Bmax and Qmax},
assume that $\mathrm{Cov}\left(L_{\psi,kh}\left(\beta^{*}\right),\,L_{\psi,\left(k-j\right)h}\right.$
~ $\left(\beta^{*}\right)\bigr)$ depends on $j$ but not on $kh$.
Then, under $H_{0}$, (i) Let $n_{T}\asymp T^{5/8}$. Then, $\widehat{\nu}_{2,h},\,\widehat{\nu}_{4,h}=\nu_{L}+O_{\mathbb{P}}\left(T_{n}^{-1/16}\log\left(T_{n}\right)\right)$;
(ii) Let $n_{T}\asymp T^{1/3}$. Then $\mathbb{E}\left(\left[\widehat{\nu}_{3,h}^{2}-\nu_{L}^{2}\right]^{2}\right)=O\left(T_{n}^{-2/3}\right)$. 
\end{thm}
Under covariance-stationarity, given Theorem \ref{Theorem: Var Est 3 in Wu and Zhao},
the results of Corollary \ref{Corollary CR Null Distrb Quadratic}-\ref{Corollary CR Null Distrb General}
are applicable after replacing $\nu_{L}$ by an appropriate consistent
estimator.

\subsubsection{Non-Stationary Forecast Losses}

We now consider estimation of the asymptotic variance in the case
the forecast losses are heterogeneous. The estimator $\widehat{\nu}$
that we introduce below depends on the specific loss function and
thus it can be used for replacing $\nu_{L}$ in Corollary \ref{Corollary CR Null Distrb Quadratic}-\ref{Corollary CR Null Distrb General}.
Non-stationarity implies that $\sigma_{e,t}^{2}$ is time-varying
and thus the results of Theorem \ref{Theorem: Var Est 3 in Wu and Zhao}
are not applicable due to the presence of many extra parameters that
account for the time-varying structure. To deal with this issue we
propose a novel block-wise self-normalization technique which simultaneously
addresses two issues. First, the block-wise self-normalization ensures
that the difference in forecast losses between two adjacent blocks
are asymptotically independent across non-adjacent blocks and that
within each block the losses are standardized so that time-varying
variances cancel out. Second, by computing an average\textemdash over
all blocks\textemdash of the self-normalized difference in forecast
losses we account for possible serial dependence. We derive asymptotic
results within a general framework based on the strong invariance
principle for stationary processes developed in \citet{wu:07} and
extended to modulated stationary processes by \citet{zhao/li:13}.

For each block $b=0,\ldots,\,m_{T}-2$, let
\begin{align*}
A_{h,b}\left(\widehat{\beta}\right) & \triangleq n_{T}^{-1}\sum_{j=1}^{n_{T}}\left(L_{\psi,T_{m}+\tau+\left(b+1\right)n_{T}+j-1}\left(\widehat{\beta}_{T_{m}+\left(b+1\right)n_{T}+j-1}\right)\right),\\
V_{h,b}\left(\widehat{\beta}\right) & \triangleq n_{T}^{-1}\sum_{j=1}^{n_{T}}\left(L_{\psi,T_{m}+\tau+\left(b+1\right)n_{T}+j-1}\left(\widehat{\beta}\right)-\overline{L}{}_{\psi,b}\left(\widehat{\beta}\right)\right)^{2},
\end{align*}
where $\overline{L}{}_{\psi,b}\left(\widehat{\beta}\right)=n_{T}^{-1}\sum_{j=1}^{n_{T}}L_{\psi,T_{m}+\tau+\left(b+1\right)n_{T}+j-1}\left(\widehat{\beta}\right)$
and define the statistic
\begin{align*}
\zeta_{h,b}\left(\widehat{\beta}\right)\triangleq & \sqrt{n_{T}}\left(A_{h,b}\left(\widehat{\beta}\right)-A_{h,b-1}\left(\widehat{\beta}\right)\right)/\sqrt{V_{h,b}}.
\end{align*}
Finally, an average\textemdash over all blocks $m_{T}$\textemdash of
the per-block self-normalized statistics $\zeta_{h,b}$'s is used
to define an estimator of the asymptotic variance: $\widehat{\nu}_{L}^{2}\triangleq2^{-1}\left(m_{T}-1\right)^{-1}\sum_{b=0}^{m_{T}-1}\zeta_{h,b}^{2}$.

Let $\sigma_{L,kh}^{2}\triangleq\mathrm{Var}\left(L_{\psi,kh}\left(\beta^{*}\right)\right)$.
We also need to introduce the following quantities,
\begin{align*}
F_{h,b}^{*} & \triangleq\left|\sigma_{L,\left(T_{m}+\tau+\left(b+2\right)n_{T}\right)h}\right|,\qquad\qquad J_{h,b}^{*}\triangleq\sigma_{L,\left(T_{m}+\tau+\left(b+2\right)n_{T}\right)h}^{2},\\
\Sigma_{h,b}^{*} & \triangleq\sum_{j=1}^{n_{T}}\sigma_{L,\left(T_{m}+\tau+\left(b+1\right)n_{T}+j\right)h}^{2},\qquad\qquad\widetilde{\Sigma}_{h,b}^{*}\triangleq\left(\sum_{j=1}^{n_{T}}\sigma_{L,\left(T_{m}+\tau+\left(b+1\right)n_{T}+j\right)h}^{4}\right)^{1/2}.
\end{align*}
\begin{thm}
\label{Theorem Long-Run Variance}Under Condition \ref{Cond The-auxiliary-sequence LR VAR}
we have $\widehat{\nu}_{L}^{2}-\nu_{L}^{2}=O_{\mathbb{P}}\left(r_{h}^{-1}\right)$,
where $r_{h}=O_{\mathbb{P}}\left(T_{n}^{\epsilon}/\left(\log\left(T_{n}\right)\right)^{2}\right)$
with $\epsilon\in\left(0,\,1/4\right)$ such that $r_{h}\rightarrow\infty$.
\end{thm}
The theorem simply states that $\widehat{\nu}_{L}$ is consistent
for $\nu_{L}$ and therefore the asymptotic results of Section \ref{Section CR Distribution Theory Test Stats}
continue to hold when we replace $\nu_{L}$ by $\widehat{\nu}_{L}$.

\section{\label{Section Ito Vol}Continuous Semimartingale Volatility and
Asymptotic Local Power}

\subsection{Asymptotic Results under Continuous Semimartingale Volatility }

In this section we relax the Lipschitz condition on $\sigma_{e,t}$
and extend the results for the quadratic loss case from Theorem \ref{Theoem Asymptotic H0 Distrbution Bmax and Qmax}
to stochastic volatility models driven by a Wiener process. Consequently,
this relaxation enables one to utilize the tests proposed in this
paper in setting involving high-frequency financial variables. More
specifically, we assume that $\sigma_{e,t}$ is an It\^o continuous
semimartingale that is almost surely bounded and strictly positive
adapted process. We replace Assumption  \ref{Assumption Lipchtitz cont of Sigma}
by the following.
\begin{assumption}
\label{Assumption Smooth of Sigma}Under $H_{0}$ the process $\left\{ \sigma_{e,t}\right\} _{t\geq0}$
satisfies $\phi_{\sigma,\eta,\tau_{h}\wedge N}\leq K_{h}\eta^{\kappa}$
for some $\kappa>0$, some sequence of stopping times $\tau_{h}\rightarrow\infty$
and some $\mathbb{P}$-a.s. finite random variable $K_{h}$. 
\end{assumption}
The assumption implies that $\sigma_{e,t}$ belongs to a rather large
class of volatility processes usually considered in financial econometrics.
The parameter $\kappa$ plays a key role in the testing framework
of this section and we refer to it as the regularity exponent. When
$\kappa=1$ we recover the case of Lipschitz volatility considered
in the previous sections while the standard stochastic volatility
model without jumps correspond to $\kappa=1/2-\epsilon$ for a sufficiently
small $\epsilon>0.$ Next, we have a slightly different version of
Condition \ref{Cond The-auxiliary-sequence}.
\begin{condition}
\label{Cond The-auxiliary-sequence Ito Vol}The sequence $\left\{ n_{T}\right\} $
satisfies for some $\epsilon>0,$ 
\begin{align}
n_{T}\rightarrow\infty\quad\mathrm{as\quad}T\rightarrow\infty & \qquad\mathrm{and}\qquad T^{\epsilon}n_{T}^{-1}+\sqrt{n_{T}}\left(n_{T}h\right)^{\kappa}\sqrt{\log\left(T\right)}\rightarrow0.\label{eq. Condition auxiliary nT Ito Vol}
\end{align}
\end{condition}
For It\^o continuous semimartingale volatility $\sigma_{e,t}$ the
condition suggests $n_{T}\propto T^{1/2-\epsilon}$ for small $\epsilon>0.$
Let $\varGamma_{t}\triangleq\mathbb{E}_{\sigma}\left[dL_{t}\left(e_{t};\,\beta^{*}\right)/dt\right]$.\footnote{For example, for the quadratic loss with $\mu_{e,t}=0$ the notation
reduces to $\varGamma_{t}=\sigma_{e,t}^{2}.$ } The more general framework considered here requires us to consider
the following null hypotheses: under quadratic loss, 
\begin{align}
H_{0}: & \quad\left\{ \varGamma_{t}\right\} _{t\in\left[N_{\mathrm{in}}+h,\,N\right]}\in\boldsymbol{C}\left(\kappa,\,K_{h}\right),\label{eq H0 Ito Vol}
\end{align}
 where $\boldsymbol{C}\left(\kappa,\,K_{h}\right)$ is a class of
continuous functions on $\left[N_{\mathrm{in}}+h,\,N\right]$, 
\begin{align*}
\boldsymbol{C}\left(\kappa,\,K_{h}\right) & \triangleq\left\{ \left\{ \varGamma_{t}\right\} _{t\in\left[N_{\mathrm{in}}+h,\,N\right]}:\,\sup_{s,t\in\left[N_{\mathrm{in}}+h,\,N\right],\left|t-s\right|<\eta}\left|\varGamma_{t}-\varGamma_{s}\right|\leq K_{h}\eta^{\kappa}\right\} ,
\end{align*}
where $\kappa>0$ and $K_{h}$ is given in Assumption \ref{Assumption Smooth of Sigma}.
Thus, we wish to discriminate between $H_{0}$ and 
\begin{align}
H_{1} & :\,\exists\lambda\in\left[N_{\mathrm{in}}+h,\,N\right]\quad\mathrm{with}\quad\left\{ \varGamma_{t}\left(\omega\right)\right\} _{t\in\left[N_{\mathrm{in}}+h,\,N\right]}\in\boldsymbol{J}_{\lambda}\left(\kappa,\,K_{h},\,d_{h}\right),\label{eq. H1 Ito vol}
\end{align}
where $\boldsymbol{J}_{\lambda}\left(\kappa,\,K_{h},\,d_{h}\right)\triangleq\left\{ \left\{ \varGamma_{t}\right\} _{t\in\left[N_{\mathrm{in}}+h,\,N\right]}:\,\left\{ \varGamma_{t}-\Delta\varGamma_{t}\right\} _{t\in\left[N_{\mathrm{in}}+h,\,N\right]}\in\boldsymbol{C}\left(\kappa,\,K_{h}\right);\,\left|\Delta\varGamma_{\lambda}\right|\geq d_{h}\right\} $,
$\Delta\varGamma_{\lambda}=\varGamma_{\lambda}-\lim_{s\uparrow\lambda}\varGamma_{s}$
and $\left\{ d_{h}\right\} $ is a decreasing sequence. The following
theorem extends Theorem \ref{Theoem Asymptotic H0 Distrbution Bmax and Qmax}
to the current setting.
\begin{thm}
\label{Theoem Asymptotic H0 Distrbution Bmax and Qmax Ito Vol}Let
$m_{T},\,\gamma_{m_{T}}$ and $\mathscr{V}$ as defined in Theorem
\ref{Theoem Asymptotic H0 Distrbution Bmax and Qmax}. Assume the
assumptions of Theorem \ref{Theoem Asymptotic H0 Distrbution Bmax and Qmax}
hold with Assumption \ref{Assumption Lipchtitz cont of Sigma} replaced
by Assumption \ref{Assumption Smooth of Sigma}. Under Condition \ref{Cond The-auxiliary-sequence Ito Vol}
and quadratic loss, the same results of Theorem \ref{Theoem Asymptotic H0 Distrbution Bmax and Qmax}
hold.
\end{thm}

\subsection{Asymptotic Local Power}

In this section we consider the behavior of $\mathrm{MQ}_{\mathrm{max},h}$
under a sequence of local alternatives.
\begin{assumption}
\label{Ass Local Power}We have the same assumptions as in Theorem
\ref{Theoem Asymptotic H0 Distrbution Bmax and Qmax Ito Vol} and
assume (i) in model \eqref{Mode for Y} we replace $\beta^{*}$ by
$\beta_{t}=\beta^{*}+\mu_{\beta,t}/\left(\log\left(T_{n}\right)n_{T}\right)^{1/4}$
where $\mu_{\beta,t}\in\mathbb{R}^{q}$ is $\mathbb{P}$-a.s. locally
bounded and adapted process; (ii) we set $\mu_{e,t}=0$ for all $t\geq0$;
(iii) we replace Assumption \ref{Assumption Roo-T Consistent beta }
by $\left\Vert \widehat{\beta}_{k}-\beta^{*}\right\Vert =\mu_{\beta,kh}/\left(\log\left(T_{n}\right)n_{T}\right)^{1/4}+O_{\mathbb{P}}\left(T^{-1/2}\right)$
uniformly in $k.$
\end{assumption}
Part (iii) is a consequence of part (i) as it can be easily verified.
Let 
\begin{align*}
\mathrm{\widetilde{MQ}}{}_{\mathrm{max},h}\left(T_{n},\,\tau\right) & \triangleq\nu_{L}^{-1}\max_{i=n_{T},\ldots,\,T_{n}-n_{T}}\left|n_{T}^{-1}\sum_{j=i+1}^{i+n_{T}}\left(SL_{\psi,T_{m}+\tau+j-1}\left(\widehat{\beta}_{T_{m}+j-1}\right)-2\zeta_{\mu,j,+}\right)\right.\\
 & \quad\left.-n_{T}^{-1}\sum_{j=i-n_{T}+1}^{i}\left(SL_{\psi,T_{m}+\tau+j-1}\left(\widehat{\beta}_{T_{m}+j-1}\right)-2\zeta_{\mu,j,-}\right)\right|,
\end{align*}
where 
\begin{align*}
\zeta_{\mu,j,+} & \triangleq\mu'_{\beta,\left(T_{m}+\tau+j-1\right)h}\Sigma_{X,\left(T_{m}+\tau+i-1\right)h}\mu_{\beta,\left(T_{m}+\tau+j-1\right)h}/\left(\log\left(T_{n}\right)n_{T}\right)^{1/2}\\
\zeta_{\mu,j,-} & \triangleq\mu'_{\beta,\left(T_{m}+\tau+j-1\right)h}\Sigma_{X,\left(T_{m}+\tau+i-n_{T}-1\right)h}\mu_{\beta,\left(T_{m}+\tau+j-1\right)h}/\left(\log\left(T_{n}\right)n_{T}\right)^{1/2}.
\end{align*}
\begin{thm}
\label{Theorem Local Asym Power}Under Assumption \ref{Ass Local Power},
\begin{align*}
\sqrt{\log\left(m_{T}\right)}\left(n_{T}^{1/2}\nu_{L}^{-1}\right)\mathrm{\widetilde{MQ}}_{\mathrm{max},h}\left(T_{n},\,\tau\right)-2\log\left(m_{T}\right)-\frac{1}{2}\log\log\left(m_{T}\right)-\log3 & \overset{}{\Rightarrow}\mathscr{V},
\end{align*}
 where $\mathscr{V},$ and $m_{T}$ are defined as in Theorem \ref{Theoem Asymptotic H0 Distrbution Bmax and Qmax}. 
\end{thm}
\begin{rem}
(i) The theorem suggests that under the local alternatives $\beta_{t}=\beta^{*}+\mu_{\beta,t}/\left(\log\left(T_{n}\right)n_{T}\right)^{1/4}$
there is a bias term arising from the presence of $\zeta_{\mu}$.
This bias term does not vanish asymptotically and results in shifting
the center of the distribution. Moreover, it depends on the second
moments of the regressors and on the function $\mu_{\beta}$; (ii)
The theorem illustrates the sensitivity of the asymptotic power to
the form of the alternative. We can attempt to compare Theorem \ref{Theorem Local Asym Power}
with the local power result regarding the sup-Wald test of \citet{andrews:93}.
Unlike Theorem 4 in \citet{andrews:93}, our result suggests that
the location of the instability should not play any special role and
the power should not be sensitive to whether the break in predictive
ability occurs at middle sample or toward the tail of the sample.
This follows because of the local nature of our test statistic and
contrasts with classical tests for parameter instability and structural
change since their performance  hinges on the location of the break
{[}see \citet{deng/perron:08}, \citet{kim/perron:09} and \citet{perron/yamamoto:16}
for additional results on the power of classical structural break
tests{]}. However, the magnitude of the break\textemdash here shrinking
at rate $\left(\log\left(T_{n}\right)n_{T}\right)^{1/4}$\textemdash under
our specification of the local alternatives is larger than the one
considered by \citet{andrews:93}\textemdash which shrinks at rate
$1/\sqrt{T}$. This implies a trade-off between location and magnitude
of the break, and it is consistent with the evidence provided in our
simulation study; (iii) Although not shown here, the local power
of the tests is the same when a subset of the vector $\beta$ is not
subject to shift. 
\end{rem}

Theorem \ref{Theorem Local Asym Power} can be used to show that our
test possesses nontrivial power against alternatives for which the
parameter $\beta_{t}$ is time-varying and non-smooth. 
\begin{cor}
\label{Corollary Local Power}Suppose the assumptions of the previous
theorem hold with $\beta_{t}=\beta^{*}+c\mu_{\beta,t}/\left(\log\left(T_{n}\right)n_{T}\right)^{1/4}$,
where $c\in\mathbb{R}$. If $\mu_{\beta,\cdot}$ and/or $\left\{ \mu_{\beta,t}\cdot\sigma_{X,t}\right\} _{t\geq0}$
is non-smooth, we have 
\begin{alignat*}{1}
\underset{c\rightarrow\infty}{\lim}\underset{h\downarrow0}{\lim} & \mathbb{P}\left(\sqrt{\log\left(m_{T}\right)n_{T}}\nu_{L}^{-1}\mathrm{MQ}_{\mathrm{max},h}\left(T_{n},\,\tau\right)-2\log\left(m_{T}\right)-\frac{1}{2}\log\log\left(m_{T}\right)-\log3>\mathrm{cv}_{1-\alpha}\right)=1,
\end{alignat*}
 where $\mathrm{cv}_{1-\alpha}$ is the level $\left(1-\alpha\right)$
critical value of the distribution of $\mathscr{V}$ and $\alpha\in\left(0,\,1\right)$. 
\end{cor}

\section{Extensions\label{Section Extensions}}

A number of extensions is treated in our companion paper \citet{casini_FFb}.
As explained above, it would be useful to ensure that there are no
instabilities in the in-sample period $\left[0,\,N_{\mathrm{in}}\right].$
We propose a procedure that involves a pre-test about instability
on $\left[0,\,N_{\mathrm{in}}\right]$ for a given $N_{\mathrm{in}}$
chosen by the forecaster. Instabilities in the in-sample $\left[0,\,N_{\mathrm{in}}\right]$
are much easier to be detected relative to instabilities in the out-of
sample because they do not face the so-called ``contamination effect''.
The latter arises, for example under the recursive and rolling scheme,
when the instability originally occurring in the out-of-sample eventually
enters the moving in-sample window {[}cf. \citet{casini/perron_Oxford_Survey}
and \citet{perron/yamamoto:18}{]}. The consequence is that existing
tests face substantial power losses. This property is not shared by
our test statistics because of their local nature. Our procedure works
very well and we show through simulations that instabilities occurring
in the in-sample only or occurring both in the in-sample and in the
out-of-sample simultaneously, lead easily to rejection of the null
hypotheses relative to instabilities occurring in the out-of-sample
only\textemdash as we consider here. 

A second issue is that, in this paper, we have considered processes
that have a continuous sample path under the null hypotheses. Thus,
it is of interest to extend the results to a setting that involves
jump processes which are important in high-frequency financial data.
This can be achieved by using techniques that are able to separate
the continuous part from the discontinuous part of a It� semimartingale
{[}see e.g. \citet{li/todorov/tauchen:17} and \citet{li/xiu:16}{]}.

Another important issue is the estimation of the time at which forecast
instability occurs. Once the null hypotheses has been rejected, a
forecaster may take into consideration the possibility of revising
the forecasting method and/or model. Hence, it becomes crucial to
learn some information about the timing of the instability. For example,
consider the case of a one-time structural change in a parameter of
the data-generating process at time $T_{b}^{0}=\left\lfloor T\lambda_{b}^{0}\right\rfloor $,
where $T_{b}^{0}$ is the \textit{break point} and $\lambda_{b}^{0}\in\left(0,\,1\right)$
is the \textit{fractional break date}. Once $H_{0}$ is rejected,
a forecaster would benefit from knowing that the forecast method originally
employed is found to statistically either under or over-perform over
part of the sample after $T_{b}^{0}$ relative to the part prior $T_{b}^{0}$.
Then, a forecaster would entertain the possibility of modifying the
forecast model in order to generate future forecasts for $Y_{t}$.
Not only the forecast model might be revised but most importantly,
knowledge of beginning of the instability at $T_{b}^{0}$ can be further
exploited to design the forecasting method for the future forecasts.
It would be inappropriate, for instance, to use a rolling scheme where
the rolling window used to construct the forecast include observations
prior to $T_{b}^{0}$ since those observations provide little informational
content for the purpose of predicting $Y_{t}$ after the change-point
$T_{b}^{0}$. On the other hand, this line of reasoning is justified
by this particular example and indeed in practice many issues arise
when dealing with the timing of the insatiability in our context.
For example, the exact form of the insatiability may be unknown. Under
the latter scenarios, there is no clear-cut break date $T_{b}^{0}$
that can be defined. Thus, it is less obvious how a forecaster should
proceed in those cases. Nonetheless, one can meaningfully think about
the timing of the instability by not just attempting to estimate $T_{b}^{0}$\textemdash which
is not clear how it is defined\textemdash but rather attempting to
detect the initial date in the sample after which the forecasts become
unstable as well as to detect the last date after which the forecasts
remain stable relative to the in-sample period. Since our test statistics
are local in nature, one can introduce a procedure which sequentially
tests the hypotheses $H_{0}$ in regions of the sample where $H_{0}$
has not yet been rejected. One then records the number of times for
which $H_{0}$ is rejected and estimates the corresponding change-point
dates. After ordering these change-point dates, one has finally access
to useful information such has the initial timing of the forecast
instability and the last part of the out-of-sample period which remains
stable. Such information can arguably be advantageous to the forecaster.

\section{\label{Section Simulation Study}Small-Sample Evaluation}

We now examine the empirical size and power of our proposed tests
and compare them to those of \citet{giacomini/rossi:09}, abbreviated
GR (2009). In particular, we consider both the uncorrected and corrected
version of the $t_{T_{m},T_{n},\tau}^{\mathrm{stat}}$ statistic of
GR (2009).\footnote{We use a superscript $\mathrm{c}$ to indicate the corrected version:
$t^{\mathrm{stat,c}}$.} Size and power properties for the Quadratic loss with fixed scheme
are reported in Section \ref{Subsection MC Size} and Section \ref{Subsection MC Power},
respectively. The Supplemental Material includes corresponding simulation
studies for the recursive and rolling schemes and for the Linex loss;
these results are not reported here because they are qualitatively
equivalent. Overall, one can draw the following conclusions from our
simulation study. In terms of size control, the statistics $\mathrm{B}_{\mathrm{max},h}$
and $\mathrm{Q}_{\mathrm{max},h}$ are comparable with the corrected
version $t^{\mathrm{stat,c}}$ proposed by GR (2009).\footnote{As shown by GR (2009), the uncorrected version of $t^{\mathrm{stat}}$
can be oversized for models that induce serial dependence in the forecast
losses. The authors then proposed a finite-sample correction and did
not consider $t^{\mathrm{stat}}$ further in their power analysis.
Similarly, GR (2009) showed that just using classical structural break
tests in this context is not very helpful as they might have statistical
power equals to the size in some cases. Moreover, simulations in
\citet{perron/yamamoto:18} confirmed that, under rolling and recursive
scheme, structural break tests suffer power losses which can be attributed
to a so-called ``contamination effect'' arising when the instability
enters the in-sample window {[}see also \citet{casini/perron_Oxford_Survey}{]}. } Moreover, the test $\mathrm{MQ}_{\mathrm{max},h}$ that uses overlapping
blocks is also comparable in terms of size. The same is not true for
$\mathrm{MB}_{\mathrm{max},h}$ because often it seems to be somewhat
liberal. Turning to the power comparison, each of our test statistics
$\mathrm{B}_{\mathrm{max},h}$, $\mathrm{Q}_{\mathrm{max},h}$ and
$\mathrm{MQ}_{\mathrm{max},h}$ displays significant power gains over
the $t_{T_{m},T_{n},\tau}^{\mathrm{stat}}$ statistics especially
as the period of instability (i) is comparatively short relative to
the total sample size and/or (ii) is not located at middle sample.
In the latter circumstances, the gains in power are, uniformly over
different data-generating processes and over parameter break magnitudes,
on the order of 30-40\%.

Throughout, we restrict attention to one-step ahead forecast horizon
(i.e., $\tau=1$), and we use the same loss function for estimation
and evaluation. We use our asymptotic results as an approximation
for the case where $h=1$ in our theoretical model in \eqref{Discretized Model Y}
and consider discrete-time DGPs. In models with serially correlated
losses (i.e., S2 and S6 below) for the statistics $\mathrm{Q}_{\mathrm{max},h}$
and $\mathrm{MQ}_{\mathrm{max},h}$ we employ the long-run variance
estimator from Theorem \ref{Theorem Long-Run Variance}. With regards
to the tests of GR (2009) we use the appropriate version of $t^{\mathrm{stat}}$
and of $t^{\mathrm{stat,c}}$.\footnote{As recommended by GR (2009) we set the truncation lag of their HAC
estimator equal to $\left\lfloor T_{n}^{1/3}\right\rfloor $; we also
the use the truncation lag $\left\lfloor 0.75T_{n}^{1/3}\right\rfloor $.} 
\begin{rem}
Implementation of our tests statistics requires to choose the number
of blocks $m_{T}$\textemdash{} satisfying Condition \ref{Cond The-auxiliary-sequence}.
The finite-sample properties can be sensitive to the choice of $m_{T}$.
This is confirmed in our numerical study, where assigning larger values
to $m_{T}$ than the smallest one allowed by the condition may result
in oversized tests. Therefore, we recommend practitioners to set $m_{T}$
equal to the smallest integer as allowed by Condition \ref{Cond The-auxiliary-sequence}.
This is the strategy we have adopted in the Monte Carlo study of this
section, and as we will show, it results in approximately correct
size and good power across different data-generating mechanisms. 
\end{rem}

\subsection{\label{Subsection MC Size}Empirical Size}

We consider discrete-time DGPs of the form
\begin{align}
Y_{t}=\mu+\beta X_{t-1}+e_{t}, & \qquad\qquad t=1,\ldots,\,T,\label{Eq. DGP Simulation Study}
\end{align}
for various in-sample and out-of-sample sizes and with a total sample
size ranging from $T=100$ to $T=500$. Note that \eqref{Eq. DGP Simulation Study}
is a special case of the theoretical model with a sampling interval
$h=1$. We consider six versions of \eqref{Eq. DGP Simulation Study},
where the first and second specification (S1 and S2 below) are calibrated
to the Phillips curve model of U.S. inflation from \citet{staiger/stock/watson:97}:
S1 involves $\mu=2.73$, $\beta=-0.44$, and where $\left\{ X_{t}\right\} $
and $\left\{ e_{t}\right\} $ are independent sequences of zero-mean
i.i.d. Gaussian disturbances with unit variance; S2 is the same as
S1 but with ARCH errors $e_{t}=\sigma_{e,t}u_{t}$, $\sigma_{e,t}=1+0.5e_{t-1}^{2}$
with $u_{t}\sim\mathscr{N}\left(0,\,1\right)$; S3 specifies $\left\{ X_{t}\right\} $
to follow a zero-mean Gaussian AR(1) with autoregressive coefficient
0.4, $\beta=1$ and $e_{t}\sim\mathscr{N}\left(0,\,0.49\right)$ independent
of $X_{t}$; S5 is a model with a lagged dependent variable $X_{t-1}=Y_{t-1}$,
$\mu=0$, $\beta=0.3$ and $e_{t}\sim\mathscr{N}\left(0,\,0.49\right)$;
S6 involves serially correlated disturbances $e_{t}=0.3e_{t-1}+u_{t}$,
$u_{t}\sim\mathscr{N}\left(0,\,1\right)$.

Table \ref{Table S1}-\ref{Table S2} report the rejection rates for
significance levels $\alpha=0.05$ and $0.10$ for model S1-S2. Results
for the other DGPs can be found in Table \ref{Table Size S3}-\ref{Table S6}
in the Supplement. We first focus on i.i.d. forecast losses (i.e.,
models S1 and S3-S5). Both $\mathrm{B}_{\mathrm{max},h}$ and $\mathrm{Q}_{\mathrm{max},h}$
are well-sized. As the sample size increases their performance improves
and we note that their rejection frequencies are closer to the nominal
level when the in-sample size is one half of the total sample. In
model S1, when the in-sample size is $0.25T,$ $\mathrm{B}_{\mathrm{max},h}$
and $\mathrm{Q}_{\mathrm{max},h}$ tend to be slightly conservative
while the opposite occurs when in-sample size is $0.75T$. The version
of $\mathrm{B}_{\mathrm{max},h}$ that uses overlapping blocks $\left(\mathrm{MB}_{\mathrm{max},h}\right)$
can be quite liberal (cf. models S1 and S3). In contrast, $\mathrm{MQ}_{\mathrm{max},h}$
seems to control the size well, though it tends to be slightly liberal
but that depends on the relative size of the in-sample and out-of-sample
windows. We observe that there is no clear pattern in size performance
for our test statistics as we raise the sample size $T$. The reason
is straightforward: as we raise $T$ we also need to adjust the choice
of $m_{T}$ (the number of blocks) in accordance with Condition \ref{Cond The-auxiliary-sequence}.
This explains why, for example, in Table \ref{Table S1}, top panel,
the empirical size of $\mathrm{Q}_{\mathrm{max},h}$ for $\left(T_{m}=100,\,T_{n}=100\right)$
is better than for $\left(T_{m}=150,\,T_{n}=150\right)$. Turning
to the $t^{\mathrm{stat}}$ statistics of \citet{giacomini/rossi:09},
the uncorrected version performs better than the corrected version
since the latter systematically displays an empirical size 2-3\% below
the nominal level. We can conclude that in models with i.i.d. errors
the statistics $\mathrm{B}_{\mathrm{max},h}$, $\mathrm{Q}_{\mathrm{max},h}$,
$\mathrm{MQ}_{\mathrm{max},h}$ and $t^{\mathrm{stat}}$ are comparable
in terms of empirical size, whereas $\mathrm{MB}_{\mathrm{max},h}$
and $t^{\mathrm{stat,c}}$ tend to over-reject and under-reject, respectively.

Let us now turn to models with serially correlated losses. When the
disturbances follow an ARCH process, (cf. model S2, Table \ref{Table S2}),
we observe that both statistics that do not use overlapping blocks,
$\mathrm{B}_{\mathrm{max},h}$ and $\mathrm{Q}_{\mathrm{max},h}$,
show reasonable size control. The same feature applies to $\mathrm{MQ}_{\mathrm{max},h}$
while $\mathrm{MB}_{\mathrm{max},h}$ displays rejection rates that
are systematically above the significance level. It also appears that
the corrected version of the statistic of GR (2009) is now regularly
below the nominal level. In contrast, the uncorrected version $t^{\mathrm{stat}}$
seems to control size well. When the errors follow an autoregressive
process (cf. model S6), Table \ref{Table S6} shows that $t^{\mathrm{stat}}$
and $\mathrm{MB}_{\mathrm{max},h}$ are arbitrarily oversized for
all sample sizes. $\mathrm{MQ}_{\mathrm{max},h}$ and $t^{\mathrm{stat,c}}$
possess rejection rates frequently below the desired nominal level.
The statistic that shows the best empirical sizes across different
$T$ is $\mathrm{Q}_{\mathrm{max},h}$.

Overall, our analysis on the size properties of the tests suggests
that when the DGP involves i.i.d. errors it is fair to compare $\mathrm{B}_{\mathrm{max},h}$,
$\mathrm{Q}_{\mathrm{max},h}$, $\mathrm{MQ}_{\mathrm{max},h}$ and
$t^{\mathrm{stat}}$ whereas the rejection rates of $\mathrm{MB}_{\mathrm{max},h}$
and $t^{\mathrm{stat,c}}$ tend to deviate systematically from the
nominal level. When there are autocorellated errors, it is difficult
to compare $t^{\mathrm{stat}}$ and $\mathrm{MB}_{\mathrm{max},h}$
with the other statistics because the former can be highly oversized.
The statistics that appear to perform better in terms of approximate
size control uniformly over different data-generating mechanisms are
$\mathrm{Q}_{\mathrm{max},h}$ and $\mathrm{MQ}_{\mathrm{max},h}$.

\subsection{\label{Subsection MC Power}Empirical Power}

We report the small sample power of the tests under various sources
of forecast instability. We consider several sample sizes $T$ as
well as several designs varying for the distribution of the total
sample between in-sample and out-of-sample window. The break date\textemdash or
the date of the first change-point when more complicated designs are
used\textemdash is denoted by $T_{b}^{0}=T\lambda_{0}$, where $\lambda_{0}\in\left(0,\,1\right)$
is the fractional break date. We shall bring special attention to
the location of $T_{b}^{0}$ in the sample as well as to the duration
of the instability (i.e., $T-T_{b}^{0}$). We shall see that both
factors are actually important for the performance of the methods
proposed by \citet{giacomini/rossi:09} while our test statistics
being local in nature possess essentially uniform power over distinct
locations $T_{b}^{0}$. Furthermore, our definition of forecast instability
does not demand any relationship between the stable and unstable period
and thus it is useful to examine the differences in power properties
when a one-time change-point is present relative to when short-lasting
instabilities arise. 

We consider both discrete shifts\textemdash a structural break\textemdash and
recurrent changes in a parameter: model P1a (break in a regression
coefficient): $Y_{t}=2.73-0.44X_{t-1}+\delta X_{t-1}\mathbf{1}\left\{ t>T_{b}^{0}\right\} +e_{t}$,
where $X_{t-1}\sim\mathrm{i.i.d.}\mathscr{N}\left(0,\,1\right)$ and
$e_{t}\sim\mathrm{i.i.d.}\mathscr{N}\left(0,\,1\right)$; model P1b:
it is the same as model P1a but with $X_{t-1}\sim\mathrm{i.i.d.}\mathscr{N}\left(1,\,1\right)$;
model P2: $Y_{t}=X_{t-1}+\delta X_{t-1}\mathbf{1}\left\{ t>T_{b}^{0}\right\} +e_{t}$,
where $X_{t-1}$ is a Gaussian AR(1) with autoregressive coefficient
0.4 and unit variance, and $e_{t}\sim\mathrm{i.i.d.}\mathscr{N}\left(0,\,0.49\right)$;
model P3 (recurrent break in mean): $Y_{t}=\beta_{t}+e_{t}$, where
$\beta_{t}$ switches between $\delta$ and $0$ every $p$ periods
and $e_{t}\sim\mathrm{i.i.d.}\mathscr{N}\left(0,\,0.64\right)$; model
P4 (single break in variance): $Y_{t}=0.5X_{t-1}+\left(1+\delta\mathbf{1}\left\{ t>T_{b}^{0}\right\} \right)e_{t}$
where $X_{t-1}\sim\mathrm{i.i.d.}\mathscr{N}\left(1,\,1\right)$ and
$e_{t}\sim\mathrm{i.i.d.}\mathscr{N}\left(0,\,1\right)$; model P5
(recurrent break in variance): $Y_{t}=\mu+\left(1+\beta_{t}\right)e_{t}$,
where $\beta_{t}$ switches between $\delta$ and $0$ every $p$
periods and $e_{t}\sim\mathrm{i.i.d.}\mathscr{N}\left(0,\,0.49\right)$;
model P6 (lagged dependent variable): $Y_{t}=\delta\mathbf{1}\left\{ t>T_{b}^{0}\right\} +0.3Y_{t-1}+e_{t}$,
$e_{t}\sim\mathrm{i.i.d.}\mathscr{N}\left(0,\,0.49\right)$; model
P7 (ARCH disturbances): $Y_{t}=2.73-0.44X_{t-1}+\delta X_{t-1}\mathbf{1}\left\{ t>T_{b}^{0}\right\} +e_{t}$,
where $X_{t-1}\sim\mathrm{i.i.d.}\mathscr{N}\left(0,\,1.5\right)$
and $e_{t}=\sigma_{t}u_{t}$, $\sigma_{t}^{2}=0.5+0.5e_{t-1}^{2}$,
$u_{t}\sim\mathrm{i.i.d.}\mathscr{N}\left(0,\,1\right)$; model P8
(autocorrelated errors): $Y_{t}=1+X_{t-1}+\delta X_{t-1}\mathbf{1}\left\{ t>T_{b}^{0}\right\} +e_{t}$,
where $X_{t-1}\sim\mathrm{i.i.d.}\mathscr{N}\left(0,\,1.4\right)$
and $e_{t}=0.4e_{t-1}+u_{t}$, $u_{t}\sim\mathrm{i.i.d.}\mathscr{N}\left(0,\,1\right)$.
For models that do not involve recurrent changes we also consider
power comparisons when the instability lasts only for some period
of time as opposed to the post-$T_{b}^{0}$ period. This requires
replacing $\mathbf{1}\left\{ t>T_{b}^{0}\right\} $ in models P1-P2,
P4 and P6-P8 with $\mathbf{1}\left\{ T_{b}^{0}<t\leq T_{b}^{0}+p\right\} $
where $p$ is the number of consecutive observations in which the
forecast model is unstable. The value of $p$ depends on the sample
size $T$. For example, when $T=100$ we set $p=10$; when $T=200$
we set $p=20$ and so on.\footnote{Note that for $\left(T_{m}=50,\,T_{n}=50\right)$ the value $p=10$
corresponds to a period of instability lasting for one-fifth of the
out-of-sample; thus, the duration of the instability is nontrivial
and consistent with forecasting applications. See the notes to each
figure for the other values of $p$. The title of a figure corresponding
to a short-lasting instability is labeled ``short-term instability''.} The case of short-term instability is the most prevalent in empirical
work because it is very unlikely that a professional forecaster would
use a poor-performing predictor or forecast model for many consecutive
years (e.g., the whole out-of-sample).

Figure \ref{Fig_P1a_150_078}-\ref{Fig_P4_450_068_st} in the Appendix
plot the power functions for models P1a, P4 and P7. Figure \ref{Fig_P1b_150}-\ref{Fig_P2_1200_078_st}
in the Supplement plot the power functions for the remaining DGPs.
They include several sample sizes ranging from $T=100$ to $T=500,$
several in-sample and out-of-sample sizes as well as different locations
$\lambda_{0}$ of the breaks. We begin with considering general instabilities
first and then move to short-term instabilities. Figure \ref{Fig_P1a_150_078}-\ref{Fig_P1a_230_078}
reports the results for model P1a. When $T=100,\,150$ Figure \ref{Fig_P1a_150_078}
shows that our tests have good power against model P1a while the tests
of GR (2009) seem to be less powerful. For example, when the break
date is at $T_{0}=0.8T$ our tests display reasonable power. However,
both $t^{\mathrm{stat}}$ statistics of GR (2009) perform significantly
worse and the associated power curve is bounded away from one even
for a very large break size $\delta=3$. This feature disappears when
we raise the sample size to $T\geq200$ and maintain the break date
at $T_{0}=0.8T$; see Figure \ref{Fig_P1a_230_078}. The latter figure
also shows that for large sample sizes and instabilities that last
for more than 50\% of the out-of-sample (top panels) all tests have
good power even though the $t^{\mathrm{stat}}$ statistics of GR (2009)
have slightly higher power. The power turns to be essentially the
same when $\lambda_{0}=0.8$ (i.e., the instability only lasts for
40\% of the out-of-sample). For model P2, Figure \ref{Fig_P2_120_078}
plots the power functions for $T=100,\,200$ and $\lambda_{0}=0.7,\,0.8.$
Except for the pair $\left(T=200,\,\lambda_{0}=0.7\right)$ (cf. top-right
panel) for which our tests and the $t^{\mathrm{stat}}$-type tests
display roughly the same power, it is clear that our tests are  more
powerful than the $t^{\mathrm{stat}}$ tests (both corrected and uncorrected
version). The power gains are substantial and range from 20\% to 40\%.
Moreover, as for model P1a and P1b when the instability lasts for
less than 50\% of the out-of-sample (cf. $\lambda_{0}=0.8$; bottom-left
panel) the statistics $\mathrm{B}_{\mathrm{max},h}$ and $\mathrm{Q}_{\mathrm{max},h}$
achieve trivial power already for a break magnitude $\delta=1.5$
whereas the $t^{\mathrm{stat}}$ tests of GR (2009) display rejection
rates below 60\% even when $\delta=2$ and yet below 70\% when $\delta=2.5$;
that is, their power function does not attain unit power even for
very large break magnitudes. These properties characterize all models
with i.i.d. errors and extend to model with lagged dependent variables
as predictors (cf. model P7; Figure \ref{Fig_P6_2300_078} in the
Supplement). 

Let us now turn to recurrent breaks in the mean. For recurrent breaks
we implement the statistics $\mathrm{MB}_{\mathrm{max},h}$ and $\mathrm{MQ}_{\mathrm{max},h}$
that use overlapping blocks. Figure \ref{Fig_P3_230_056} plots the
power curves for model P3. All tests have power and their performance
is essentially the same. Figure \ref{Fig_P4_230_068} corresponds
to model P4 (single break in the variance) and shows that when the
instability begins in the second half of the out-of-sample (cf. $\lambda_{0}=0.8$;
bottom panels) our tests $\mathrm{MB}_{\mathrm{max},h}$ and $\mathrm{MQ}_{\mathrm{max},h}$
achieves  good power while the $t^{\mathrm{stat}}$-type tests have
little power that does not attain unity even for a large break magnitude
$\delta=1.5$. When there are recurrent breaks in the variance as
in model P5, Figure \ref{Fig_P5_230_056} shows that the our tests
$\mathrm{MB}_{\mathrm{max},h}$ and $\mathrm{MQ}_{\mathrm{max},h}$
and the $t^{\mathrm{stat}}$-type tests have all good power and their
performance is analogous.

Let us now consider models with either ARCH errors or autocorrelated
errors. Observe that the latter models both imply that the forecast
losses are serially correlated. Figure \ref{Fig_P7_2300_078} shows
that when the errors follow an ARCH(1) process the statistic $\mathrm{Q}_{\mathrm{max},h}$
based on the asymptotic variance estimator $\widehat{\nu}_{L}^{2}$
performs well in terms of empirical power. In contrast, the $t^{\mathrm{stat}}$-type
tests fail as their power is non-monotonic, never reaches 20\% and
it decreases to zero as the magnitude of the break rises.\footnote{We actually implemented the $t^{\mathrm{stat}}$ by using either \citeauthor{andrews:91}'s
(1991) or \citeauthor{newey/west:87}'s (1987) estiamtor of the long-run
variance. We also experimented different choices for the truncation
lag. The results, however, were unchanged. We suspect that this property
depends on the estimation of the long-run variance in our forecasting
context which can be challenging due to small sample sizes and to
the presence of breaks. The same issues were found in \citet{martins/perron:16}
and \citet{fossati:17}.} We note that the version of $\widehat{\nu}_{L}$ that uses more blocks
is less precises. The same results hold true when the disturbances
are autocorrelated; see Figure \ref{Fig_P8_2300_078}.

Finally, we consider short-term instabilities in Figure \ref{Fig_P1_150_078_st}-\ref{Fig_P4_450_068_st}.
It is straightforward to recognize a general pattern: the tests of
GR (2009) have little  power whereas our tests possess good empirical
power against all data-generating processes, break locations and sample
sizes. Furthermore, the small sample power properties are uniform
over the location of the instability and over the relative size of
the in-sample and out-of-sample windows. The latter property is important
in practice because forecast instabilities are frequently short-lived.

To sum up, our test statistics perform well in controlling the size,
even though the versions that use overlapping blocks are somewhat
liberal. For our tests, empirical size being close to the significance
level is a feature that holds over different DGPs and sample sizes.
Turning to power comparison, there is clear evidence that our tests
are reliable in that they have good power against different form of
instabilities. There appears to be substantial power gains relative
to existing methods especially when the instability (i) is short-lasting
and/or (ii) is located toward the tail of the out-of-sample. These
properties characterize both statistics using non-overlapping and
overlapping blocks.

\section{\label{Section Conclusions}Conclusions}

We have formalized the concepts of forecast instability and forecast
failure. Our definition poses at the center the economic forecaster
and emphasizes the importance of the time duration of the instability.
We assume the data arise as an outcome of an underlying system of
stochastic differential equations which then implies that we can approximate
the sequence of forecast losses by a continuous-time stochastic process.
We have built a testing framework based on the local pathwise properties
of that process and have adopted an infill asymptotics to derive the
null distribution of the test statistics. The null distribution follows
an extreme value distribution. Our results can be used to test whether
the predictive ability of a given forecast model changes over time
and can be applied in forecasting exercises involving either low-frequency
as well as high-frequency macroeconomic and financial variables. The
simulation study confirms that there are substantial power gains especially
when the instability (i) is short-lasting and/or (ii) is located toward
the tail of the out-of-sample. Our framework allows for misspecification,
different types of parameter instability and arbitrary forms of non-stationarity
such as heteroskedasticity and serial correlation. Our continuous-time
specification and associated continuous record asymptotic scheme can
provide a promising complementary framework to the classical approach
for forecasting in economics.

\newpage{}

\begin{singlespace}

\bibliographystyle{econometrica}
\bibliography{References}
\addcontentsline{toc}{section}{References}

\end{singlespace}

\newpage{}

%\lipsum
\clearpage 
\pagenumbering{arabic}% resets `page` counter to 1 
\renewcommand*{\thepage}{A-\arabic{page}}
\appendix
%\begin{appendices}

\section{Appendix}

\subsection{Tables}

\begin{table}[H]
\caption{\label{Table S1}Empirical small sample size of forecast instability
tests based on model S1}
\begin{centering}
\begin{tabular}{cccccccccccc}
\hline 
 &  &  & \multicolumn{2}{c}{GR (2009)} & $\mathrm{B}_{\mathrm{max},h}$ & \multicolumn{2}{c}{$\mathrm{Q}_{\mathrm{max},h}$} & \multicolumn{2}{c}{$\mathrm{MB}_{\mathrm{max},h}$} & \multicolumn{2}{c}{$\mathrm{MQ}_{\mathrm{max},h}$}\tabularnewline
 &  &  & $t^{\mathrm{stat}}$ (uncorrected) & $t^{\mathrm{stat,c}}$ (corrected) &  & \multicolumn{6}{c}{}\tabularnewline
\hline 
\multicolumn{3}{l}{$\alpha=0.05$ } & \multicolumn{2}{c}{} & \tabularnewline
 & $T_{m}$ & $T_{n}$ &  &  &  & \multicolumn{2}{c}{} & \multicolumn{2}{c}{} & \multicolumn{2}{c}{}\tabularnewline
$T=100$ & 25 & 75 & 0.052 & 0.038 & 0.044 & \multicolumn{2}{c}{0.037} & \multicolumn{2}{c}{0.112} & \multicolumn{2}{c}{0.140}\tabularnewline
 & 50 & 50 & 0.063 & 0.030 & 0.019 & \multicolumn{2}{c}{0.011} & \multicolumn{2}{c}{0.078} & \multicolumn{2}{c}{0.064}\tabularnewline
 & 75 & 25 & 0.051 & 0.046 & 0.056 & \multicolumn{2}{c}{0.050} & \multicolumn{2}{c}{0.096} & \multicolumn{2}{c}{0.095}\tabularnewline
$T=200$ & 50 & 150 & 0.047 & 0.036 & 0.029 & \multicolumn{2}{c}{0.032} & \multicolumn{2}{c}{0.136} & \multicolumn{2}{c}{0.083}\tabularnewline
 & 100 & 100 & 0.052 & 0.035 & 0.032 & \multicolumn{2}{c}{0.030} & \multicolumn{2}{c}{0.110} & \multicolumn{2}{c}{0.070}\tabularnewline
 & 150 & 50 & 0.078 & 0.028 & 0.060 & \multicolumn{2}{c}{0.058} & \multicolumn{2}{c}{0.059} & \multicolumn{2}{c}{0.058}\tabularnewline
$T=300$ & 75 & 225 & 0.045 & 0.041 & 0.040 & \multicolumn{2}{c}{0.049} & \multicolumn{2}{c}{0.106} & \multicolumn{2}{c}{0.046}\tabularnewline
 & 150 & 150 & 0.054 & 0.034 & 0.061 & \multicolumn{2}{c}{0.057} & \multicolumn{2}{c}{0.145} & \multicolumn{2}{c}{0.086}\tabularnewline
 & 225 & 75 & 0.072 & 0.028 & 0.104 & \multicolumn{2}{c}{0.095} & \multicolumn{2}{c}{0.092} & \multicolumn{2}{c}{0.076}\tabularnewline
$T=400$ & 100 & 300 & 0.050  & 0.048 & 0.054 & \multicolumn{2}{c}{0.068} & \multicolumn{2}{c}{0.129} & \multicolumn{2}{c}{0.055}\tabularnewline
 & 200 & 200 & 0.056 & 0.042 & 0.063 & \multicolumn{2}{c}{0.063} & \multicolumn{2}{c}{0.122} & \multicolumn{2}{c}{0.059}\tabularnewline
 & 300 & 100 & 0.059 & 0.030 & 0.069 & \multicolumn{2}{c}{0.067} & \multicolumn{2}{c}{0.108} & \multicolumn{2}{c}{0.068}\tabularnewline
 &  &  &  &  &  &  &  & \multicolumn{2}{c}{} &  & \tabularnewline
\multicolumn{3}{l}{$\alpha=0.10$ } &  &  &  &  &  &  &  &  & \tabularnewline
 & $T_{m}$ & $T_{n}$ & \multicolumn{2}{c}{} &  & \multicolumn{2}{c}{} & \multicolumn{2}{c}{} & \multicolumn{2}{c}{}\tabularnewline
$T=100$ & 25 & 75 & 0.154 & 0.112 & 0.099 & \multicolumn{2}{c}{0.142} & \multicolumn{2}{c}{0.186} & \multicolumn{2}{c}{0.163}\tabularnewline
 & 50 & 50 & 0.102 & 0.087 & 0.115 & \multicolumn{2}{c}{0.096} & \multicolumn{2}{c}{0.111} & \multicolumn{2}{c}{0.093}\tabularnewline
 & 75 & 25 & 0.137 & 0.071 & 0.053 & \multicolumn{2}{c}{0.067} & \multicolumn{2}{c}{0.128} & \multicolumn{2}{c}{0.130}\tabularnewline
$T=200$ & 50 & 150 & 0.103 & 0.106 & 0.095 & \multicolumn{2}{c}{0.117} & \multicolumn{2}{c}{0.130} & \multicolumn{2}{c}{0.068}\tabularnewline
 & 100 & 100 & 0.116 & 0.096 & 0.100 & \multicolumn{2}{c}{0.096} & \multicolumn{2}{c}{0.157} & \multicolumn{2}{c}{0.105}\tabularnewline
 & 150 & 50 & 0.114 & 0.076 & 0.128 & \multicolumn{2}{c}{0.150} & \multicolumn{2}{c}{0.103} & \multicolumn{2}{c}{0.093}\tabularnewline
$T=300$ & 75 & 225 & 0.108 & 0.110 & 0.077 & \multicolumn{2}{c}{0.109} & \multicolumn{2}{c}{0.167} & \multicolumn{2}{c}{0.085}\tabularnewline
 & 150 & 150 & 0.103 & 0.094 & 0.116 & \multicolumn{2}{c}{0.106} & \multicolumn{2}{c}{0.204} & \multicolumn{2}{c}{0.130}\tabularnewline
 & 225 & 75 & 0.135 & 0.132 & 0.142 & \multicolumn{2}{c}{0.118} & \multicolumn{2}{c}{0.194} & \multicolumn{2}{c}{0.163}\tabularnewline
$T=400$ & 100 & 300 & 0.098 & 0.108 & 0.108 & \multicolumn{2}{c}{0.108} & \multicolumn{2}{c}{0.192} & \multicolumn{2}{c}{0.096}\tabularnewline
 & 200 & 200 & 0.105 & 0.091 & 0.087 & \multicolumn{2}{c}{0.139} & \multicolumn{2}{c}{0.167} & \multicolumn{2}{c}{0.088}\tabularnewline
 & 300 & 100 & 0.112 & 0.079 & 0.114 & \multicolumn{2}{c}{0.109} & \multicolumn{2}{c}{0.166} & \multicolumn{2}{c}{0.112}\tabularnewline
\hline 
\end{tabular}
\par\end{centering}
\noindent\begin{minipage}[t]{1\columnwidth}%
{\small{}The table reports the rejection probabilities of $100\alpha\%$-level
tests proposed in the paper and those proposed by \citet{giacomini/rossi:09}
{[}(abbreviated GR (2009){]} for model S1. For all methods we use
the fixed forecasting scheme. $T=T_{m}+T_{n}$, where $T$ is the
total sample size, $T_{m}$ is the size of the in-sample window and
$T_{n}$ is the size of the out-of-sample window. $m_{T}$ is set
equal to the smallest integer allowed by Condition \ref{Cond The-auxiliary-sequence}.
Based on 5,000 replications.}%
\end{minipage}
\end{table}

\begin{table}[H]
\caption{\label{Table S2}Empirical small sample size of forecast instability
tests based on model S2}
\begin{centering}
\begin{tabular}{ccccccccc||ccc}
\hline 
 &  &  & \multicolumn{2}{c}{GR (2009)} & $\mathrm{B}_{\mathrm{max},h}$ & \multicolumn{2}{c}{$\mathrm{Q}_{\mathrm{max},h}$} & \multicolumn{2}{c}{$\mathrm{MB}_{\mathrm{max},h}$} & \multicolumn{2}{c}{$\mathrm{MQ}_{\mathrm{max},h}$}\tabularnewline
 &  &  & $t^{\mathrm{stat}}$ (uncorrected) & $t^{\mathrm{stat,c}}$ (corrected) &  & \multicolumn{6}{c}{}\tabularnewline
\hline 
\multicolumn{3}{l}{$\alpha=0.05$ } & \multicolumn{2}{c}{} & \tabularnewline
 & $T_{m}$ & $T_{n}$ &  &  &  & \multicolumn{2}{c}{} & \multicolumn{2}{c}{} & \multicolumn{2}{c}{}\tabularnewline
$T=100$ & 25 & 75 & 0.049 & 0.019 & 0.086 & \multicolumn{2}{c}{0.098} & \multicolumn{2}{c}{0.090} & \multicolumn{2}{c}{0.089}\tabularnewline
\multirow{2}{*}{} & 50 & 50 & 0.069 & 0.024 & 0.058 & \multicolumn{2}{c}{0.072} & \multicolumn{2}{c}{0.083} & \multicolumn{2}{c}{0.067}\tabularnewline
 & 75 & 25 & 0.039 & 0.016 & 0.081 & \multicolumn{2}{c}{0.111} & \multicolumn{2}{c}{0.092} & \multicolumn{2}{c}{0.091}\tabularnewline
$T=200$ & 50 & 150 & 0.049 & 0.025 & 0.076 & \multicolumn{2}{c}{0.072} & \multicolumn{2}{c}{0.138} & \multicolumn{2}{c}{0.089}\tabularnewline
\multirow{2}{*}{} & 100 & 100 & 0.057 & 0.026 & 0.070 & \multicolumn{2}{c}{0.073} & \multicolumn{2}{c}{0.106} & \multicolumn{2}{c}{0.068}\tabularnewline
 & 150 & 50 & 0.075 & 0.020 & 0.055 & \multicolumn{2}{c}{0.070} & \multicolumn{2}{c}{0.082} & \multicolumn{2}{c}{0.070}\tabularnewline
$T=300$ & 75 & 225 & 0.050 & 0.029 & 0.058 & \multicolumn{2}{c}{0.036} & \multicolumn{2}{c}{0.102} & \multicolumn{2}{c}{0.044}\tabularnewline
\multirow{2}{*}{} & 150 & 150 & 0.059 & 0.032 & 0.077 & \multicolumn{2}{c}{0.072} & \multicolumn{2}{c}{0.144} & \multicolumn{2}{c}{0.086}\tabularnewline
 & 225 & 75 & 0.065 & 0.025 & 0.096 & \multicolumn{2}{c}{0.103} & \multicolumn{2}{c}{0.152} & \multicolumn{2}{c}{0.123}\tabularnewline
$T=400$ & 100 & 300 & 0.054 & 0.032 & 0.061 & \multicolumn{2}{c}{0.041} & \multicolumn{2}{c}{0.123} & \multicolumn{2}{c}{0.046}\tabularnewline
\multirow{2}{*}{} & 200 & 200 & 0.051 & 0.035 & 0.065 & \multicolumn{2}{c}{0.048} & \multicolumn{2}{c}{0.111} & \multicolumn{2}{c}{0.052}\tabularnewline
 & 300 & 100 & 0.068 & 0.031 & 0.067 & \multicolumn{2}{c}{0.063} & \multicolumn{2}{c}{0.115} & \multicolumn{2}{c}{0.074}\tabularnewline
 &  &  &  &  &  &  &  & \multicolumn{2}{c}{} &  & \tabularnewline
\multicolumn{3}{l}{$\alpha=0.10$ } &  &  &  & \multicolumn{2}{c}{} & \multicolumn{2}{c}{} & \multicolumn{2}{c}{}\tabularnewline
 & $T_{m}$ & $T_{n}$ & \multicolumn{2}{c}{} &  & \multicolumn{2}{c}{} & \multicolumn{2}{c}{} & \multicolumn{2}{c}{}\tabularnewline
$T=100$ & 25 & 75 & 0.109 & 0.069 & 0.136 & \multicolumn{2}{c}{0.152} & \multicolumn{2}{c}{0.191} & \multicolumn{2}{c}{0.165}\tabularnewline
\multirow{2}{*}{} & 50 & 50 & 0.107 & 0.069 & 0.095 & \multicolumn{2}{c}{0.118} & \multicolumn{2}{c}{0.118} & \multicolumn{2}{c}{0.095}\tabularnewline
 & 75 & 25 & 0.134 & 0.060 & 0.112 & \multicolumn{2}{c}{0.152} & \multicolumn{2}{c}{0.125} & \multicolumn{2}{c}{0.128}\tabularnewline
$T=200$ & 50 & 150 & 0.100 & 0.078 & 0.125 & \multicolumn{2}{c}{0.113} & \multicolumn{2}{c}{0.199} & \multicolumn{2}{c}{0.133}\tabularnewline
\multirow{2}{*}{} & 100 & 100 & 0.106 & 0.073 & 0.108 & \multicolumn{2}{c}{0.111} & \multicolumn{2}{c}{0.160} & \multicolumn{2}{c}{0.108}\tabularnewline
 & 150 & 50 & 0.101 & 0.078 & 0.101 & \multicolumn{2}{c}{0.105} & \multicolumn{2}{c}{0.112} & \multicolumn{2}{c}{0.091}\tabularnewline
$T=300$ & 75 & 225 & 0.102 & 0.081 & 0.103 & \multicolumn{2}{c}{0.071} & \multicolumn{2}{c}{0.159} & \multicolumn{2}{c}{0.077}\tabularnewline
\multirow{2}{*}{} & 150 & 150 & 0.111 & 0.079 & 0.119 & \multicolumn{2}{c}{0.112} & \multicolumn{2}{c}{0.189} & \multicolumn{2}{c}{0.129}\tabularnewline
 & 225 & 75 & 0.114 & 0.068 & 0.144 & \multicolumn{2}{c}{0.159} & \multicolumn{2}{c}{0.197} & \multicolumn{2}{c}{0.170}\tabularnewline
$T=400$ & 100 & 300 & 0.097 & 0.082 & 0.109 & \multicolumn{2}{c}{0.079} & \multicolumn{2}{c}{0.193} & \multicolumn{2}{c}{0.096}\tabularnewline
\multirow{2}{*}{} & 200 & 200 & 0.089 & 0.106 & 0.075 & \multicolumn{2}{c}{0.079} & \multicolumn{2}{c}{0.171} & \multicolumn{2}{c}{0.088}\tabularnewline
 & 300 & 100 & 0.112 & 0.079 & 0.104 & \multicolumn{2}{c}{0.110} & \multicolumn{2}{c}{0.164} & \multicolumn{2}{c}{0.104}\tabularnewline
\hline 
\end{tabular}
\par\end{centering}
\noindent\begin{minipage}[t]{1\columnwidth}%
{\small{}Model S2. We use the estimator $\widehat{\nu}_{L}$ from
Theorem \ref{Theorem Long-Run Variance} for the asymptotic variance
of $\mathrm{Q}_{\mathrm{max},h}$ and $\mathrm{MQ}_{\mathrm{max},h}$.
For the statistics $t^{\mathrm{stat}}$ and $t^{\mathrm{stat,c}}$
we use the Newey-West estimator with truncation lags $\left\lfloor T_{n}^{1/3}\right\rfloor $
as recommended by \citet{giacomini/rossi:09}. The notes of Table
\ref{Table S1} applies.}%
\end{minipage}
\end{table}

\clearpage{}\pagebreak{}

\subsection{\label{Section Figures}Figures}

\subsubsection{General Instability}

\noindent \begin{center}
\begin{figure}[H]
\includegraphics[width=18cm,height=8.5cm]{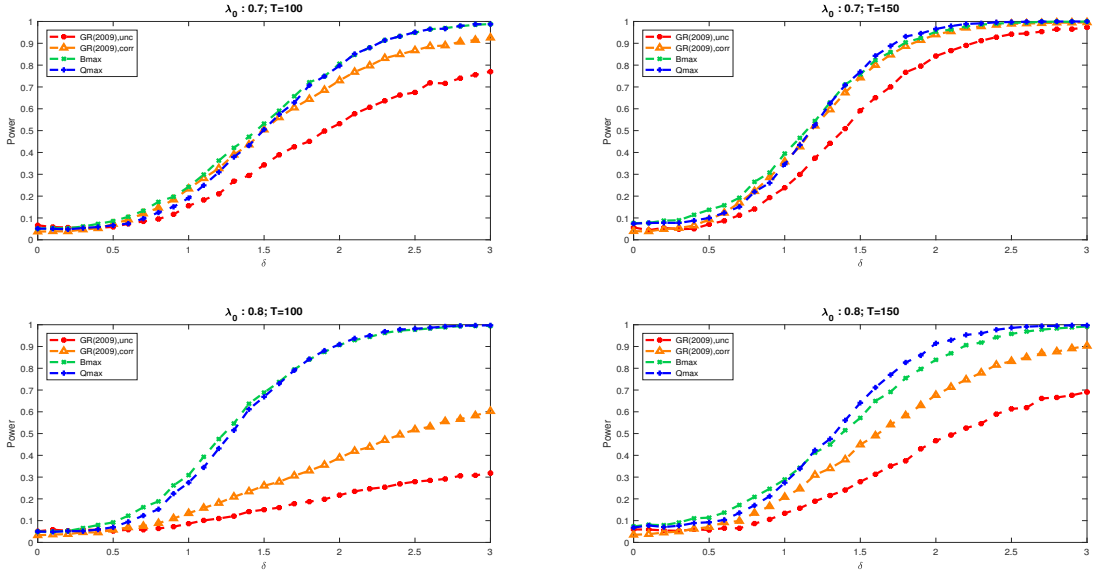}

{\footnotesize{}\caption{{\footnotesize{}\label{Fig_P1a_150_078}Power functions for model
P1a: $Y_{t}=2.73-0.44X_{t-1}+\delta X_{t-1}\mathbf{1}\left\{ t>T_{b}^{0}\right\} +e_{t}$},{\footnotesize{}
where $X_{t-1}\sim\mathrm{i.i.d.}\mathscr{N}\left(0,\,1\right)$,
$e_{t}\sim\mathrm{i.i.d.}\mathscr{N}\left(0,\,1\right)$, and $T_{b}^{0}=T\lambda_{0}$.
$T=100$ (left panels) and $T=150$ (right panels). $\lambda_{0}=0.7$
(top panels) and $\lambda_{0}=0.8$ (bottom panels). In-sample size
is $T_{m}=0.4T$ while out-of-sample size is $T_{n}=0.6T$. The green
and blue broken lines correspond to $\mathrm{B}_{\mathrm{max},h}$
and $\mathrm{Q}_{\mathrm{max},h}$, respectively. The red and orange
broken lines correspond to the $t^{\mathrm{stat}}$ of \citet{giacomini/rossi:09},
respectively, the uncorrected and corrected version.}}
}{\footnotesize \par}
\end{figure}
\end{center}

\begin{singlespace}
\noindent 
\begin{figure}[H]
\includegraphics[width=18cm,height=8.5cm]{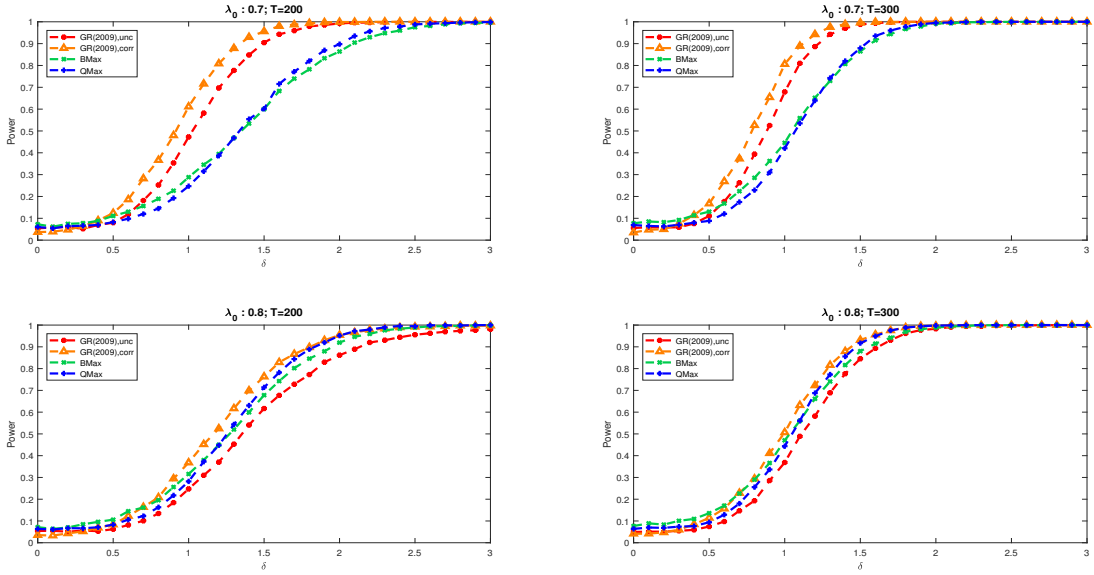}

{\footnotesize{}\caption{{\footnotesize{}\label{Fig_P1a_230_078}Power functions for model
P1a. $T=200$ (left panels) and $T=300$ (right panels). The notes
of Figure \ref{Fig_P1a_150_078} apply.}}
}{\footnotesize \par}
\end{figure}

\end{singlespace}

\begin{singlespace}
\noindent 
\begin{figure}[H]
\includegraphics[width=18cm,height=8.5cm]{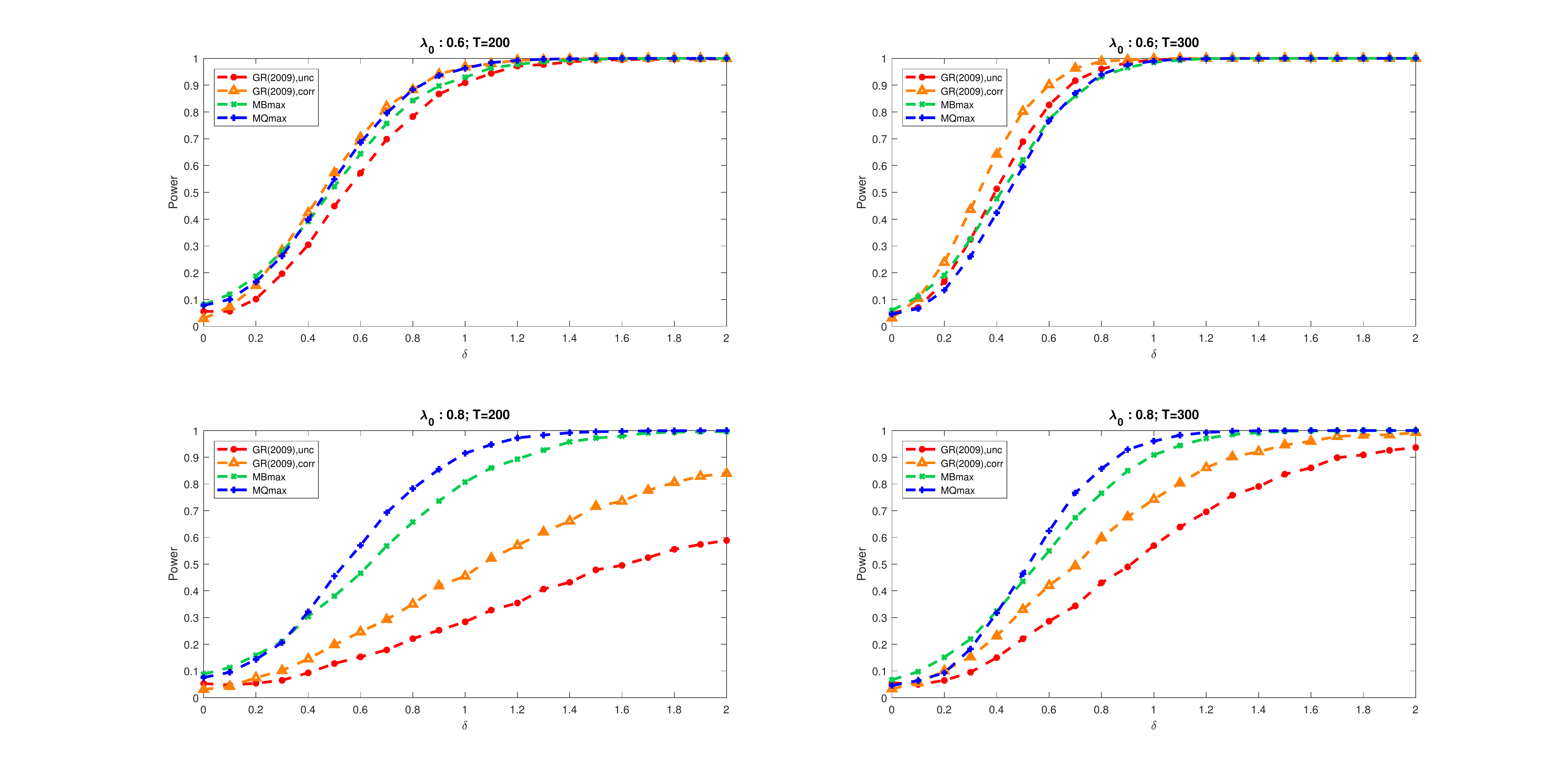}

{\footnotesize{}\caption{{\footnotesize{}\label{Fig_P4_230_068}Power functions for model
P4 (single break in variance): $Y_{t}=0.5X_{t-1}+\left(1+\delta\mathbf{1}\left\{ t>T_{b}^{0}\right\} \right)e_{t}$
where $X_{t-1}\sim\mathrm{i.i.d.}\mathscr{N}\left(1,\,1\right)$ and
$e_{t}\sim\mathrm{i.i.d.}\mathscr{N}\left(0,\,1\right)$. $T=200$
(left panels) and $T=300$ (right panels). $\lambda_{0}=0.6$ (top
panels) and $\lambda_{0}=0.8$ (bottom panels). In-sample size is
$T_{m}=0.3T$ while out-of-sample size is $T_{n}=0.7T$. The green
and blue broken lines correspond to $\mathrm{B}_{\mathrm{max},h}$
and $\mathrm{Q}_{\mathrm{max},h}$, respectively. The red and orange
broken lines correspond to the $t^{\mathrm{stat}}$ of \citet{giacomini/rossi:09},
respectively, the uncorrected and corrected version.}}
}{\footnotesize \par}
\end{figure}

\end{singlespace}

\begin{singlespace}
\noindent 
\begin{figure}[H]
\includegraphics[width=18cm,height=8.5cm]{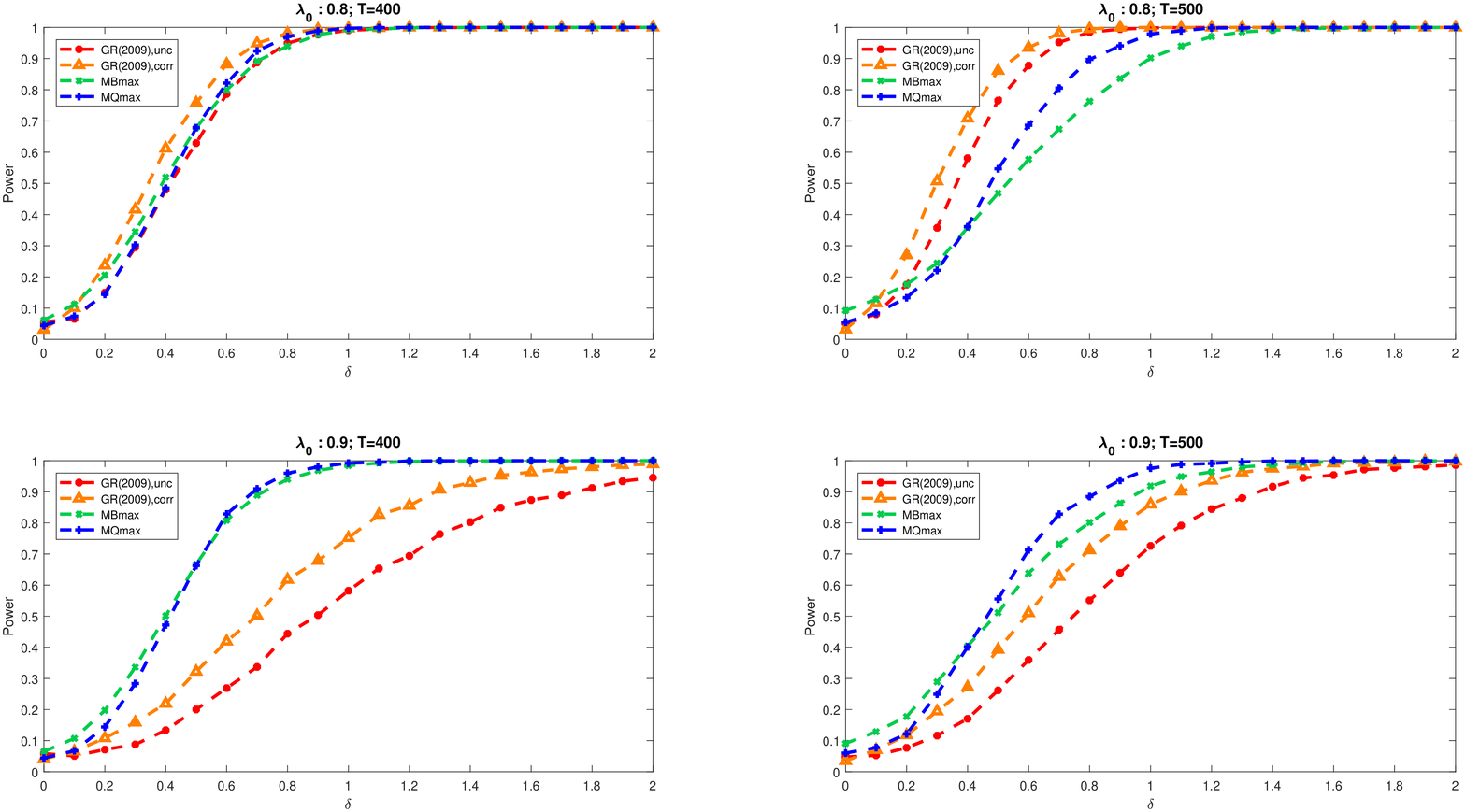}

{\footnotesize{}\caption{{\footnotesize{}\label{Fig_P4_450_089}Power functions for model
P4. $T=400$ (left panels) and $T=500$ (right panels). $\lambda_{0}=0.8$
(top panels) and $\lambda_{0}=0.9$ (bottom panels). The notes of
Figure \ref{Fig_P4_230_068} apply.}}
}{\footnotesize \par}
\end{figure}

\end{singlespace}

\begin{singlespace}
\noindent 
\begin{figure}[H]
\includegraphics[width=18cm,height=8.5cm]{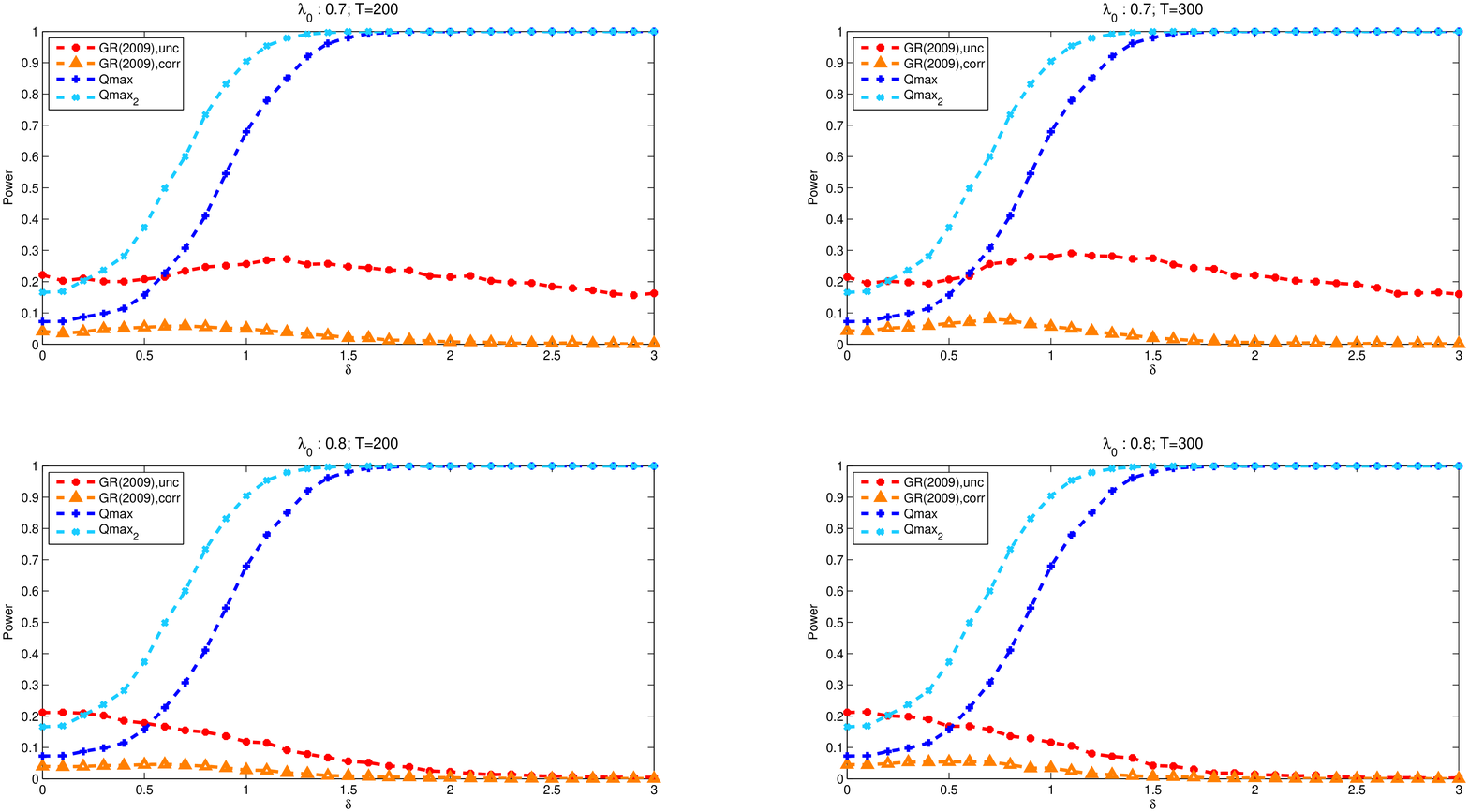}

{\footnotesize{}\caption{{\footnotesize{}\label{Fig_P7_2300_078}Power functions for model
P7 (ARCH errors): $Y_{t}=2.73-0.44X_{t-1}+\delta X_{t-1}\mathbf{1}\left\{ t>T_{b}^{0}\right\} +e_{t}$,
where $X_{t-1}\sim\mathrm{i.i.d.}\mathscr{N}\left(0,\,1.5\right)$
and $e_{t}=\sigma_{t}u_{t}$, $\sigma_{t}^{2}=0.5+0.5e_{t-1}^{2}$,
$u_{t}\sim\mathrm{i.i.d.}\mathscr{N}\left(0,\,1\right)$. $T=200$
(left panels) and $T=300$ (right panels). $\lambda_{0}=0.7$ (top
panels) and $\lambda_{0}=0.8$ (bottom panels). $T_{m}=0.5T$ and
$T_{n}=0.5T$. The light-blue and blue broken lines correspond to
a version of $\mathrm{Q}_{\mathrm{max},h}$ that uses $\widehat{\nu}_{L}$
but with different choices of $m_{T}$ (for the light-blue broken
line we increase the number of blocks by one relative to the recommended
value of $m_{T}$).}}
}{\footnotesize \par}
\end{figure}

\end{singlespace}

\subsubsection{Short-Term Instability}

\begin{singlespace}
\noindent 
\begin{figure}[H]
\includegraphics[width=18cm,height=8.5cm]{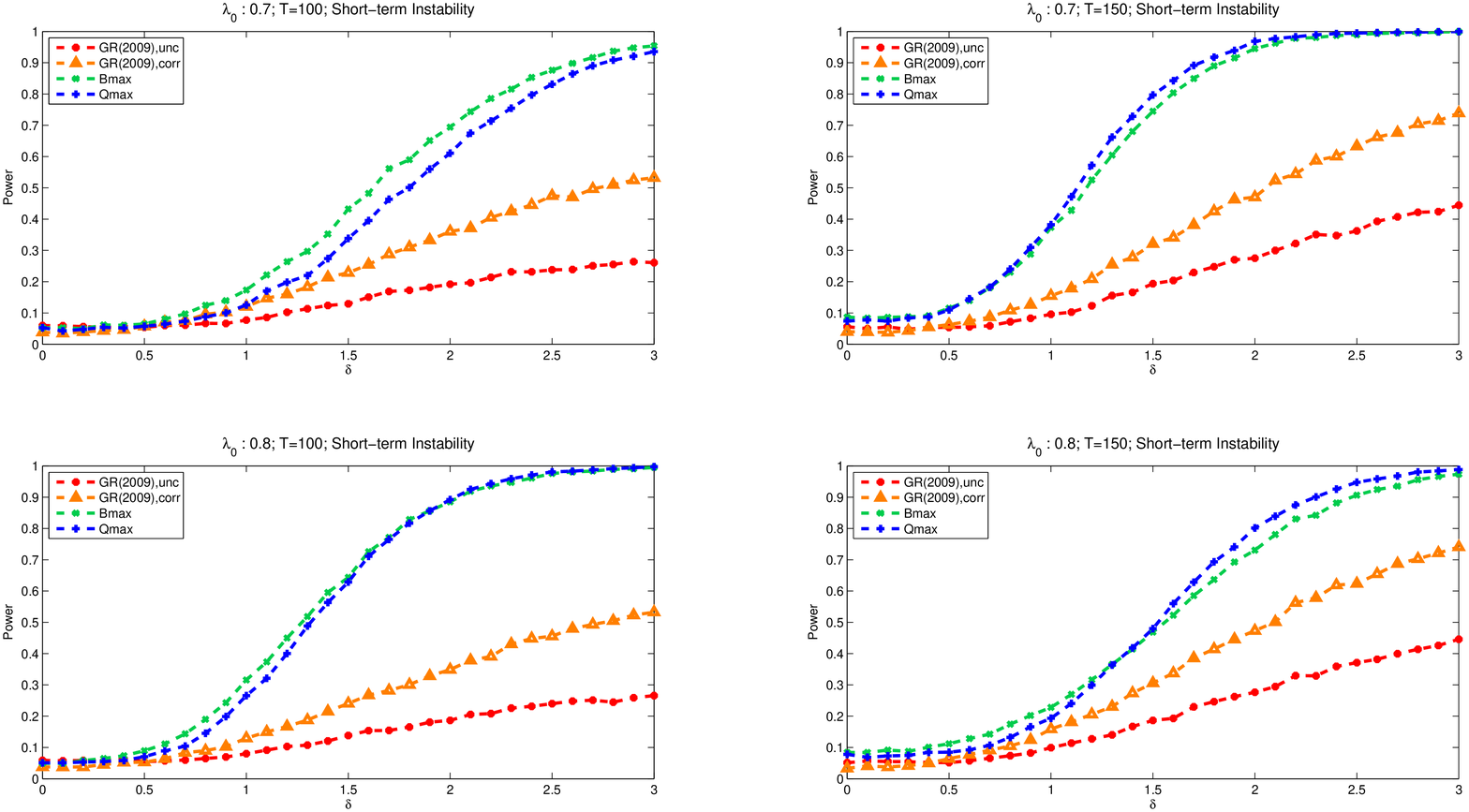}

{\footnotesize{}\caption{{\footnotesize{}\label{Fig_P1_150_078_st}Power functions for model
P1a with short-term instability: $Y_{t}=2.73-0.44X_{t-1}+\delta X_{t-1}\mathbf{1}\left\{ T_{b}^{0}<t\leq T_{b}^{0}+p\right\} +e_{t}$},{\footnotesize{}
where $X_{t-1}\sim\mathrm{i.i.d.}\mathscr{N}\left(0,\,1\right)$,
$e_{t}\sim\mathrm{i.i.d.}\mathscr{N}\left(0,\,1\right)$, and $T_{b}^{0}=T\lambda_{0}$.
We set $\left(T,\,p\right)=\left\{ \left(100,\,20\right),\,\left(150,\,25\right)\right\} $.
$\lambda_{0}=0.7$ (top panels) and $\lambda_{0}=0.8$ (bottom panels).
$T_{m}=0.4T$ and $T_{n}=0.6T$. The green and blue broken lines correspond
to $\mathrm{B}_{\mathrm{max},h}$ and $\mathrm{Q}_{\mathrm{max},h}$,
respectively. The red and orange broken lines correspond to the $t^{\mathrm{stat}}$
of \citet{giacomini/rossi:09}, respectively, the uncorrected and
corrected version.}}
}{\footnotesize \par}
\end{figure}

\end{singlespace}

\begin{singlespace}
\noindent 
\begin{figure}[H]
\includegraphics[width=18cm,height=7.5cm]{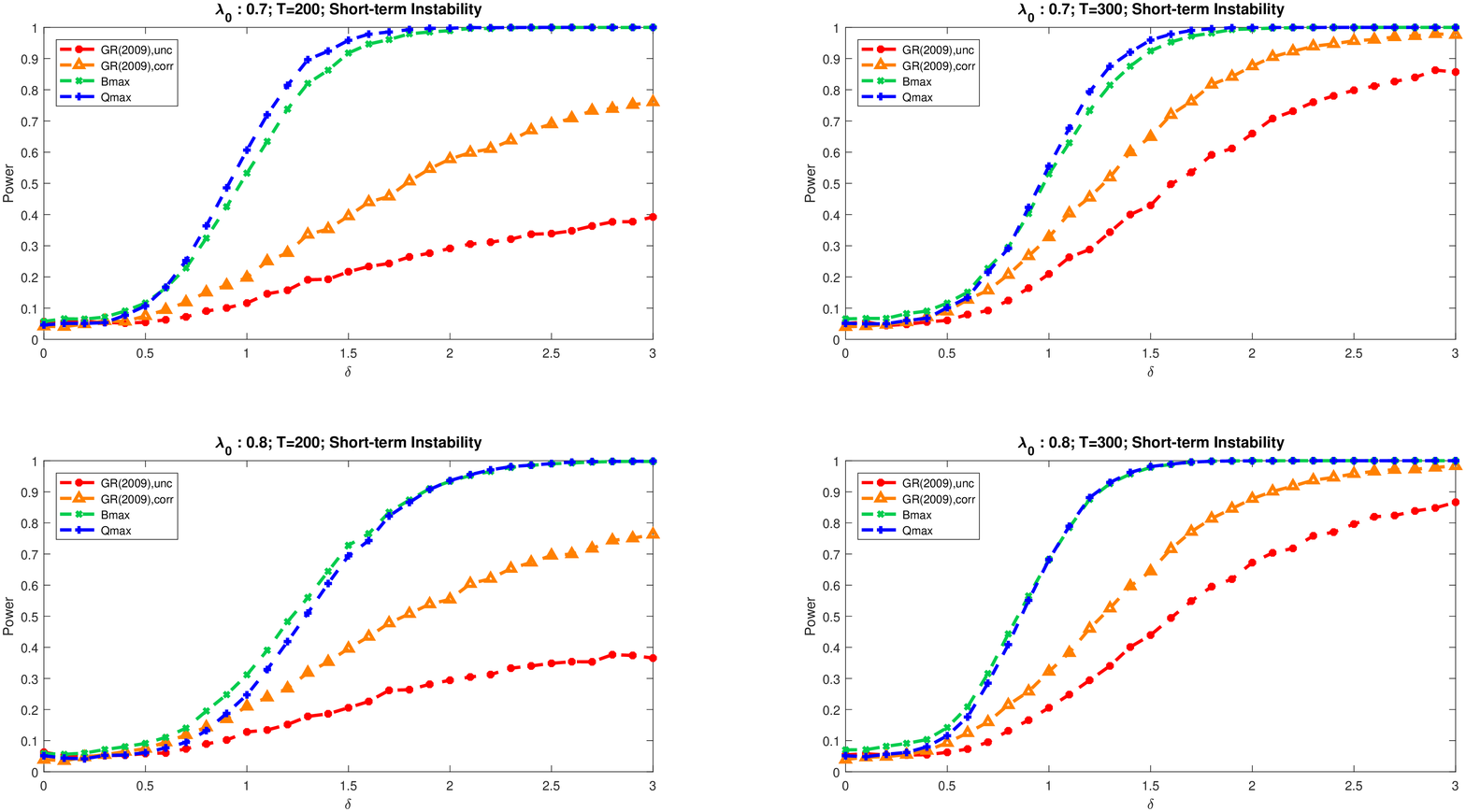}

{\footnotesize{}\caption{{\footnotesize{}\label{Fig_P1a_2300_st}Power functions for model
P1a. We set $\left(T,\,p\right)=\left\{ \left(200,\,20\right),\,\left(300,\,30\right)\right\} $.
The notes of Figure \ref{Fig_P1_150_078_st} apply.}}
}{\footnotesize \par}
\end{figure}

\end{singlespace}

\begin{singlespace}
\noindent 

\noindent 
\begin{figure}[H]
\includegraphics[width=18cm,height=8.5cm]{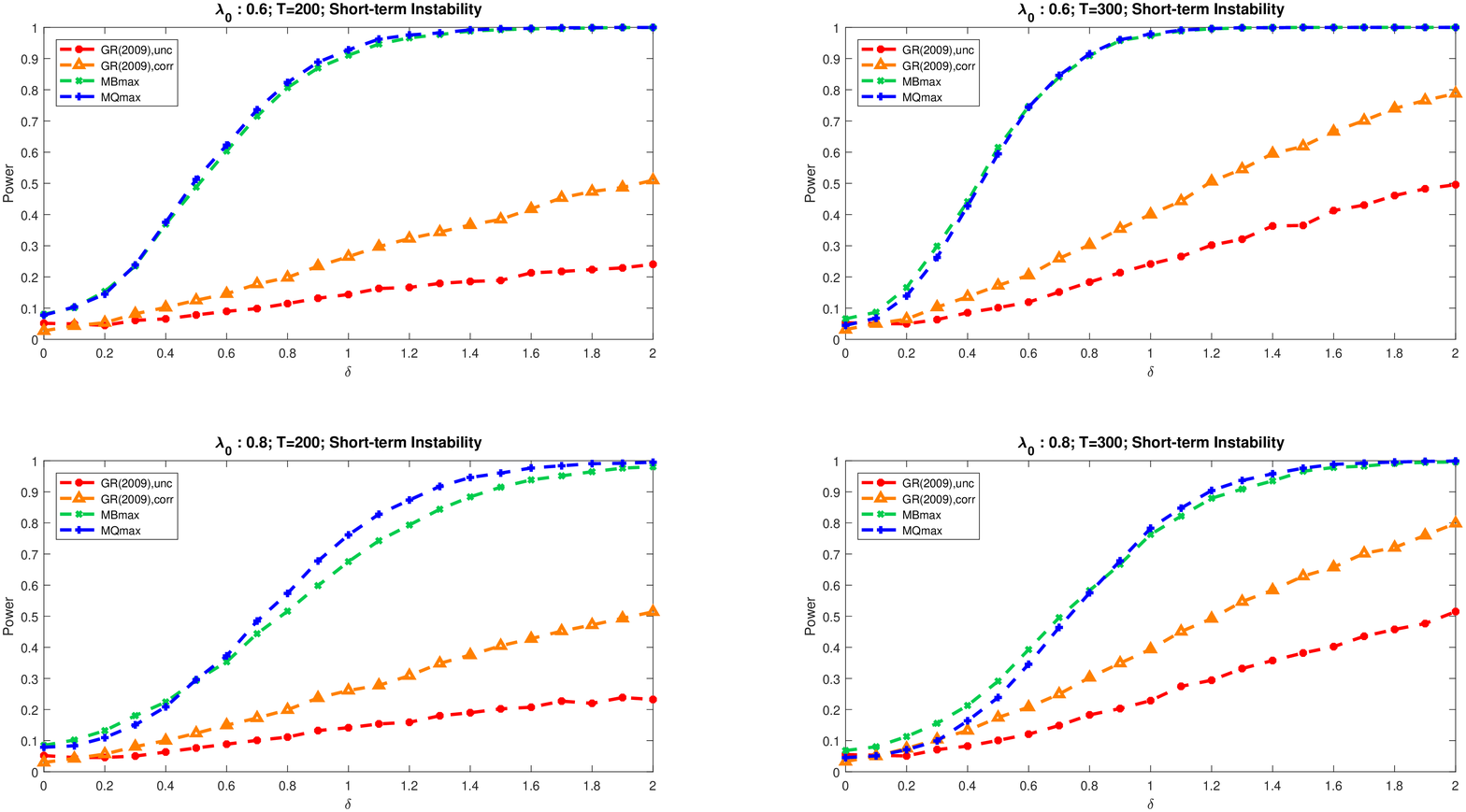}

{\footnotesize{}\caption{{\footnotesize{}\label{Fig_P4_230_068_st}Power functions for model
P4 (single break in variance) with short-term instability: $Y_{t}=0.5X_{t-1}+\left(1+\delta\mathbf{1}\left\{ T_{b}^{0}<t\leq T_{b}^{0}+p\right\} \right)e_{t}$
where $X_{t-1}\sim\mathrm{i.i.d.}\mathscr{N}\left(1,\,1\right)$ and
$e_{t}\sim\mathrm{i.i.d.}\mathscr{N}\left(0,\,1\right)$. We set $\left(T,\,p\right)=\left\{ \left(200,\,30\right),\,\left(300,\,30\right)\right\} $.
$\lambda_{0}=0.6$ (top panels) and $\lambda_{0}=0.8$ (bottom panels).
$T_{m}=0.3T$ and $T_{n}=0.7T$. The green and blue broken lines correspond
to $\mathrm{B}_{\mathrm{max},h}$ and $\mathrm{Q}_{\mathrm{max},h}$,
respectively. The red and orange broken lines correspond to the $t^{\mathrm{stat}}$
of \citet{giacomini/rossi:09}, respectively, the uncorrected and
corrected version.}}
}{\footnotesize \par}
\end{figure}

\end{singlespace}

\begin{singlespace}
\noindent 
\begin{figure}[H]
\includegraphics[width=18cm,height=8.5cm]{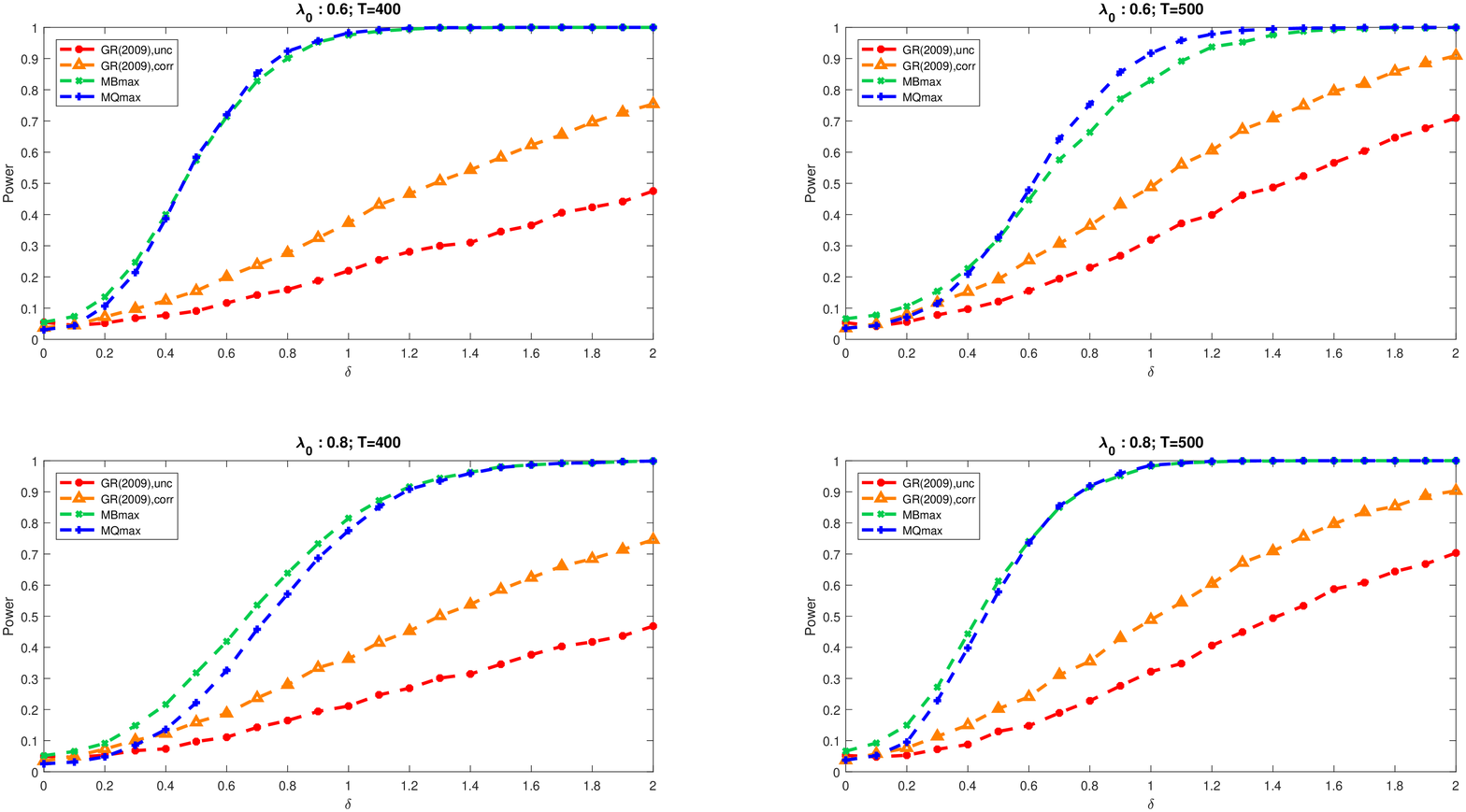}

{\footnotesize{}\caption{{\footnotesize{}\label{Fig_P4_450_068_st}Power functions for model
P4 (single break in variance) with short-term instability We set $\left(T,\,p\right)=\left\{ \left(400,\,30\right),\,\left(500,\,30\right)\right\} $.
The notes of Figure \ref{Fig_P4_230_068_st} apply.}}
}{\footnotesize \par}
\end{figure}

\end{singlespace}

\clearpage{}

\newpage{}

\pagebreak{}

\section*{}
\addcontentsline{toc}{part}{Supplemental Material}
\begin{center}
\Large{{Supplemental Material} to} 
\end{center}

\begin{center}
\title{\textbf{\Large{Tests for Forecast Instability and Forecast Failure under a Continuous Record Asymptotic Framework}}} 
\maketitle
\end{center}
\medskip{} 
\medskip{} 
\medskip{} 
\thispagestyle{empty}

\begin{center}
\author{\textsc{\textcolor{MyBlue}{Alessandro Casini}}}\\ 
\medskip{}
\medskip{} 
\medskip{} 

\small{{Department of Economics}}\\
\small{{Boston University}}\\
\medskip{}
\medskip{} 
\medskip{} 
\medskip{} 
\date{\small{\today}} %\\
%\footnotesize{First Vesrion: \printdate{28.10.2015}}}
\medskip{} 
\medskip{} 
\medskip{} 
\end{center}
\begin{abstract}
{\footnotesize{}This supplemental material is structured as follows.
Section \ref{Section Mathematical-Appendix} contains the Mathematical
Appendix which includes all proofs of the results in the paper. Section
\ref{Section Additional Figures Supp} reports additional figures
and tables from the simulation study of Section \ref{Section Simulation Study}.
In Section \ref{Sec Additional-Monte-Carlo} we collect additional
simulation results pertaining to recursive and rolling schemes and
to the Linex loss function.}{\footnotesize \par}
\end{abstract}
\setcounter{page}{0}
\setcounter{section}{0}
\renewcommand*{\theHsection}{\the\value{section}}

\newpage{}

\begin{singlespace} 
\noindent 
%\fontsize{10}{15}
\small

\allowdisplaybreaks

%\appendix

\renewcommand{\thepage}{S-\arabic{page}}   %\renewcommand{\thesection}{S.\arabic{section}}
\renewcommand{\thesection}{S.\Alph{section}}   
\renewcommand{\theequation}{S.\arabic{equation}}

%\appendixpagenumbering

%\begin{appendices}

%\simpleheading

\section{\label{Section Mathematical-Appendix}Mathematical Proofs }

The Mathematical Appendix is structured as follows. The proofs of
the results in Section \ref{Section CR Distribution Theory Test Stats}
and \ref{Section Estimation-of-Asymptotic Variance} are collected
in Section \ref{Subsection Proofs-of-Section CR Asym Distrb} and
\ref{Subsection Proofs-of-Section Var Est}, respectively. The results
of Section \ref{Section Ito Vol} are covered in Section \ref{subSection Proofs-of-Section Ito Vol}. 

\subsection{Additional Notation}

Throughout the proofs, $C$ is a generic constant that may vary from
line to line; we may sometime write $C_{r}$ to emphasize the dependence
of $C$ on a scalar $r.$ For brevity, we indicate that a sequence
$\left\{ U_{k}\right\} $ is formed by independent and non-identically
distributed random variables by labeling it as i.n.d. For the variables
$\Delta_{h}e_{k}$ and $\Delta_{h}X_{k}$ we use a tilde notation
to denote their normalized version: $\Delta_{h}\widetilde{e}_{k}=h^{-1/2}\Delta_{h}e_{k}$
and $\Delta_{h}\widetilde{X}_{k}=h^{-1/2}\Delta_{h}X_{k}$. We use
a star superscript $\left(*\right)$ on $\Delta_{h}e_{k}$ to indicate
the residuals obtained when $\beta=\beta^{*}$: $\Delta_{h}\widetilde{e}_{k}^{*}=h^{-1/2}\left(\Delta_{h}Y_{k}-\left(\beta^{*}\right)'\Delta_{h}X_{k-\tau}\right)$.
We sometime omit the index from $\widehat{\beta}_{k}$ and simply
use $\widehat{\beta}$ when it is clear from the context.  

\subsection{Localization}

As it is typical in the high-frequency statistics literature, we use
a localization argument {[}cf. Section I.1.d in \citet{jacod/shiryaev:03}{]}.
Thus, we replace Assumption \ref{Assumption 1, CT} and Assumption
\ref{Assumption Lipchtitz cont of Sigma} by the following stronger
assumption which basically turns the local restrictions into global. 
\begin{assumption}
\label{Assumption Localization Condition 1}Let Assumption \ref{Assumption 1, CT}-\ref{Assumption Lipchtitz cont of Sigma},
Assumption \ref{Ass the process}-\ref{Assumption Roo-T Consistent beta }
and Condition \ref{Cond The-auxiliary-sequence} hold. When $\mu_{e,t}=0$
for all $t\geq0$ the process $\left\{ Z_{t}\right\} _{t\geq0}$ takes
value in some compact set; the processes $\left\{ \sigma_{X,t},\,\sigma_{e,t}\right\} _{t\geq0}$
are bounded c�dl�g and $\left\{ \mu_{X,t},\,\mu_{e,t}\right\} _{t\geq0}$
are bounded c�dl�g. Furthermore, $\phi_{\sigma,\eta,N}\leq C\eta$
for some $C<\infty.$ 
\end{assumption}

\subsection{Preliminary Lemmas}
\begin{lem}
\label{Lemma Prelim A1}For any $1\leq r,\,l\leq q$, and $1\leq i\leq n_{T}$,
we have 

(i) $\sup_{b=0,\ldots,\,\left\lfloor T_{n}/n_{T}\right\rfloor -2}\sum_{j=1}^{T_{m}+bn_{T}+i-1}\Delta_{h}X_{k}^{\left(r\right)}\Delta_{h}e_{k}^{*}\overset{\mathbb{P}}{\rightarrow}0$;

(ii) $\sup_{b=0,\ldots,\,\left\lfloor T_{n}/n_{T}\right\rfloor -2}\left\Vert \sum_{j=1}^{T_{m}+bn_{T}+i-1}\Delta_{h}X_{k}^{\left(r\right)}\Delta_{h}X_{k}^{\left(l\right)\prime}-\int_{0}^{N_{\mathrm{in}}+bn_{T}h}\Sigma_{X,s}^{\left(r,l\right)}ds\right\Vert \overset{\mathbb{P}}{\rightarrow}0$;

(iii) the central limit theorem in Lemma S.A.5 in \citet{casini/perron_CR_Single_Break}
holds for $X_{t}$.
\end{lem}

\noindent \textit{Proof.} Part (i)-(ii) are a consequence of the
law of large numbers for quadratic variation; see Section S.A.3 in
\citet{casini/perron_CR_Single_Break}. For part (iii) see the above
referenced theorem. $\square$

\subsection{\label{Subsection Proofs-of-Section CR Asym Distrb}Proofs of Section
\ref{Section CR Distribution Theory Test Stats}}

Throughout this section we maintain Assumption \ref{Assumption Localization Condition 1}.

\subsubsection{Proof of Theorem \ref{Theoem Asymptotic H0 Distrbution Bmax and Qmax}}

The idea behind the proof of both Theorem \ref{Theoem Asymptotic H0 Distrbution Bmax and Qmax}-\ref{Theoem Asymptotic H0 Distrbution Gmax QGmax}
is the same. Thus, the quadratic loss case serves as a guide and
we then use some of these derivations for the general loss case. All
the results in this section are proved under $H_{0}$. 

\paragraph{Proof of part (i) of Theorem \ref{Theoem Asymptotic H0 Distrbution Bmax and Qmax}}

The theorem is proved through several lemmas. The first step involves
showing that the error in replacing $\widehat{\beta}$ by $\beta^{*}$
is asymptotically negligible. We provide the proof of this first step
by assuming that $\mu_{e,t}=0$ in \eqref{Mode for Y}. That is, in
Lemma \ref{Lemma QA1}-\ref{Lemma QA2} we have\textbf{ }$\mu_{e,t}=0$
and we show how these results continue to hold without this restriction
in Section \ref{par Negligibility-of-the}. We focus for simplicity
on the recursive scheme only; the proofs for the other cases are similar
and omitted. Let $U_{h,b}\triangleq n_{T}^{-1}\sum_{j=1}^{n_{T}}SL_{\psi,T_{m}+\tau+bn_{T}+j-1}\left(\beta^{*}\right)$,
$\overline{U}_{h,b}\triangleq n_{T}^{-1}\sum_{j=1}^{n_{T}}L_{\psi,T_{m}+\tau+bn_{T}+j-1}\left(\beta^{*}\right)$
and $\mathrm{U}_{\max,h}\left(T_{n},\,\tau\right)\triangleq\max_{b=0,\ldots,\,\left\lfloor T_{n}/n_{T}\right\rfloor -2}\left|\left(U_{h,b+1}-U_{h,b}\right)/\overline{U}_{h,b+1}\right|$.
In some steps of the proof, we will use the following simple result.
For any integer $m\geq1$, let $c_{1,b}$ and $c_{2,b}$ $\left(b=1,\ldots,\,m\right)$
be arbitrary real numbers, then
\begin{align}
\left|c_{1,b}\right| & \leq\left|c_{1,b}-c_{2,b}\right|+\left|c_{2,b}\right|\leq\max_{b=1,\ldots,m}\left|c_{1,b}-c_{2,b}\right|+\max_{b=1,\ldots,m}\left|c_{2,b}\right|.\label{eq. Inequality in Prop A1}
\end{align}
\begin{lem}
\label{Lemma QA1}As $h\downarrow0$, $\left(\log\left(T_{n}\right)n_{T}\right)^{1/2}\left(\mathrm{U}_{\max,h}\left(T_{n},\,\tau\right)-\mathrm{B}_{\mathrm{max},h}\left(T_{n},\,\tau\right)\right)\overset{\mathbb{P}}{\rightarrow}0$. 
\end{lem}

\noindent \textit{Proof.} By the reverse triangle inequality, inequality
\eqref{eq. Inequality in Prop A1} and Lemma \ref{Lemma QA2} below,
for some $C_{1},\,C_{2}<\infty$, 
\begin{align}
\left|\right. & \left.\mathrm{U}_{\max,h}\left(T_{n},\,\tau\right)-\mathrm{B}_{\mathrm{max},h}\left(T_{n},\,\tau\right)\right|\nonumber \\
 & \leq\max_{b=0,\ldots,\,\left\lfloor T_{n}/n_{T}\right\rfloor -2}\left|n_{T}^{-1}\sum_{j=1}^{n_{T}}\left(SL_{\psi,T_{m}+\tau+\left(b+1\right)n_{T}+j-1}\left(\beta^{*}\right)/\overline{B}_{h,b+1}-SL_{\psi,T_{m}+\tau+\left(b+1\right)n_{T}+j-1}\left(\widehat{\beta}\right)/\overline{U}_{h,b+1}\right)\right|\nonumber \\
 & \quad+\max_{b=0,\ldots,\,\left\lfloor T_{n}/n_{T}\right\rfloor -2}\left|n_{T}^{-1}\left(\sum_{j=1}^{n_{T}}SL_{\psi,T_{m}+\tau+bn_{T}+j-1}\left(\beta^{*}\right)/\overline{B}_{h,b+1}-SL_{\psi,T_{m}+\tau+bn_{T}+j-1}\left(\widehat{\beta}\right)/\overline{U}_{h,b+1}\right)\right|\nonumber \\
 & \leq C_{1}\max_{b=0,\ldots,\,\left\lfloor T_{n}/n_{T}\right\rfloor -2}\left|n_{T}^{-1}\left(\frac{\sum_{j=1}^{n_{T}}SL_{\psi,T_{m}+\tau+\left(b+1\right)n_{T}+j-1}\left(\beta^{*}\right)-SL_{\psi,T_{m}+\tau+\left(b+1\right)n_{T}+j-1}\left(\widehat{\beta}\right)}{\overline{U}_{h,b+1}^{2}}\right)\right|\label{eq (Lemma A1 A1)}\\
 & \quad+C_{2}\max_{b=0,\ldots,\,\left\lfloor T_{n}/n_{T}\right\rfloor -2}\left|n_{T}^{-1}\left(\frac{\sum_{j=1}^{n_{T}}SL_{\psi,T_{m}+\tau+bn_{T}+j-1}\left(\beta^{*}\right)-SL_{\psi,T_{m}+\tau+bn_{T}+j-1}\left(\widehat{\beta}\right)}{\overline{U}_{h,b+1}^{2}}\right)\right|.\nonumber 
\end{align}
Note that for any $j=1,\ldots,\,n_{T}$, 
\begin{align*}
SL_{\psi,T_{m}+\tau+bn_{T}+j-1} & \left(\beta^{*}\right)-SL_{\psi,T_{m}+\tau+bn_{T}+j-1}\left(\widehat{\beta}\right)\\
 & =L_{\psi,T_{m}+\tau+bn_{T}+j-1}\left(\beta^{*}\right)-L_{\psi,T_{m}+\tau+bn_{T}+j-1}\left(\widehat{\beta}\right)+o_{\mathbb{P}}\left(T^{-1/2}\right),
\end{align*}
where the $o_{\mathbb{P}}\left(T^{-1/2}\right)$ term arises from
the proof of Lemma \ref{Lemma QA2}. Recall that for $1\leq j\leq n_{T}$,
\begin{align*}
\Delta_{h}\widetilde{e}_{T_{m}+\tau+bn_{T}+j}^{*} & =\sigma_{e,\left(T_{m}+\tau+bn_{T}\right)h}\left(h^{-1/2}\Delta_{h}W_{e,T_{m}+\tau+bn_{T}+j}\right),
\end{align*}
so that 
\begin{align}
L_{\psi,T_{m}+\tau+bn_{T}+j} & \left(\beta^{*}\right)-L_{\psi,T_{m}+\tau+bn_{T}+j}\left(\widehat{\beta}\right)\nonumber \\
 & =-\left(\widehat{\beta}-\beta^{*}\right)'\Delta_{h}\widetilde{X}_{T_{m}+bn_{T}+j}\Delta_{h}\widetilde{X}'_{T_{m}+bn_{T}+j}\left(\widehat{\beta}-\beta^{*}\right)\nonumber \\
 & \quad+2\sigma_{e,\left(T_{m}+\tau+bn_{T}\right)h}\left(h^{-1/2}\Delta_{h}W_{e,T_{m}+\tau+bn_{T}+j}\right)\left(\widehat{\beta}-\beta^{*}\right)'\Delta_{h}\widetilde{X}_{T_{m}+bn_{T}+j}.\label{eq (A2)}
\end{align}
Recall that $\Delta_{h}\widetilde{X}_{k}=h^{-1/2}\Delta_{h}X_{k}$
and thus 
\begin{align*}
n_{T}^{-1}\sum_{j=1}^{n_{T}}\Delta_{h}\widetilde{X}_{T_{m}+bn_{T}+j}\Delta_{h}\widetilde{X}'_{T_{m}+bn_{T}+j}-\Sigma_{X,\left(T_{m}+bn_{T}\right)h} & =o_{\mathbb{P}}\left(1\right),
\end{align*}
which follows from Theorem 9.3.2 part (i) in \citet{jacod/protter:12}.
This implies that 
\begin{align*}
n_{T}^{-1}\sum_{j=1}^{n_{T}}\Delta_{h}\widetilde{X}_{T_{m}+bn_{T}+j-1}\Delta_{h}\widetilde{X}'_{T_{m}+bn_{T}+j-1} & =O_{\mathbb{P}}\left(1\right),
\end{align*}
 by Assumption \ref{Assumption 1, CT}-(iv). By Assumption \ref{Assumption 1, CT}-(v)
and the aforementioned theorem,
\begin{align*}
n_{T}^{-1}\sum_{j=1}^{n_{T}}\left(h^{-1/2}\Delta_{h}W_{e,T_{m}+\tau+bn_{T}+j-1}\right)\Delta_{h}\widetilde{X}_{T_{m}+bn_{T}+j-1} & \overset{\mathbb{P}}{\rightarrow}0.
\end{align*}
Note that by Assumption \ref{Assumption Roo-T Consistent beta },
$\widehat{\beta}_{k}-\beta^{*}=O_{\mathbb{P}}\left(1/\sqrt{T}\right)$
uniformly in $k\geq T_{m}$. Therefore, from these arguments we deduce
that
\begin{align}
n_{T}^{-1}\sum_{j=1}^{n_{T}}\left(SL_{\psi,T_{m}+\tau+bn_{T}+j-1}\left(\beta^{*}\right)-SL_{\psi,T_{m}+\tau+bn_{T}+j-1}\left(\widehat{\beta}\right)\right) & =o_{\mathbb{P}}\left(1/\sqrt{T}\right).\label{eq S* - Shat}
\end{align}
Then, for any $\varepsilon>0$ and any constant $K>0$, the first
term on the right-hand side of \eqref{eq (Lemma A1 A1)} is 
\begin{align}
\mathbb{P} & \left(\max_{b=0,\ldots,\,\left\lfloor T_{n}/n_{T}\right\rfloor -2}\left|\frac{\left(\log\left(T_{n}\right)n_{T}\right)^{1/2}\left(U_{h,b+1}-B_{h,b+1}\right)}{\overline{U}_{h,b+1}^{2}}\right|>\varepsilon\right)\nonumber \\
 & \leq\mathbb{P}\left(\max_{b=0,\ldots,\,\left\lfloor T_{n}/n_{T}\right\rfloor -2}\left|\left(\log\left(T_{n}\right)n_{T}\right)^{1/2}\left(U_{h,b+1}-B_{h,b+1}\right)\right|>\varepsilon/K\right)\nonumber \\
 & \quad+\mathbb{P}\left(\max_{b=0,\ldots,\,\left\lfloor T_{n}/n_{T}\right\rfloor -2}1/\left|\overline{U}_{h,b+1}^{2}\right|>K\right).\label{eq (A3)}
\end{align}
 Given the result on the negligibility of the drift term from Section
\ref{par Negligibility-of-the}, we can apply Lemma \ref{Lemma eq(22) in Vetter (2012)}
to $\overline{U}_{h,b}$. Then, the second probability term above
is equal to $\mathbb{P}\left(\min_{b=0,\ldots,\,\left\lfloor T_{n}/n_{T}\right\rfloor -2}\left|\overline{U}_{h,b+1}^{2}\right|<1/K\right)$
which converges to zero by letting $K=4/\sigma_{-}^{4}$. As for
the first probability term, we use \eqref{eq S* - Shat} and choose
$r>0$ sufficiently large to deduce that, 
\begin{align*}
\mathbb{P} & \left(\max_{b=0,\ldots,\,\left\lfloor T_{n}/n_{T}\right\rfloor -2}\left|\left(\log\left(T_{n}\right)n_{T}\right)^{1/2}\left(U_{h,b+1}-B_{h,b+1}\right)\right|>\varepsilon/K\right)\\
 & \leq\left(\frac{K}{\varepsilon}\right)^{r}\sum_{b=0}^{\left\lfloor T_{n}/n_{T}\right\rfloor -2}\mathbb{E}\left[\left|\left(\log\left(T_{n}\right)n_{T}\right)^{1/2}\left(U_{h,b+1}-B_{h,b+1}\right)\right|^{r}\right]\\
 & =\left(\frac{K}{\varepsilon}\right)^{r}\left(\log\left(T_{n}\right)\right)^{r/2}n_{T}^{r/2-1}O_{\mathbb{P}}\left(1/T^{r/2-1}\right)\rightarrow0,
\end{align*}
in view of Condition \ref{Cond The-auxiliary-sequence} and $T_{n}=O\left(T\right)$.
We can repeat the same argument for the second term of \eqref{eq (Lemma A1 A1)}.
Altogether, this establishes the claim of the lemma. $\square$
\begin{lem}
\label{Lemma QA2}As $h\downarrow0$,
\begin{align*}
\max_{b=0,\ldots,\,\left\lfloor T_{n}/n_{T}\right\rfloor -2}\left(\log\left(T_{n}\right)n_{T}\right)^{1/2}\left|n_{T}^{-1}\sum_{j=1}^{n_{T}}\left(\overline{L}_{\psi,\left(T_{m}+\left(b+1\right)n_{T}+j-1\right)h}\left(\widehat{\beta}\right)-\overline{L}_{\psi,\left(T_{m}+bn_{T}+j-1\right)h}\left(\widehat{\beta}\right)\right)\right| & \overset{\mathbb{P}}{\rightarrow}0,
\end{align*}
and the same result holds with $\beta^{*}$ in place $\widehat{\beta}$.
Furthermore, as $h\downarrow0$,
\begin{align*}
\max_{b=0,\ldots,\,\left\lfloor T_{n}/n_{T}\right\rfloor -2}\left(\log\left(T_{n}\right)n_{T}\right)^{1/2}\left|n_{T}^{-1}\sum_{j=1}^{n_{T}}\left(\overline{L}_{\psi,\left(T_{m}+bn_{T}+j-1\right)h}\left(\widehat{\beta}\right)-\overline{L}_{\psi,\left(T_{m}+bn_{T}+j-1\right)h}\left(\beta^{*}\right)\right)\right| & \overset{\mathbb{P}}{\rightarrow}0.
\end{align*}
\end{lem}

\noindent \textit{Proof.} By definition, 
\begin{align*}
\left|n_{T}^{-1}\sum_{j=1}^{n_{T}}\right. & \left.\left(\overline{L}_{\psi,\left(T_{m}+\left(b+1\right)n_{T}+j-1\right)h}\left(\beta^{*}\right)-\overline{L}_{\psi,\left(T_{m}+bn_{T}+j-1\right)h}\left(\beta^{*}\right)\right)\right|\\
 & =\left|n_{T}^{-1}\sum_{j=1}^{n_{T}}\left(\sum_{l=1}^{T_{m}+\left(b+1\right)n_{T}+j-1}\frac{\left(\Delta_{h}\widetilde{e}_{l}^{*}\right)^{2}}{T_{m}+\left(b+1\right)n_{T}+j-1}-\sum_{l=1}^{T_{m}+bn_{T}+j-1}\frac{\left(\Delta_{h}\widetilde{e}_{l}^{*}\right)^{2}}{T_{m}+bn_{T}+j-1}\right)\right|\\
 & =\left|n_{T}^{-1}\sum_{j=1}^{n_{T}}\left(\sum_{l=1}^{T_{m}+\left(b+1\right)n_{T}+j-1}\left(\Delta_{h}\widetilde{e}_{l}^{*}\right)^{2}\left(\frac{1}{T_{m}+\left(b+1\right)n_{T}+j-1}-\frac{1}{T_{m}+bn_{T}+j-1}\right)\right.\right.\\
 & \quad\left.\left.+\sum_{l=T_{m}+bn_{T}+j}^{T_{m}+\left(b+1\right)n_{T}+j-1}\frac{\left(\Delta_{h}\widetilde{e}_{l}^{*}\right)^{2}}{T_{m}+bn_{T}+j-1}\right)\right|.
\end{align*}
 By a law of large numbers for a sequence of i.n.d. random variables
{[}see \citet{white:01}, Section 3.2{]} and the boundedness of $\left\{ \sigma_{t}\right\} _{t\geq0}$,
we have $\left(T_{m}+\left(b+1\right)n_{T}+j-1\right)^{-1}\sum_{l=1}^{T_{m}+\left(b+1\right)n_{T}+j-1}\left(\Delta_{h}\widetilde{e}_{l}^{*}\right)^{2}=O_{\mathbb{P}}\left(1\right)$.
On the other hand, the second term is negligible because there are
$n_{T}-1$ summands and $T_{m}=O\left(T\right)$. Altogether,
\begin{align*}
\left|n_{T}^{-1}\sum_{j=1}^{n_{T}}\left(\overline{L}_{\psi,\left(T_{m}+\left(b+1\right)n_{T}+j-1\right)h}\left(\beta^{*}\right)-\overline{L}_{\psi,\left(T_{m}+bn_{T}+j-1\right)h}\left(\beta^{*}\right)\right)\right| & \leq C\left(O_{\mathbb{P}}\left(\frac{n_{T}}{T_{m}}\right)+O_{\mathbb{P}}\left(\frac{n_{T}}{T_{m}+m_{T}n_{T}}\right)\right).
\end{align*}
Thus, for any $\varepsilon>0$, 
\begin{align}
\mathbb{P} & \left(\max_{b=0,\ldots,\,\left\lfloor T_{n}/n_{T}\right\rfloor -2}\left(\log\left(T_{n}\right)n_{T}\right)^{1/2}\left|n_{T}^{-1}\sum_{j=1}^{n_{T}}\left(\overline{L}_{\psi,\left(T_{m}+\left(b+1\right)n_{T}+j-1\right)h}\left(\beta^{*}\right)-\overline{L}_{\psi,\left(T_{m}+bn_{T}+j-1\right)h}\left(\beta^{*}\right)\right)\right|>\varepsilon\right)\nonumber \\
 & \leq\sum_{b=0}^{\left\lfloor T_{n}/n_{T}\right\rfloor -2}\mathbb{P}\left(\left(\log\left(T_{n}\right)n_{T}\right)^{1/2}\left|n_{T}^{-1}\sum_{j=1}^{n_{T}}\left(\overline{L}_{\psi,\left(T_{m}+\left(b+1\right)n_{T}+j-1\right)h}\left(\beta^{*}\right)-\overline{L}_{\psi,\left(T_{m}+bn_{T}+j-1\right)h}\left(\beta^{*}\right)\right)\right|>\varepsilon\right)\nonumber \\
 & \leq\varepsilon^{-r}\sum_{b=0}^{\left\lfloor T_{n}/n_{T}\right\rfloor -2}\mathbb{E}\left[\left(\log\left(T_{n}\right)n_{T}\right)^{r/2}\left|n_{T}^{-1}\sum_{j=1}^{n_{T}}\left(\overline{L}_{\psi,\left(T_{m}+\left(b+1\right)n_{T}+j-1\right)h}\left(\beta^{*}\right)-\overline{L}_{\psi,\left(T_{m}+bn_{T}+j-1\right)h}\left(\beta^{*}\right)\right)\right|^{r}\right]\nonumber \\
 & \leq\varepsilon^{-r}C\left(\log\left(T_{n}\right)n_{T}\right)^{r/2}O_{\mathbb{P}}\left(n_{T}^{r-1}T_{n}^{1-r}\right)\rightarrow0,\label{eq (A1a0)}
\end{align}
for $r>0$ sufficiently large and in view of Condition \ref{Cond The-auxiliary-sequence}
since $T_{m}$ is of the same order as $T_{n}$. For the last claim
of the lemma, note that
\begin{align}
\overline{L}_{\psi,\left(T_{m}+bn_{T}+j-1\right)h}\left(\widehat{\beta}\right) & -\overline{L}_{\psi,\left(T_{m}+bn_{T}+j-1\right)h}\left(\beta^{*}\right)\nonumber \\
 & =\sum_{l=1}^{T_{m}+bn_{T}+j-1}\frac{\left(\Delta_{h}\widetilde{e}_{l}\right)^{2}}{T_{m}+bn_{T}+j-1}-\sum_{l=1}^{T_{m}+bn_{T}+j-1}\frac{\left(\Delta_{h}\widetilde{e}_{l}^{*}\right)^{2}}{T_{m}+bn_{T}+j-1}\nonumber \\
 & =\frac{1}{T_{m}+bn_{T}+j-1}\sum_{l=1}^{T_{m}+bn_{T}+j-1}\left(\widehat{\beta}-\beta^{*}\right)'\Delta_{h}\widetilde{X}_{l}\Delta_{h}\widetilde{X}'_{l}\left(\widehat{\beta}-\beta^{*}\right)\label{eq (A1a)}\\
 & \quad-\frac{2}{T_{m}+bn_{T}+j-1}\sum_{l=1}^{T_{m}+bn_{T}+j-1-\tau}\Delta_{h}\widetilde{e}_{l}^{*}\left(\widehat{\beta}-\beta^{*}\right)'\Delta_{h}\widetilde{X}{}_{l}.\label{eq (A1b)}
\end{align}
By Lemma \ref{Lemma Prelim A1}, $\left(T_{m}+bn_{T}+j-1\right)^{-1}\sum_{l=1}^{T_{m}+bn_{T}+j-1-\tau}\Delta_{h}\widetilde{X}{}_{l}\Delta_{h}\widetilde{X}{}_{l}'=O_{\mathbb{P}}\left(1\right)$.
Since $\widehat{\beta}_{k}-\beta^{*}=O_{\mathbb{P}}\left(1/\sqrt{T}\right)$
uniformly in $k\geq T_{m}$ by Assumption \ref{Assumption Roo-T Consistent beta },
the term in \eqref{eq (A1a)} is $O_{\mathbb{P}}\left(T^{-1}\right)$
whereas the term \eqref{eq (A1b)} is $o_{\mathbb{P}}\left(T^{-1/2}\right)$
by Lemma \ref{Lemma Prelim A1}. Therefore, upon using Condition \ref{Cond The-auxiliary-sequence}\textbf{
}and the same argument that led to \eqref{eq (A1a0)} we show the
last claim of the lemma. The proof of the second claim then follows
from combining the result of the first and last claim. $\square$
\begin{lem}
\label{Lemma eq(22) in Vetter (2012)}Let $B_{h,b}^{0}=\left(n_{T}h\right)^{-1}\sum_{j=1}^{n_{T}}\sigma_{e,\left(T_{m}+\tau+bn_{T}-1\right)h}^{2}\left(\Delta_{h}W_{e,T_{m}+\tau+bn_{T}+j-1}\right)^{2}$.
For any $\varepsilon>0$ and some constant $K>0$, $\mathbb{P}\left(\max_{b=0,\ldots,\,\left\lfloor T_{n}/n_{T}\right\rfloor -2}\left|1/B_{h,b}^{0}\right|>K\right)\rightarrow0$.
\end{lem}

\noindent \textit{Proof.} Note that 
\begin{align*}
\mathbb{P} & \left(\max_{b=0,\ldots,\,\left\lfloor T_{n}/n_{T}\right\rfloor -2}\left|1/B_{h,b}^{0}\right|>K\right)\\
 & =\mathbb{P}\left(\min_{b=0,\ldots,\,\left\lfloor T_{n}/n_{T}\right\rfloor -2}\left|B_{h,b}^{0}\right|<K^{-1}\right)\\
 & =\mathbb{P}\left(\min_{b=0,\ldots,\,\left\lfloor T_{n}/n_{T}\right\rfloor -2}\frac{1}{n_{T}h}\sum_{j=1}^{n_{T}}\left(\sigma_{e,\left(T_{m}+\tau+bn_{T}-1\right)h}^{2}\left(\Delta_{h}W_{e,T_{m}+\tau+bn_{T}+j-1}\right)^{2}\right)<K^{-1}\right)\\
 & \leq\sum_{b=0}^{\left\lfloor T/n_{T}\right\rfloor -2}\mathbb{P}\left(\frac{1}{n_{T}}\sum_{j=1}^{n_{T}}\left(\sigma_{e,\left(T_{m}+\tau+bn_{T}-1\right)h}\left(h^{-1/2}\Delta_{h}W_{e,T_{m}+\tau+bn_{T}+j-1}\right)\right)^{2}<K^{-1}\right).
\end{align*}
With $K=2/\sigma_{-}^{2}$ {[}with $\sigma_{-}$ defined in Assumption
\ref{Assumption 1, CT}-(iii){]}, we can use Markov's inequality to
deduce, for any $r>0,$
\begin{align*}
\mathbb{P} & \left(\frac{1}{n_{T}}\sum_{j=1}^{n_{T}}\left(\sigma_{e,\left(T_{m}+\tau+bn_{T}-1\right)h}\Delta_{h}W_{e,T_{m}+\tau+bn_{T}+j-1}\right)^{2}<\sigma_{-}^{2}/2\right)\\
 & \leq\mathbb{P}\left(\frac{1}{n_{T}}\sum_{j=1}^{n_{T}}\sigma_{e,\left(T_{m}+\tau+bn_{T}-1\right)h}^{2}\left(\left(h^{-1/2}\Delta_{h}W_{e,T_{m}+\tau+bn_{T}+j-1}\right)^{2}-1\right)<-\sigma_{-}^{2}/2\right)\\
 & \leq\left(\frac{2}{\sigma_{-}^{2}}\right)^{r}n_{T}^{-r/2}\mathbb{E}\left[\left|n_{T}^{-1/2}\sum_{j=1}^{n_{T}}\sigma_{e,\left(T_{m}+\tau+bn_{T}-1\right)h}^{2}\left(\left(h^{-1/2}\Delta_{h}W_{e,T_{m}+\tau+bn_{T}+j-1}\right)^{2}-1\right)\right|^{r}\right].
\end{align*}
  From a standard central limit theorem for i.i.d. observations
we have 
\begin{align*}
\mathbb{E}\left[\left|n_{T}^{-1/2}\sum_{j=1}^{n_{T}}\left(\left(h^{-1/2}\Delta_{h}W_{T_{m}+\tau+bn_{T}+j-1}\right)^{2}-1\right)\right|^{r}\right] & <C_{2,r},
\end{align*}
 where $C_{2,r}<\infty$. Thus, since we can choose $r$ sufficiently
large we can deduce, 
\begin{align*}
\sum_{b=0}^{\left\lfloor T/n_{T}\right\rfloor -2}\mathbb{P} & \left(\frac{1}{n_{T}}\sum_{j=1}^{n_{T}}\left(\sigma_{e,\left(T_{m}+\tau+bn_{T}-1\right)h}h^{-1/2}\Delta_{h}W_{T_{m}+\tau+bn_{T}+j-1}\right)^{2}<K\right)\\
 & \leq C_{r}\left(\frac{2}{\sigma_{-}^{2}}\right)^{r}\left(T_{n}/n_{T}\right)n_{T}^{-r/2}\rightarrow0,
\end{align*}
where we have also used Condition \ref{Cond The-auxiliary-sequence}.
This concludes the proof. $\square$

\medskip{}

Next, let 
\begin{align}
\mathrm{B}_{\mathrm{max},h}^{0}\left(T_{n},\,\tau\right) & \triangleq\max_{b=0,\ldots,\,\left\lfloor T_{n}/n_{T}\right\rfloor -2}\left|\left(B_{h,b+1}^{0}-B_{h,b}^{0}\right)/B_{h,b+1}^{0}\right|\label{eq: B^0 and B^*}\\
\mathrm{B}_{\mathrm{max},h}^{*}\left(T_{n},\,\tau\right) & \triangleq\max_{b=0,\ldots,\,\left\lfloor T_{n}/n_{T}\right\rfloor -2}\left|\left(B_{h,b+1}^{*}-\overline{U}_{h,b}\right)/\overline{U}_{h,b+1}\right|,\nonumber 
\end{align}
where $B_{h,b}^{0}=\left(n_{T}h\right)^{-1}\sum_{j=1}^{n_{T}}\sigma_{e,\left(T_{m}+\tau+bn_{T}-1\right)h}^{2}\left(\Delta_{h}W_{e,T_{m}+\tau+bn_{T}+j-1}\right)^{2}$
and $B_{h,b}^{*}=n_{T}^{-1}\sum_{j=1}^{n_{T}}\left(\Delta_{h}\widetilde{e}_{T_{m}+\tau+bn_{T}+j-1}^{*}\right)^{2}$.
The following lemma shows that, under $H_{0}$, the difference in
the in-sample losses $\overline{L}_{\psi,kh}\left(\widehat{\beta}_{k}\right)$
across adjacent blocks is negligible asymptotically. 
\begin{lem}
\label{Lemma QA3}As $h\downarrow0$, $\left(\log\left(T_{n}\right)n_{T}\right)^{1/2}\left(\mathrm{B}_{\mathrm{max},h}^{*}\left(T_{n},\,\tau\right)-\mathrm{U}_{\mathrm{max},h}\left(T_{n},\,\tau\right)\right)\overset{\mathbb{P}}{\rightarrow}0$.
\end{lem}

\noindent \textit{Proof.} We begin with the inequality, 
\begin{align*}
\left|\mathrm{B}_{\mathrm{max},h}^{*}\left(T_{n},\,\tau\right)\right. & \left.-\mathrm{U}_{\mathrm{max},h}\left(T_{n},\,\tau\right)\right|\\
 & =\left|\max_{b=0,\ldots,\left\lfloor T_{n}/n_{T}\right\rfloor -2}\left|\left(B_{h,b+1}^{*}-B_{h,b}^{*}\right)/\overline{U}_{h,b+1}\right|-\max_{i=0,\ldots,\left\lfloor n/k_{n}\right\rfloor -2}\left|\left(U_{h,b+1}-U_{h,b}\right)/\overline{U}_{h,b+1}\right|\right|\\
 & \leq\max_{b=0,\ldots,\left\lfloor T_{n}/n_{T}\right\rfloor -2}\left|\left(\sum_{j=1}^{n_{T}}\left(\overline{L}_{\psi,\left(T_{m}+\left(b+1\right)n_{T}+j-1\right)h}\left(\beta^{*}\right)-\overline{L}_{\psi,\left(T_{m}+bn_{T}+j-1\right)h}\left(\beta^{*}\right)\right)\right)/\overline{U}_{h,b+1}\right|.
\end{align*}
 For any $\varepsilon>0$ and any $K>0,$ 
\begin{align*}
\mathbb{P} & \left(\max_{b=0,\ldots,\,\left\lfloor T_{n}/n_{T}\right\rfloor -2}\left|\frac{\left(\log\left(T_{n}\right)n_{T}\right)^{1/2}\sum_{j=1}^{n_{T}}\left(\overline{L}_{\psi,\left(T_{m}+\left(b+1\right)n_{T}+j-1\right)h}\left(\beta^{*}\right)-\overline{L}_{\psi,\left(T_{m}+bn_{T}+j-1\right)h}\left(\beta^{*}\right)\right)}{\overline{U}_{h,b}}\right|>\varepsilon\right)\\
 & \leq\mathbb{P}\left(\max_{b=0,\ldots,\,\left\lfloor T_{n}/n_{T}\right\rfloor -2}\left|\left(\log\left(T_{n}\right)n_{T}\right)^{1/2}\sum_{j=1}^{n_{T}}\left(\overline{L}_{\psi,\left(T_{m}+\left(b+1\right)n_{T}+j-1\right)h}\left(\beta^{*}\right)-\overline{L}_{\psi,\left(T_{m}+bn_{T}+j-1\right)h}\left(\beta^{*}\right)\right)\right|>\varepsilon/\sqrt{K}\right)\\
 & \quad+\mathbb{P}\left(\max_{b=0,\ldots,\,\left\lfloor T_{n}/n_{T}\right\rfloor -2}1/\left|\overline{U}_{h,b+1}\right|>\sqrt{K}\right).
\end{align*}
 By the second result in Lemma \ref{Lemma QA2} the first term converges
to zero. As for the second term, it was already treated in \eqref{eq (A3)}
with $\overline{U}_{h,b+1}^{2}$ in place of $\overline{U}_{h,b+1}$,
and a similar argument can be applied to yield the same result.
$\square$

\medskip{}

  Lemma \ref{Lemma QA3} implies that the asymptotic behavior of
the test statistics under $H_{0}$ is determined by the sequence of
out-of-sample losses only. Next, let us define the following quantity
which has the volatility shifted back by one block of time-length
$n_{T}h$:  
\begin{align*}
\widetilde{B}_{h,b}^{0} & =\left(n_{T}h\right)^{-1}\sum_{j=1}^{n_{T}}\sigma_{e,\left(T_{m}+\tau+\left(b-1\right)n_{T}-1\right)h}^{2}\left(\Delta_{h}W_{e,T_{m}+\tau+bn_{T}+j-1}\right)^{2},
\end{align*}
and use it to define the statistic 
\begin{align*}
\widetilde{\mathrm{B}}_{\mathrm{max},h}^{0}\left(T_{n},\,\tau\right) & \triangleq\max_{b=0,\ldots,\,\left\lfloor T_{n}/n_{T}\right\rfloor -2}\left|\left(\widetilde{B}_{h,b+1}^{0}-B_{h,b}^{0}\right)/\widetilde{B}_{h,b+1}^{0}\right|.
\end{align*}
Our final goal is to show that $\left(\log\left(T_{n}\right)n_{T}\right)^{1/2}\left(\mathrm{V}_{\mathrm{max},h}\left(T_{n},\,\tau\right)-\widetilde{\mathrm{B}}_{\mathrm{max},h}^{0}\left(T_{n},\,\tau\right)\right)$
converges to zero in probability, where 
\begin{align}
\mathrm{V}_{\mathrm{max},h}\left(T_{n},\,\tau\right) & \triangleq\max_{b=0,\ldots,\,\left\lfloor T_{n}/n_{T}\right\rfloor -2}\left|\frac{\widetilde{B}_{h,b+1}^{0}-B_{h,b}^{0}}{\sigma_{e,\left(T_{m}+\tau+bn_{T}-1\right)h}^{2}}\right|.\label{eq. Def Vmax (QB)}
\end{align}
We deduce this result from several small lemmas. We begin by replacing
$\mathrm{B}_{\mathrm{max},h}^{*}\left(T_{n},\,\tau\right)$ by $\mathrm{B}_{\mathrm{max},h}^{0}\left(T_{n},\,\tau\right)$.
\begin{lem}
\label{Prop QA1 BJV}As $h\downarrow0$, $\left(\log\left(T_{n}\right)n_{T}\right)^{1/2}\left(\mathrm{B}_{\mathrm{max},h}^{*}\left(T_{n},\,\tau\right)-\mathrm{B}_{\mathrm{max},h}^{0}\left(T_{n},\,\tau\right)\right)\overset{\mathbb{P}}{\rightarrow}0$. 
\end{lem}

\noindent \textit{Proof.} We begin by using inequality \eqref{eq. Inequality in Prop A1},
   
\begin{align}
\left|\mathrm{B}_{\mathrm{max},h}^{*}\left(T_{n},\,\tau\right)\right. & \left.-\mathrm{B}_{\mathrm{max},h}^{0}\left(T_{n},\,\tau\right)\right|\nonumber \\
 & =\left|\max_{b=0,\ldots,\left\lfloor T_{n}/n_{T}\right\rfloor -2}\left|\left(B_{h,b+1}^{*}-B{}_{h,b}^{*}\right)/\overline{U}_{h,b+1}\right|-\max_{i=0,\ldots,\left\lfloor T_{n}/n_{T}\right\rfloor -2}\left|B_{h,b}^{0}/B_{h,b+1}^{0}-1\right|\right|\nonumber \\
 & \leq\max_{b=0,\ldots,\left\lfloor T_{n}/n_{T}\right\rfloor -2}\left|B_{h,b}^{*}/\overline{U}_{h,b+1}-1-\left(B_{h,b}^{0}/B_{h,b+1}^{0}-1\right)\right|\nonumber \\
 & \leq\max_{b=0,\ldots,\,\left\lfloor T_{n}/n_{T}\right\rfloor -2}\left|B_{h,b}^{*}\left(\frac{1}{\overline{U}_{h,b+1}}-\frac{1}{B_{h,b+1}^{0}}\right)\right|+\max_{b=0,\ldots,\,\left\lfloor T_{n}/n_{T}\right\rfloor -2}\left|\frac{B_{h,b}^{*}-B{}_{h,b}^{0}}{B_{h,b+1}^{0}}\right|.\label{eq. 47 BJV}
\end{align}
Consider the second term of \eqref{eq. 47 BJV}. Let $K>0$. For any
$\varepsilon>0$, 
\begin{align}
\mathbb{P} & \left(\max_{b=0,\ldots,\,\left\lfloor T_{n}/n_{T}\right\rfloor -2}\left|\frac{\left(\log\left(T_{n}\right)n_{T}\right)^{1/2}\left(B_{h,b}^{*}-B{}_{h,b}^{0}\right)}{B_{h,b+1}^{0}}\right|>\varepsilon\right)\nonumber \\
 & \leq\mathbb{P}\left(\max_{b=0,\ldots,\,\left\lfloor T_{n}/n_{T}\right\rfloor -2}\left|\left(\log\left(T_{n}\right)n_{T}\right)^{1/2}\left(B_{h,b}^{*}-B{}_{h,b}^{0}\right)\right|>\varepsilon/K\right)\nonumber \\
 & \quad+\mathbb{P}\left(\max_{b=0,\ldots,\,\left\lfloor T_{n}/n_{T}\right\rfloor -2}\left|1/B_{h,b+1}^{0}\right|>K\right).\label{eq (48 or 69) BJV}
\end{align}
 By Lemma \ref{Lemma eq(22) in Vetter (2012)}, $\mathbb{P}\left(\max_{b=0,\ldots,\,\left\lfloor T_{n}/n_{T}\right\rfloor -2}\left|1/B_{h,b+1}^{0}\right|>K\right)=o_{\mathbb{P}}\left(1\right)$
if we set for instance $K=2/\sigma_{-}^{2}$.\textbf{} Let us consider
the first term of \eqref{eq (48 or 69) BJV}. By It\^o's formula,
 
\begin{align}
B_{h,b}^{*} & -B_{h,b}^{0}\nonumber \\
 & =\left(n_{T}h\right)^{-1}\sum_{j=1}^{n_{T}}\left(\Delta_{h}e_{T_{m}+\tau+bn_{T}+j-1}^{*}\right)^{2}-\left(n_{T}h\right)^{-1}\sum_{j=1}^{n_{T}}\sigma_{e,\left(T_{m}+\tau+bn_{T}-1\right)h}^{2}\left(\Delta_{h}W_{e,T_{m}+\tau+bn_{T}+j-1}\right)^{2}\nonumber \\
 & =\left(n_{T}h\right)^{-1}\sum_{j=1}^{n_{T}}2\int_{\left(T_{m}+\tau+bn_{T}+j-2\right)h}^{\left(T_{m}+\tau+bn_{T}+j-1\right)h}\left(\sigma_{s}^{2}-\sigma_{e,\left(T_{m}+\tau+bn_{T}-1\right)h}^{2}\right)ds\nonumber \\
 & \quad+\left(n_{T}h\right)^{-1}\sum_{j=1}^{n_{T}}2\int_{\left(T_{m}+\tau+bn_{T}+j-2\right)h}^{\left(T_{m}+\tau+bn_{T}+j-1\right)h}\Biggl(\left(e_{s}-e_{\left(T_{m}+\tau+bn_{T}+j-2\right)h}\right)\sigma_{e,s}\nonumber \\
 & \quad-\left(W_{e,s}-W_{e,\left(T_{m}+\tau+bn_{T}+j-2\right)h}\right)\sigma_{e,\left(T_{m}+\tau+bn_{T}-1\right)h}^{2}\Biggl)dW_{e,s}\nonumber \\
 & \quad+\left(n_{T}h\right)^{-1}\sum_{j=1}^{n_{T}}2\int_{\left(T_{m}+\tau+bn_{T}+j-2\right)h}^{\left(T_{m}+\tau+bn_{T}+j-1\right)h}\left(e_{s}-e_{\left(T_{m}+\tau+bn_{T}+j-2\right)h}\right)\mu_{e,s}h^{-\vartheta}ds.\label{eq. (49) BJV}
\end{align}
Consider the first term of \eqref{eq. (49) BJV},
\begin{align}
\max_{b=0,\ldots,\,\left\lfloor T_{n}/n_{T}\right\rfloor -2} & \left|\left(\log\left(T_{n}\right)n_{T}\right)^{1/2}\left(n_{T}h\right)^{-1}\sum_{j=1}^{n_{T}}2\int_{\left(T_{m}+\tau+bn_{T}+j-2\right)h}^{\left(T_{m}+\tau+bn_{T}+j-1\right)h}\left(\sigma_{e,s}^{2}-\sigma_{e,\left(T_{m}+\tau+bn_{T}-1\right)h}^{2}\right)ds\right|\nonumber \\
 & \leq\max_{b=0,\ldots,\,\left\lfloor T_{n}/n_{T}\right\rfloor -2}\left(\log\left(T_{n}\right)n_{T}\right)^{1/2}\left(n_{T}h\right)^{-1}\sum_{j=1}^{n_{T}}2\int_{\left(T_{m}+\tau+bn_{T}+j-2\right)h}^{\left(T_{m}+\tau+bn_{T}+j-1\right)h}\left|\sigma_{e,s}^{2}-\sigma_{e,\left(T_{m}+\tau+bn_{T}-1\right)h}^{2}\right|ds\nonumber \\
 & \leq\left(\log\left(T_{n}\right)n_{T}\right)^{1/2}\left(n_{T}h\right)^{-1}2\phi_{\sigma,n_{T}h,N}\cdot n_{T}h\nonumber \\
 & \leq C\left(\log\left(T_{n}\right)n_{T}\right)^{1/2}n_{T}h\rightarrow0,\label{eq (49b)}
\end{align}
 by Condition \ref{Cond The-auxiliary-sequence}. Let us now turn
to the last term. We have for any integer $r>0,$ 
\begin{align*}
\mathbb{P} & \left(\max_{b=0,\ldots,\,\left\lfloor T_{n}/n_{T}\right\rfloor -2}\right.\left(\log\left(T_{n}\right)n_{T}\right)^{1/2}\left(n_{T}h\right)^{-1}\\
 & \quad\left.\times\left|\sum_{j=1}^{n_{T}}2\int_{\left(T_{m}+\tau+bn_{T}+j-2\right)h}^{\left(T_{m}+\tau+bn_{T}+j-1\right)h}\left(e_{s}-e_{\left(T_{m}+\tau+bn_{T}+j-2\right)h}\right)\mu_{e,s}h^{-\vartheta}ds\right|>\varepsilon/\left(4K\right)\right)\\
 & \leq\sum_{b=0}^{\left\lfloor T_{n}/n_{T}\right\rfloor -2}\mathbb{P}\left(\left(\log\left(T_{n}\right)n_{T}\right)^{1/2}\left(n_{T}h\right)^{-1}\right.\\
 & \quad\times\left.\left|\sum_{j=1}^{n_{T}}2\int_{\left(T_{m}+\tau+bn_{T}+j-2\right)h}^{\left(T_{m}+\tau+bn_{T}+j-1\right)h}\left(e_{s}-e_{\left(T_{m}+\tau+bn_{T}+j-2\right)h}\right)\mu_{e,s}h^{-\vartheta}ds\right|>\varepsilon/\left(4K\right)\right)\\
 & \leq\left(\frac{4K}{\varepsilon}\right)^{r}\sum_{b=0}^{\left\lfloor T_{n}/n_{T}\right\rfloor -2}\\
 & \quad\times\mathbb{E}\left[\left|\left(\log\left(T_{n}\right)n_{T}\right)^{1/2}\left(n_{T}h\right)^{-1}\sum_{j=1}^{n_{T}}2\int_{\left(T_{m}+\tau+bn_{T}+j-2\right)h}^{\left(T_{m}+\tau+bn_{T}+j-1\right)h}\left(e_{s}-e_{\left(T_{m}+\tau+bn_{T}+j-2\right)h}\right)\mu_{e,s}h^{-\vartheta}ds\right|^{r}\right]\\
 & \leq C_{r}\left(\frac{4K}{\varepsilon}\right)^{r}\sum_{b=0}^{\left\lfloor T_{n}/n_{T}\right\rfloor -2}\\
 & \quad\times\left[\left|\left(\log\left(T_{n}\right)n_{T}\right)^{1/2}\left(n_{T}h\right)^{-1}\sum_{j=1}^{n_{T}}2\int_{\left(T_{m}+\tau+bn_{T}+j-2\right)h}^{\left(T_{m}+\tau+bn_{T}+j-1\right)h}\left(\mathbb{E}\left[\left|\left(e_{s}-e_{\left(T_{m}+\tau+bn_{T}+j-2\right)h}\right)\mu_{e,s}\right|^{r}h^{-\vartheta r}\right]\right)^{1/r}ds\right|^{r}\right],
\end{align*}
where the last inequality follows from using Jensen's and Minkowski's
inequalities. By the Burkh\"{o}lder-Davis-Gundy inequality, for any
$s\in\left[\left(T_{m}+\tau+bn_{T}+j-2\right)h,\,\left(T_{m}+\tau+bn_{T}+j-1\right)h\right],$
we have
\begin{align*}
\mathbb{E}\left[\left|\left(e_{s}-e_{\left(T_{m}+\tau+bn_{T}+j-2\right)h}\right)\mu_{e,s}\right|^{r}h^{-\vartheta r}\right] & \leq C_{r}h^{r/2-\vartheta r},
\end{align*}
and therefore since $\vartheta\in[0,\,1/8)$, 
\begin{align*}
\mathbb{P} & \left(\max_{b=0,\ldots,\,\left\lfloor T_{n}/n_{T}\right\rfloor -2}\right.\left(\log\left(T_{n}\right)n_{T}\right)^{1/2}\left(n_{T}h\right)^{-1}\\
 & \left.\times\left|\sum_{j=1}^{n_{T}}2\int_{\left(T_{m}+\tau+bn_{T}+j-2\right)h}^{\left(T_{m}+\tau+bn_{T}+j-1\right)h}\left(e_{s}-e_{\left(T_{m}+\tau+bn_{T}+j-2\right)h}\right)\mu_{e,s}h^{-\vartheta}ds\right|>\varepsilon/\left(4K\right)\right)\\
 & \leq C_{r}\left(\frac{8K}{\varepsilon}\right)^{r}\sum_{b=0}^{\left\lfloor T_{n}/n_{T}\right\rfloor -2}\left(\left(\log\left(T_{n}\right)n_{T}\right)^{1/2}\left(n_{T}h\right)^{-1}n_{T}h^{1+3/8}\right)^{r}\\
 & \leq C_{r}\left(\frac{8K}{\varepsilon}\right)^{r}\sqrt{\log\left(T_{n}\right)}T_{n}^{1/3+\epsilon}\left(h^{1/24+\epsilon}\right)^{r}\rightarrow0,
\end{align*}
 for $r>0$ sufficiently large and where $\epsilon>0$ is a small
real number. Next, consider the second term of \eqref{eq. (49) BJV},
\begin{align}
\left(e_{s}-e_{\left(T_{m}+\tau+bn_{T}+j-2\right)h}\right) & \sigma_{e,s}-\left(W_{e,s}-W_{e,\left(T_{m}+\tau+bn_{T}+j-2\right)h}\right)\sigma_{e,\left(T_{m}+\tau+bn_{T}-1\right)h}^{2}\label{eq. (51, 72) BJV}\\
 & =\sigma_{e,s}\int_{\left(T_{m}+\tau+bn_{T}+j-2\right)h}^{s}\sigma_{e,v}dW_{e,v}\nonumber \\
 & \quad-\sigma_{e,\left(T_{m}+\tau+bn_{T}-1\right)h}^{2}\int_{\left(T_{m}+\tau+bn_{T}+j-2\right)h}^{s}dW_{e,v}\nonumber \\
 & \quad+\sigma_{e,s}\int_{\left(T_{m}+\tau+bn_{T}+j-2\right)h}^{s}\mu_{e,v}h^{-\vartheta}dv\nonumber \\
 & =\left(\sigma_{e,s}-\sigma_{e,\left(T_{m}+\tau+bn_{T}-1\right)h}\right)\int_{\left(T_{m}+\tau+bn_{T}+j-2\right)h}^{s}\sigma_{e,v}dW_{e,v}\nonumber \\
 & \quad+\sigma_{e,\left(T_{m}+\tau+bn_{T}-1\right)h}\int_{\left(T_{m}+\tau+bn_{T}+j-2\right)h}^{s}\left(\sigma_{e,v}-\sigma_{e,\left(T_{m}+\tau+bn_{T}-1\right)h}\right)dW_{e,v}\nonumber \\
 & \quad+\sigma_{e,s}\int_{\left(T_{m}+\tau+bn_{T}+j-2\right)h}^{s}\mu_{e,v}h^{-\vartheta}dv.\nonumber 
\end{align}
For any integer $r>2,$ 
\begin{align*}
\mathbb{P} & \left(\max_{b=0,\ldots,\,\left\lfloor T_{n}/n_{T}\right\rfloor -2}\right.\left(\log\left(T_{n}\right)n_{T}\right)^{1/2}\left(n_{T}h\right)^{-1}\\
 & \quad\left.\times\left|\sum_{j=1}^{n_{T}}2\int_{\left(T_{m}+\tau+bn_{T}+j-2\right)h}^{\left(T_{m}+\tau+bn_{T}+j-1\right)h}\left(\sigma_{e,s}-\sigma_{e,\left(T_{m}+\tau+bn_{T}-1\right)h}\right)\int_{\left(T_{m}+\tau+bn_{T}+j-2\right)h}^{s}\sigma_{e,v}dW_{e,v}dW_{e,s}\right|>\varepsilon/\left(12K\right)\right)\\
 & \leq\left(\frac{\varepsilon}{12K}\right)^{-r}\sum_{b=1}^{\left\lfloor T_{n}/n_{T}\right\rfloor -2}\left(\log\left(T_{n}\right)n_{T}\right)^{1/2}\left(n_{T}h\right)^{-1}\\
 & \quad\times\mathbb{E}\left[\left|\sum_{j=1}^{n_{T}}2\int_{\left(T_{m}+\tau+bn_{T}+j-2\right)h}^{\left(T_{m}+\tau+bn_{T}+j-1\right)h}\left(\sigma_{e,s}-\sigma_{e,\left(T_{m}+\tau+bn_{T}-1\right)h}\right)\int_{\left(T_{m}+\tau+bn_{T}+j-2\right)h}^{s}\sigma_{e,v}dW_{e,v}dW_{e,s}\right|^{r}\right].
\end{align*}
Then, by H\"older's inequality, 

\begin{align}
\mathbb{E} & \left[\left|\left(\log\left(T_{n}\right)n_{T}\right)^{1/2}\left(n_{T}h\right)^{-1}\right.\right.\nonumber \\
 & \quad\left.\left.\times\sum_{j=1}^{n_{T}}2\int_{\left(T_{m}+\tau+bn_{T}+j-2\right)h}^{\left(T_{m}+\tau+bn_{T}+j-1\right)h}\left(\sigma_{e,s}-\sigma_{e,\left(T_{m}+\tau+bn_{T}-1\right)h}\right)\int_{\left(T_{m}+\tau+bn_{T}+j-2\right)h}^{s}\sigma_{e,v}dW_{e,v}dW_{e,s}\right|^{r}\right]\nonumber \\
 & \leq C_{r}\left(\frac{\sqrt{\log\left(T_{n}\right)}}{h\sqrt{n_{T}}}\right)^{r}\nonumber \\
 & \quad\times\left(\int_{\left(T_{m}+\tau+bn_{T}-1\right)h}^{\left(T_{m}+\tau+\left(b+1\right)n_{T}-1\right)h}\sum_{j=1}^{n_{T}}\mathbb{E}\left[\left(\sigma_{e,s}-\sigma_{e,\left(T_{m}+\tau+bn_{T}-1\right)h}\right)^{r}\left(\int_{\left(T_{m}+\tau+bn_{T}+j-2\right)h}^{s}\sigma_{e,v}dW_{e,v}\right)^{r}\right.\right.\nonumber \\
 & \quad\left.\left.\left.\times\mathbf{1}_{\left\{ \left[\left(T_{m}+\tau+bn_{T}+j-2\right)h,\,\left(T_{m}+\tau+bn_{T}+j-1\right)h\right]\right\} }\left(s\right)\right]\right)^{2/r}ds\right)^{r/2}\nonumber \\
 & \leq C_{r}\left(\frac{\sqrt{\log\left(T_{n}\right)}}{h\sqrt{n_{T}}}\right)^{r}\nonumber \\
 & \quad\times\left(\int_{\left(T_{m}+\tau+bn_{T}-1\right)h}^{\left(T_{m}+\tau+\left(b+1\right)n_{T}-1\right)h}\sum_{j=1}^{n_{T}}\mathbb{E}\left[\phi_{\sigma,n_{T}h,N}^{r}\left(\int_{\left(T_{m}+\tau+bn_{T}+j-2\right)h}^{s}\sigma_{e,v}dW_{e,v}\right)^{r}\right.\right.\nonumber \\
 & \quad\times\left.\left.\left.\mathbf{1}_{\left\{ \left[\left(T_{m}+\tau+bn_{T}+j-2\right)h,\,\left(T_{m}+\tau+bn_{T}+j-1\right)h\right]\right\} }\left(s\right)\right]\right)^{2/r}ds\right)^{r/2}\nonumber \\
 & \leq C_{r}\left(\frac{\sqrt{\log\left(T_{n}\right)}}{h\sqrt{n_{T}}}\right)^{r}\left(\int_{\left(T_{m}+\tau+bn_{T}-1\right)h}^{\left(T_{m}+\tau+\left(b+1\right)n_{T}-1\right)h}\left(\left(n_{T}h\right)^{r}h^{r/2}\right)^{2/r}ds\right)^{r/2}\nonumber \\
 & \leq C_{r}\left(\sqrt{\log\left(T_{n}\right)}\right)^{r}h^{r}n_{T}^{r}\rightarrow0.\label{eq. (49c)}
\end{align}
 The same bound holds for the second term in \eqref{eq. (51, 72) BJV}.
Finally, the last term of \eqref{eq. (51, 72) BJV} is such that
\begin{align}
\mathbb{P} & \left(\max_{b=0,\ldots,\,\left\lfloor T_{n}/n_{T}\right\rfloor -2}\right.\left(\log\left(T_{n}\right)n_{T}\right)^{1/2}\left(n_{T}h\right)^{-1}\nonumber \\
 & \times\left.\left|\sum_{j=1}^{n_{T}}2\int_{\left(T_{m}+\tau+bn_{T}+j-2\right)h}^{\left(T_{m}+\tau+bn_{T}+j-1\right)h}\sigma_{e,s}\int_{\left(T_{m}+\tau+bn_{T}+j-2\right)h}^{s}\mu_{e,v}h^{-\vartheta}dvdW_{e,s}\right|>\varepsilon/\left(12K\right)\right)\nonumber \\
 & \leq\sum_{b=0}^{\left\lfloor T_{n}/n_{T}\right\rfloor -2}\mathbb{P}\left(\left(\log\left(T_{n}\right)n_{T}\right)^{1/2}\left(n_{T}h\right)^{-1}\right.\nonumber \\
 & \quad\times\left.\left|\sum_{j=1}^{n_{T}}2\int_{\left(T_{m}+\tau+bn_{T}+j-2\right)h}^{\left(T_{m}+\tau+bn_{T}+j-1\right)h}\sigma_{e,s}\int_{\left(T_{m}+\tau+bn_{T}+j-2\right)h}^{s}\mu_{e,v}h^{-1/8}dvdW_{e,v}\right|>\varepsilon/\left(12K\right)\right)\nonumber \\
 & \leq\left(\frac{12K}{\varepsilon}\right)^{r}\sum_{b=0}^{\left\lfloor T_{n}/n_{T}\right\rfloor -2}\nonumber \\
 & \quad\times\mathbb{E}\left[\left|\left(\log\left(T_{n}\right)n_{T}\right)^{1/2}\left(n_{T}h\right)^{-1}\sum_{j=1}^{n_{T}}2\int_{\left(T_{m}+\tau+bn_{T}+j-2\right)h}^{\left(T_{m}+\tau+bn_{T}+j-1\right)h}\sigma_{e,s}\int_{\left(T_{m}+\tau+bn_{T}+j-2\right)h}^{s}\mu_{e,v}h^{-1/8}dvdW_{e,v}\right|^{r}\right].\label{eq (73) BJV}
\end{align}
Let 
\begin{align*}
f_{s} & \triangleq2\sum_{j=1}^{n_{T}}\sigma_{e,s}\int_{\left(T_{m}+\tau+bn_{T}+j-2\right)h}^{s}\mu_{e,v}h^{-\vartheta}du\times\mathbf{1}_{\left[\left(T_{m}+\tau+bn_{T}+j-2\right)h,\,\left(T_{m}+\tau+bn_{T}+j-1\right)h\right]}\left(s\right),
\end{align*}
and observe that for any integer $r>1$, 
\begin{align*}
\mathbb{E}\left(f_{s}^{r}\right) & =2^{r}\sum_{j=1}^{n_{T}}\mathbb{E}\left[\sigma_{e,s}^{r}\left(\int_{\left(T_{m}+\tau+bn_{T}+j-2\right)h}^{s}\mu_{e,v}h^{-\vartheta}dv\right)^{r}\right]\mathbf{1}_{\left[\left(T_{m}+\tau+bn_{T}+j-2\right)h,\,\left(T_{m}+\tau+bn_{T}+j-1\right)h\right]}\left(s\right)\\
 & \leq2^{r}C_{r}h^{r\left(1-\vartheta\right)}.
\end{align*}
Therefore, the right-hand side of \eqref{eq (73) BJV} is less than
\begin{align*}
\left(\frac{12K}{\varepsilon}\right)^{r}\sum_{b=0}^{\left\lfloor T_{n}/n_{T}\right\rfloor -2} & \left(\left(\log\left(T_{n}\right)n_{T}\right)^{1/2}\left(n_{T}h\right)^{-1}\right)^{r}\mathbb{E}\left[\left|\int_{\left(T_{m}+\tau+bn_{T}-1\right)h}^{\left(T_{m}+\tau+\left(b+1\right)n_{T}-1\right)h}f_{s}dW_{e,s}\right|^{r}\right]\\
 & \leq\left(\frac{12K}{\varepsilon}\right)^{r}\sum_{b=0}^{\left\lfloor T_{n}/n_{T}\right\rfloor -2}\left(\left(\log\left(T_{n}\right)n_{T}\right)^{1/2}\left(n_{T}h\right)^{-1}\right)^{r}\mathbb{E}\left[\left|\int_{\left(T_{m}+\tau+bn_{T}-1\right)h}^{\left(T_{m}+\tau+\left(b+1\right)n_{T}-1\right)h}f_{s}^{2}ds\right|^{r/2}\right]\\
 & \leq\left(\frac{12K}{\varepsilon}\right)^{r}\sum_{b=0}^{\left\lfloor T_{n}/n_{T}\right\rfloor -2}\left(\left(\log\left(T_{n}\right)n_{T}\right)^{1/2}\left(n_{T}h\right)^{-1}\right)^{r}\left(\int_{\left(T_{m}+\tau+bn_{T}-1\right)h}^{\left(T_{m}+\tau+\left(b+1\right)n_{T}-1\right)h}\left(\mathbb{E}\left(f_{s}^{r}\right)\right)^{2/r}ds\right)^{r/2}\\
 & \leq C_{r}\left(\frac{12K}{\varepsilon}\right)^{r}\sum_{b=0}^{\left\lfloor T_{n}/n_{T}\right\rfloor -2}\left(\left(\log\left(T_{n}\right)n_{T}\right)^{1/2}\left(n_{T}h\right)^{-1}\right)^{r}\left(\int_{\left(T_{m}+\tau+bn_{T}-1\right)h}^{\left(T_{m}+\tau+\left(b+1\right)n_{T}-1\right)h}h^{2\left(1-\vartheta\right)}ds\right)^{r/2}\\
 & \leq C_{r}\left(\frac{12K}{\varepsilon}\right)^{r}\left(\left(\log\left(T_{n}\right)\right)^{1/2}\right)^{r}\left(h^{r/24-4/3}\right)\rightarrow0,
\end{align*}
 for $r$ sufficiently large. This leads to
\begin{align}
\mathbb{P}\left(\max_{b=0,\ldots,\,\left\lfloor T_{n}/n_{T}\right\rfloor -2}\left|\left(\log\left(T_{n}\right)n_{T}\right)^{1/2}\left(B_{h,b}^{*}-B{}_{h,b}^{0}\right)\right|>\varepsilon/K\right) & \rightarrow0.\label{eq (49d)}
\end{align}
We now turn to the first term on the right hand side of \eqref{eq. 47 BJV}.
Choose any $\varepsilon>0$ and positive $K<\infty,$ and note that
\begin{align}
\mathbb{P} & \left(\max_{b=0,\ldots,\,\left\lfloor T_{n}/n_{T}\right\rfloor -2}\left(\log\left(T_{n}\right)n_{T}\right)^{1/2}\left|B_{h,b}^{*}\left(\frac{1}{\overline{U}_{h,b+1}}-\frac{1}{B_{h,b+1}^{0}}\right)\right|>\varepsilon\right)\label{eq. (53b) BJV}\\
 & \leq\mathbb{P}\left(\max_{b=0,\ldots,\,\left\lfloor T_{n}/n_{T}\right\rfloor -2}\left(\log\left(T_{n}\right)n_{T}\right)^{1/2}\left|B_{h,b}^{*}\left(\overline{U}_{h,b+1}-B{}_{h,b+1}^{0}\right)\right|>\varepsilon/K\right)\nonumber \\
 & +\mathbb{P}\left(\max_{b=0,\ldots,\,\left\lfloor T_{n}/n_{T}\right\rfloor -2}1/\left|\overline{U}_{h,b+1}B{}_{h,b+1}^{0}\right|>K\right).\nonumber 
\end{align}
We can manipulate the second term as follows: 
\begin{align*}
\mathbb{P} & \left(\max_{b=0,\ldots,\,\left\lfloor T_{n}/n_{T}\right\rfloor -2}1/\left|\overline{U}_{h,b+1}B{}_{h,b+1}^{0}\right|>K\right)\\
 & =\mathbb{P}\left(\min_{b=0,\ldots,\,\left\lfloor T_{n}/n_{T}\right\rfloor -2}\left|\overline{U}_{h,b+1}B{}_{h,b+1}^{0}\right|<1/K\right)\\
 & \leq\mathbb{P}\left(\min_{b=0,\ldots,\,\left\lfloor T_{n}/n_{T}\right\rfloor -2}\left|\overline{U}_{h,b+1}\right|<1/\sqrt{K}\right)+\mathbb{P}\left(\min_{b=0,\ldots,\,\left\lfloor T_{n}/n_{T}\right\rfloor -2}\left|B_{h,b+1}^{0}\right|<1/\sqrt{K}\right)\\
 & \leq\mathbb{P}\left(\max_{b=0,\ldots,\,\left\lfloor T_{n}/n_{T}\right\rfloor -2}\left|\overline{U}_{h,b+1}-B{}_{h,b+1}^{0}\right|>1/\sqrt{K}\right)+\mathbb{P}\left(\min_{b=0,\ldots,\,\left\lfloor T_{n}/n_{T}\right\rfloor -2}\left|B_{h,b+1}^{0}\right|<2/\sqrt{K}\right)\\
 & \quad+\mathbb{P}\left(\min_{b=0,\ldots,\,\left\lfloor T_{n}/n_{T}\right\rfloor -2}\left|B_{h,b+1}^{0}\right|<1/\sqrt{K}\right).
\end{align*}
 The second and third term on the right-hand side of the the last
inequality converge to zero in view of Lemma \ref{Lemma eq(22) in Vetter (2012)}.
Noting that $\overline{U}_{h,b}$ coincides with $B_{h,b}^{*}$, we
can use the same arguments that led to \eqref{eq (49d)} which shows
a tighter bound since it involves $\max_{b=0,\ldots,\,\left\lfloor T_{n}/n_{T}\right\rfloor -2}\left|B_{h,b+1}^{*}-B{}_{h,b+1}^{0}\right|$
multiplied by $\left(\log\left(T_{n}\right)n_{T}\right)^{1/2}$.
Turning to the first term in \eqref{eq. (53b) BJV}, note that for
any $K_{2}>0,$ 
\begin{align*}
\mathbb{P} & \left(\max_{b=0,\ldots,\,\left\lfloor T_{n}/n_{T}\right\rfloor -2}\left(\log\left(T_{n}\right)n_{T}\right)^{1/2}\left|B_{h,b}^{*}\left(\overline{U}_{h,b+1}-B{}_{h,b+1}^{0}\right)\right|>\varepsilon/K\right)\\
 & \leq\mathbb{P}\left(\max_{b=0,\ldots,\,\left\lfloor T_{n}/n_{T}\right\rfloor -2}\left|B_{h,b}^{*}\right|>K_{2}\right)+\mathbb{P}\left(\max_{b=0,\ldots,\,\left\lfloor T_{n}/n_{T}\right\rfloor -2}\left(\log\left(T_{n}\right)n_{T}\right)^{1/2}\left|\overline{U}_{h,b+1}-B{}_{h,b+1}^{0}\right|>\varepsilon/\left(K\cdot K_{2}\right)\right).
\end{align*}
Noting that $\overline{U}_{h,b+1}$ coincides with $B_{h,b+1}^{*}$,
we can use the same arguments as in \eqref{eq (49d)}. The second
term converges to zero by the same argument as above. Then,
\begin{align*}
\mathbb{P} & \left(\max_{b=0,\ldots,\,\left\lfloor T_{n}/n_{T}\right\rfloor -2}\left|B_{h,b}^{*}\right|>K_{2}\right)\\
 & \leq\mathbb{P}\left(\max_{b=0,\ldots,\,\left\lfloor T_{n}/n_{T}\right\rfloor -2}\left|B_{h,b}^{*}-B_{h,b}^{0}\right|>K_{2}/2\right)+\mathbb{P}\left(\max_{b=0,\ldots,\,\left\lfloor T_{n}/n_{T}\right\rfloor -2}\left|B_{h,b}^{0}\right|>K_{2}/2\right).
\end{align*}
The first term was already discussed above whereas the second term
converges to zero by invoking again Lemma \ref{Lemma eq(22) in Vetter (2012)}
together with the localization assumption {[}cf. Assumption \ref{Assumption 1, CT}-(iii){]}
which implies the $\sigma_{e,t}$ is bounded from above for all $t\geq0$.
$\square$
\begin{lem}
\label{Lemma Prop QA2 BJV }As $h\downarrow0$, $\left(\log\left(T_{n}\right)n_{T}\right)^{1/2}\left(\mathrm{B}_{\mathrm{max},h}^{0}\left(T_{n},\,\tau\right)-\widetilde{\mathrm{B}}_{\mathrm{max},h}^{0}\left(T_{n},\,\tau\right)\right)\overset{\mathbb{P}}{\rightarrow}0$.
\end{lem}

\noindent \textit{Proof.} By simple rearrangements, 
\begin{align*}
\left|\mathrm{B}_{\mathrm{max},h}^{0}\left(T_{n},\,\tau\right)-\widetilde{\mathrm{B}}_{\mathrm{max},h}^{0}\left(T_{n},\,\tau\right)\right| & \leq\max_{b=0,\ldots,\,\left\lfloor T_{n}/n_{T}\right\rfloor -2}\left|\frac{B_{h,b}^{0}\left(\widetilde{B}_{h,b+1}^{0}-B_{h,b+1}^{0}\right)}{\widetilde{B}_{h,b+1}^{0}B_{h,b+1}^{0}}\right|.
\end{align*}
We shall show that 
\begin{align}
\max_{b=0,\ldots,\,\left\lfloor T_{n}/n_{T}\right\rfloor -2}\left(\log\left(T_{n}\right)n_{T}\right)^{1/2}\left|\frac{B_{h,b}^{0}\left(\widetilde{B}_{h,b+1}^{0}-B_{h,b+1}^{0}\right)}{\widetilde{B}_{h,b+1}^{0}B_{h,b+1}^{0}}\right| & =o_{\mathbb{P}}\left(1\right).\label{eq. (54b) in Prop A.2}
\end{align}
 By Lemma \ref{Lemma eq(22) in Vetter (2012)}, $\mathbb{P}\left(\min_{b=0,\ldots,\,\left\lfloor T_{n}/n_{T}\right\rfloor -2}\left|\widetilde{B}_{h,b+1}^{0}B_{h,b+1}^{0}\right|<1/K\right)\rightarrow0$
for some $K>0$; for example, set $\sqrt{K}=2/\sigma_{-}^{2}$. Turning
to the numerator of \eqref{eq. (54b) in Prop A.2}, for any $\varepsilon>0$
and any $K>0,$ 
\begin{align}
\mathbb{P} & \left(\max_{b=0,\ldots,\,\left\lfloor T_{n}/n_{T}\right\rfloor -2}\left(\log\left(T_{n}\right)n_{T}\right)^{1/2}\left|B_{h,b}^{0}\left(\widetilde{B}_{h,b+1}^{0}-B_{h,b+1}^{0}\right)\right|>\varepsilon\right)\nonumber \\
 & \leq\mathbb{P}\left(\max_{b=0,\ldots,\,\left\lfloor T_{n}/n_{T}\right\rfloor -2}\left|B_{h,b}^{0}\right|>K\right)+\mathbb{P}\left(\max_{b=0,\ldots,\,\left\lfloor T_{n}/n_{T}\right\rfloor -2}\left(\log\left(T_{n}\right)n_{T}\right)^{1/2}\left|\widetilde{B}_{h,b+1}^{0}-B_{h,b+1}^{0}\right|>\varepsilon/K\right),\label{eq. (55) BJV}
\end{align}
 where the first term converges to zero by the same argument as in
the last part of the proof of Lemma \ref{Prop QA1 BJV}. Therefore,
it remains to deal with the second term for which

\begin{align*}
\mathbb{P} & \left(\max_{b=0,\ldots,\,\left\lfloor T_{n}/n_{T}\right\rfloor -2}\left(\log\left(T_{n}\right)n_{T}\right)^{1/2}\left|\widetilde{B}_{h,b+1}^{0}-B_{h,b+1}^{0}\right|>\varepsilon/K\right)\\
 & =\mathbb{P}\left(\max_{b=0,\ldots,\left\lfloor T_{n}/n_{T}\right\rfloor -2}\frac{\left(\log\left(T_{n}\right)n_{T}\right)^{1/2}}{n_{T}h}\right.\\
 & \quad\times\left.\left|\sigma_{e,\left(T_{m}+\tau+bn_{T}-1\right)h}^{2}-\sigma_{e,\left(T_{m}+\tau+\left(b+1\right)n_{T}-1\right)h}^{2}\right|\sum_{j=1}^{n_{T}}\left(\Delta_{h}W_{e,T_{m}+\tau+\left(b+1\right)n_{T}+j-1}\right)^{2}>\varepsilon/K\right)\\
 & \leq\mathbb{P}\left(\max_{b=0,\ldots,\,\left\lfloor T_{n}/n_{T}\right\rfloor -2}\left(\log\left(T_{n}\right)n_{T}\right)^{1/2}\left|\sigma_{e,\left(T_{m}+\tau+bn_{T}-1\right)h}^{2}-\sigma_{e,\left(T_{m}+\tau+\left(b+1\right)n_{T}-1\right)h}^{2}\right|>\varepsilon/2K\right)\\
 & \quad+\mathbb{P}\left(\max_{b=0,\ldots,\,\left\lfloor T_{n}/n_{T}\right\rfloor -2}\left|\left(n_{T}h\right)^{-1}\sum_{j=1}^{n_{T}}\left(\Delta_{h}W_{e,T_{m}+\tau+\left(b+1\right)n_{T}+j-1}\right)^{2}\right|>2\right).
\end{align*}
 By Assumption \ref{Assumption Lipchtitz cont of Sigma}, Markov's
inequality and sufficiently large $r>0$,
\begin{align}
\mathbb{P} & \left(\max_{b=0,\ldots,\,\left\lfloor T_{n}/n_{T}\right\rfloor -2}\left(\log\left(T_{n}\right)n_{T}\right)^{1/2}\left|\sigma_{e,\left(T_{m}+\tau+bn_{T}-1\right)h}^{2}-\sigma_{e,\left(T_{m}+\tau+\left(b+1\right)n_{T}-1\right)h}^{2}\right|>\varepsilon/2K\right)\nonumber \\
 & \leq C_{r}\left(\frac{2K}{\varepsilon}\right)^{r}\left(\log\left(T_{n}\right)n_{T}\right)^{r/2}\left(T_{n}/n_{T}\right)\phi_{\sigma,n_{T}h,N}^{r}\rightarrow0.\label{eq (55b)}
\end{align}
 Finally, for all integers $r>0$, 
\begin{align}
\mathbb{P} & \left(\max_{b=0,\ldots,\,\left\lfloor T_{n}/n_{T}\right\rfloor -2}\left|\left(n_{T}h\right)^{-1}\sum_{j=1}^{n_{T}}\left(\Delta_{h}W_{e,T_{m}+\tau+\left(b+1\right)n_{T}+j-1}\right)^{2}\right|>2\right)\label{eq. (56) BJV}\\
 & \leq\sum_{b=0}^{\left\lfloor T_{n}/n_{T}\right|-2}\mathbb{P}\left(\left|n_{T}^{-1}\sum_{j=1}^{n_{T}}\left(\left(h^{-1/2}\Delta_{h}W_{e,T_{m}+\tau+\left(b+1\right)n_{T}+j-1}\right)^{2}-1\right)\right|^{r}>1\right)\nonumber \\
 & \leq\sum_{b=0}^{\left\lfloor T_{n}/n_{T}\right|-2}\mathbb{E}\left(\left|n_{T}^{-1}\sum_{j=1}^{n_{T}}\left(\left(h^{-1/2}\Delta_{h}W_{e,T_{m}+\tau+\left(b+1\right)n_{T}+j-1}\right)^{2}-1\right)\right|^{r}\right).\nonumber \\
 & \leq C_{r}\left(T_{n}/n_{T}\right)n_{T}^{-r/2}=C_{r}T_{n}n_{T}^{-1-r/2}\rightarrow0,\nonumber 
\end{align}
in view of Condition \ref{Cond The-auxiliary-sequence} by choosing
$r$ sufficiently large. Using this together with \eqref{eq (55b)}
into \eqref{eq. (55) BJV} we deduce \eqref{eq. (54b) in Prop A.2}.
$\square$
\begin{lem}
\label{Lemma Prop A.3 in BJV}As $h\downarrow0$, $\left(\log\left(T_{n}\right)n_{T}\right)^{1/2}\left(\mathrm{V}_{\mathrm{max},h}\left(T_{n},\,\tau\right)-\widetilde{\mathrm{B}}_{\mathrm{max},h}^{0}\left(T_{n},\,\tau\right)\right)\overset{\mathbb{P}}{\rightarrow}0$.
\end{lem}

\noindent \textit{Proof.} Note that 
\begin{align*}
\mathrm{V}_{\mathrm{max},h}\left(T_{n},\,\tau\right)-\widetilde{\mathrm{B}}_{\mathrm{max},h}^{0}\left(T_{n},\,\tau\right) & =\max_{b=0,\ldots,\,\left\lfloor T_{n}/n_{T}\right\rfloor -2}\left|\frac{\widetilde{B}_{h,b+1}^{0}-B_{h,b}^{0}}{\sigma_{e,\left(T_{m}+\tau+bn_{T}-1\right)h}^{2}}\right|-\max_{b=0,\ldots,\,\left\lfloor T_{n}/n_{T}\right\rfloor -2}\left|B_{h,b}^{0}/\widetilde{B}_{h,b+1}^{0}-1\right|\\
 & \leq\max_{b=0,\ldots,\,\left\lfloor T_{n}/n_{T}\right\rfloor -2}\left|\frac{\left(\widetilde{B}_{h,b+1}^{0}-B_{h,b}^{0}\right)\left(\widetilde{B}_{h,b+1}^{0}-\sigma_{e,\left(T_{m}+\tau+bn_{T}-1\right)h}^{2}\right)}{\sigma_{e,\left(T_{m}+\tau+bn_{T}-1\right)h}^{2}\widetilde{B}_{h,b+1}^{0}}\right|,
\end{align*}
and thus we show 
\begin{align}
\left(\log\left(T_{n}\right)n_{T}\right)^{1/2}\max_{b=0,\ldots,\,\left\lfloor T_{n}/n_{T}\right\rfloor -2}\left|\frac{\left(B_{h,b+1}^{0}-\widetilde{B}_{h,b}^{0}\right)\left(\widetilde{B}_{h,b+1}^{0}-\sigma_{e,\left(T_{m}+\tau+bn_{T}-1\right)h}^{2}\right)}{\sigma_{e,\left(T_{m}+\tau+bn_{T}-1\right)h}^{2}\widetilde{B}_{h,b+1}^{0}}\right| & =o_{\mathbb{P}}\left(1\right).\label{eq. Eq. in Prop A.3}
\end{align}
By the boundedness of $\sigma_{e,t},\,t\geq0$, and upon using the
same arguments as in the previous lemmas for $\widetilde{B}_{h,b}^{0},$
the denominator is $O_{\mathbb{P}}\left(1\right)$. Since for any
$\varepsilon>0,$
\begin{align*}
\mathbb{P} & \left(\left(\log\left(T_{n}\right)n_{T}\right)^{1/4}\max_{b=0,\ldots,\,\left\lfloor T_{n}/n_{T}\right\rfloor -2}\left|\widetilde{B}_{h,b+1}^{0}-B_{h,b}^{0}\right|>\sqrt{\varepsilon}\right)\\
 & \leq\mathbb{P}\left(\max_{b=0,\ldots,\left\lfloor T_{n}/n_{T}\right\rfloor -2}\frac{\left(\log\left(T_{n}\right)n_{T}\right)^{1/4}}{n_{T}h}\right.\\
 & \quad\times\left.\sigma_{e,\left(T_{m}+\tau+bn_{T}-1\right)h}^{2}\left|\sum_{j=1}^{n_{T}}\left(\left(\Delta_{h}W_{e,T_{m}+\tau+bn_{T}+j-1}\right)^{2}-\left(\Delta_{h}W_{e,T_{m}+\tau+\left(b+1\right)n_{T}+j-1}\right)^{2}\right)\right|>\sqrt{\varepsilon}\right)\\
 & \leq\mathbb{P}\left(\max_{b=0,\ldots,\left\lfloor T_{n}/n_{T}\right\rfloor -2}\frac{\left(\log\left(T_{n}\right)n_{T}\right)^{1/4}}{n_{T}h}\sigma_{e,\left(T_{m}+\tau+bn_{T}-1\right)h}^{2}\left|\sum_{j=1}^{n_{T}}\left(\Delta_{h}W_{e,T_{m}+\tau+bn_{T}+j-1}\right)^{2}-1\right|>\sqrt{\varepsilon}/2\right)\\
 & \quad+\mathbb{P}\left(\max_{b=0,\ldots,\left\lfloor T_{n}/n_{T}\right\rfloor -2}\frac{\left(\log\left(T_{n}\right)n_{T}\right)^{1/4}}{n_{T}h}\sigma_{e,\left(T_{m}+\tau+bn_{T}-1\right)h}^{2}\left|\sum_{j=1}^{n_{T}}\left(\Delta_{h}W_{e,T_{m}+\tau+\left(b+1\right)n_{T}+j-1}\right)^{2}-1\right|>\sqrt{\varepsilon}/2\right).
\end{align*}
We consider the first probability term; the argument for the second
is analogous. By using a similar argument as in \eqref{eq. (56) BJV},
the first term is less than
\begin{align*}
\sum_{b=0}^{\left\lfloor T_{n}/n_{T}\right|-2}\mathbb{P} & \left(\left(\log\left(T_{n}\right)n_{T}\right)^{1/4}\sigma_{e,\left(T_{m}+\tau+bn_{T}-1\right)h}^{2}n_{T}^{-1}\left|\sum_{j=1}^{n_{T}}\left(h^{-1/2}\Delta_{h}W_{e,T_{m}+\tau+\left(b+1\right)n_{T}+j-1}\right)^{2}-1\right|>\sqrt{\varepsilon}/2\right)\\
 & \leq C_{r}\left(\frac{2}{\sqrt{\varepsilon}}\right)^{r}\left(\log\left(T_{n}\right)n_{T}\right)^{r/4}\sum_{b=0}^{\left\lfloor T_{n}/n_{T}\right|-2}\mathbb{E}\left(\left|n_{T}^{-1}\sum_{j=1}^{n_{T}}\left(\left(h^{-1/2}\Delta_{h}W_{e,T_{m}+\tau+\left(b+1\right)n_{T}+j-1}\right)^{2}-1\right)\right|^{r}\right)\\
 & =\left(\frac{2}{\sqrt{\varepsilon}}\right)^{r}\left(\log\left(T_{n}\right)\right)^{r/4}\left(T_{n}/n_{T}\right)n_{T}^{-r/4},
\end{align*}
 which goes to zero by choosing $r>0$ sufficiently large. It remains
to show that
\begin{align}
\mathbb{P}\left(\left(\log\left(T_{n}\right)n_{T}\right)^{1/4}\max_{b=0,\ldots,\,\left\lfloor T_{n}/n_{T}\right\rfloor -2}\left|\widetilde{B}_{h,b+1}^{0}-\sigma_{e,\left(T_{m}+\tau+bn_{T}-1\right)h}^{2}\right|>\varepsilon^{1/2}\right) & \rightarrow0.\label{eq Eq. Prop A.3b}
\end{align}
Simple manipulations yield for some $C<\infty,$ with $\sigma_{+}\leq\sqrt{C}$,
\begin{align*}
\mathbb{P} & \left(\left(\log\left(T_{n}\right)n_{T}\right)^{1/4}\max_{b=0,\ldots,\,\left\lfloor T_{n}/n_{T}\right\rfloor -2}\left|\widetilde{B}_{h,b+1}^{0}-\sigma_{e,\left(T_{m}+\tau+bn_{T}-1\right)h}^{2}\right|>\varepsilon^{1/2}\right)\\
 & \leq\mathbb{P}\left(\left(\log\left(T_{n}\right)n_{T}\right)^{1/4}\max_{b=0,\ldots,\,\left\lfloor T_{n}/n_{T}\right\rfloor -2}\sigma_{e,\left(T_{m}+\tau+bn_{T}-1\right)h}^{2}\left|\frac{1}{n_{T}}\sum_{j=1}^{n_{T}}\left(\left(h^{-1/2}\Delta_{h}W_{e,T_{m}+\tau+\left(b+1\right)n_{T}+j-1}\right)^{2}-1\right)\right|>\sqrt{\varepsilon}\right)\\
 & \leq C^{r}\left(\frac{1}{\sqrt{\varepsilon}}\right)^{r}\left(\log\left(T_{n}\right)n_{T}\right)^{r/4}\sum_{b=0}^{\left\lfloor T_{n}/n_{T}\right|-2}\mathbb{E}\left(\left|n_{T}^{-1}\sum_{j=1}^{n_{T}}\left(\left(h^{-1/2}\Delta_{h}W_{e,T_{m}+\tau+\left(b+1\right)n_{T}+j-1}\right)^{2}-1\right)\right|^{r}\right)\\
 & \leq C_{r}\left(\frac{2}{\sqrt{\varepsilon}}\right)^{r}\left(\log\left(T_{n}\right)\right)^{r/4}\left(T_{n}/n_{T}\right)n_{T}^{-r/4}\rightarrow0.
\end{align*}
We have \eqref{eq Eq. Prop A.3b} and thus \eqref{eq. Eq. in Prop A.3},
which concludes the proof. $\square$ 

\medskip{}
From Lemma \ref{Lemma QA1}-\ref{Lemma Prop A.3 in BJV} we deduce
$\sqrt{\log\left(T_{n}\right)n_{T}}\left(\mathrm{B}_{\mathrm{max},h}\left(T_{n},\,\tau\right)-\mathrm{V}_{\mathrm{max},h}\left(T_{n},\,\tau\right)\right)=o_{\mathbb{P}}\left(1\right),$
where $\mathrm{V}_{\mathrm{max},h}\left(T_{n},\,\tau\right)$ was
defined in \eqref{eq. Def Vmax (QB)}. By the properties of the Wiener
process, for each block $b$ the variables $\left(\Delta_{h}W_{e,T_{m}+\tau+bn_{T}+j-1}\right)^{2}$
are a sequence of $\chi_{1}^{2}$ random variables which are independent
over $j$. After centering these variables, (i.e., $\left\{ \left(\Delta_{h}W_{e,T_{m}+\tau+bn_{T}+j-1}\right)^{2}-1\right\} $),
we can apply the results in Lemma 1-2 in \citet{wu/zhao:07}. This
leads us to a limit theorem for the statistic $\mathrm{V}_{\mathrm{max},h}\left(T_{n},\,\tau\right)$
which takes a similar form to the statistic in equation (13) of \citet{wu/zhao:07}.
Therefore, in Lemma \ref{Lemma QB Prop A.5.} we provide a limit theorem
which adapts Theorem 1 of \citet{wu/zhao:07} to our context. The
difference hinges on (i) the dependence structure of the variables
$\left\{ \left(\Delta_{h}W_{e,T_{m}+\tau+bn_{T}+j-1}\right)^{2}-1\right\} _{i=1}^{n_{T}}$
relative to the sequence $\left\{ X_{k}\right\} _{k\geq1}$ appearing
in \citet{wu/zhao:07}, and on (ii) the form of our test statistics
which allow both for additive and multiplicative structure. For the
quadratic loss case, our problem is then similar to that of \citet{bibinger/jirak/vetter:16}
who also uses Lemma 1-2 in \citet{wu/zhao:07}; yet even in the quadratic
loss case our context differs from that of \citet{bibinger/jirak/vetter:16}
because we allow for model misspecification via the additional term
$\mu_{e,t}$ in \eqref{Discretized Model Y} and estimation of $\widehat{\beta}_{k}$.
\begin{assumption}
\label{Assumption 4th moment fo forecast error}The sequence of rescaled
forecasts errors $\left\{ \Delta_{h}\widetilde{e}_{k}^{*}\right\} _{k\geq1}$
satisfies, for some $p\geq4$, $\mathbb{E}\left[\left|\Delta_{h}\widetilde{e}_{k}^{*}\right|^{p}\right]<\infty$
for all $k\geq1$. Furthermore, the sequence of forecast losses $\left\{ L_{\psi,kh}\right\} _{k\geq1}$
satisfies the same assumption. 
\end{assumption}
We now explain how to verify Assumption \ref{Assumption 4th moment fo forecast error}. 
\begin{lem}
\label{Lemma verify Assumption 4th moments}Given the model in \eqref{Discretized Model Y},
Assumption \ref{Assumption 4th moment fo forecast error} holds. 
\end{lem}

\noindent \textit{Proof.} We know that $\Delta_{h}e_{k}^{*}=\int_{\left(k-1\right)h}^{kh}\mu_{e,s}h^{-\vartheta}ds+\int_{\left(k-1\right)h}^{kh}\sigma_{e,s}dW_{e,s}$.
Note that conditional on $\left\{ \mu_{e,t}\right\} _{t\geq0}$ and
$\left\{ \sigma_{e,t}\right\} _{t\geq0}$, 
\begin{align}
\left(\Delta_{h}e_{k}^{*}\right)^{2} & =\left(\int_{\left(k-1\right)h}^{kh}\mu_{e,s}h^{-\vartheta}ds\right)^{2}+\left(\int_{\left(k-1\right)h}^{kh}\sigma_{e,s}dW_{e,s}\right)^{2}+2\int_{\left(k-1\right)h}^{kh}\int_{\left(k-1\right)h}^{kh}\mu_{e,v}h^{-\vartheta}\sigma_{e,s}dvdW_{e,s}\nonumber \\
 & =O\left(h^{2\left(1-\vartheta\right)}\right)+\left(\int_{\left(k-1\right)h}^{kh}\sigma_{e,s}dW_{e,s}\right)^{2}+O_{\mathbb{P}}\left(h^{3/2-\vartheta}\right)\nonumber \\
 & =o\left(h\right)+\left(\int_{\left(k-1\right)h}^{kh}\sigma_{e,s}dW_{e,s}\right)^{2}+o_{\mathbb{P}}\left(h\right).\label{eq (1) Neg Drfit}
\end{align}
Hence, $\mathbb{E}\left[\left|\Delta_{h}e_{k}^{*}\right|^{p}|\,\mathscr{F}_{\left(k-1\right)h}\right]=\mathbb{E}\left[\left|\int_{\left(k-1\right)h}^{kh}\sigma_{e,s}dW_{e,s}\right|^{p}|\,\mathscr{F}_{\left(k-1\right)h}\right]+C_{p}o_{\mathbb{P}}\left(h^{p/2}\right)$
and Assumption \ref{Assumption 4th moment fo forecast error} is verified
given the properties of the Wiener process and $\psi_{h}=h^{1/2}$.
$\square$
\begin{lem}
\label{Lemma QB Prop A.5.}For $n=1,\ldots,\,T_{n}$, let $\mu_{n}=\mu\left(n_{T}/T_{n}\right)$
with $\mu\in\boldsymbol{Lip}\left(\left[0,\,1\right]\right)$. Let
$\left\{ U_{n}\right\} _{n\geq1}$ denote a sequence of i.n.d. random
variables with $U_{n}=\mu_{n}+\widetilde{U}_{n}$, $\mathbb{E}\left(\widetilde{U}_{n}\right)=0$,
$\mathrm{Var}\left(\widetilde{U}_{n}\right)=\sigma_{\widetilde{U}}^{2}$
and $\mathbb{E}\left[\left|\widetilde{U}_{n}\right|^{p}\right]<\infty$
for some $p\geq4.$ Set $m_{T}=\left\lfloor T_{n}/n_{T}\right\rfloor $
and define 
\begin{align*}
\mathrm{B}_{\mathrm{max},T_{n}} & \triangleq\frac{1}{n_{T}}\max_{0\leq b\leq\left\lfloor T_{n}/n_{T}\right\rfloor -2}\left|\sum_{j=1}^{n_{T}}\left(U_{\left(b+1\right)n_{T}+j}-U_{bn_{T}+j}\right)\right|,
\end{align*}
 and 
\begin{align*}
\mathrm{MB}_{\mathrm{max},T_{n}} & \triangleq\frac{1}{n_{T}}\max_{n_{T}\leq i\leq T_{n}-n_{T}}\left|\sum_{j=i+1}^{n_{T}+i}U_{j}-\sum_{j=i-n_{T}+1}^{i}U_{j}\right|.
\end{align*}
If the following condition holds,
\begin{align}
n_{T}^{-p/2}T_{n} & =o\left(\left(\log\left(T_{n}\right)\right)^{-p/2}\right),\label{eq. 66 BJV}
\end{align}
 then
\begin{align}
\sqrt{\log\left(m_{T}\right)}\left(\frac{\sqrt{n_{T}}}{\sigma_{\widetilde{U}}}\mathrm{B}_{\mathrm{max},T_{n}}-\gamma_{m_{T}}\right) & \overset{}{\Rightarrow}\mathscr{V},\label{eq QB Prop A5 result}
\end{align}
 and 
\begin{align}
\sqrt{\log\left(m_{T}\right)}\left(\frac{\sqrt{n_{T}}}{\sigma_{\widetilde{U}}^ {}}\mathrm{MB}_{\mathrm{max},T_{n}}-2\log\left(m_{T}\right)-\frac{1}{2}\log\log\left(m_{T}\right)-\log3\right) & \overset{}{\Rightarrow}\mathscr{V},\label{QB Prop A5 result MB}
\end{align}
where $\gamma_{m_{T}}=\left[4\log\left(m_{T}\right)-2\log\left(\log\left(m_{T}\right)\right)\right]^{1/2}$
and $\mathscr{V}$ satisfies $\mathbb{P}\left(\mathscr{V}\leq v\right)=\exp\left(-\pi^{-1/2}\exp\left(-v\right)\right).$ 
\end{lem}

\noindent \textit{Proof.} Without loss of generality, we set $\sigma_{\widetilde{U}}=1$.
By the Donsker-Prokhorov invariance principle $T_{n}^{-1/2}\sum_{j=1}^{\left\lfloor sT_{n}\right\rfloor }\widetilde{U}_{j}\Rightarrow\mathbb{B}\left(s\right)$,
where $\left\{ \mathbb{B}\left(s\right)\right\} _{s\in\left[0,\,1\right]}$
is a standard Wiener process on $\left[0,\,1\right]$. Then, we have
by definition that $Z_{b+1}\triangleq n_{T}^{-1/2}\left(\mathbb{B}\left(\left(b+1\right)n_{T}\right)-\mathbb{B}\left(bn_{T}\right)\right)$,
$b=0,\ldots m_{T}-1$, are i.i.d. standard normal random variables.
We have the decomposition 
\begin{align*}
n_{T}^{-1}\sum_{j=1}^{n_{T}}U_{\left(b+1\right)n_{T}+j} & =\frac{Z_{b+1}}{\sqrt{n_{T}}}+\frac{1}{n_{T}}\sum_{j=1}^{n_{T}}\mu_{\left(b+1\right)n_{T}+j}+\frac{R_{b+1,n_{T}}}{n_{T}},
\end{align*}
 where $R_{b,T_{n}}\triangleq\sum_{j=1}^{bn_{T}}\widetilde{U}_{j}-\mathbb{B}\left(bn_{T}\right)-\left(\sum_{j=1}^{\left(b-1\right)n_{T}}\widetilde{U}_{j}-\mathbb{B}\left(\left(b-1\right)n_{T}\right)\right)$
and recall $\widetilde{U}_{j}=U_{j}-\mu_{j}$. By the strong invariance
principle of \citet{komlos:75}, $\max_{b\leq m_{T}-1}\left|R_{b+1,T_{n}}\right|=o_{\mathrm{a.s.}}\left(T_{n}^{1/p}\right)$,
where we have used the independence structure of $\left\{ \widetilde{U}_{j}\right\} $.
Since $\mu\in\boldsymbol{Lip}\left(\left[0,\,1\right]\right)$, we
have uniformly over $b$ and $j$, $n_{T}^{-1}\sum_{j=1}^{n_{T}}\left(\mu_{\left(b+1\right)n_{T}+j}-\mu_{bn_{T}+j}\right)=O\left(n_{T}/T_{n}\right)$.
Altogether, 
\begin{align*}
n_{T}^{-1}\sum_{j=1}^{n_{T}}\left(U_{\left(b+1\right)n_{T}+j}-U_{bn_{T}+j}\right) & =Z_{b+1}-Z_{b}+O_{\mathrm{a.s.}}\left(n_{T}^{3/2}/T_{n}+n_{T}^{-1/2}T_{n}^{1/p}\right)\\
 & =Z_{b+1}-Z_{b}+o_{\mathrm{a.s.}}\left(\left(\log\left(m_{T}\right)\right)^{-1/2}\right).
\end{align*}
 The result in equation \eqref{eq QB Prop A5 result} then follows
from Lemma 1 in \citet{wu/zhao:07}. 

We now turn to the corresponding result for the overlapping case.
We redefine $\left\{ Z_{j}\right\} _{j\geq1}$ as being a sequence
of standard normal random variables. Then, 
\begin{align*}
\max_{n_{T}\leq i\leq T_{n}-n_{T}} & \left|\sum_{j=i+1}^{n_{T}+i}\left(U_{j}-Z_{j}\right)-\sum_{j=n_{T}-i+1}^{i}\left(U_{j}-Z_{j}\right)\right|\\
 & =\max_{n_{T}\leq i\leq T_{n}-n_{T}}\left|\sum_{j=i+1}^{n_{T}+i}\left(U_{j}-\mu_{j}-Z_{j}\right)-\sum_{j=n_{T}-i+1}^{i}\left(U_{j}-\mu_{j}-Z_{j}\right)+\sum_{j=i+1}^{n_{T}+i}\mu_{j}-\sum_{j=n_{T}-i+1}^{i}\mu_{j}\right|\\
 & \leq4\max_{n_{T}\leq i\leq T_{n}-n_{T}}\left|\sum_{j=1}^{i}\left(\widetilde{U}_{j}-Z_{j}\right)\right|+\max_{n_{T}\leq i\leq T_{n}-n_{T}}\left|\sum_{j=i+1}^{n_{T}+i}\mu_{j}-\sum_{j=n_{T}-i+1}^{i}\mu_{j}\right|\\
 & =4\max_{n_{T}\leq i\leq T_{n}-n_{T}}\left|\sum_{j=1}^{i}\left(\widetilde{U}_{j}-Z_{j}\right)\right|+O\left(n_{T}^{2}/T_{n}\right),
\end{align*}
where the last equality follows from $\mu\in\boldsymbol{Lip}\left(\left[0,\,1\right]\right)$
and $O\left(n_{T}^{2}/T_{n}\right)$ being uniform. Next, we use Theorem
4 of \citet{komlos:76} to derive a bound on the approximation error
for the first term above. Let $\left\{ a_{T_{n}}\right\} _{T_{n}\in\mathbb{N}}$
be a positive sequence. By Markov's inequality,
\begin{align*}
\mathbb{P}\left(\max_{n_{T}\leq i\leq T_{n}}\left|\sum_{j=1}^{i}\left(\widetilde{U}_{j}-Z_{j}\right)\right|\geq a_{T_{n}}\right) & \leq C_{1,p}\frac{1}{a_{T_{n}}^{p}}\sum_{j=1}^{T_{n}}\mathbb{E}\left(\left|\widetilde{U}_{j}\right|^{p}\right)\leq C_{2,p}\frac{T_{n}}{a_{T_{n}}^{p}},
\end{align*}
where $C_{1,p},\,C_{2,p}<\infty$. The conditions of Theorem 4 in
\citet{komlos:76} are satisfied if we set $a_{T_{n}}=\sqrt{n_{T}/\log\left(T_{n}\right)}$.
This leads to
\begin{align}
\max_{n_{T}\leq i\leq T_{n}-n_{T}}\left|\sum_{j=i+1}^{n_{T}+i}\left(\widetilde{U}_{j}-Z_{j}\right)-\sum_{j=n_{T}-i+1}^{i}\left(\widetilde{U}_{j}-Z_{j}\right)\right| & =o_{\mathbb{P}}\left(\sqrt{n_{T}}\left(\log\left(T_{n}\right)\right)^{-1/2}\right),\label{eq (67) Prop A.5 BJV}
\end{align}
 where we have used \eqref{eq. 66 BJV}. Let $\mathbb{B}\left(i\right)=\sum_{j=1}^{i}Z_{j}$
and define $H\left(u\right)\triangleq\left(\mathbf{1}\left(0\leq u<1\right)-\mathbf{1}\left(-1<u<0\right)\right)/\sqrt{2}$.
Use \eqref{eq (67) Prop A.5 BJV} to deduce that, 
\begin{align*}
\frac{\sqrt{n_{T}}\mathrm{MB}_{\mathrm{max},T_{n}}}{\sqrt{2}} & =\frac{1}{\sqrt{2n_{T}}}\max_{n_{T}\leq i\leq T_{n}-n_{T}}\left|\sum_{j=i+1}^{n_{T}+i}\widetilde{U}_{j}-\sum_{j=i-n_{T}+1}^{i}\widetilde{U}_{j}\right|+O\left(n_{T}^{3/2}/T_{n}\right)\\
 & =\frac{1}{\sqrt{2n_{T}}}\max_{n_{T}\leq i\leq T_{n}-n_{T}}\left|\mathbb{B}\left(i+n_{T}\right)-\mathbb{B}\left(i\right)-\left(\mathbb{B}\left(i\right)-\mathbb{B}\left(i-n_{T}\right)\right)\right|+o_{\mathbb{P}}\left(\frac{\left(\log\left(T_{n}\right)\right)^{-1/2}}{\sqrt{2}}\right).
\end{align*}
 Therefore, letting $\varrho_{n}=\sup\left\{ \left|\mathbb{B}\left(u\right)-\mathbb{B}\left(u'\right)\right|:\,u,\,u'\in\left[0,\,T_{n}\right],\,\left|u-u'\right|\leq1\right\} $,
we have 
\begin{align*}
\frac{\sqrt{n_{T}}\mathrm{MB}_{\mathrm{max},T_{n}}}{\sqrt{2}} & =\frac{1}{\sqrt{2n_{T}}}\max_{n_{T}\leq i\leq T_{n}-n_{T}}\left|\int_{\mathbb{R}}H\left(\frac{s-u}{n_{T}}\right)d\mathbb{B}\left(u\right)\right|+\frac{O\left(\varrho{}_{n}\right)}{\sqrt{n_{T}}}+\frac{o_{\mathbb{P}}\left(1\right)}{\sqrt{\log\left(T_{n}\right)}}.
\end{align*}
By the global modulus of continuity of the standard Wiener process
{[}cf. Theorem 2.9.25 in \citet{karatzas/shreve:96}{]}, we know that
$\varrho_{n}=o_{\mathbb{P}}\left(\sqrt{\log\left(T_{n}\right)}\right)$.
The result for the overlapping case then follows from Lemma 2 in \citet{wu/zhao:07}
with $\alpha=1$, $D_{H,1}=3$, bandwidth $b_{n}=m_{T}^{-1}$ and
$n=T_{n}$; see their Definition 1 as well and note that their lemma
can be applied because $\left(\log\left(T_{n}\right)\right)^{6}=o\left(n_{T}\right)$
holds by condition \eqref{eq. 66 BJV}. $\square$ 

\medskip{}

\noindent\textit{Proof of Theorem \ref{Theoem Asymptotic H0 Distrbution Bmax and Qmax}-(i).}
From Lemma \ref{Lemma QA1}-\ref{Lemma Prop A.3 in BJV}, $\sqrt{\log\left(T_{n}\right)n_{T}}\left(\mathrm{B}_{\mathrm{max},h}\left(T_{n},\,\tau\right)-\mathrm{V}_{\mathrm{max},h}\left(T_{n},\,\tau\right)\right)=o_{\mathbb{P}}\left(1\right)$.
Lemma \ref{Lemma verify Assumption 4th moments} shows that Assumption
\ref{Assumption 4th moment fo forecast error} holds. Then, under
Condition \ref{Cond The-auxiliary-sequence}, we can apply Lemma \ref{Lemma QB Prop A.5.}
to $\mathrm{V}_{\mathrm{max},h}\left(T_{n},\,\tau\right)$ which in
turn leads to the result for $\mathrm{B}_{\mathrm{max},h}\left(T_{n},\,\tau\right)$
in part (i) of Theorem \ref{Theoem Asymptotic H0 Distrbution Bmax and Qmax}.
$\square$

\paragraph{Proof of part (ii) of Theorem \ref{Theoem Asymptotic H0 Distrbution Bmax and Qmax}}

The proof can be simplified considerably by using arguments similar
to those of part (i) of Theorem \ref{Theoem Asymptotic H0 Distrbution Bmax and Qmax}.
Let 
\begin{align}
\mathrm{MB}_{\mathrm{max},h}^{*}\left(T_{n},\,\tau\right) & =\max_{i=n_{T},\ldots,\,T_{n}-n_{T}}\left|\frac{n_{T}^{-1}\sum_{j=i-n_{T}+1}^{i}\left(\Delta_{h}\widetilde{e}_{T_{m}+\tau+j-1}^{*}\right)^{2}}{n_{T}^{-1}\sum_{j=i+1}^{i+n_{T}}\left(\Delta_{h}\widetilde{e}_{T_{m}+\tau+j-1}^{*}\right)^{2}}-1\right|,\label{eq (44) BJV}
\end{align}
 and 
\begin{align}
\mathrm{MB}_{\mathrm{max},h}^{0}\left(T_{n},\,\tau\right) & =\max_{i=n_{T},\ldots,\,T_{n}-n_{T}}\left|\frac{n_{T}^{-1}\sum_{j=i-n_{T}+1}^{i}\sigma_{e,\left(T_{m}+\tau+i-n_{T}-1\right)h}^{2}\left(h^{-1/2}\Delta_{h}W_{e,T_{m}+\tau+j-1}\right)^{2}}{n_{T}^{-1}\sum_{j=i+1}^{i+n_{T}}\sigma_{e,\left(T_{m}+\tau+i-1\right)h}^{2}\left(h^{-1/2}\Delta_{h}W_{e,T_{m}+\tau+j-1}\right)^{2}}-1\right|.\label{eq MB*}
\end{align}
\begin{lem}
\label{Prop A.4 BJV}$\sqrt{\log\left(T_{n}\right)n_{T}}\left(\mathrm{MB}_{\mathrm{max},h}\left(T_{n},\,\tau\right)-\mathrm{MB}{}_{\mathrm{max},h}^{0}\left(T_{n},\,\tau\right)\right)\overset{\mathbb{P}}{\rightarrow}0.$ 
\end{lem}

\noindent \textit{Proof.} Note that the choice of overlapping blocks
does not alter the results of Lemma \ref{Lemma QA1}-\ref{Lemma QA3},
which in turn give $\left(\log\left(T_{n}\right)n_{T}\right)^{1/2}\left(\mathrm{MB}_{\mathrm{max},h}\left(T_{n},\,\tau\right)-\mathrm{MB}_{\mathrm{max},h}^{*}\left(T_{n},\,\tau\right)\right)\overset{\mathbb{P}}{\rightarrow}0$.
Thus, we can begin by proving a result analogous to Lemma \ref{Prop QA1 BJV}:
$\left(\log\left(T_{n}\right)n_{T}\right)^{1/2}\left(\mathrm{MB}_{\mathrm{max},h}^{*}\left(T_{n},\,\tau\right)-\mathrm{MB}_{\mathrm{max},h}^{0}\left(T_{n},\,\tau\right)\right)\overset{\mathbb{P}}{\rightarrow}0$.
Note that proceeding as in \eqref{eq. 47 BJV}, we have 
\begin{align*}
\left|\right. & \left.\mathrm{MB}_{\mathrm{max},h}^{*}\left(T_{n},\,\tau\right)-\mathrm{MB}_{\mathrm{max},h}^{0}\left(T_{n},\,\tau\right)\right|\\
 & \leq\max_{i=n_{T},\ldots,\,T_{n}-n_{T}}\left|n_{T}^{-1}\sum_{j=i-n_{T}+1}^{i}\left(\Delta_{h}\widetilde{e}_{T_{m}+\tau+j-1}^{*}\right)^{2}\right.\\
 & \quad\times\left.\left(\frac{1}{n_{T}^{-1}\sum_{j=i+1}^{i+n_{T}}\left(\Delta_{h}\widetilde{e}_{T_{m}+\tau+j-1}^{*}\right)^{2}}-\frac{1}{n_{T}^{-1}\sum_{j=i+1}^{i+n_{T}}\sigma_{e,\left(T_{m}+\tau+i-1\right)h}^{2}\left(h^{-1/2}\Delta_{h}W_{e,T_{m}+\tau+j-1}\right)^{2}}\right)\right|\\
 & \quad+\max_{i=n_{T},\ldots,\,T_{n}-n_{T}}\left|\frac{n_{T}^{-1}\sum_{j=i-n_{T}+1}^{i}\left(\left(\Delta_{h}\widetilde{e}_{T_{m}+\tau+j-1}^{*}\right)^{2}-\sigma_{e,\left(T_{m}+\tau+i-n_{T}-1\right)h}^{2}\left(h^{-1/2}\Delta_{h}W_{e,T_{m}+\tau+j-1}\right)^{2}\right)}{n_{T}^{-1}\sum_{j=i+1}^{i+n_{T}}\sigma_{e,\left(T_{m}+\tau+i-1\right)h}^{2}\left(h^{-1/2}\Delta_{h}W_{e,T_{m}+\tau+j-1}\right)^{2}}\right|.
\end{align*}
Then, we can use the same decomposition as in \eqref{eq (48 or 69) BJV},
\begin{align*}
\mathbb{P} & \left(\max_{i=n_{T},\ldots,\,T_{n}-n_{T}}\left(\log\left(T_{n}\right)n_{T}\right)^{1/2}\right.\\
 & \quad\times\left.\left|\frac{n_{T}^{-1}\sum_{j=i-n_{T}+1}^{i}\left(\left(\Delta_{h}\widetilde{e}_{T_{m}+\tau+j-1}^{*}\right)^{2}-\sigma_{e,\left(T_{m}+\tau+i-n_{T}-1\right)h}^{2}\left(h^{-1/2}\Delta_{h}W_{e,T_{m}+\tau+j-1}\right)^{2}\right)}{n_{T}^{-1}\sum_{j=i+1}^{i+n_{T}}\sigma_{e,\left(T_{m}+\tau+i-1\right)h}^{2}\left(h^{-1/2}\Delta_{h}W_{e,T_{m}+\tau+j-1}\right)^{2}}\right|>\varepsilon\right)\\
 & \leq\mathbb{P}\left(\max_{i=n_{T},\ldots,\,T-n_{T}}\left(\log\left(T_{n}\right)n_{T}\right)^{1/2}\left|n_{T}^{-1}\sum_{j=i-n_{T}+1}^{i}\left(\left(\Delta_{h}\widetilde{e}_{T_{m}+\tau+j-1}^{*}\right)^{2}\right.\right.\right.\\
 & \quad\left.\left.\left.-\sigma_{e,\left(T_{m}+\tau+i-n_{T}-1\right)h}^{2}\left(h^{-1/2}\Delta_{h}W_{e,T_{m}+\tau+j-1}\right)^{2}\right)\right|>\varepsilon/K\right)\\
 & \quad+\mathbb{P}\left(\min_{i=n_{T},\ldots,\,T-n_{T}}\left|n_{T}^{-1}\sum_{j=i+1}^{i+n_{T}}\sigma_{e,\left(T_{m}+\tau+i-1\right)h}^{2}\left(h^{-1/2}\Delta_{h}W_{e,T_{m}+\tau+j-1}\right)^{2}\right|<1/K\right),
\end{align*}
 which holds for any $\varepsilon>0$ and any constant $K>0$. Using
the same reasoning as in the proof involving the second term of \eqref{eq. 47 BJV}
and choosing $K$ appropriately, we have for the second term, 
\begin{align*}
\mathbb{P}\left(\max_{i=n_{T},\ldots,\,T-n_{T}}\left|n_{T}^{-1}\sum_{j=i+1}^{i+n_{T}}\sigma_{e,\left(T_{m}+\tau+i-1\right)h}^{2}\left(h^{-1/2}\Delta_{h}W_{e,T_{m}+\tau+j-1}\right)^{2}\right|>K\right) & \rightarrow0.
\end{align*}
Thus, it remains to consider the first term on the right-hand side
above. For the non-overlapping case it was treated in \eqref{eq. (49) BJV}
and its final bound can be obtained from \eqref{eq (49b)}-\eqref{eq (49d)}.
However, for the overlapping block case, the maximum is over a larger
number of arguments. Indeed, the final bound is an order $O\left(n_{T}\right)$
larger than the one for the non-overlapping case. Nonetheless, the
same conclusion holds upon choosing $r$ large enough there: 
\begin{align}
\mathbb{P} & \left(\max_{i=n_{T},\ldots,\,T-n_{T}}\left(\log\left(T_{n}\right)n_{T}\right)^{1/2}\right.\label{eq (p.32)}\\
 & \quad\times\left.\left|n_{T}^{-1}\sum_{j=i-n_{T}+1}^{i}\left(\left(\Delta_{h}\widetilde{e}_{T_{m}+\tau+j-1}^{*}\right)^{2}-\sigma_{e,\left(T_{m}+\tau+i-n_{T}-1\right)h}^{2}\left(h^{-1/2}\Delta_{h}W_{e,T_{m}+\tau+j-1}\right)^{2}\right)\right|>\varepsilon/K\right)\nonumber \\
 & \quad\rightarrow0.\nonumber 
\end{align}
Generalizing the arguments that led to \eqref{eq (p.32)} and noting
that the bounds involving the Lipschitz continuity of $\left\{ \sigma_{e,t}\right\} _{t\geq0}$
remain the same as in the non-overlapping case, the corresponding
results in Lemma \ref{Prop QA1 BJV}-\ref{Lemma Prop A.3 in BJV}
can be verified. This together with Lemma \ref{Lemma QA1}-\ref{Lemma QA3}\textemdash which
are valid for both cases with minor changes in notation\textemdash yield
the conclusion of the lemma. $\square$

\medskip{}

\noindent\textit{Proof of Theorem \ref{Theoem Asymptotic H0 Distrbution Bmax and Qmax}-(ii).}
From Lemma \ref{Prop A.4 BJV}, $\sqrt{\log\left(T_{n}\right)n_{T}}\left(\mathrm{MB}_{\mathrm{max},h}\left(T_{n},\,\tau\right)-\mathrm{MB}{}_{\mathrm{max},h}^{0}\left(T_{n},\,\tau\right)\right)=o_{\mathbb{P}}\left(1\right)$.
For the non-overlapping case, Lemma \ref{Lemma verify Assumption 4th moments}
shows that Assumption \ref{Assumption 4th moment fo forecast error}
is satisfied. Given Condition \ref{Cond The-auxiliary-sequence},
Lemma \ref{Lemma QB Prop A.5.} {[}cf. the result pertaining to $\mathrm{MB}_{\mathrm{max},T_{n}}$
there{]} applied to $\mathrm{MB}{}_{\mathrm{max},h}^{0}\left(T_{n},\,\tau\right)$
gives part (ii) of the theorem. $\square$

\paragraph{\label{par Negligibility-of-the}Negligibility of the $\mu_{e,t}$
term}

The negligibility of the drift term can be proven by using similar
arguments to those in Section A.3.3 in \citet{casini/perron_CR_Single_Break}.
From the decomposition in \eqref{eq (1) Neg Drfit} we have for any
$b=0,\ldots,\,\left\lfloor T_{n}/n_{T}\right\rfloor -2$, 
\begin{align}
\sum_{j=1}^{n_{T}}\left(\Delta_{h}e_{T_{m}+\tau+bn_{T}+j-1}^{*}\right)^{2} & =\sum_{j=1}^{n_{T}}\left(\int_{\left(T_{m}+\tau+bn_{T}+j-2\right)h}^{\left(T_{m}+\tau+bn_{T}+j-1\right)h}\mu_{e,s}h^{-\vartheta}ds\right)^{2}+\sum_{j=1}^{n_{T}}\left(\int_{\left(T_{m}+\tau+bn_{T}+j-2\right)h}^{\left(T_{m}+\tau+bn_{T}+j-1\right)h}\sigma_{e,s}dW_{e,s}\right)^{2}\nonumber \\
 & \quad+2\sum_{j=1}^{n_{T}}\int_{\left(T_{m}+\tau+bn_{T}+j-2\right)h}^{\left(T_{m}+\tau+bn_{T}+j-1\right)h}\int_{\left(T_{m}+\tau+bn_{T}+j-2\right)h}^{\left(T_{m}+\tau+bn_{T}+j-1\right)h}\mu_{e,v}h^{-\vartheta}\sigma_{e,s}dvdW_{e,s}\nonumber \\
 & =o\left(n_{T}h^{2\left(1-\vartheta\right)}\right)+\sum_{j=1}^{n_{T}}\left(\int_{\left(T_{m}+\tau+bn_{T}+j-2\right)h}^{\left(T_{m}+\tau+bn_{T}+j-1\right)h}\sigma_{e,s}dW_{e,s}\right)^{2}+o_{\mathbb{P}}\left(n_{T}h^{3/2-\vartheta}\right),\label{eq(2) Neg Drift}
\end{align}
for small $\epsilon>0.$ The limit theorems involve normalizing the
above sums by the factor $\sqrt{\log\left(T_{n}\right)n_{T}}/\left(n_{T}h\right)=h^{-2/3-\epsilon/2}$.
Then the first term is $o\left(h^{5/12+\epsilon}\right)$. The bound
can be extended to hold for the maximum over blocks $b=0,\ldots,\,\left\lfloor T_{n}/n_{T}\right\rfloor -2$
by using the same argument as in \eqref{eq (73) BJV}. The latter
bound also applies to the third term of \eqref{eq(2) Neg Drift} which
is even of higher order. Therefore, the results of Lemma \ref{Lemma QA1}-\ref{Lemma QA3}
still holds when $\mu_{e,t}$ is not restricted to be null for all
$t\geq0$. 

\subsubsection{Proof of Corollary \ref{Corollary CR Null Distrb Quadratic}}

\noindent\textit{Proof}. The proof follows easily from Lemma \ref{Lemma QB Prop A.5.}
with $\sigma_{\widetilde{U}}=\nu_{L}$. That is, we have now $R_{b,T_{n}}\triangleq\sum_{j=1}^{bn_{T}}\widetilde{U}_{j}-\nu_{L}\mathbb{B}\left(bn_{T}\right)-\left(\sum_{j=1}^{\left(b-1\right)n_{T}}\widetilde{U}_{j}-\nu_{L}\mathbb{B}\left(\left(b-1\right)n_{T}\right)\right)$
which satisfies the same bound as above. Then, proceeding as above,
\begin{align*}
\nu_{L}^{-1}n_{T}^{-1/2}\sum_{j=1}^{n_{T}}\left(U_{\left(b+1\right)n_{T}+j}-U_{bn_{T}+j}\right) & =Z_{b+1}-Z_{b}+o_{\mathrm{a.s.}}\left(\left(\log\left(m_{T}\right)\right)^{-1/2}\right),
\end{align*}
and the final result for $\mathrm{Q}_{\mathrm{max},h}$ can be deduced
again from Lemma 1 in \citet{wu/zhao:07}. $\square$

\subsubsection{Proof of Theorem \ref{Theoem Asymptotic H0 Distrbution Gmax QGmax}}

\paragraph{Proof of part (i) of Theorem \ref{Theoem Asymptotic H0 Distrbution Gmax QGmax}}

Recall the notation for the normalized forecast error $\Delta_{h}\widetilde{e}_{k}\triangleq\Delta_{h}e_{k}/\psi_{h}$
and for the normalized forecast loss $L_{\psi,T_{m}+\tau+bn_{T}+j}\left(\beta^{*}\right)=g\left(\Delta_{h}\widetilde{e}_{T_{m}+\tau+bn_{T}+j-1};\,\beta^{*}\right)$.
We use the quantity $\mathrm{U}_{\max,h}\left(T_{n},\,\tau\right)$
as defined in the proof for the quadratic case. However, $\overline{U}_{h,b}$
is now defined as $D_{h,b}$ but with $\beta^{*}$ in place of $\widehat{\beta}$.
Let $B_{h,b}^{*}=n_{T}^{-1}\sum_{j=1}^{n_{T}}g\left(\Delta_{h}\widetilde{e}_{T_{m}+\tau+bn_{T}+j-1}^{*};\,\beta^{*}\right)$.
We only provide the proof for the recursive forecasting scheme. As
in the proof of Theorem \ref{Theoem Asymptotic H0 Distrbution Bmax and Qmax},
we first assume that $\mu_{e,t}=0$ in \eqref{Mode for Y} and relax
such restriction in Section \ref{Section Negligibility-of-the drift }.
We again omit the index from $\widehat{\beta}$ when it is clear from
the context.
\begin{lem}
\label{Lemma GB A2}For any $L\in\boldsymbol{L}_{e}$, the results
of Lemma \ref{Lemma QA2} hold.
\end{lem}
\noindent\textit{Proof}. By definition and upon using basic manipulations,
\begin{align*}
\left|n_{T}^{-1}\right. & \left.\sum_{j=1}^{n_{T}}\left(\overline{L}_{\psi,\left(T_{m}+\left(b+1\right)n_{T}+j-1\right)h}\left(\beta^{*}\right)-\overline{L}_{\psi,\left(T_{m}+bn_{T}+j-1\right)h}\left(\beta^{*}\right)\right)\right|\\
 & =\left|n_{T}^{-1}\sum_{j=1}^{n_{T}}\left(\sum_{l=1}^{T_{m}+\left(b+1\right)n_{T}+j-1}\frac{g\left(\Delta_{h}\widetilde{e}_{l}^{*};\,\beta^{*}\right)}{T_{m}+\left(b+1\right)n_{T}+j-1}-\sum_{l=1}^{T_{m}+bn_{T}+j-1}\frac{g\left(\Delta_{h}\widetilde{e}_{l}^{*};\,\beta^{*}\right)}{T_{m}+bn_{T}+j-1}\right)\right|\\
 & =\left|n_{T}^{-1}\sum_{j=1}^{n_{T}}\left(\sum_{l=1}^{T_{m}+\left(b+1\right)n_{T}+j-1}g\left(\Delta_{h}\widetilde{e}_{l}^{*};\,\beta^{*}\right)\left(\frac{1}{T_{m}+\left(b+1\right)n_{T}+j-1}-\frac{1}{T_{m}+bn_{T}+j-1}\right)\right.\right.\\
 & \quad\left.\left.+\sum_{l=T_{m}+bn_{T}+j}^{T_{m}+\left(b+1\right)n_{T}+j-1}\frac{g\left(\Delta_{h}\widetilde{e}_{l}^{*};\,\beta^{*}\right)}{T_{m}+bn_{T}+j-1}\right)\right|\\
 & =O_{\mathbb{P}}\left(\frac{n_{T}}{T_{m}}\right)+O_{\mathbb{P}}\left(\frac{n_{T}}{T_{m}}\right),
\end{align*}
where the latter bounds are implied by basic law of large numbers
given Assumption \ref{Assumption Moments of Losses}. Then use the
same arguments as in the proof of Lemma \ref{Lemma QA2} to yield
a bound similar to \eqref{eq (A1a0)}. Finally, consider a mean-value
expansion of $g\left(\Delta_{h}\widetilde{e}_{l};\,\widehat{\beta}\right)$
around $\beta^{*}$,
\begin{align*}
g\left(\Delta_{h}\widetilde{e}_{l};\,\widehat{\beta}\right) & =g\left(\Delta_{h}\widetilde{e}_{l}^{*};\,\beta^{*}\right)+\frac{\partial g\left(\Delta_{h}\widetilde{e}_{l}^{*};\,\beta^{*}\right)}{\partial\beta}\left(\widehat{\beta}-\beta^{*}\right)+\frac{1}{2}\left(\widehat{\beta}-\beta^{*}\right)'\frac{\partial^{2}g\left(\Delta_{h}\widetilde{e}_{l};\,\overline{\beta}\right)}{\partial\beta\partial\beta'}\left(\widehat{\beta}-\beta^{*}\right),
\end{align*}
 where $\overline{\beta}$ is an intermediate point between $\beta^{*}$
and $\widehat{\beta}.$ It follows that 
\begin{align*}
\overline{L} & _{\psi,\left(T_{m}+bn_{T}+j-1\right)h}\left(\widehat{\beta}\right)-\overline{L}_{\psi,\left(T_{m}+bn_{T}+j-1\right)h}\left(\beta^{*}\right)\\
 & =\sum_{l=1}^{T_{m}+bn_{T}+j-1}\frac{g\left(\Delta_{h}\widetilde{e}_{l};\,\widehat{\beta}\right)}{T_{m}+bn_{T}+j-1}-\sum_{l=1}^{T_{m}+bn_{T}+j-1}\frac{g\left(\Delta_{h}\widetilde{e}_{l}^{*};\,\beta^{*}\right)}{T_{m}+bn_{T}+j-1}\\
 & =\frac{1}{T_{m}+bn_{T}+j-1}\sum_{l=1}^{T_{m}+bn_{T}+j-1}\left(\frac{\partial g\left(\Delta_{h}\widetilde{e}_{l};\,\beta^{*}\right)}{\partial\beta}\left(\widehat{\beta}-\beta^{*}\right)+\frac{1}{2}\left(\widehat{\beta}-\beta^{*}\right)'\frac{\partial^{2}g\left(\Delta_{h}\widetilde{e}_{l};\,\overline{\beta}\right)}{\partial\beta\partial\beta'}\left(\widehat{\beta}-\beta^{*}\right)\right).
\end{align*}
By Assumption  \ref{Assumption Differentiability Loss}, $\left|\partial^{2}g\left(\Delta_{h}\widetilde{e}_{l};\,\overline{\beta}\right)/\partial\beta\partial\beta'\right|<C$
and thus the second term is $O_{\mathbb{P}}\left(1/T\right)$ uniformly
in $l$. By Assumption \ref{Assumption Moments of Losses}, $\mathbb{E}\left(\left\Vert \partial g\left(\Delta_{h}\widetilde{e}_{l};\,\beta^{*}\right)/\partial\beta\right\Vert \right)^{2+\varpi}<\infty$
uniformly in $l$. Since 
\begin{align*}
\left\{ \partial g\left(\Delta_{h}\widetilde{e}_{l};\,\beta^{*}\right)/\partial\beta-\mathbb{E}\left(\partial g\left(\Delta_{h}\widetilde{e}_{l};\,\beta^{*}\right)/\partial\beta\right)\right\} _{l\geq T_{m}} & ,
\end{align*}
forms a martingale difference sequence we can use classical bounds
on averages of m.d.s. By Assumption \ref{Assumption Roo-T Consistent beta },
$\widehat{\beta}-\beta^{*}=O_{\mathbb{P}}\left(1/\sqrt{T}\right)$
because $\widehat{\beta}_{l}-\beta^{*}=O_{\mathbb{P}}\left(1/\sqrt{T}\right)$
uniformly in $l\geq T_{m}$. Thus, $\overline{L}{}_{\psi,\left(T_{m}+bn_{T}+j-1\right)h}\left(\widehat{\beta}\right)-\overline{L}_{\psi,\left(T_{m}+bn_{T}+j-1\right)h}\left(\beta^{*}\right)=O_{\mathbb{P}}\left(1/\sqrt{T}\right)$.
Proceeding as in \eqref{eq (A1a0)}-\eqref{eq (A1b)} one verifies,
\begin{align*}
\max_{b=0,\ldots,\,\left\lfloor T_{n}/n_{T}\right\rfloor -2}\left(\log\left(T_{n}\right)n_{T}\right)^{1/2}\left|n_{T}^{-1}\sum_{j=1}^{n_{T}}\left(\overline{L}_{\psi,\left(T_{m}+bn_{T}+j-1\right)h}\left(\widehat{\beta}\right)-\overline{L}_{\psi,\left(T_{m}+bn_{T}+j-1\right)h}\left(\beta^{*}\right)\right)\right| & \overset{\mathbb{P}}{\rightarrow}0.\,\square
\end{align*}

We now have a corresponding result to Lemma \ref{Lemma QA1}.
\begin{lem}
\label{Lemma GB A1}As $h\downarrow0$, $\left(\log\left(T_{n}\right)n_{T}\right)^{1/2}\left(\mathrm{U}_{\max,h}\left(T_{n},\,\tau\right)-\mathrm{G}_{\mathrm{max},h}\left(T_{n},\,\tau\right)\right)\overset{\mathbb{P}}{\rightarrow}0$. 
\end{lem}
\noindent\textit{Proof}. The same manipulations as in Lemma \ref{Lemma QA1}
yield, 
\begin{align}
\left|\mathrm{U}_{\max,h}\right. & \left.\left(T_{n},\,\tau\right)-\mathrm{G}_{\mathrm{max},h}\left(T_{n},\,\tau\right)\right|\label{eq (1) Lemma GB A1}\\
 & \leq\max_{b=0,\ldots,\,\left\lfloor T_{n}/n_{T}\right\rfloor -2}\left|\frac{n_{T}^{-1}\sum_{j=1}^{n_{T}}\left(SL_{\psi,T_{m}+\tau+\left(b+1\right)n_{T}+j-1}\left(\beta^{*}\right)-SL_{\psi,T_{m}+\tau+\left(b+1\right)n_{T}+j-1}\left(\widehat{\beta}\right)\right)}{\sqrt{\overline{U}_{b+1,h}D_{b+1,h}}}\right|\nonumber \\
 & \quad+\max_{b=0,\ldots,\,\left\lfloor T_{n}/n_{T}\right\rfloor -2}\left|n_{T}^{-1}\frac{\sum_{j=1}^{n_{T}}\left(SL_{\psi,T_{m}+\tau+bn_{T}+j-1}\left(\beta^{*}\right)-SL_{\psi,T_{m}+\tau+bn_{T}+j-1}\left(\widehat{\beta}\right)\right)}{\sqrt{\overline{U}_{b+1,h}D_{b+1,h}}}\right|\nonumber \\
 & \leq C_{1}\max_{b=0,\ldots,\,\left\lfloor T_{n}/n_{T}\right\rfloor -2}\left|\frac{n_{T}^{-1}\sum_{j=1}^{n_{T}}\left(SL_{\psi,T_{m}+\tau+\left(b+1\right)n_{T}+j-1}\left(\beta^{*}\right)-SL_{\psi,T_{m}+\tau+\left(b+1\right)n_{T}+j-1}\left(\widehat{\beta}\right)\right)}{\overline{U}_{b+1,h}}\right|\nonumber \\
 & \quad+C_{2}\max_{b=0,\ldots,\,\left\lfloor T_{n}/n_{T}\right\rfloor -2}\left|n_{T}^{-1}\left(\frac{\sum_{j=1}^{n_{T}}\left(SL_{\psi,T_{m}+\tau+bn_{T}+j-1}\left(\beta^{*}\right)-SL_{\psi,T_{m}+\tau+bn_{T}+j-1}\left(\widehat{\beta}\right)\right)}{\overline{U}_{b+1,h}}\right)\right|.\nonumber 
\end{align}
 From Lemma \ref{Lemma GB A2}, for any $j=1,\ldots,\,n_{T}$,
\begin{align*}
SL_{T_{m}+\tau+bn_{T}+j-1}^{\psi} & \left(\beta^{*}\right)-SL_{T_{m}+\tau+bn_{T}+j-1}^{\psi}\left(\widehat{\beta}\right)\\
 & =L_{\psi,T_{m}+\tau+bn_{T}+j-1}\left(\beta^{*}\right)-L_{\psi,T_{m}+\tau+bn_{T}+j-1}\left(\widehat{\beta}\right)+o_{\mathbb{P}}\left(T^{-1/2}\right).
\end{align*}
Note that, 
\begin{align*}
L_{\psi,T_{m}+\tau+bn_{T}+j-1} & \left(\widehat{\beta}\right)-L_{\psi,T_{m}+\tau+bn_{T}+j-1}\left(\beta^{*}\right)\\
 & =g\left(\Delta_{h}\widetilde{e}_{T_{m}+\tau+bn_{T}+j-1};\,\widehat{\beta}\right)-g\left(\sigma_{e,\left(T_{m}+\tau+bn_{T}-1\right)h}\left(h^{-1/2}\Delta_{h}W_{e,T_{m}+\tau+bn_{T}+j-1}\right);\,\beta^{*}\right),
\end{align*}
 and taking a mean-value expansion of $g\left(\Delta_{h}\widetilde{e}_{T_{m}+\tau+bn_{T}+j-1};\,\widehat{\beta}\right)$
around $\beta^{*}$ we have 
\begin{align*}
g\left(\Delta_{h}\widetilde{e}_{T_{m}+\tau+bn_{T}+j-1};\,\widehat{\beta}\right) & =g\left(\sigma_{e,\left(T_{m}+\tau+bn_{T}-1\right)h}\left(h^{-1/2}\Delta_{h}W_{e,T_{m}+\tau+bn_{T}+j-1}\right);\,\beta^{*}\right)\left(\widehat{\beta}-\beta^{*}\right)\\
 & \quad+\frac{\partial g\left(\sigma_{e,\left(T_{m}+\tau+bn_{T}-1\right)h}\left(h^{-1/2}\Delta_{h}W_{e,T_{m}+\tau+bn_{T}+j-1}\right);\,\beta^{*}\right)}{\partial\beta}\left(\widehat{\beta}-\beta^{*}\right)\\
 & \quad+\frac{1}{2}\left(\widehat{\beta}-\beta^{*}\right)'\frac{\partial^{2}g\left(\Delta_{h}\widetilde{e}_{T_{m}+\tau+bn_{T}+j-1};\,\overline{\beta}\right)}{\partial\beta\partial\beta'}\left(\widehat{\beta}-\beta^{*}\right).
\end{align*}
 Therefore, using the last three relationships above, Assumption \ref{Assumption Differentiability Loss}-\ref{Assumption Moments of Losses}
and Assumption \ref{Assumption Roo-T Consistent beta } we have for
the numerator of \eqref{eq (1) Lemma GB A1},
\begin{align*}
\left|\frac{1}{n_{T}}\sum_{j=1}^{n_{T}}\right. & \left.\left(SL_{\psi,T_{m}+\tau+bn_{T}+j-1}\left(\beta^{*}\right)-SL_{\psi,T_{m}+\tau+bn_{T}+j-1}\left(\widehat{\beta}\right)\right)\right|\\
 & \leq\left|\frac{1}{n_{T}}\sum_{j=1}^{n_{T}}\left(\frac{\partial g\left(\sigma_{e,\left(T_{m}+\tau+bn_{T}-1\right)h}\left(h^{-1/2}\Delta_{h}W_{e,T_{m}+\tau+bn_{T}+j-1}\right);\,\beta^{*}\right)}{\partial\beta}\left(\widehat{\beta}-\beta^{*}\right)\right.\right.\\
 & \quad+\left.\left.\frac{1}{2}\left(\widehat{\beta}-\beta^{*}\right)'\frac{\partial^{2}g\left(\Delta_{h}\widetilde{e}_{T_{m}+\tau+bn_{T}+j-1};\,\overline{\beta}\right)}{\partial\beta\partial\beta'}\left(\widehat{\beta}-\beta^{*}\right)\right)\right|\\
 & \leq\left\Vert \widehat{\beta}-\beta^{*}\right\Vert \frac{1}{n_{T}}\sum_{j=1}^{n_{T}}\left\Vert \frac{\partial g\left(\sigma_{e,\left(T_{m}+\tau+bn_{T}-1\right)h}\left(h^{-1/2}\Delta_{h}W_{e,T_{m}+\tau+bn_{T}+j-1}\right);\,\beta^{*}\right)}{\partial\beta}\right\Vert \\
 & \quad+\left\Vert \widehat{\beta}-\beta^{*}\right\Vert ^{2}\frac{1}{2n_{T}}\sum_{j=1}^{n_{T}}\left\Vert \frac{\partial^{2}g\left(\Delta_{h}\widetilde{e}_{T_{m}+\tau+bn_{T}+j-1};\,\overline{\beta}\right)}{\partial\beta\partial\beta'}\right\Vert .
\end{align*}
Since $\widehat{\beta}_{k}-\beta^{*}=O_{\mathbb{P}}\left(1/\sqrt{T}\right)$
uniformly, the first term on the right-hand side above is $CO_{\mathbb{P}}\left(T^{-1/2}\right)$
by Assumption \ref{Assumption Moments of Losses} while the second
term is $O_{\mathbb{P}}\left(T^{-1}\right)$ by Assumption \ref{Assumption Differentiability Loss}.
Both bounds are uniform in $b$. Combining the latter two results
we have 
\begin{align}
\left|\frac{1}{n_{T}}\sum_{j=1}^{n_{T}}\left(SL_{\psi,T_{m}+\tau+bn_{T}+j-1}\left(\beta^{*}\right)-SL_{\psi,T_{m}+\tau+bn_{T}+j-1}\left(\widehat{\beta}\right)\right)\right| & =KO_{\mathbb{P}}\left(T^{-1/2}\right).\label{eq (2) GB A2}
\end{align}
Thus, for any $\varepsilon>0$ and any constant $K>0$, the first
term on the right-hand side of \eqref{eq (1) Lemma GB A1} is such
that 
\begin{align}
\mathbb{P} & \left(\max_{b=0,\ldots,\,\left\lfloor T_{n}/n_{T}\right\rfloor -2}\left|\left(\log\left(T_{n}\right)n_{T}\right)^{1/2}\frac{n_{T}^{-1}\sum_{j=1}^{n_{T}}\left(SL_{\psi,T_{m}+\tau+\left(b+1\right)n_{T}+j-1}\left(\beta^{*}\right)-SL_{\psi,T_{m}+\tau+\left(b+1\right)n_{T}+j-1}\left(\widehat{\beta}\right)\right)}{\sqrt{\overline{U}_{b+1,h}D_{b+1,h}}}\right|>\varepsilon\right)\nonumber \\
 & \leq\mathbb{P}\left(\max_{b=0,\ldots,\,\left\lfloor T_{n}/n_{T}\right\rfloor -2}\left|\left(\log\left(T_{n}\right)n_{T}\right)^{1/2}n_{T}^{-1}\sum_{j=1}^{n_{T}}\left(SL_{\psi,T_{m}+\tau+\left(b+1\right)n_{T}+j-1}\left(\beta^{*}\right)-SL_{\psi,T_{m}+\tau+\left(b+1\right)n_{T}+j-1}\left(\widehat{\beta}\right)\right)\right|>\varepsilon/K\right)\nonumber \\
 & \quad+\mathbb{P}\left(\max_{b=0,\ldots,\,\left\lfloor T_{n}/n_{T}\right\rfloor -2}1/\left|\overline{U}_{b+1,h}\right|<K\right).\label{eq(1) GB A2}
\end{align}
By Lemma \ref{Lemma eq(22) in Vetter (2012) for GB } below, $\mathbb{P}\left(\min_{b=0,\ldots,\,\left\lfloor T_{n}/n_{T}\right\rfloor -2}\left|\overline{U}_{b+1,h}\right|<1/K\right)\rightarrow0$
by letting, for example, $K=2/\sigma_{L,-}^{2}$, where $\sigma_{L,-}$
was introduced in that proof. As for the first probability term, by
using Markov's inequality and the relationship in \eqref{eq (2) GB A2},
we have for any $r>0$, 
\begin{align*}
\mathbb{P} & \left(\max_{b=0,\ldots,\,\left\lfloor T_{n}/n_{T}\right\rfloor -2}\left|\left(\log\left(T_{n}\right)n_{T}\right)^{1/2}n_{T}^{-1}\sum_{j=1}^{n_{T}}\left(SL_{\psi,T_{m}+\tau+\left(b+1\right)n_{T}+j-1}\left(\beta^{*}\right)-SL_{\psi,T_{m}+\tau+\left(b+1\right)n_{T}+j-1}\left(\widehat{\beta}\right)\right)\right|>\varepsilon/K\right)\\
 & \leq\left(\frac{K}{\varepsilon}\right)^{r}\sum_{b=0}^{\left\lfloor T_{n}/n_{T}\right\rfloor -2}\mathbb{E}\left[\left|\left(\log\left(T_{n}\right)n_{T}\right)^{1/2}n_{T}^{-1}\sum_{j=1}^{n_{T}}\left(SL_{\psi,T_{m}+\tau+\left(b+1\right)n_{T}+j-1}\left(\beta^{*}\right)-SL_{\psi,T_{m}+\tau+\left(b+1\right)n_{T}+j-1}\left(\widehat{\beta}\right)\right)\right|^{r}\right]\\
 & =\left(\frac{K}{\varepsilon}\right)^{r}\left(\log\left(T_{n}\right)n_{T}\right)^{r/2}O_{\mathbb{P}}\left(T_{n}/\left(T^{r/2}n_{T}\right)\right)\\
 & =\left(\frac{K}{\varepsilon}\right)^{r}\left(\log\left(T_{n}\right)\right)^{r/2}O_{\mathbb{P}}\left(1/\left(T_{n}^{r/6+1/2}\right)\right)\rightarrow0,
\end{align*}
using Condition \ref{Cond The-auxiliary-sequence}. The argument for
the second term of \eqref{eq (1) Lemma GB A1} is equivalent. This
concludes the proof of the lemma. $\square$

\medskip{}

Let $u_{T_{m}+\tau+bn_{T}+j-1}\triangleq g\left(\sigma_{e,\left(T_{m}+\tau+bn_{T}-1\right)h}h^{-1/2}\left(\Delta_{h}W_{e,T_{m}+\tau+bn_{T}+j-1};\,\beta^{*}\right)\right)$
and define $\mathrm{B}_{\mathrm{max},h}^{0}\left(T_{n},\,\tau\right)\triangleq\max_{b=0,\ldots,\,\left\lfloor T_{n}/n_{T}\right\rfloor -2}\left|\left(B_{h,b+1}^{0}-B_{h,b}^{0}\right)/\sqrt{D_{h,b+1}^{0}}\right|,$
where $B_{h,b}^{0}=n_{T}^{-1}\sum_{j=1}^{n_{T}}u_{T_{m}+\tau+bn_{T}+j-1}$
and
\begin{align*}
D_{h,b}^{0} & \triangleq n_{T}^{-1}\sum_{j=1}^{n_{T}}\left(u_{T_{m}+\tau+bn_{T}+j-1}-\overline{g}_{b}\right)^{2},
\end{align*}
with $\overline{g}_{b}\triangleq n_{T}^{-1}\sum_{j=1}^{n_{T}}u_{T_{m}+\tau+bn_{T}+j-1}$.

Similarly, define $B_{h,b}^{*}=n_{T}^{-1}\sum_{j=1}^{n_{T}}u_{T_{m}+\tau+bn_{T}+j-1}^{*}$,
where $u_{T_{m}+\tau+bn_{T}+j-1}^{*}\triangleq g\left(\Delta_{h}\widetilde{e}_{T_{m}+\tau+bn_{T}+j-1}^{*};\,\beta^{*}\right)$.
  The next quantity that we define is similar to $B_{h,b}^{0}$
but has all the parameters shifted back by one block of time length
$n_{T}h$: $\widetilde{B}_{h,b}^{0}=n_{T}^{-1}\sum_{j=1}^{n_{T}}\widetilde{u}_{T_{m}+\tau+bn_{T}+j-1},$
where
\begin{align*}
\widetilde{u}_{T_{m}+\tau+bn_{T}+j-1} & \triangleq g\left(\sigma_{e,\left(T_{m}+\tau+\left(b-1\right)n_{T}-1\right)h}h^{-1/2}\Delta_{h}W_{e,T_{m}+\tau+bn_{T}+j-1};\,\beta^{*}\right).
\end{align*}
 With this notation we can define the statistic $\widetilde{\mathrm{B}}_{\mathrm{max},h}^{0}\left(T_{n},\,\tau\right)\triangleq\max_{b=0,\ldots,\,\left\lfloor T_{n}/n_{T}\right\rfloor -2}\left|\left(\widetilde{B}_{h,b+1}^{0}-B_{h,b}^{0}\right)/\sqrt{\widetilde{D}_{h,b+1}^{0}}\right|$,
where $\widetilde{D}_{h,b}^{0}\triangleq n_{T}^{-1}\sum_{j=1}^{n_{T}}\left(\widetilde{u}_{T_{m}+\tau+bn_{T}+j-1}-\overline{\widetilde{g}}_{b}\right)^{2}$
with $\overline{\widetilde{g}}_{b}\triangleq n_{T}^{-1}\sum_{j=1}^{n_{T}}\widetilde{u}_{T_{m}+\tau+bn_{T}+j-1}$.
We want to show that 
\begin{align*}
\mathbb{P}\left(\left(\log\left(T_{n}\right)n_{T}\right)^{1/2}\left(\mathrm{V}_{\mathrm{max},h}\left(T_{n},\,\tau\right)-\widetilde{\mathrm{B}}_{\mathrm{max},h}^{0}\left(T_{n},\,\tau\right)\right)>\varepsilon\right) & \rightarrow0,
\end{align*}
for any $\varepsilon>0,$ where 
\begin{align*}
\mathrm{V}_{\mathrm{max},h}\left(T_{n},\,\tau\right) & \triangleq\max_{b=0,\ldots,\,\left\lfloor T_{n}/n_{T}\right\rfloor -2}\left|\frac{\widetilde{B}_{h,b+1}^{0}-B_{h,b}^{0}}{\sigma_{u,\left(T_{m}+\tau+bn_{T}-1\right)h}}\right|,
\end{align*}
with $\sigma_{u,\left(T_{m}+\tau+bn_{T}-1\right)h}^{2}\triangleq\mathrm{Var}\left(u_{T_{m}+\tau+bn_{T}}\right)$.
The normalization by $\sigma_{u,\left(T_{m}+\tau+bn_{T}-1\right)h}$
ensures that we obtain a distribution-free limit theory. Note that
the localization assumption implies that there exist $0<\sigma_{u,-}<\sigma_{u,+}<\infty$
defined by $\sigma_{u,-}\triangleq\inf_{k\geq1}\left\{ \sigma_{u,kh}\right\} $
and $\sigma_{u,+}\triangleq\sup_{k\geq1}\left\{ \sigma_{u,kh}\right\} $.
Furthermore, under $H_{0}$, $\sigma_{u,\left(T_{m}+\tau+bn_{T}-1\right)h}$
is a smooth function of Lipschitz parameters and therefore Condition
\ref{eq. Condition auxiliary nT} applies to $\sigma_{u,\left(T_{m}+\tau+bn_{T}-1\right)h}$
as well: $\phi_{\sigma_{u},\eta,N}\leq K\eta.$ Finally, let 
\begin{align}
\mathrm{B}_{\mathrm{max},h}^{*}\left(T_{n},\,\tau\right) & \triangleq\max_{b=0,\ldots,\,\left\lfloor T_{n}/n_{T}\right\rfloor -2}\left|\left(B_{h,b+1}^{*}-B{}_{h,b}^{*}\right)/\overline{U}_{h,b+1}\right|.\label{eq: B^0 and B^*-1}
\end{align}
We proceed via small lemmas which parallel Lemma \ref{Prop QA1 BJV}-\ref{Lemma Prop A.3 in BJV}.
The following lemma shows that, under $H_{0}$, the difference in
the in-sample losses $\overline{L}_{\psi,kh}\left(\widehat{\beta}\right)$
between adjacent blocks is negligible asymptotically. 
\begin{lem}
\label{Lemma GL A3}As $h\downarrow0$, $\left(\log\left(T_{n}\right)n_{T}\right)^{1/2}\left(\mathrm{B}_{\mathrm{max},h}^{*}\left(T_{n},\,\tau\right)-\mathrm{U}_{\mathrm{max},h}\left(T_{n},\,\tau\right)\right)\overset{\mathbb{P}}{\rightarrow}0$.
\end{lem}
\noindent\textit{Proof}. Apply \eqref{eq. Inequality in Prop A1}
to yield 
\begin{align*}
\left|\right. & \left.\mathrm{B}_{\mathrm{max},h}^{*}\left(T_{n},\,\tau\right)-\mathrm{U}_{\mathrm{max},h}\left(T_{n},\,\tau\right)\right|\\
 & =\left|\max_{b=0,\ldots,\left\lfloor T_{n}/n_{T}\right\rfloor -2}\left|\left(B_{h,b+1}^{*}-B_{h,b}^{*}\right)/\sqrt{\overline{U}_{h,b+1}}\right|-\max_{b=0,\ldots,\left\lfloor T_{n}/n_{T}\right\rfloor -2}\left|\left(U_{h,b+1}-U_{h,b}\right)/\sqrt{\overline{U}_{h,b+1}}\right|\right|\\
 & \leq\max_{b=0,\ldots,\left\lfloor T_{n}/n_{T}\right\rfloor -2}\left|\sum_{j=1}^{n_{T}}\left(\overline{L}_{\psi,\left(T_{m}+\left(b+1\right)n_{T}+j-1\right)h}\left(\beta^{*}\right)-\overline{L}_{\psi,\left(T_{m}+bn_{T}+j-1\right)h}\left(\beta^{*}\right)\right)/\sqrt{\overline{U}_{h,b+1}}\right|.
\end{align*}
 For any $\varepsilon>0$ and any $K>0,$ 
\begin{align*}
\mathbb{P} & \left(\max_{b=0,\ldots,\,\left\lfloor T_{n}/n_{T}\right\rfloor -2}\left|\frac{\left(\log\left(T_{n}\right)n_{T}\right)^{1/2}\sum_{j=1}^{n_{T}}\left(\overline{L}_{\psi,\left(T_{m}+\left(b+1\right)n_{T}+j-1\right)h}\left(\beta^{*}\right)-\overline{L}_{\psi,\left(T_{m}+bn_{T}+j-1\right)h}\left(\beta^{*}\right)\right)}{\sqrt{\overline{U}_{h,b+1}}}\right|>\varepsilon\right)\\
 & \leq\mathbb{P}\left(\max_{b=0,\ldots,\,\left\lfloor T_{n}/n_{T}\right\rfloor -2}\left|\left(\log\left(T_{n}\right)n_{T}\right)^{1/2}\sum_{j=1}^{n_{T}}\left(\overline{L}_{\psi,\left(T_{m}+\left(b+1\right)n_{T}+j-1\right)h}\left(\beta^{*}\right)-\overline{L}_{\psi,\left(T_{m}+bn_{T}+j-1\right)h}\left(\beta^{*}\right)\right)\right|>\varepsilon/K\right)\\
 & \quad+\mathbb{P}\left(\max_{b=0,\ldots,\,\left\lfloor T_{n}/n_{T}\right\rfloor -2}1/\left|\sqrt{\overline{U}_{h,b+1}}\right|>K\right).
\end{align*}
 By Lemma \ref{Lemma GB A2} the first term on the right-hand size
converges to zero. As for the second term, use the same argument as
in \eqref{eq(1) GB A2}. The result then follows. $\square$
\begin{lem}
\label{Prop GB A1 BJV}As $h\downarrow0$, $\left(\log\left(T_{n}\right)n_{T}\right)^{1/2}\left(\mathrm{B}_{\mathrm{max},h}^{*}\left(T_{n},\,\tau\right)-\mathrm{B}_{\mathrm{max},h}^{0}\left(T_{n},\,\tau\right)\right)\overset{\mathbb{P}}{\rightarrow}0$. 
\end{lem}
\noindent\textit{Proof}. Note that 
\begin{align}
\left|\right. & \left.\mathrm{B}_{\mathrm{max},h}^{*}\left(T_{n},\,\tau\right)-\mathrm{B}_{\mathrm{max},h}^{0}\left(T_{n},\,\tau\right)\right|\nonumber \\
 & \leq\max_{b=0,\ldots,\,\left\lfloor T_{n}/n_{T}\right\rfloor -2}\left|\left(B_{h,b+1}^{*}-B_{h,b}^{*}\right)\left(\frac{1}{\sqrt{\overline{U}_{h,b+1}}}-\frac{1}{\sqrt{D_{h,b+1}^{0}}}\right)\right|+\max_{b=0,\ldots,\,\left\lfloor T_{n}/n_{T}\right\rfloor -2}\left|\frac{B_{h,b+1}^{*}-B{}_{h,b+1}^{0}}{\sqrt{D_{h,b+1}^{0}}}\right|\label{eq. 47 GB BJV}\\
 & \quad+\max_{b=0,\ldots,\,\left\lfloor T_{n}/n_{T}\right\rfloor -2}\left|\frac{B_{h,b}^{*}-B{}_{h,b}^{0}}{\sqrt{D_{h,b+1}^{0}}}\right|.\nonumber 
\end{align}
Consider the first term of \eqref{eq. 47 GB BJV}. We can write for
any $\varepsilon>0$, any $0<K,\,C<\infty$, and some small positive
number $\varpi<1/2$, 
\begin{align}
\mathbb{P} & \left(\max_{b=0,\ldots,\,\left\lfloor T_{n}/n_{T}\right\rfloor -2}\left|\left(\log\left(T_{n}\right)n_{T}\right)^{1/2}\left(B_{h,b+1}^{*}-B_{h,b}^{*}\right)\left(\frac{1}{\sqrt{\overline{U}_{h,b+1}}}-\frac{1}{\sqrt{D_{h,b+1}^{0}}}\right)\right|>\varepsilon\right)\nonumber \\
 & \leq\mathbb{P}\left(\max_{b=0,\ldots,\,\left\lfloor T_{n}/n_{T}\right\rfloor -2}\left|n_{T}^{\varpi}\left(B_{h,b+1}^{*}-B_{h,b}^{*}\right)\right|>\varepsilon/K\right)\nonumber \\
 & \quad+\mathbb{P}\left(\max_{b=0,\ldots,\,\left\lfloor T_{n}/n_{T}\right\rfloor -2}\left|\sqrt{\log\left(T_{n}\right)}n_{T}^{1/2-\varpi}\left(\sqrt{\overline{U}_{h,b+1}}-\sqrt{D_{h,b+1}^{0}}\right)\right|>K/C\right)\nonumber \\
 & \quad+\mathbb{P}\left(\max_{b=0,\ldots,\,\left\lfloor T_{n}/n_{T}\right\rfloor -2}\left|\frac{1}{\sqrt{D_{h,b+1}^{0}\overline{U}_{h,b+1}}}\right|>C\right)\triangleq A_{1,h}+A_{2,h}+A_{3,h}.\label{eq (A1A2A3) GB}
\end{align}
We first discuss $A_{1,h}$: 
\begin{align}
\mathbb{P} & \left(\max_{b=0,\ldots,\,\left\lfloor T_{n}/n_{T}\right\rfloor -2}\left|n_{T}^{\varpi}\left(B_{h,b+1}^{*}-B_{h,b}^{*}\right)\right|>\varepsilon/K\right)\nonumber \\
 & \leq\sum_{b=0}^{\left\lfloor T_{n}/n_{T}\right\rfloor -2}\mathbb{P}\left(\left|n_{T}^{\varpi}\left(B_{h,b+1}^{*}-B_{h,b}^{*}\right)\right|>\varepsilon/K\right)\nonumber \\
 & \leq\sum_{b=0}^{\left\lfloor T_{n}/n_{T}\right\rfloor -2}\mathbb{P}\left(\left|n_{T}^{\varpi}\left(B_{h,b+1}^{*}-\mu_{T_{m}+\tau+\left(b+1\right)n_{T}-1}\right)\right|>\varepsilon/\left(3K\right)\right)\nonumber \\
 & \quad+\sum_{b=0}^{\left\lfloor T_{n}/n_{T}\right\rfloor -2}\mathbb{P}\left(\left|n_{T}^{\varpi}\left(B_{h,b}^{*}-\mu_{T_{m}+\tau+bn_{T}-1}\right)\right|>\varepsilon/\left(3K\right)\right)\nonumber \\
 & \quad+\sum_{b=0}^{\left\lfloor T_{n}/n_{T}\right\rfloor -2}\mathbb{P}\left(\left|n_{T}^{\varpi}\left(\mu_{T_{m}+\tau+\left(b+1\right)n_{T}-1}-\mu_{T_{m}+\tau+bn_{T}-1}\right)\right|>\varepsilon/\left(3K\right)\right).\label{eq (A1) GB}
\end{align}
Since $B_{h,b}^{*}\overset{\mathbb{P}}{\rightarrow}\mu_{T_{m}+\tau+bn_{T}-1}\triangleq\mathbb{E}\left(u_{T_{m}+\tau+bn_{T}}\right)$
and $\mathbb{E}\left[\sqrt{n_{T}}\left(B_{h,b}^{*}-\mu_{T_{m}+\tau+bn_{T}-1}\right)\right]<\infty$
by a standard CLT, we have for $r>0$ sufficiently large and by choosing
$\varpi$ sufficiently small,
\begin{align*}
\sum_{b=0}^{\left\lfloor T_{n}/n_{T}\right\rfloor -2}\mathbb{P}\left(\left|n_{T}^{\varpi}\left(B_{h,b}^{*}-\mu_{T_{m}+\tau+bn_{T}-1}\right)\right|>\varepsilon/\left(3K\right)\right) & \leq K_{r}\left(\frac{3K}{\varepsilon}\right)^{r}\sum_{b=0}^{\left\lfloor T_{n}/n_{T}\right\rfloor -2}\mathbb{E}\left(\left|n_{T}^{\varpi}\left(B_{h,b}^{*}-\mu_{T_{m}+\tau+bn_{T}-1}\right)\right|^{r}\right)\\
 & \leq K_{r}\left(\frac{3K}{\varepsilon}\right)^{r}T_{n}n_{T}^{r\left(\varpi-1/2\right)-1}\rightarrow0.
\end{align*}
 The term involving $B_{h,b+1}^{*}$ admits a similar bound. For the
last term of \eqref{eq (A1) GB}, we use the Lipschitz continuity
of $\mu_{\cdot}$ to yield 
\begin{align*}
\sum_{b=0}^{\left\lfloor T_{n}/n_{T}\right\rfloor -2} & \mathbb{P}\left(\left|n_{T}^{\varpi}\left(\mu_{T_{m}+\tau+\left(b+1\right)n_{T}-1}-\mu_{T_{m}+\tau+bn_{T}-1}\right)\right|>\varepsilon/\left(3K\right)\right)\\
 & \leq\left(\frac{3K}{\varepsilon}\right)^{r}\sum_{b=0}^{\left\lfloor T_{n}/n_{T}\right\rfloor -2}\mathbb{E}\left(\left|n_{T}^{\varpi}\left(\mu_{T_{m}+\tau+\left(b+1\right)n_{T}-1}-\mu_{T_{m}+\tau+bn_{T}-1}\right)\right|^{r}\right)\\
 & \leq\left(\frac{3K}{\varepsilon}\right)^{r}\left(T_{n}/n_{T}-2\right)n_{T}^{r\left(\varpi+1\right)}h^{r}\rightarrow0,
\end{align*}
 for $r>0$ sufficiently large since $\varpi$ is chosen to be small.
Thus, $A_{1,h}\rightarrow0$ while Lemma \ref{Lemma eq(22) in Vetter (2012) for GB }
implies that $A_{3,h}\rightarrow0$ by setting $\sqrt{C}=1/\left(2\sigma_{u,-}\right)$.
Further, note that $\sigma_{u,\left(T_{m}+\tau+\left(b+1\right)n_{T}-1\right)h}^{2}$
is the limit of both $\overline{U}_{h,b+1}$ and $D_{h,b+1}^{0}$.
Thus, given the i.i.d. structure, we can use a standard CLT  to yield
\begin{align}
\mathbb{P} & \left(\max_{b=0,\ldots,\,\left\lfloor T_{n}/n_{T}\right\rfloor -2}\left|\sqrt{\log\left(T_{n}\right)}n_{T}^{1/2-\varpi}\left(\sqrt{\overline{U}_{h,b+1}}-\sqrt{D_{h,b+1}^{0}}\right)\right|>K/C\right)\nonumber \\
 & \leq\sum_{b=0}^{\left\lfloor T_{n}/n_{T}\right\rfloor -2}\mathbb{P}\left(\left|\sqrt{\log\left(T_{n}\right)}n_{T}^{1/2-\varpi}\left(\sqrt{\overline{U}_{h,b+1}}-\sigma_{u,\left(T_{m}+\tau+\left(b+1\right)n_{T}-1\right)h}\right)\right|^{r}>\left(K/2C\right)^{r}\right)\nonumber \\
 & \quad+\sum_{b=0}^{\left\lfloor T_{n}/n_{T}\right\rfloor -2}\mathbb{P}\left(\left|\sqrt{\log\left(T_{n}\right)}n_{T}^{1/2-\varpi}\left(\sqrt{D_{h,b+1}^{0}}-\sigma_{u,\left(T_{m}+\tau+\left(b+1\right)n_{T}-1\right)h}\right)\right|^{r}>\left(K/2C\right)^{r}\right)\nonumber \\
 & \leq2\left(K/2C\right)^{-r}\left(T_{n}/n_{T}\right)O_{\mathbb{P}}\left(\left(\sqrt{\log\left(T_{n}\right)}n_{T}^{-\varpi}\right)^{r}\right)\rightarrow0,\label{eq (A2) GB}
\end{align}
for $r>1/\varpi$ sufficiently large. This shows that $A_{2,h}\rightarrow0$.
It remains to discuss the second term of \eqref{eq. 47 GB BJV}; the
argument for the third term is equivalent and omitted. Recall the
definition of $u_{T_{m}+\tau+\left(b+1\right)n_{T}-1}^{*}$ and $u_{T_{m}+\tau+\left(b+1\right)n_{T}-1}$.
By a mean-value expansion, 
\begin{align}
B_{h,b+1}^{*} & -B_{h,b+1}^{0}\nonumber \\
 & =n_{T}^{-1}\sum_{j=1}^{n_{T}}\left(u_{T_{m}+\tau+\left(b+1\right)n_{T}+j-1}^{*}-u_{T_{m}+\tau+\left(b+1\right)n_{T}+j-1}\right)\nonumber \\
 & =n_{T}^{-1}\sum_{j=1}^{n_{T}}\left(g\left(\Delta_{h}\widetilde{e}_{T_{m}+\tau+\left(b+1\right)n_{T}+j-1}^{*};\,\beta^{*}\right)-g\left(\sigma_{e,\left(T_{m}+\tau+\left(b+1\right)n_{T}-1\right)h}h^{-1/2}\left(\Delta_{h}W_{e,T_{m}+\tau+\left(b+1\right)n_{T}+j-1}\right);\,\beta^{*}\right)\right)\nonumber \\
 & =n_{T}^{-1}\sum_{j=1}^{n_{T}}\left[\partial_{e}g\left(\sigma_{e,\left(T_{m}+\tau+\left(b+1\right)n_{T}-1\right)h}h^{-1/2}\left(\Delta_{h}W_{e,T_{m}+\tau+\left(b+1\right)n_{T}+j-1}\right);\,\beta^{*}\right)\right.\nonumber \\
 & \quad\times\left(\Delta_{h}\widetilde{e}_{T_{m}+\tau+\left(b+1\right)n_{T}+j-1}-\sigma_{e,\left(T_{m}+\tau+\left(b+1\right)n_{T}-1\right)h}h^{-1/2}\Delta_{h}W_{e,T_{m}+\tau+\left(b+1\right)n_{T}+j-1}\right)\nonumber \\
 & \quad+\partial_{e}^{2}g\left(\sigma_{e,\left(T_{m}+\tau+\left(b+1\right)n_{T}-1\right)h}h^{-1/2}\left(\Delta_{h}\overline{W}_{e,T_{m}+\tau+\left(b+1\right)n_{T}+j-1}\right);\,\beta^{*}\right)\nonumber \\
 & \quad\left.\times\left(\Delta_{h}\widetilde{e}_{T_{m}+\tau+\left(b+1\right)n_{T}+j-1}-\sigma_{e,\left(T_{m}+\tau+\left(b+1\right)n_{T}-1\right)h}h^{-1/2}\Delta_{h}W_{e,T_{m}+\tau+\left(b+1\right)n_{T}+j-1}\right)^{2}\right].\label{eq (GB A1) -1}
\end{align}
Since for $r=1,\,2$, $\left|\partial_{e}^{r}g\left(e;\,\beta\right)\right|<C_{r}$
for some $C_{r}<\infty$ by Assumption \ref{Assumption Bounded on Bounded Sets LX (2-iv)},
the right-hand side above is less than 
\begin{align}
\left(\log\left(T_{n}\right)n_{T}\right)^{1/2} & C_{1}\left(n_{T}\sqrt{h}\right)^{-1}\nonumber \\
 & \quad\times\sum_{j=1}^{n_{T}}\left(\Delta_{h}e_{T_{m}+\tau+\left(b+1\right)n_{T}+j-1}^{*}-\sigma_{e,\left(T_{m}+\tau+\left(b+1\right)n_{T}-1\right)h}\Delta_{h}W_{e,T_{m}+\tau+\left(b+1\right)n_{T}+j-1}\right)\nonumber \\
 & +\left(\log\left(T_{n}\right)n_{T}\right)^{1/2}C_{2}\left(n_{T}h\right)^{-1}\nonumber \\
 & \quad\times\sum_{j=1}^{n_{T}}\left(\Delta_{h}e_{T_{m}+\tau+\left(b+1\right)n_{T}+j-1}^{*}-\sigma_{e,\left(T_{m}+\tau+\left(b+1\right)n_{T}-1\right)h}\Delta_{T_{m}+\tau+\left(b+1\right)n_{T}+j-1}W_{e}\right)^{2}.\label{eq (49) GB A.1}
\end{align}
Let us consider the first term of \eqref{eq (49) GB A.1}. By It\^o's
formula, 
\begin{align}
\Delta_{h}e_{T_{m}+\tau+\left(b+1\right)n_{T}+j-1} & -\sigma_{e,\left(T_{m}+\tau+\left(b+1\right)n_{T}-1\right)h}\Delta_{T_{m}+\tau+\left(b+1\right)n_{T}+j-1}W_{e}\label{eq (1) Taylor GB e - e*}\\
 & =\int_{\left(T_{m}+\tau+\left(b+1\right)n_{T}+j-1\right)h}^{\left(T_{m}+\tau+\left(b+1\right)n_{T}+j\right)h}\left(\sigma_{e,s}-\sigma_{e,\left(T_{m}+\tau+\left(b+1\right)n_{T}-1\right)h}\right)dW_{e,s}\nonumber \\
 & \quad+\int_{\left(T_{m}+\tau+\left(b+1\right)n_{T}+j-1\right)h}^{\left(T_{m}+\tau+\left(b+1\right)n_{T}+j\right)h}\mu_{e,s}h^{-\vartheta}ds.\nonumber 
\end{align}
 Then, for an integer $r>2$, by Jensen's inequality,
\begin{align}
\mathbb{E} & \left[\left|\left(\log\left(T_{n}\right)n_{T}\right)^{1/2}\left(n_{T}\sqrt{h}\right)^{-1}\sum_{j=1}^{n_{T}}\int_{\left(T_{m}+\tau+\left(b+1\right)n_{T}+j-1\right)h}^{\left(T_{m}+\tau+\left(b+1\right)n_{T}+j\right)h}\left(\sigma_{e,s}-\sigma_{e,\left(T_{m}+\tau+\left(b+1\right)n_{T}-1\right)h}\right)dW_{e,s}\right|^{r}\right]\nonumber \\
 & \leq K_{r}\left(\left(\log\left(T_{n}\right)n_{T}\right)^{1/2}\left(n_{T}\sqrt{h}\right)^{-1}\right)^{r}\left(\sum_{j=1}^{n_{T}}\left(\mathbb{E}\left[\int_{\left(T_{m}+\tau+\left(b+1\right)n_{T}+j-1\right)h}^{\left(T_{m}+\tau+\left(b+1\right)n_{T}+j\right)h}\left(\sigma_{e,s}-\sigma_{e,\left(T_{m}+\tau+\left(b+1\right)n_{T}-2\right)h}\right)^{2r}\right]\right)^{1/r}ds\right)^{r/2}\nonumber \\
 & \leq K_{r}\left(\left(\log\left(T_{n}\right)n_{T}\right)^{1/2}\left(n_{T}\sqrt{h}\right)^{-1}\right)^{r}\left(\int_{\left(T_{m}+\tau+\left(b+1\right)n_{T}\right)h}^{\left(T_{m}+\tau+\left(b+2\right)n_{T}\right)h}\left(\left(\mathbb{E}\left[\phi_{\sigma,n_{T}h,N}^{2r}\right]\right)^{1/r}\right)ds\right)^{r/2}\nonumber \\
 & \leq K_{r}\left(\left(\log\left(T_{n}\right)n_{T}\right)^{1/2}\left(n_{T}\sqrt{h}\right)^{-1}\right)^{r}\left(\left(n_{T}h\right)^{2}n_{T}h\right)^{r/2}\nonumber \\
 & \leq K_{r}\left(\left(\log\left(T_{n}\right)\right)^{1/2}\right)^{r}h^{r/3-\epsilon}\rightarrow0,\label{eq (2) GB 2 - e*-3}
\end{align}
 by choosing $r$ large enough. Next, we consider the second term
of \eqref{eq (1) Taylor GB e - e*},
\begin{align}
\mathbb{E} & \left[\left|\left(\log\left(T_{n}\right)n_{T}\right)^{1/2}\left(n_{T}\sqrt{h}\right)^{-1}\sum_{j=1}^{n_{T}}\int_{\left(T_{m}+\tau+\left(b+1\right)n_{T}+j-1\right)h}^{\left(T_{m}+\tau+\left(b+1\right)n_{T}+j\right)h}\mu_{e,s}h^{-\vartheta}ds\right|^{r}\right]\nonumber \\
 & \leq K_{r}\left(\left(\log\left(T_{n}\right)n_{T}\right)^{1/2}\left(n_{T}\sqrt{h}\right)^{-1}\right)^{r}\left(n_{T}h^{1-\vartheta}\right)^{r}\nonumber \\
 & \leq K_{r}\left(\left(\log\left(T_{n}\right)\right)^{1/2}\right)^{r}h^{21r/24-\epsilon}\rightarrow0.\label{eq (2) GB 2 - e*-3-1}
\end{align}
 For the term in the second line of \eqref{eq (49) GB A.1} apply
the same arguments as in \eqref{eq (2) GB 2 - e*-3}-\eqref{eq (2) GB 2 - e*-3-1}
with $m=r/2$ in place of $r$ above. Choosing $m$ large enough yields
the same result. Thus, using the latter results into the second term
of \eqref{eq. 47 GB BJV} via \eqref{eq (GB A1) -1} we have
\begin{align}
\mathbb{P}\left(\max_{b=0,\ldots,\,\left\lfloor T_{n}/n_{T}\right\rfloor -2}\left|\left(\log\left(T_{n}\right)n_{T}\right)^{1/2}\left(B_{h,b+1}^{*}-B{}_{h,b+1}^{0}\right)\right|>\varepsilon/K\right) & \rightarrow0.\label{eq (49) GB A1}
\end{align}
Note that the same result holds for $B_{h,b}^{*}-B{}_{h,b}^{0}$.
The first term of \eqref{eq. 47 GB BJV} has been treated above and
so the claim of the lemma follows. $\square$
\begin{lem}
\label{Lemma eq(22) in Vetter (2012) for GB }Assume $\mu_{e,t}=0$
for all $t\geq0$. Then, $\mathbb{P}\left(\max_{b=0,\ldots,\,\left\lfloor T_{n}/n_{T}\right\rfloor -2}\left|1/\overline{U}_{h,b}\right|>K\right)\rightarrow0$
for some constant $K>0$.
\end{lem}
\noindent\textit{Proof}. Note that 
\begin{align*}
\mathbb{P} & \left(\max_{b=0,\ldots,\,\left\lfloor T_{n}/n_{T}\right\rfloor -2}\left|1/\overline{U}_{h,b}\right|>K\right)\\
 & =\mathbb{P}\left(\min_{b=0,\ldots,\,\left\lfloor T_{n}/n_{T}\right\rfloor -2}\left|\overline{U}_{h,b}\right|<K^{-1}\right)\\
 & =\mathbb{P}\left(\min_{b=0,\ldots,\,\left\lfloor T_{n}/n_{T}\right\rfloor -2}n_{T}^{-1}\sum_{j=1}^{n_{T}}\left(L_{\psi,\left(T_{m}+\tau+bn_{T}+j-1\right)h}\left(\beta^{*}\right)-\overline{L}_{\psi,b}\left(\beta^{*}\right)\right)^{2}<K^{-1}\right)\\
 & \leq\sum_{b=0}^{\left\lfloor T/n_{T}\right\rfloor -2}\mathbb{P}\left(n_{T}^{-1}\sum_{j=1}^{n_{T}}\left(L_{\psi,\left(T_{m}+\tau+bn_{T}+j-1\right)h}\left(\beta^{*}\right)-\overline{L}_{\psi,b}\left(\beta^{*}\right)\right)^{2}<K^{-1}\right).
\end{align*}
The rest of the proof continues by setting $K^{-1}=\sigma_{L,-}^{2}/2$
where $\sigma_{L,-}^{2}\triangleq\inf_{k\geq1}\sigma_{L,kh}^{2}$
with $\sigma_{L,kh}^{2}\triangleq\mathrm{Var}\left(L_{\psi,kh}\left(\beta^{*}\right)\right)$.
We can use Markov's inequality to deduce for any $r>0,$  
\begin{align*}
\mathbb{P} & \left(n_{T}^{-1}\sum_{j=1}^{n_{T}}\left(L_{\psi,\left(T_{m}+\tau+bn_{T}+j-1\right)h}\left(\beta^{*}\right)-\overline{L}_{\psi,b}\left(\beta^{*}\right)\right)^{2}<\sigma_{L,-}^{2}/2\right)\\
 & \leq\mathbb{P}\left(\left|n_{T}^{-1}\sum_{j=1}^{n_{T}}\left(\left(L_{\psi,\left(T_{m}+\tau+bn_{T}+j-1\right)h}\left(\beta^{*}\right)-\overline{L}_{\psi,b}\left(\beta^{*}\right)\right)^{2}-\sigma_{L,\left(T_{m}+\tau+bn_{T}-1\right)h}^{2}\right)\right|>\sigma_{L,-}^{2}/2\right)\\
 & \leq C_{r}\left(\frac{2}{\sigma_{L,-}^{2}}\right)^{r}\mathbb{E}\left[\left|n_{T}^{-1}\sum_{j=1}^{n_{T}}\left(\left(L_{\psi,\left(T_{m}+\tau+bn_{T}+j-1\right)h}\left(\beta^{*}\right)-\overline{L}_{\psi,b}\left(\beta^{*}\right)\right)^{2}-\sigma_{L,\left(T_{m}+\tau+bn_{T}-1\right)h}^{2}\right)\right|^{r}\right].
\end{align*}
Observe that, conditional on $\left\{ \sigma_{e,t}\right\} _{t\geq0}$,
$\mathrm{Var}_{\sigma}\left[L_{\psi,\left(T_{m}+\tau+bn_{T}+j-1\right)h}\left(\beta^{*}\right)\right]$
is constant across $j=1,\ldots,\,n_{T}$ for a given $b$. Then, Assumption
\ref{Assumption 4th moment fo forecast error} implies that we can
rely on a basic CLT for i.i.d. observations to yield, $\mathbb{E}\left[\left|n_{T}^{-1/2}\sum_{j=1}^{n_{T}}\left(\left(L_{\psi,\left(T_{m}+\tau+bn_{T}+j-1\right)h}\left(\beta^{*}\right)-\overline{L}_{\psi,b}\left(\beta^{*}\right)\right)^{2}-\sigma_{L,\left(T_{m}+\tau+bn_{T}-1\right)h}^{2}\right)\right|^{r}\right]<C_{r},$
where $C_{r}<\infty$. Thus, choose $r$ sufficiently large  so that
\begin{align*}
\sum_{b=0}^{\left\lfloor T/n_{T}\right\rfloor -2} & \mathbb{P}\left(n_{T}^{-1}\sum_{j=1}^{n_{T}}\left(L_{\psi,\left(T_{m}+\tau+bn_{T}+j-1\right)h}\left(\beta^{*}\right)-\overline{L}_{\psi,b}\left(\beta^{*}\right)\right)^{2}<K^{-1}\right)\\
 & \leq C_{r}\left(\frac{2}{\sigma_{L,-}^{2}}\right)^{r}O_{\mathbb{P}}\left(T_{n}/n_{T}\right)n_{T}^{-r/2}\rightarrow0,
\end{align*}
and the proof is concluded. $\square$
\begin{lem}
\label{Lemma Prop GB A2 BJV}As $h\downarrow0$, $\left(\log\left(T_{n}\right)n_{T}\right)^{1/2}\left(\mathrm{B}_{\mathrm{max},h}^{0}\left(T_{n},\,\tau\right)-\widetilde{\mathrm{B}}_{\mathrm{max},h}^{0}\left(T_{n},\,\tau\right)\right)\overset{\mathbb{P}}{\rightarrow}0$.
\end{lem}
\noindent\textit{Proof}. By basic manipulations, 
\begin{align}
\left|\right. & \left.\mathrm{B}_{\mathrm{max},h}^{0}\left(T_{n},\,\tau\right)-\widetilde{\mathrm{B}}_{\mathrm{max},h}^{0}\left(T_{n},\,\tau\right)\right|\nonumber \\
 & \leq\max_{b=0,\ldots,\,\left\lfloor T_{n}/n_{T}\right\rfloor -2}\left|\frac{\sqrt{\widetilde{D}_{h,b+1}^{0}}\left(B_{h,b+1}^{0}-B_{h,b}^{0}\right)-\sqrt{D_{h,b+1}^{0}}\left(\widetilde{B}_{h,b+1}^{0}-B_{h,b}^{0}\right)}{\sqrt{D_{h,b+1}^{0}\widetilde{D}_{h,b+1}^{0}}}\right|\nonumber \\
 & \leq\max_{b=0,\ldots,\,\left\lfloor T_{n}/n_{T}\right\rfloor -2}\left|\frac{\sqrt{\widetilde{D}_{h,b+1}^{0}}\left(B_{h,b+1}^{0}-\widetilde{B}_{h,b+1}^{0}\right)+\left(\widetilde{B}_{h,b+1}^{0}-B_{h,b}^{0}\right)\left(\sqrt{\widetilde{D}_{h,b+1}^{0}}-\sqrt{D_{h,b+1}^{0}}\right)}{\sqrt{D_{h,b+1}^{0}\widetilde{D}_{h,b+1}^{0}}}\right|\nonumber \\
 & \leq\max_{b=0,\ldots,\,\left\lfloor T_{n}/n_{T}\right\rfloor -2}\left|\frac{\sqrt{\widetilde{D}_{h,b+1}^{0}}\left(B_{h,b+1}^{0}-\widetilde{B}_{h,b+1}^{0}\right)}{\sqrt{D_{h,b+1}^{0}\widetilde{D}_{h,b+1}^{0}}}\right|+\max_{b=0,\ldots,\,\left\lfloor T_{n}/n_{T}\right\rfloor -2}\left|\frac{\left(\widetilde{B}_{h,b+1}^{0}-B_{h,b}^{0}\right)\left(\sqrt{\widetilde{D}_{h,b+1}^{0}}-\sqrt{D_{h,b+1}^{0}}\right)}{\sqrt{D_{h,b+1}^{0}\widetilde{D}_{h,b+1}^{0}}}\right|\nonumber \\
 & \triangleq R_{1,h}+R_{2,h}.\label{eq Lemma A2 GB R1 R2 R3}
\end{align}
We begin with showing that $\left(\log\left(T_{n}\right)n_{T}\right)^{1/2}R_{1,h}\overset{\mathbb{P}}{\rightarrow}0,$
or 
\begin{align}
\max_{b=0,\ldots,\,\left\lfloor T_{n}/n_{T}\right\rfloor -2}\left(\log\left(T_{n}\right)n_{T}\right)^{1/2}\left|\frac{\sqrt{\widetilde{D}_{h,b+1}^{0}}\left(B_{h,b+1}^{0}-\widetilde{B}_{h,b+1}^{0}\right)}{\sqrt{D_{h,b+1}^{0}\widetilde{D}_{h,b+1}^{0}}}\right| & =o_{\mathbb{P}}\left(1\right).\label{eq. (54b) in Prop A.2 GB}
\end{align}
By Lemma \ref{Lemma eq(22) in Vetter (2012) for GB }, $\mathbb{P}\left(\min_{b=0,\ldots,\,\left\lfloor T_{n}/n_{T}\right\rfloor -2}\left|\sqrt{D_{h,b+1}^{0}}\right|<K^{-1/2}\right)\rightarrow0$,
where, for example, $\sqrt{K}=2/\sigma_{u,-}$. A similar argument
can be used for $\widetilde{D}_{h,b+1}^{0}$ and therefore it remains
to consider the first term of the following decomposition which is
valid for any $\varepsilon>0$ and any $K>0,$ 
\begin{align}
\mathbb{P} & \left(\max_{b=0,\ldots,\,\left\lfloor T_{n}/n_{T}\right\rfloor -2}\left(\log\left(T_{n}\right)n_{T}\right)^{1/2}\left|\frac{\sqrt{\widetilde{D}_{h,b+1}^{0}}\left(B_{h,b+1}^{0}-\widetilde{B}_{h,b+1}^{0}\right)}{\sqrt{D_{h,b+1}^{0}\widetilde{D}_{h,b+1}^{0}}}\right|>\varepsilon/K\right)\nonumber \\
 & \leq\mathbb{P}\left(\max_{b=0,\ldots,\,\left\lfloor T_{n}/n_{T}\right\rfloor -2}\left(\log\left(T_{n}\right)n_{T}\right)^{1/2}\left|\sqrt{\widetilde{D}_{h,b+1}^{0}}\left(B_{h,b+1}^{0}-\widetilde{B}_{h,b+1}^{0}\right)\right|>\varepsilon/K\right)\nonumber \\
 & \quad+\mathbb{P}\left(\max_{b=0,\ldots,\,\left\lfloor T_{n}/n_{T}\right\rfloor -2}\sqrt{D_{h,b+1}^{0}\widetilde{D}_{h,b+1}^{0}}>K\right).\label{eq. (55) BJV-2}
\end{align}
We have for any positive $K_{2}<\infty$,
\begin{align*}
\mathbb{P} & \left(\max_{b=0,\ldots,\,\left\lfloor T_{n}/n_{T}\right\rfloor -2}\left(\log\left(T_{n}\right)n_{T}\right)^{1/2}\left|\sqrt{\widetilde{D}_{h,b+1}^{0}}\left(B_{h,b+1}^{0}-\widetilde{B}_{h,b+1}^{0}\right)\right|>\varepsilon/K\right)\\
 & \leq\mathbb{P}\left(\max_{b=0,\ldots,\,\left\lfloor T_{n}/n_{T}\right\rfloor -2}\left(\log\left(T_{n}\right)n_{T}\right)^{1/2}\left|B_{h,b+1}^{0}-\widetilde{B}_{h,b+1}^{0}\right|>\varepsilon/\left(K\cdot K_{2}\right)\right)\\
 & \quad+\mathbb{P}\left(\max_{b=0,\ldots,\,\left\lfloor T_{n}/n_{T}\right\rfloor -2}\sqrt{\widetilde{D}_{h,b+1}^{0}}>K_{2}\right).
\end{align*}
 It is straightforward to see that Lemma \ref{Lemma eq(22) in Vetter (2012) for GB }
can be applied also to the second term on the right-hand side above.
Hence, it is sufficient to focus on the first term only. Recall the
definition of $\widetilde{u}_{T_{m}+\tau+bn_{T}-1}$ and $u_{T_{m}+\tau+bn_{T}-1}$
introduced before Lemma \ref{Lemma GL A3}. We write  

\begin{align}
\mathbb{P} & \left(\max_{b=0,\ldots,\,\left\lfloor T_{n}/n_{T}\right\rfloor -2}\left(\log\left(T_{n}\right)n_{T}\right)^{1/2}\left|B_{h,b+1}^{0}-\widetilde{B}_{h,b+1}^{0}\right|>\varepsilon/K\right)\nonumber \\
 & =\mathbb{P}\left(\max_{b=0,\ldots,\,\left\lfloor T_{n}/n_{T}\right\rfloor -2}\left(\log\left(T_{n}\right)n_{T}\right)^{1/2}n_{T}^{-1}\sum_{j=1}^{n_{T}}\left(u_{T_{m}+\tau+\left(b+1\right)n_{T}+j-1}-\widetilde{u}_{T_{m}+\tau+\left(b+1\right)n_{T}+j-1}\right)>\varepsilon/K\right).\label{eq (GB) A2}
\end{align}
By a mean-value expansion (omitting the second argument of $g\left(\cdot;\,\cdot\right)$
which is for both terms here equal to $\beta^{*}$), 
\begin{align*}
g & \left(\sigma_{e,\left(T_{m}+\tau+\left(b+1\right)n_{T}-1\right)h}h^{-1/2}\left(\Delta_{h}W_{e,T_{m}+\tau+\left(b+1\right)n_{T}+j-1}\right)\right)\\
 & \quad-g\left(\sigma_{e,\left(T_{m}+\tau+bn_{T}-1\right)h}h^{-1/2}\left(\Delta_{h}W_{e,T_{m}+\tau+\left(b+1\right)n_{T}+j-1}\right)\right)\\
 & =g_{e}\left(\sigma_{e,\left(T_{m}+\tau+bn_{T}-1\right)h}h^{-1/2}\Delta_{h}W_{e,T_{m}+\tau+\left(b+1\right)n_{T}+j-1}\right)\\
 & \quad\times\left[\left(\sigma_{e,\left(T_{m}+\tau+bn_{T}-1\right)h}-\sigma_{e,\left(T_{m}+\tau+\left(b+1\right)n_{T}-1\right)h}\right)\left(h^{-1/2}\Delta_{h}W_{e,T_{m}+\tau+\left(b+1\right)n_{T}+j-1}\right)\right]\\
 & \quad+2^{-1}g_{ee,b,j}\left(\overline{e}\right)\left[\left(\sigma_{e,\left(T_{m}+\tau+bn_{T}-1\right)h}-\sigma_{e,\left(T_{m}+\tau+\left(b+1\right)n_{T}-1\right)h}\right)\left(h^{-1/2}\Delta_{h}W_{e,T_{m}+\tau+\left(b+1\right)n_{T}+j-1}\right)\right]^{2}.
\end{align*}
 In view of Assumption \ref{Assumption Lipchtitz cont of Sigma},
for $r=1,\,2$, 
\begin{align}
\left|\sigma_{e,\left(T_{m}+\tau+bn_{T}-1\right)h}-\sigma_{e,\left(T_{m}+\tau+\left(b+1\right)n_{T}-1\right)h}\right|^{r} & \leq C_{r}\left(n_{T}h\right)^{r},\label{eq (3) GB A2}
\end{align}
uniformly in $b$ where $C_{r}<\infty$. Let
\begin{align*}
\overline{C}_{1}\triangleq2\sup_{k\geq1}\sup_{t\geq0}\left|g_{e}\left(\sigma_{e,t}h^{-1/2}\Delta_{h}W_{e,k}\right)\right|, & \qquad\overline{C}_{2}\triangleq2\sup_{k\geq1}\sup_{t\geq0}\left|g_{ee}\left(\sigma_{e,t}h^{-1/2}\Delta_{h}W_{e,k}\right)\right|.
\end{align*}
Then, the right-hand side of \eqref{eq (GB) A2} can be decomposed
as follows with $K_{1}=\sqrt{2\overline{C}_{1}}$ and $K_{2}=\sqrt{2\overline{C}_{2}}$,
\begin{align*}
\mathbb{P} & \left(\max_{b=0,\ldots,\,\left\lfloor T_{n}/n_{T}\right\rfloor -2}\left(\log\left(T_{n}\right)n_{T}\right)^{1/2}n_{T}^{-1}\sum_{j=1}^{n_{T}}\left(\widetilde{u}_{T_{m}+\tau+\left(b+1\right)n_{T}-1}-u_{T_{m}+\tau+\left(b+1\right)n_{T}-1}\right)>\varepsilon/K\right)\\
 & \leq\mathbb{P}\left(\max_{b=0,\ldots,\,\left\lfloor T_{n}/n_{T}\right\rfloor -2}\left|\left(\log\left(T_{n}\right)n_{T}\right)^{1/2}\left(\sigma_{e,\left(T_{m}+\tau+bn_{T}-1\right)h}-\sigma_{e,\left(T_{m}+\tau+\left(b+1\right)n_{T}-1\right)h}\right)\right|>\varepsilon/\left(K_{1}\cdot K\right)\right)\\
 & \quad+\mathbb{P}\left(\max_{b=0,\ldots,\,\left\lfloor T_{n}/n_{T}\right\rfloor -2}\left|n_{T}^{-1}\sum_{j=1}^{n_{T}}g_{e}\left(\sigma_{e,\left(T_{m}+\tau+bn_{T}-1\right)h}h^{-1/2}\Delta_{h}W_{e,T_{m}+\tau+\left(b+1\right)n_{T}+j-1}\right)\right.\right.\\
 & \left.\quad\times\left.\left(h^{-1/2}\Delta_{h}W_{e,T_{m}+\tau+\left(b+1\right)n_{T}+j-1}\right)\right|>K_{1}\right)\\
 & \quad+\mathbb{P}\left(\max_{b=0,\ldots,\,\left\lfloor T_{n}/n_{T}\right\rfloor -2}\left(\log\left(T_{n}\right)n_{T}\right)^{1/2}\left(\sigma_{e,\left(T_{m}+\tau+bn_{T}-1\right)h}-\sigma_{e,\left(T_{m}+\tau+\left(b+1\right)n_{T}-1\right)h}\right)^{2}>\varepsilon/\left(2K_{2}\cdot K\right)\right)\\
 & \quad+\mathbb{P}\left(\max_{b=0,\ldots,\,\left\lfloor T_{n}/n_{T}\right\rfloor -2}\left|2^{-1}n_{T}^{-1}\sum_{j=1}^{n_{T}}g_{ee,b,j}\left(\overline{e}\right)\left(h^{-1/2}\Delta_{h}W_{e,T_{m}+\tau+\left(b+1\right)n_{T}+j-1}\right)^{2}\right|>K_{2}\right)\\
 & \triangleq A_{1,h}+A_{2,h}+A_{3,h}+A_{4,h}.
\end{align*}
The relationship in \eqref{eq (3) GB A2} implies that $A_{1,h},\,A_{3,h}\rightarrow0$
using Condition \ref{Cond The-auxiliary-sequence} because 
\begin{align*}
\max_{b=0,\ldots,\,\left\lfloor T_{n}/n_{T}\right\rfloor -2} & \left|\left(\log\left(T_{n}\right)n_{T}\right)^{1/2}\left(\sigma_{e,\left(T_{m}+\tau+bn_{T}-1\right)h}-\sigma_{e,\left(T_{m}+\tau+\left(b+1\right)n_{T}-1\right)h}\right)\right|\\
 & \leq\left(\log\left(T_{n}\right)n_{T}\right)^{1/2}\phi_{\sigma,n_{T}h,N}\rightarrow0.
\end{align*}
The boundedness of $g_{e}\left(\cdot,\cdot\right)$ {[}cf. Assumption
\ref{Assumption Bounded on Bounded Sets LX (2-iv)}{]}, implies that
for $r>0$ large enough, 
\begin{align*}
\mathbb{P} & \left(\max_{b=0,\ldots,\,\left\lfloor T_{n}/n_{T}\right\rfloor -2}n_{T}^{-1}\sum_{j=1}^{n_{T}}\left|g_{e}\left(\sigma_{e,\left(T_{m}+\tau+bn_{T}-1\right)h}h^{-1/2}\Delta_{h}W_{e,T_{m}+\tau+\left(b+1\right)n_{T}+j-1}\right)\right.\right.\\
 & \quad\left.\left.\times h^{-1/2}\Delta_{h}W_{e,T_{m}+\tau+\left(b+1\right)n_{T}+j-1}\right|>K_{1}\right)\\
 & \leq\mathbb{P}\left(\max_{b=0,\ldots,\,\left\lfloor T_{n}/n_{T}\right\rfloor -2}\left|n_{T}^{-2}\overline{C}_{1}^{2}\sum_{j=1}^{n_{T}}\left(h^{-1/2}\Delta_{h}W_{e,T_{m}+\tau+\left(b+1\right)n_{T}+j-1}\right)^{2}-1\right|^{r/2}>\left(K_{1}^{2}/2\right)^{r/2}\right)\\
 & \leq\left(2/K_{1}^{2}\right)^{r/2}C_{r}\sum_{b=0}^{\left\lfloor T_{n}/n_{T}\right|-2}\mathbb{E}\left[\left|n_{T}^{-2}\sum_{j=1}^{n_{T}}\left(\left(h^{-1/2}\Delta_{h}W_{e,T_{m}+\tau+\left(b+1\right)n_{T}+j-1}\right)^{2}-1\right)\right|^{r/2}\right]\\
 & \leq\left(2/K_{1}^{2}\right)^{r}C_{r}T_{n}n_{T}^{-1-3r/2}\rightarrow0.
\end{align*}
 We can apply the same argument with $K_{2}=\sqrt{2}\overline{C}_{2}$
to $A_{4,h}$ to show that 
\begin{align*}
\mathbb{P} & \left(\max_{b=0,\ldots,\,\left\lfloor T_{n}/n_{T}\right\rfloor -2}n_{T}^{-1}\sum_{i=1}^{n_{T}}g_{ee,b,j}\left(\overline{e}\right)\left(h^{-1/2}\Delta_{T_{m}+\tau+bn_{T}+i-1}W_{e}\right)^{2}>K_{2}\right)\\
 & \leq K_{2}^{2r}C_{r}T_{n}n_{T}^{-1-r}\rightarrow0,
\end{align*}
and so $A_{4,h}\rightarrow0$. This gives \eqref{eq. (55) BJV-2}
and thus \eqref{eq. (54b) in Prop A.2 GB}. Next, we consider $R_{2,h}$
and want to show $\left(\log\left(T_{n}\right)n_{T}\right)^{1/2}R_{2,h}\overset{\mathbb{P}}{\rightarrow}0,$
or 
\begin{align}
\max_{b=0,\ldots,\,\left\lfloor T_{n}/n_{T}\right\rfloor -2}\left(\log\left(T_{n}\right)n_{T}\right)^{1/2}\left|\frac{\left(\widetilde{B}_{h,b+1}^{0}-B_{h,b}^{0}\right)\left(\sqrt{\widetilde{D}_{h,b+1}^{0}}-\sqrt{D_{h,b+1}^{0}}\right)}{\sqrt{D_{h,b+1}^{0}\widetilde{D}_{h,b+1}^{0}}}\right| & =o_{\mathbb{P}}\left(1\right).\label{eq. (54b) in Prop A.2 GB R2}
\end{align}
 Proceeding as in \eqref{eq. (55) BJV-2}, it is sufficient show 
\begin{align*}
\mathbb{P}\left(\max_{b=0,\ldots,\,\left\lfloor T_{n}/n_{T}\right\rfloor -2}\left(\log\left(T_{n}\right)n_{T}\right)^{1/2}\left|\left(\widetilde{B}_{h,b+1}^{0}-B_{h,b}^{0}\right)\left(\sqrt{\widetilde{D}_{h,b+1}^{0}}-\sqrt{D_{h,b+1}^{0}}\right)\right|>\varepsilon/K\right) & \rightarrow0.
\end{align*}
 The argument for $\left(\log\left(T_{n}\right)n_{T}\right)^{1/2}\left(\widetilde{B}_{h,b+1}^{0}-B_{h,b}^{0}\right)$
is similar to the one used above, but now one needs an additional
step using a Taylor series expansion of $g$; we omit the details.
Thus, we have to show 
\begin{align*}
\mathbb{P}\left(\max_{b=0,\ldots,\,\left\lfloor T_{n}/n_{T}\right\rfloor -2}\left|\sqrt{\widetilde{D}_{h,b+1}^{0}}-\sqrt{D_{h,b+1}^{0}}\right|>C\varepsilon\right) & \rightarrow0,
\end{align*}
 for some finite $C>0$. Note that 
\begin{align*}
\sqrt{\widetilde{D}_{h,b}^{0}}-\sqrt{D_{h,b}^{0}} & =\sigma_{u,\left(T_{m}+\tau+\left(b-1\right)n_{T}-1\right)h}-\sigma_{u,\left(T_{m}+\tau+bn_{T}-1\right)h}+O_{\mathbb{P}}\left(n_{T}^{-1/2}\right)\\
 & =\phi_{\sigma_{u},n_{T}h,N}+O_{\mathbb{P}}\left(n_{T}^{-1/2}\right).
\end{align*}
 Since $\phi_{\sigma_{u},n_{T}h,N}\leq Cn_{T}h$ uniformly over $h,\ldots,\,Th=N,$
we can show using the same arguments employed above that $\mathbb{P}\left(\max_{b=0,\ldots,\,\left\lfloor T_{n}/n_{T}\right\rfloor -2}\left|\left(\widetilde{D}_{h,b+1}^{0}-D_{h,b+1}^{0}\right)\right|>C\varepsilon\right)\rightarrow0$.
Therefore, we have $\left(\log\left(T_{n}\right)n_{T}\right)^{1/2}R_{2,h}\overset{\mathbb{P}}{\rightarrow}0$.
The claim of the lemma follows. $\square$
\begin{lem}
\label{Lemma GB Prop A.3 in BJV}As $h\downarrow0$, $\left(\log\left(T_{n}\right)n_{T}\right)^{1/2}\left(\mathrm{V}_{\mathrm{max},h}\left(T_{n},\,\tau\right)-\widetilde{\mathrm{B}}_{\mathrm{max},h}^{0}\left(T_{n},\,\tau\right)\right)\overset{\mathbb{P}}{\rightarrow}0$.
\end{lem}
\noindent\textit{Proof}. We have the inequality, 
\begin{align*}
\mathrm{V}_{\mathrm{max},h}\left(T_{n},\,\tau\right) & -\widetilde{\mathrm{B}}_{\mathrm{max},h}^{0}\left(T_{n},\,\tau\right)\\
 & =\max_{b=0,\ldots,\,\left\lfloor T_{n}/n_{T}\right\rfloor -2}\left|\frac{\widetilde{B}_{h,b+1}^{0}-B_{h,b}^{0}}{\sigma_{u,\left(T_{m}+\tau+bn_{T}-1\right)h}}\right|-\max_{b=0,\ldots,\,\left\lfloor T_{n}/n_{T}\right\rfloor -2}\left|\frac{\widetilde{B}_{h,b+1}^{0}-B_{h,b}^{0}}{\sqrt{\widetilde{D}_{h,b+1}^{0}}}\right|\\
 & \leq\max_{b=0,\ldots,\,\left\lfloor T_{n}/n_{T}\right\rfloor -2}\left|\frac{\left(\widetilde{B}_{h,b+1}^{0}-B_{h,b}^{0}\right)\left(\widetilde{D}_{h,b+1}^{0}-\sigma_{u,\left(T_{m}+\tau+bn_{T}-1\right)h}\right)}{\sigma_{u,\left(T_{m}+\tau+bn_{T}-1\right)h}\sqrt{\widetilde{D}_{h,b+1}^{0}}}\right|.
\end{align*}
Thus, we want to show that 
\begin{align}
\left(\log\left(T_{n}\right)n_{T}\right)^{1/2}\max_{b=0,\ldots,\,\left\lfloor T_{n}/n_{T}\right\rfloor -2}\left|\frac{\left(\widetilde{B}_{h,b+1}^{0}-B_{h,b}^{0}\right)\left(\sqrt{\widetilde{D}_{h,b+1}^{0}}-\sigma_{u,\left(T_{m}+\tau+bn_{T}-1\right)h}\right)}{\sigma_{u,\left(T_{m}+\tau+bn_{T}-1\right)h}\sqrt{\widetilde{D}_{h,b+1}^{0}}}\right| & =o_{\mathbb{P}}\left(1\right).\label{eq. Eq. in GB Prop A.3}
\end{align}
By Assumption \ref{Assumption Moments of Losses}, $0<\sigma_{u,\left(T_{m}+\tau+bn_{T}-1\right)h}<\infty$
for all $b\geq0$ while $\widetilde{D}_{h,b+1}^{0}$ was already shown
to bounded from below and above. Thus, basic manipulations as in the
previous lemmas show that the denominator is also $O_{\mathbb{P}}\left(1\right)$.
Turning to the numerator, we have
\begin{align*}
\mathbb{P} & \left(\left(\log\left(T_{n}\right)n_{T}\right)^{1/2}\max_{b=0,\ldots,\,\left\lfloor T_{n}/n_{T}\right\rfloor -2}\left|\left(\widetilde{B}_{h,b+1}^{0}-B_{h,b}^{0}\right)\left(\sqrt{\widetilde{D}_{h,b+1}^{0}}-\sigma_{u,\left(T_{m}+\tau+bn_{T}-1\right)h}\right)\right|>\varepsilon\right)\\
 & \leq\mathbb{P}\left(\left(\log\left(T_{n}\right)n_{T}\right)^{1/2}\max_{b=0,\ldots,\,\left\lfloor T_{n}/n_{T}\right\rfloor -2}\left|\widetilde{B}_{h,b+1}^{0}-B_{h,b}^{0}\right|>\sqrt{\varepsilon}\right)\\
 & +\mathbb{P}\left(\max_{b=0,\ldots,\,\left\lfloor T_{n}/n_{T}\right\rfloor -2}\left|\sqrt{\widetilde{D}_{h,b+1}^{0}}-\sigma_{u,\left(T_{m}+\tau+bn_{T}-1\right)h}\right|>\sqrt{\varepsilon}\right).
\end{align*}
 In view of the proof of the last part of Lemma \ref{Lemma Prop GB A2 BJV},
\begin{align*}
\mathbb{P}\left(\left(\log\left(T_{n}\right)n_{T}\right)^{1/2}\max_{b=0,\ldots,\,\left\lfloor T_{n}/n_{T}\right\rfloor -2}\left|\widetilde{B}_{h,b+1}^{0}-B_{h,b}^{0}\right|>\sqrt{\varepsilon}\right) & \rightarrow0.
\end{align*}
To conclude the proof of the lemma it remains to show that
\begin{align}
\mathbb{P}\left(\max_{b=0,\ldots,\,\left\lfloor T_{n}/n_{T}\right\rfloor -2}\left|\sqrt{\widetilde{D}_{h,b+1}^{0}}-\sigma_{u,\left(T_{m}+\tau+bn_{T}-1\right)h}\right|>\sqrt{\varepsilon}\right) & \rightarrow0.\label{eq Eq. GB Prop A.3b}
\end{align}
By the definition of $\widetilde{D}_{h,b+1}^{0}$, the summands $\widetilde{u}_{T_{m}+\tau+bn_{T}+j-1},\,\left(j=1,\ldots,\,n_{T}\right)$
are independent and each satisfies $\mathrm{Var}\left[\widetilde{u}_{T_{m}+\tau+bn_{T}+j-1}\right]=\sigma_{u,\left(T_{m}+\tau+bn_{T}-1\right)h}^{2}$.
Then,
\begin{align*}
\mathbb{P} & \left(\max_{b=0,\ldots,\,\left\lfloor T_{n}/n_{T}\right\rfloor -2}\left|\sqrt{\widetilde{D}_{h,b+1}^{0}}-\sigma_{u,\left(T_{m}+\tau+bn_{T}-1\right)h}\right|>\sqrt{\varepsilon}\right)\\
 & \leq\mathbb{P}\left(\max_{b=0,\ldots,\,\left\lfloor T_{n}/n_{T}\right\rfloor -2}\right.\\
 & \quad\times\left.\left|\sqrt{n_{T}^{-1}\sum_{j=1}^{n_{T}}\left(g\left(\sigma_{e,\left(T_{m}+\tau+bn_{T}-1\right)h}h^{-1/2}\left(\Delta_{h}W_{e,T_{m}+\tau+\left(b+1\right)n_{T}+j-1}\right)\right)-\overline{\widetilde{g}}_{b+1}\right)^{2}}-\sigma_{u,\left(T_{m}+\tau+bn_{T}-1\right)h}\right|>\sqrt{\varepsilon}\right).
\end{align*}
Note that the variables $g\left(\sigma_{e,\left(T_{m}+\tau+bn_{T}-1\right)h}h^{-1/2}\left(\Delta_{h}W_{e,T_{m}+\tau+\left(b+1\right)n_{T}+j-1}\right)\right)$
are independent over $j$ and their variances are constant and equal
to $\sigma_{u,\left(T_{m}+\tau+bn_{T}-1\right)h}^{2}$. Due to the
i.i.d. structure we can rely on a basic CLT for the sample variance
which, given Assumption \ref{Assumption 4th moment fo forecast error},
yields 
\begin{align*}
\mathbb{E}\left[\left|\sqrt{n_{T}^{-1/2}\sum_{j=1}^{n_{T}}\left(g\left(\sigma_{e,\left(T_{m}+\tau+bn_{T}-1\right)h}h^{-1/2}\left(\Delta_{h}W_{e,T_{m}+\tau+\left(b+1\right)n_{T}+j-1}\right)\right)-\overline{\widetilde{g}}_{b+1}\right)^{2}}-\sigma_{u,\left(T_{m}+\tau+bn_{T}-1\right)h}\right|\right] & <\infty,
\end{align*}
and thus for $r>0$ sufficiently large, we have by Condition \ref{Cond The-auxiliary-sequence},
\begin{align*}
\mathbb{P} & \left(\max_{b=0,\ldots,\,\left\lfloor T_{n}/n_{T}\right\rfloor -2}\left|\sqrt{\widetilde{D}_{h,b+1}^{0}}-\sigma_{u,\left(T_{m}+\tau+bn_{T}-1\right)h}\right|>\sqrt{\varepsilon}\right)\\
 & \leq C_{r}\varepsilon^{-r/2}\sum_{b=0}^{\left\lfloor T_{n}/n_{T}\right\rfloor -2}\\
 & \quad\times\mathbb{E}\left[\left|\sqrt{n_{T}^{-1}\sum_{j=1}^{n_{T}}\left(g\left(\sigma_{e,\left(T_{m}+\tau+bn_{T}-1\right)h}h^{-1/2}\left(\Delta_{T_{m}+\tau+\left(b+1\right)n_{T}+j-1}W_{e}\right)\right)-\overline{\widetilde{g}}_{b+1}\right)^{2}}-\sigma_{u,\left(T_{m}+\tau+bn_{T}-1\right)h}\right|^{r}\right]\\
 & \leq C_{r}\varepsilon^{-r/2}O_{\mathbb{P}}\left(T_{n}/n_{T}\right)n_{T}^{-r/2}\rightarrow0.
\end{align*}
 Altogether, we have \eqref{eq Eq. GB Prop A.3b} and thus \eqref{eq. Eq. in GB Prop A.3},
which concludes the proof. $\square$ 

\medskip{}

\noindent \textit{Proof of Theorem \ref{Theoem Asymptotic H0 Distrbution Gmax QGmax}-(i).}
By Lemma \ref{Lemma GB A1}-\ref{Lemma GB Prop A.3 in BJV},
\begin{align*}
\left(\log\left(T_{n}\right)n_{T}\right)^{1/2}\left(\mathrm{G}_{\mathrm{max},h}\left(T_{n},\,\tau\right)-\mathrm{V}_{\mathrm{max},h}\left(T_{n},\,\tau\right)\right) & \overset{\mathbb{P}}{\rightarrow}0.
\end{align*}
 We now apply Lemma \ref{Lemma QB Prop A.5.} to $\mathrm{V}_{\mathrm{max},h}\left(T_{n},\,\tau\right)$.
Let
\begin{align*}
\widetilde{U}_{T_{m}+\tau+bn_{T}+j-1} & \triangleq\frac{\widetilde{u}_{T_{m}+\tau+bn_{T}+j-1}-\mu_{u,T_{m}+\tau+\left(b-1\right)n_{T}-1}}{\sigma_{u,\left(T_{m}+\tau+\left(b-1\right)n_{T}-1\right)h}}\\
U_{T_{m}+\tau+bn_{T}+j-1} & \triangleq\frac{u_{T_{m}+\tau+bn_{T}+j-1}-\mu_{u,T_{m}+\tau+bn_{T}-1}}{\sigma_{u,\left(T_{m}+\tau+bn_{T}-1\right)h}},
\end{align*}
with $\mu_{u,T_{m}+\tau+bn_{T}-1}\triangleq\mathbb{E}\left(u_{T_{m}+\tau+bn_{T}-1}\right)$.
Then, write 
\begin{align*}
\mathrm{V}_{\mathrm{max},h}\left(T_{n},\,\tau\right) & =\max_{b=0,\ldots,\,\left\lfloor T_{n}/n_{T}\right\rfloor -2}\left|n_{T}^{-1}\sum_{j=1}^{n_{T}}\left(\widetilde{U}_{T_{m}+\tau+\left(b+1\right)n_{T}+j-1}-U_{T_{m}+\tau+bn_{T}+j-1}\right)\right|.
\end{align*}
Observe that the variables $\widetilde{U}_{T_{m}+\tau+\left(b+1\right)n_{T}+j-1}$
and $U_{T_{m}+\tau+bn_{T}+j-1}$, $\left(j=1,\ldots,\,n_{T}\right)$
have both zero mean, unit variance and are independent over $b$ and
$j$. Thus, $\mathrm{V}_{\mathrm{max},h}\left(T_{n},\,\tau\right)$
corresponds to $\mathrm{B}_{\mathrm{max},T_{n}}$ from Lemma \ref{Lemma QB Prop A.5.}.
In addition, under Assumption \ref{Assumption 4th moment fo forecast error}
and Condition \ref{Cond The-auxiliary-sequence} the final result
can be deduced from the same lemma. $\square$

\paragraph{Proof of part (ii) of Theorem \ref{Theoem Asymptotic H0 Distrbution Gmax QGmax}}

The proof follows similar steps as those used for $\mathrm{MQ}_{\mathrm{max},h}$.
More specifically, we can repeat the same proof as in Lemma \ref{Lemma GB A2}
so that corresponding results for a general loss function are still
valid. Let $\overline{U}_{h,i}\triangleq n_{T}^{-1}\sum_{j=i+1}^{i+n_{T}}\left(L_{\psi,\left(T_{m}+\tau+j-1\right)h}\left(\beta^{*}\right)-\overline{L}_{\psi,i}\left(\beta^{*}\right)\right)^{2}$
and define 
\begin{align*}
\mathrm{U}_{\max,h}\left(T_{n},\,\tau\right) & \triangleq\max_{i=n_{T},\ldots,\,T_{n}-n_{T}}\left|\frac{n_{T}^{-1}\sum_{j=i+1}^{i+n_{T}}SL_{\psi,T_{m}+\tau+j-1}\left(\beta^{*}\right)-n_{T}^{-1}\sum_{j=i-n_{T}+1}^{i}SL_{\psi,T_{m}+\tau+j-1}\left(\beta^{*}\right)}{\sqrt{\overline{U}_{h,i}}}\right|.
\end{align*}
\begin{lem}
\label{Lemma GB A2 MB}For any $L\in\boldsymbol{L}_{e}$, we have
the results of Lemma \ref{Lemma GB A2} and
\begin{align*}
\left(\log\left(T_{n}\right)n_{T}\right)^{1/2}\left(\mathrm{U}_{\max,h}\left(T_{n},\,\tau\right)-\mathrm{MG}_{\mathrm{max},h}\left(T_{n},\,\tau\right)\right) & \overset{\mathbb{P}}{\rightarrow}0.
\end{align*}
\end{lem}
\noindent \textit{Proof}. The first claim can be proven in the same
fashion as in Lemma \ref{Lemma GB A2} with minor changes in notations.
Proceeding as in Lemma \ref{Lemma GB A1}, 
\begin{align}
\left|\mathrm{U}_{\max,h}\right. & \left.\left(T_{n},\,\tau\right)-\mathrm{MG}_{\mathrm{max},h}\left(T_{n},\,\tau\right)\right|\label{eq (1) Lemma GB MB A1}\\
 & \leq\max_{i=n_{T},\ldots,\,T_{n}-n_{T}}\left|\frac{n_{T}^{-1}\sum_{j=i+1}^{i+n_{T}}\left(SL_{\psi,T_{m}+\tau+j-1}\left(\beta^{*}\right)-SL_{\psi,T_{m}+\tau+j-1}\left(\widehat{\beta}\right)\right)}{\sqrt{\overline{U}_{h,i}D_{h,i}}}\right|\nonumber \\
 & \quad+\max_{i=n_{T},\ldots,\,T_{n}-n_{T}}\left|\frac{n_{T}^{-1}\sum_{j=i-n_{T}+1}^{i}\left(SL_{\psi,T_{m}+\tau+j-1}\left(\beta^{*}\right)-SL_{\psi,T_{m}+\tau+j-1}\left(\widehat{\beta}\right)\right)}{\sqrt{\overline{U}_{h,i}D_{h,i}}}\right|\nonumber \\
 & \leq C_{1}\max_{i=n_{T},\ldots,\,T_{n}-n_{T}}\left|\frac{n_{T}^{-1}\sum_{j=i+1}^{i+n_{T}}\left(SL_{\psi,T_{m}+\tau+j-1}\left(\beta^{*}\right)-SL_{\psi,T_{m}+\tau+j-1}\left(\widehat{\beta}\right)\right)}{\overline{U}_{h,i}}\right|\nonumber \\
 & \quad+C_{2}\max_{i=n_{T},\ldots,\,T_{n}-n_{T}}\left|\frac{n_{T}^{-1}\sum_{j=i-n_{T}+1}^{i}\left(SL_{\psi,T_{m}+\tau+j-1}\left(\beta^{*}\right)-SL_{\psi,T_{m}+\tau+j-1}\left(\widehat{\beta}\right)\right)}{\overline{U}_{h,i}^ {}}\right|.\nonumber 
\end{align}
Following the same derivations as in the non-overlapping case we have
a result corresponding to equation \eqref{eq (2) GB A2}, 
\begin{align}
\left|n_{T}^{-1}\sum_{j=i+1}^{i+n_{T}}\left(SL_{\psi,T_{m}+\tau+j-1}\left(\beta^{*}\right)-SL_{\psi,T_{m}+\tau+j-1}\left(\widehat{\beta}\right)\right)\right| & =KO_{\mathbb{P}}\left(T^{-1/2}\right).\label{eq (2) GB A2-1}
\end{align}
For any $\varepsilon>0$ and any constant $K>0$, we then have the
decomposition, 
\begin{align}
\mathbb{P} & \left(\max_{i=n_{T},\ldots,\,T_{n}-n_{T}}\left|\left(\log\left(T_{n}\right)n_{T}\right)^{1/2}\frac{n_{T}^{-1}\sum_{j=i+1}^{i+n_{T}}\left(SL_{\psi,T_{m}+\tau+j-1}\left(\beta^{*}\right)-SL_{\psi,T_{m}+\tau+j-1}\left(\widehat{\beta}\right)\right)}{\overline{U}_{h,i}}\right|>\varepsilon\right)\nonumber \\
 & \leq\mathbb{P}\left(\max_{i=n_{T},\ldots,\,T_{n}-n_{T}}\left|\left(\log\left(T_{n}\right)n_{T}\right)^{1/2}n_{T}^{-1}\sum_{j=i+1}^{i+n_{T}}\left(SL_{\psi,T_{m}+\tau+j-1}\left(\beta^{*}\right)-SL_{\psi,T_{m}+\tau+j-1}\left(\widehat{\beta}\right)\right)\right|>\varepsilon/K\right)\nonumber \\
 & \quad+\mathbb{P}\left(\max_{i=n_{T},\ldots,\,T_{n}-n_{T}}1/\left|\overline{U}_{h,i}\right|>K\right).\label{eq(1) GB A2-1}
\end{align}
Observe that Lemma \ref{Lemma eq(22) in Vetter (2012) for GB } remains
valid when blocks overlap and so $\mathbb{P}\left(\max_{i=n_{T},\ldots,\,T_{n}-n_{T}}1/\left|\overline{U}_{h,i}\right|>K\right)\rightarrow0$
by setting, for example, $K=1/\sigma_{L,-}^{2}$. Upon using Markov's
inequality, \eqref{eq (2) GB A2-1} (which holds uniformly in $i$)
and Condition \ref{Cond The-auxiliary-sequence} we can conclude the
proof with 
\begin{align*}
\mathbb{P} & \left(\max_{i=n_{T},\ldots,\,T_{n}-n_{T}}\left|\left(\log\left(T_{n}\right)n_{T}\right)^{1/2}n_{T}^{-1}\sum_{j=i+1}^{i+n_{T}}\left(SL_{\psi,T_{m}+\tau+j-1}\left(\beta^{*}\right)-SL_{\psi,T_{m}+\tau+j-1}\left(\widehat{\beta}\right)\right)\right|>\varepsilon/K\right)\\
 & \leq\frac{K}{\varepsilon}\mathbb{E}\left[\left(\log\left(T_{n}\right)n_{T}\right)^{1/2}n_{T}^{-1}\sum_{i=1}^{n_{T}}\left(SL_{\psi,T_{m}+\tau+j-1}\left(\beta^{*}\right)-SL_{\psi,T_{m}+\tau+j-1}\left(\widehat{\beta}\right)\right)\right]\\
 & =\frac{K}{\varepsilon}\left(\log\left(T_{n}\right)n_{T}\right)^{1/2}O_{\mathbb{P}}\left(1/\sqrt{T}\right)\rightarrow0.\,\square
\end{align*}

\medskip{}

Define
\begin{align*}
\mathrm{MB}_{\mathrm{max},h}^{0}\left(T_{n},\,\tau\right) & \triangleq\max_{i=n_{T},\ldots,\,T_{n}-n_{T}}\left|\left(n_{T}^{-1}\sum_{j=i+1}^{i+n_{T}}g\left(\sigma_{e,\left(T_{m}+\tau+i-1\right)h}h^{-1/2}\left(\Delta_{h}W_{e,T_{m}+\tau+j-1}\right)\right)\right.\right.\\
 & \left.\left.\quad-n_{T}^{-1}\sum_{j=i-n_{T}+1}^{i}g\left(\sigma_{e,\left(T_{m}+\tau+i-n_{T}-1\right)h}h^{-1/2}\left(\Delta_{h}W_{e,T_{m}+\tau+j-1}\right)\right)\right)/\sqrt{D_{h,i}^{0}}\right|.
\end{align*}
where
\begin{align*}
D_{h,i}^{0} & \triangleq n_{T}^{-1}\sum_{j=i+1}^{i+n_{T}}\left(g\left(\sigma_{e,\left(T_{m}+\tau+i-1\right)h}h^{-1/2}\left(\Delta_{h}W_{e,T_{m}+\tau+j-1}\right)\right)-\overline{g}_{i}\right)^{2},
\end{align*}
with $\overline{g}_{i}\triangleq n_{T}^{-1}\sum_{j=i+1}^{i+n_{T}}g\left(\sigma_{e,\left(T_{m}+\tau+i-1\right)h}h^{-1/2}\left(\Delta_{h}W_{e,T_{m}+\tau+j-1}\right)\right)$.
Next, let 
\begin{align}
\mathrm{MB}_{\mathrm{max},h}^{*}\left(T_{n},\,\tau\right) & \triangleq\max_{i=n_{T},\ldots,\,T_{n}-n_{T}}\left|\frac{n_{T}^{-1}\sum_{j=i+1}^{i+n_{T}}u_{T_{m}+\tau+j-1}^{*}-n_{T}^{-1}\sum_{j=i-n_{T}+1}^{i}u_{T_{m}+\tau+j-1}^{*}}{\sqrt{\overline{U}_{h,i}}}\right|,\label{eq: B^0 and B^*-1-1}
\end{align}
 where $u_{T_{m}+\tau+j-1}^{*}\triangleq g\left(\Delta_{h}\widetilde{e}_{T_{m}+\tau+j-1};\,\beta^{*}\right)$.
Then, define 
\begin{align*}
\widetilde{\mathrm{MB}}_{\mathrm{max},h}^{0}\left(T_{n},\,\tau\right) & \triangleq\max_{i=n_{T},\ldots,\,T_{n}-n_{T}}\left|\frac{n_{T}^{-1}\sum_{j=i+1}^{i+n_{T}}\widetilde{u}_{T_{m}+\tau+j-1}-n_{T}^{-1}\sum_{j=i-n_{T}+1}^{i}u_{T_{m}+\tau+j-1}}{\sqrt{\widetilde{D}_{h,b}^{0}}}\right|
\end{align*}
where $\widetilde{D}_{h,i}^{0}\triangleq n_{T}^{-1}\sum_{j=i+1}^{i+n_{T}}\left(\widetilde{u}_{T_{m}+\tau+j-1}-\overline{\widetilde{g}}_{i}\right)^{2}$,
with $\overline{\widetilde{g}}_{i}\triangleq n_{T}^{-1}\sum_{j=i+1}^{i+n_{T}}\widetilde{u}_{T_{m}+\tau+j-1}$
and 
\begin{align*}
\widetilde{u}_{T_{m}+\tau+j-1}\triangleq g\left(\sigma_{e,\left(T_{m}+\tau+i-n_{T}-1\right)h}h^{-1/2}\Delta_{h}W_{e,T_{m}+\tau+j-1};\,\beta^{*}\right) & .
\end{align*}
In the final step we shall show that $\mathbb{P}\left(\left(\log\left(T_{n}\right)n_{T}\right)^{1/2}\left(\mathrm{MV}_{\mathrm{max},h}\left(T_{n},\,\tau\right)-\widetilde{\mathrm{MB}}_{\mathrm{max},h}^{0}\left(T_{n},\,\tau\right)\right)>\varepsilon\right)\rightarrow0$
for any $\varepsilon>0,$ where 
\begin{align*}
\mathrm{MV}_{\mathrm{max},h}\left(T_{n},\,\tau\right) & \triangleq\max_{i=n_{T},\ldots,\,T_{n}-n_{T}}\left|\frac{n_{T}^{-1}\sum_{j=i+1}^{i+n_{T}}\widetilde{u}_{T_{m}+\tau+j-1}-n_{T}^{-1}\sum_{j=i-n_{T}+1}^{i}u_{T_{m}+\tau+j-1}}{\sigma_{u,\left(T_{m}+\tau+i-n_{T}-1\right)h}}\right|,
\end{align*}
with $\sigma_{u,\left(T_{m}+\tau-i-n_{T}-1\right)h}\triangleq\left(\mathrm{Var}\left(u_{T_{m}+\tau+i-n_{T}}|\,\mathscr{F}_{\left(T_{m}+\tau+i-n_{T}-1\right)h}\right)\right)^{1/2}$.
By Assumption \ref{Assumption Localization Condition 1} there exist
$0<\sigma_{u,-}<\sigma_{u,+}<\infty$ defined by $\sigma_{u,-}\triangleq\inf_{k\geq1}\left\{ \sigma_{u,kh}\right\} $
and $\sigma_{u,+}\triangleq\sup_{k\geq1}\left\{ \sigma_{u,kh}\right\} $.
In parts of the derivations below we shall use some of the results
from the non-overlapping case. In particular, the only difference
arises from the fact that now the maximum is over a larger set and
therefore the bounds should be adjusted accordingly.
\begin{lem}
\label{Prop GB A1 BJV-1}As $h\downarrow0$, $\left(\log\left(T_{n}\right)n_{T}\right)^{1/2}\left(\mathrm{U}_{\mathrm{max},h}\left(T_{n},\,\tau\right)-\mathrm{MB}_{\mathrm{max},h}^{0}\left(T_{n},\,\tau\right)\right)\overset{\mathbb{P}}{\rightarrow}0$. 
\end{lem}
\noindent \textit{Proof}. First, given Lemma \ref{Lemma GB A2 MB}
it follows that $\left(\log\left(T_{n}\right)n_{T}\right)^{1/2}\left(\mathrm{MB}_{\mathrm{max},h}^{*}\left(T_{n},\,\tau\right)-\mathrm{U}_{\mathrm{max},h}\left(T_{n},\,\tau\right)\right)\overset{\mathbb{P}}{\rightarrow}0$.
Thus, we have to show
\begin{align*}
\left(\log\left(T_{n}\right)n_{T}\right)^{1/2}\left(\mathrm{MB}_{\mathrm{max},h}^{*}\left(T_{n},\,\tau\right)-\mathrm{MB}_{\mathrm{max},h}^{0}\left(T_{n},\,\tau\right)\right) & \overset{\mathbb{P}}{\rightarrow}0.
\end{align*}
 Note that the result of Lemma \ref{Lemma eq(22) in Vetter (2012) for GB }
still holds. Thus, we have decompositions similar to \eqref{eq. 47 GB BJV}
and \eqref{eq (A1A2A3) GB} and then one can follow the same steps
as above. However, the bounds in \eqref{eq (A1) GB} and \eqref{eq (A2) GB}
are now different because the maximum is over $i=n_{T},\ldots,\,T_{n}-n_{T}$.
The bound in \eqref{eq (A1) GB} is now $\left(K/\varepsilon\right)^{r}n_{T}^{\varpi r+r}\left(2h^{r-1}\right)$
which converges to zero by choosing $r$ sufficiently large and $\varpi$
small. The bound corresponding to \eqref{eq (A2) GB} also goes to
zero for large enough $r>0$. All the steps leading to \eqref{eq (2) GB 2 - e*-3}
can be repeated with minor changes. Indeed, the bound \eqref{eq (2) GB 2 - e*-3}
also remains the same because it involves using the condition on Lipschitz
continuity, which gives for $r>0$ large enough,
\begin{align}
\mathbb{E} & \left[\left|\left(\log\left(T_{n}\right)n_{T}\right)^{1/2}\left(n_{T}\sqrt{h}\right)^{-1}\sum_{j=1}^{n_{T}}\int_{\left(T_{m}+\tau+i+j-2\right)h}^{\left(T_{m}+\tau+i+j-1\right)h}\left(\sigma_{e,s}-\sigma_{e,\left(T_{m}+\tau+i-2\right)h}\right)dW_{e,s}\right|^{r}\right]\nonumber \\
 & \leq K_{r}\left(\left(\log\left(T_{n}\right)n_{T}\right)^{1/2}\left(n_{T}\sqrt{h}\right)^{-1}\right)^{r}\left(\int_{\left(T_{m}+\tau+i-1\right)h}^{\left(T_{m}+\tau+i+n_{T}-1\right)h}\left(\mathbb{E}\left[\phi_{\sigma,n_{T}h,N}^{r}\right]\right)^{2/r}ds\right)^{r/2}\nonumber \\
 & \leq K_{r}\left(\left(\log\left(T_{n}\right)\right)^{1/2}\right)^{r}h^{-1/3+r/3+\epsilon}\rightarrow0.\label{eq (2) GB 2 - e*-1 MB}
\end{align}
 Altogether, these arguments can be used to verify the result of the
lemma. $\square$
\begin{lem}
\label{Lemma Prop GB MB A2 BJV}As $h\downarrow0$, $\left(\log\left(T_{n}\right)n_{T}\right)^{1/2}\left(\mathrm{MB}_{\mathrm{max},h}^{0}\left(T_{n},\,\tau\right)-\mathrm{MV}_{\mathrm{max},h}\left(T_{n},\,\tau\right)\right)\overset{\mathbb{P}}{\rightarrow}0$.
\end{lem}
\noindent \textit{Proof}. The proof follows exactly the same steps
as in the proof of Lemma \ref{Lemma Prop GB A2 BJV}-\ref{Lemma GB Prop A.3 in BJV}.
Since some of the bounds need to be adjusted to account for the maximum
being over $i=n_{T},\ldots,\,T_{n}-n_{T}$, we can use the same argument
as in the previous lemma. Then, all the quantities generalizing the
expressions in the proofs of Lemma \ref{Lemma Prop GB A2 BJV}-\ref{Lemma GB Prop A.3 in BJV}
are controlled thereby yielding $\left(\log\left(T_{n}\right)n_{T}\right)^{1/2}\left(\mathrm{MB}_{\mathrm{max},h}^{0}\left(T_{n},\,\tau\right)-\widetilde{\mathrm{MB}}_{\mathrm{max},h}^{0}\left(T_{n},\,\tau\right)\right)\overset{\mathbb{P}}{\rightarrow}0$
and $\left(\log\left(T_{n}\right)n_{T}\right)^{1/2}\left(\mathrm{MV}_{\mathrm{max},h}\left(T_{n},\,\tau\right)-\widetilde{\mathrm{MB}}_{\mathrm{max},h}^{0}\left(T_{n},\,\tau\right)\right)\overset{\mathbb{P}}{\rightarrow}0$.
$\square$

\medskip{}

\noindent\textit{Proof of Theorem \ref{Theoem Asymptotic H0 Distrbution Gmax QGmax}-(ii).}
From Lemma \ref{Lemma GB A2 MB}-\ref{Lemma Prop GB MB A2 BJV}, $\sqrt{\log\left(T\right)n_{T}}\left(\mathrm{MG}_{\mathrm{max},h}\left(T_{n},\,\tau\right)-\mathrm{MV}{}_{\mathrm{max},h}\left(T_{n},\,\tau\right)\right)=o_{\mathbb{P}}\left(1\right)$.
As for the non-overlapping case, we deduce the limit distribution
of $\mathrm{MV}{}_{\mathrm{max},h}$ from that of $\mathrm{MB}{}_{\mathrm{max},T_{n}}$
derived in Lemma \ref{Lemma QB Prop A.5.}. Let 
\begin{align*}
\widetilde{U}_{T_{m}+\tau+j-1} & \triangleq\begin{cases}
\frac{\widetilde{u}_{T_{m}+\tau+j-1}-\mu_{u,T_{m}+\tau+i-n_{T}-1}}{\sigma_{u,\left(T_{m}+\tau+i-n_{T}-1\right)h}}, & \quad\mathrm{for}\,j=i+1,\ldots,\,i+n_{T}\\
\frac{u_{T_{m}+\tau+j-1}-\mu_{u,T_{m}+\tau+i-n_{T}-1}}{\sigma_{u,\left(T_{m}+\tau+i-n_{T}-1\right)h}} & \quad\mathrm{for}\,j=i-n_{T}+1,\ldots,\,i.
\end{cases}
\end{align*}
Then, we have $\mathbb{E}\left(\widetilde{U}_{T_{m}+\tau+j-1}\right)=0$,
$\mathrm{Var}\left(\widetilde{U}_{T_{m}+\tau+j-1}\right)=1$ and the
$\widetilde{U}_{T_{m}+\tau+j-1}$'s are independent across $j$. $\mathrm{MV}_{\mathrm{max},h}\left(T_{n},\,\tau\right)$
now corresponds to $\mathrm{MB}_{\mathrm{max},T_{n}}$ from Lemma
\ref{Lemma QB Prop A.5.}. Thus, we can deduce the final result from
Lemma \ref{Lemma QB Prop A.5.} since Assumption \ref{Assumption 4th moment fo forecast error}
and Condition \ref{eq. Condition auxiliary nT} holds. $\square$

\paragraph{\label{Section Negligibility-of-the drift }Negligibility of the
drift term under general loss functions}

The reasoning is similar to the quadratic loss case. We only show
that the drift component $\mu_{e,t}$ is of higher order. Without
estimation uncertainty our tests statistics are simply functions of
local averages of $g\left(\Delta_{h}\widetilde{e}_{k}^{*};\,\beta^{*}\right)$,
where $g\left(\cdot,\cdot\right)$ is smooth. Note that conditional
on $\left\{ \mu_{e,t}\right\} _{t\geq0}$ and $\left\{ \sigma_{e,t}\right\} _{t\geq0}$,
\begin{align*}
h^{-1/2}\Delta_{h}e_{T_{m}+\tau+bn_{T}+j-1}^{*} & =h^{-1/2}\int_{\left(T_{m}+\tau+bn_{T}+j-2\right)h}^{\left(T_{m}+\tau+bn_{T}+j-1\right)h}\mu_{e,s}h^{-\vartheta}ds+h^{-1/2}\int_{\left(T_{m}+\tau+bn_{T}+j-2\right)h}^{\left(T_{m}+\tau+bn_{T}+j-1\right)h}\sigma_{e,s}dW_{e,s}\\
 & =O\left(h^{1-\vartheta-1/2}\right)+h^{-1/2}\int_{\left(T_{m}+\tau+bn_{T}+j-2\right)h}^{\left(T_{m}+\tau+bn_{T}+j-1\right)h}\sigma_{e,s}dW_{e,s}.
\end{align*}
 Since $\vartheta\in[0,\,1/8)$ and $\int_{\left(T_{m}+\tau+bn_{T}+j-2\right)h}^{\left(T_{m}+\tau+bn_{T}+j-1\right)h}\sigma_{e,s}dW_{e,s}\approx\mathscr{N}\left(0,\,\int_{\left(T_{m}+\tau+bn_{T}+j-2\right)h}^{\left(T_{m}+\tau+bn_{T}+j-1\right)h}\sigma_{e,s}^{2}ds\right)$
it follows that the first term above is of higher order and should
not play any role for the asymptotic results of Lemma \ref{Lemma GB A2}-\ref{Lemma GB A1}.

\subsubsection{Proof of Corollary \ref{Corollary CR Null Distrb General}}

\noindent\textit{Proof.} It follows the same arguments as for Corollary
\ref{Corollary CR Null Distrb Quadratic}. $\square$

\subsection{\label{Subsection Proofs-of-Section Var Est}Proofs of Section \ref{Section Estimation-of-Asymptotic Variance}}

\subsubsection{Proof of Theorem \ref{Theorem: Var Est 3 in Wu and Zhao}}

\noindent\textit{Proof.} See Theorem 3 in \citet{wu/zhao:07}. $\square$

\subsubsection{Proof of Theorem \ref{Theorem Long-Run Variance}}

The initial step in the proof uses a uniform strong approximation
result which essentially extends the strong invariance principle of
\citet{wu:07} to our setting. The idea behind the proof is similar
to that of Theorem 2.1 in \citet{zhao/li:13}. Before giving the result,
we need to recall the more general framework of \citet{wu:07}. 

Let $\left\{ \xi_{k}\right\} _{k=1}^{T_{n}}$ be a sequence of zero-mean
independent random variables with $\mathrm{Var}\left(\xi_{k}\right)=\sigma_{k}^{2}$
satisfying $c_{-}\leq\min_{k\geq1}\left\{ \sigma_{k}\right\} $ and
$c_{+}\geq\max_{k\geq1}\left\{ \sigma_{k}\right\} $ with $0<c_{-}<c_{+}<\infty$.
Let $\overline{\xi}_{j}\triangleq j^{-1}\sum_{k=1}^{j}\xi_{k}$, $G_{\xi,j}\triangleq j\overline{\xi}_{j}$
and $V_{\xi,j}\triangleq\sum_{k=1}^{j}\left(\xi_{k}-\overline{\xi}_{j}\right)^{2}$.
Let $\left\{ \mathbb{B}_{t}\right\} _{t\geq0}$ and $\left\{ \widetilde{\mathbb{B}}_{t}\right\} _{t\geq0}$
denote two independent one-dimensional standard Wiener processes which
need not be defined on the same probability space. Finally, let $a_{T_{n}}\triangleq\left|\sigma_{T_{n}}\right|+\sum_{k=2}^{T_{n}}\left|\sigma_{k}-\sigma_{k-1}\right|$,
$c_{T_{n}}\triangleq\left|\sigma_{T_{n}}^{2}\right|+\sum_{k=2}^{T_{n}}\left|\sigma_{k}^{2}-\sigma_{k-1}^{2}\right|$,
$\Xi_{j}\triangleq\sum_{k=1}^{j}\sigma_{k}^{2}$ and $\widetilde{\Xi}_{j}^{2}\triangleq\sum_{k=1}^{j}\sigma_{k}^{4}$.
We begin with the following lemma involving a strong invariance principle
for the process $\left\{ \xi_{k}\right\} $ and an uniform approximation
for $\left\{ V_{\xi,k}\right\} $. Without loss of generality assume
that $\xi_{k}=\sigma_{k}\epsilon_{k},$ with $\left\{ \epsilon_{k}\right\} $
being a zero-mean stationary process with $\mathbb{E}\left(\epsilon_{k}^{2}\right)=1$.
Further, denote by $\varrho^{2}$ the long-tun variance of $\epsilon_{k}:$
 $\varrho^{2}\triangleq\gamma_{0}+2T_{n}^{-1}\sum_{i=1}^{T_{n}}\gamma_{i},$
where $\gamma_{i}\triangleq\mathrm{Cov}\left(\epsilon_{k+i},\,\epsilon_{k}\right)$.
Define similarly the long-run variance of $\left\{ \epsilon_{k}^{2}-1\right\} $
and denote it by $\widetilde{\varrho}^{2}$.  Next, let
\begin{align*}
S_{k}=\sum_{j=1}^{k}\epsilon_{j} & ,\qquad\widetilde{S}_{k}=\sum_{j=1}^{k}\left(\epsilon_{j}^{2}-1\right),\qquad k=1,\ldots,\,T_{n},
\end{align*}
with the convention $S_{0}=S_{0}^{*}=0$. Then we have the following
strong invariance principles {[}cf. \citet{wu:07}{]}:
\begin{align}
\max_{1\leq k\leq T_{n}}\left|S_{k}-\varrho\mathbb{B}_{k}\right|=o_{\mathrm{a.s.}}\left(\Delta_{T_{n}}\right) & \qquad\mathrm{and}\qquad\max_{1\leq k\leq T_{n}}\left|\widetilde{S}_{k}-\widetilde{\varrho}\mathbb{\widetilde{B}}_{k}\right|=o_{\mathrm{a.s.}}\left(\Delta_{T_{n}}\right),\label{eq. SIP approx err}
\end{align}
 where $\Delta_{T_{n}}$ is an approximation error that satisfies
$\Delta_{T_{n}}\rightarrow\infty$. Under our context, the order of
$\Delta_{T_{n}}$ is given by the following assumption 
\begin{assumption}
\label{Assumttion SIP}Assume $0<\varrho,\,\widetilde{\varrho}<\infty.$
The relationships in \eqref{eq. SIP approx err} holds with $\Delta_{T_{n}}=T_{n}^{1/4}\log\left(T_{n}\right)$.
\end{assumption}
\begin{lem}
\label{Lemma: Theorem 3.1 in Zhao and Li, eq. 2.7}Given Assumption
\ref{Assumttion SIP}, for any $\eta\in(0,\,1],$ \\
(i) $\max_{T_{n}\eta\leq j\leq T_{n}}\left|G_{\xi,j}-\varrho\sum_{k=1}^{j}\sigma_{k}\left(\mathbb{B}_{k}-\mathbb{B}_{k-1}\right)\right|=O_{\mathrm{a.s.}}\left(\Delta_{T_{n}}\right);$
\\
(ii) $\max_{T_{n}\eta\leq j\leq T_{n}}\left|V_{\xi,j}-\Xi_{j}\right|=O_{\mathrm{a.s.}}\left(\Delta_{T_{n}}+\widetilde{\Xi}_{j}+\left(\Delta_{T_{n}}^{2}+\Xi_{j}\right)/T_{n}\right)$.
\end{lem}
\noindent\textit{Proof.} To prove part (ii) one needs part (i). However,
the same steps in the initial part in the proof of (ii) can be used
to prove part (i) as we explain below. Thus, we only prove part (ii).
After some simple algebraic manipulations one can verify the decomposition
$V_{\xi,j}-\Xi_{j}=U_{j}-G_{\xi,j}^{2}/j$ where $U_{j}=\sum_{k=1}^{j}\sigma_{k}^{2}\left(\epsilon_{k}^{2}-1\right)$.
By Abel's formula and $\epsilon_{k}^{2}-1=S_{k}^{*}-S_{k-1}^{*}$,
we have 
\begin{align}
U_{j} & =\sum_{k=1}^{j}\sigma_{k}^{2}\left(\widetilde{S}_{k}-\widetilde{S}_{k-1}\right)=\left(\sigma_{k}^{2}\widetilde{S}_{j}-\sigma_{0}^{2}\widetilde{S}_{0}\right)-\sum_{k=1}^{j-1}\left(\sigma_{k+1}^{2}-\sigma_{k}^{2}\right)\widetilde{S}_{k}\nonumber \\
 & =\sigma_{k}^{2}\widetilde{S}_{j}-\sum_{k=1}^{j-1}\left(\sigma_{k+1}^{2}-\sigma_{k}^{2}\right)\widetilde{S}_{k},\label{eq. A1}
\end{align}
 and by the rightmost approximation in \eqref{eq. SIP approx err}
it follows that
\begin{align}
U_{j} & =\sigma_{j}^{2}\widetilde{\varrho}\widetilde{\mathbb{B}}_{j}-\sum_{k=1}^{j-1}\left(\sigma_{k+1}^{2}-\sigma_{k}^{2}\right)\widetilde{\varrho}\widetilde{\mathbb{B}}_{k}+O_{\mathrm{a.s.}}\left(\Delta_{T_{n}}\right)\nonumber \\
 & =\widetilde{\varrho}\sum_{k=1}^{j}\sigma_{k}^{2}\left(\widetilde{\mathbb{B}}_{k}-\widetilde{\mathbb{B}}_{k-1}\right)+O_{\mathrm{a.s.}}\left(\Delta_{T_{n}}\right).\label{eq. A2}
\end{align}
Next, by Kolmogorov's maximal inequality for independent random variables
{[}cf. Theorem 22.4 in \citet{billingsley:95}{]}, we have for $C>0$,
\begin{align*}
\mathbb{P}\left[\max_{1\leq j\leq T_{n}}\left|\widetilde{\varrho}\sum_{k=1}^{j}\sigma_{k}^{2}\left(\widetilde{\mathbb{B}}_{k}-\widetilde{\mathbb{B}}_{k-1}\right)\right|\geq C\widetilde{\Xi}_{T_{n}}\right] & \leq\left(C\widetilde{\Xi}_{T_{n}}/\widetilde{\varrho}\right)^{-2}\mathbb{E}\left[\left(\sum_{k=1}^{T_{n}}\sigma_{k}^{2}\left(\widetilde{\mathbb{B}}_{k}-\widetilde{\mathbb{B}}_{k-1}\right)\right)^{2}\right]=\left(\frac{\widetilde{\varrho}}{C}\right)^{-2}.
\end{align*}
Thus, choosing $C$ large enough shows that $\max_{1\leq k\leq T_{n}}\widetilde{\Xi}_{T_{n}}^{-1}\left|\widetilde{\varrho}\sum_{k=1}^{T_{n}}\sigma_{k}^{2}\left(\widetilde{\mathbb{B}}_{k}-\widetilde{\mathbb{B}}_{k-1}\right)\right|<\infty$.
Use this result into \eqref{eq. A2} to verify that $\max_{1\leq j\leq T_{n}}\left|U_{j}\right|=O_{\mathbb{P}}\left(\widetilde{\Xi}_{T_{n}}+\Delta_{T_{n}}\right)$.
Using the same steps as in \eqref{eq. A1}-\eqref{eq. A2}, one
verifies $\max_{1\leq j\leq T_{n}}\left|G_{\xi,j}\right|=O_{\mathbb{P}}\left(\sqrt{\Xi_{T_{n}}}+\Delta_{T_{n}}\right)$.
Hence, 
\begin{align*}
V_{\xi,j}-\Xi_{j}=U_{j}-G_{\xi,j}^{2}/j & =O_{\mathrm{a.s.}}\left(\Delta_{T_{n}}+\widetilde{\Xi}_{j}+\left(\Delta_{T_{n}}^{2}+\Xi_{T_{n}}\right)/T_{n}\right),
\end{align*}
 uniformly in $1\leq j\leq T_{n}$, which proves part (ii). $\square$

\medskip{}

The first part of the proof uses Lemma \ref{Lemma: Theorem 3.1 in Zhao and Li, eq. 2.7}
applied to the sequence of normalized forecast losses $\left\{ L_{\psi,kh}\left(\beta^{*}\right)\right\} _{k=T_{m}}^{T_{m}+T_{n}}$.
We provide the proof directly for a general loss function; the case
of the quadratic loss function follows as a special case. Since we
have already dealt with the discretization error above and have shown
that $\mu_{e,s}h^{-\vartheta}$ is negligible for $\vartheta\in[0,\,1/8)$,
in this section we assume for simplicitly that $L_{\psi,kh}\left(\beta^{*}\right)=g\left(\Delta_{h}\widetilde{e}_{k}^{*};\,\beta^{*}\right),$
where $\Delta_{h}\widetilde{e}_{k}^{*}=\sigma_{e,\left(k-1\right)h}h^{-1/2}\Delta{}_{h}W_{e,k}$.
Let $\mu_{T_{m}+\tau+\left(b+1\right)n_{T}-1}\triangleq\mathbb{E}\left(L_{\psi,\left(T_{m}+\tau+\left(b+1\right)n_{T}+j-1\right)h}\left(\beta^{*}\right)\right)$
for $j=1,\ldots,\,n_{T}$, which is justified by the fact that these
variables are in the same window. 

\noindent\textit{Proof of Theorem} \ref{Theorem Long-Run Variance}\textit{.}
We shall use Lemma \ref{Lemma: Theorem 3.1 in Zhao and Li, eq. 2.7}-(ii).
Let 
\begin{align*}
\xi_{j} & \triangleq g\left(\sigma_{e,\left(T_{m}+\tau+\left(b+1\right)n_{T}-1\right)h}h^{-1/2}\Delta_{h}W{}_{e,T_{m}+\tau+\left(b+1\right)n_{T}+j-1}\right)-\mu_{T_{m}+\tau+\left(b+1\right)n_{T}-1}
\end{align*}
for $j=1,\ldots,\,n_{T}$. Using basic arguments, we also have $V_{h,b}\left(\widehat{\beta}\right)-V_{h,b}^{*}=O_{\mathbb{P}}\left(1/\sqrt{n_{T}}\right)$,
where
\begin{align*}
V_{h,b}^{*} & \triangleq n_{T}^{-1}\sum_{j=1}^{n_{T}}\left(g\left(\sigma_{e,\left(T_{m}+\tau+\left(b+1\right)n_{T}-1\right)h}h^{-1/2}\Delta{}_{T_{m}+\tau+\left(b+1\right)n_{T}+j-1}W_{e}\right)-\mu_{T_{m}+\tau+\left(b+1\right)n_{T}-1}\right)^{2};
\end{align*}
e.g., use the initial lemmas from the proof of Theorem \ref{Theoem Asymptotic H0 Distrbution Gmax QGmax}
and note that $\overline{L}{}_{\psi,b}\left(\beta^{*}\right)-\mu_{\left(T_{m}+\tau+\left(b+1\right)n_{T}-1\right)h}=O_{\mathbb{P}}\left(1/\sqrt{n_{T}}\right)$
by a basic central limit theorem for i.i.d. variables. Then, by Lemma
\ref{Lemma: Theorem 3.1 in Zhao and Li, eq. 2.7}-(ii) we have $\max_{0\leq b\leq\left\lfloor T_{n}/n_{T}\right\rfloor -2}\left|n_{T}V_{h,b}\left(\beta^{*}\right)-\Sigma_{h,b}^{*}\right|=O_{\mathbb{P}}\left(\Delta_{T_{n}}+\widetilde{\Sigma}_{h,b}^{*}\right)$,
where $\Delta_{T_{n}}=T_{n}^{1/4}\log\left(T_{n}\right)$. Let
\begin{align*}
d_{h,b} & \triangleq n_{T}V_{h,b}\left(\widehat{\beta}\right)/\Sigma_{h,b}^{*}-1,
\end{align*}
and note that\textbf{ }$d_{h,b}=O_{\mathbb{P}}\left(\left(\Delta_{T_{n}}+\widetilde{\Sigma}_{h,b}^{*}\right)/\Sigma_{h,b}^{*}\right)$
by proceeding as in the proof of Lemma \ref{Lemma GB A2} and thus
using $\widehat{\beta}_{k}-\beta^{*}=O_{\mathbb{P}}\left(T_{n}^{-1/2}\right)$
uniformly by Assumption \ref{Assumption Roo-T Consistent beta }.
By definition $\widetilde{\Sigma}_{h,b}^{*}/\Sigma_{h,b}^{*}=O_{\mathbb{P}}\left(n_{T}^{-1/2}\right)$
while by Condition \ref{Cond The-auxiliary-sequence LR VAR} $\Delta_{T_{n}}/\Sigma_{b}^{*}\rightarrow0$
so that we deduce $d_{h,b}=o_{\mathbb{P}}\left(1\right)$ uniformly
over $b$. Let $\left\{ \mathbb{B}_{t}\right\} _{t\geq0}$ be a standard
Wiener process and define 
\begin{align*}
z_{h,b} & \triangleq\left(\Sigma_{b}^{*}\right)^{-1/2}\sum_{j=1}^{n_{T}}\sigma_{L,\left(T_{m}+\tau+\left(b+1\right)n_{T}+j-1\right)h}\left(\mathbb{B}_{\left(b+1\right)n_{T}+j}-\mathbb{B}_{\left(b+1\right)n_{T}+j-1}\right).
\end{align*}
By the law of iterated logarithms {[}cf. \citet{billingsley:95},
Theorem 9.5 in Ch. 1{]}, $\max_{0\leq b\leq\left\lfloor T_{n}/n_{T}\right\rfloor -2}\left|z_{h,b}\right|=O_{\mathbb{P}}\left(\sqrt{\log\left(T_{n}\right)}\right)$.
Under $H_{0}$, by the Lipschitz continuity of $\mu$ we have $\mu_{T_{m}+\tau+\left(b+1\right)n_{T}-1}-\mu_{T_{m}+\tau+bn_{T}-1}=O_{\mathbb{P}}\left(n_{T}h\right)$.
This together with applying multiple times the bounds used at the
beginning of the proof concerning terms involving $\widehat{\beta}$
and $\beta^{*}$ allows us to that $\zeta_{h,b}\left(\widehat{\beta}\right)$
can be approximated by
\begin{align*}
\zeta_{h,b}^{*} & \triangleq\sqrt{n_{T}}\left(A_{h,b}\left(\beta^{*}\right)-\mu_{T_{m}+\tau+\left(b+1\right)n_{T}-1}-\left(A_{h,b-1}\left(\beta^{*}\right)-\mu_{T_{m}+\tau+bn_{T}-1}\right)\right)/\sqrt{V_{h,b}}
\end{align*}
because for small $\epsilon>0$,  $\sqrt{n_{T}}\left(\mu_{T_{m}+\tau+\left(b+1\right)n_{T}-1}-\mu_{T_{m}+\tau+bn_{T}-1}\right)=h^{\epsilon}\rightarrow0$.
Let 
\begin{align*}
\widetilde{A}_{h,b}\left(\beta^{*}\right)\triangleq n_{T}\left(A_{h,b}\left(\beta^{*}\right)-\mu_{T_{m}+\tau+\left(b+1\right)n_{T}-1}\right) & .
\end{align*}
Then, 
\begin{align*}
\zeta_{b,h}^{*} & =\frac{\widetilde{A}_{b,h}\left(\beta^{*}\right)}{\sqrt{\Sigma_{h,b}^{*}}\left(\sqrt{n_{T}V_{b,h}}/\sqrt{\Sigma_{h,b}^{*}}\right)}-\frac{\widetilde{A}_{b-1,h}\left(\beta^{*}\right)}{\sqrt{\Sigma_{h,b}^{*}}\left(\sqrt{n_{T}V_{b,h}}/\sqrt{\Sigma_{h,b}^{*}}\right)}\\
 & =\frac{\widetilde{A}_{b,h}\left(\beta^{*}\right)}{\sqrt{\Sigma_{h,b}^{*}\left(1+d_{h,b}\right)}}-\frac{\widetilde{A}_{b-1,h}\left(\beta^{*}\right)}{\sqrt{\Sigma_{h,b}^{*}\left(1+d_{h,b}\right)\left(\Sigma_{h,b-1}^{*}/\Sigma_{h,b-1}^{*}\right)}},
\end{align*}
and given $d_{h,b}\overset{\mathbb{P}}{\rightarrow}0$ we know that
$\sqrt{1+d_{h,b}}=1+O_{\mathbb{P}}\left(d_{h,b}\right)$. In view
of Lemma \ref{Lemma: Theorem 3.1 in Zhao and Li, eq. 2.7}-(i) we
have 
\begin{align*}
\left|\widetilde{A}_{h,b}\right| & \leq O_{\mathrm{a.s.}}\left(\Delta_{T_{n}}\right)
\end{align*}
Therefore, the last inequality leads to 
\begin{align}
\frac{\widetilde{A}_{h,b}\left(\beta^{*}\right)}{\sqrt{\Sigma_{h,b}^{*}\left(1+d_{h,b}\right)}} & =\left(1+d_{h,b}\right)\nonumber \\
 & \quad\times\left[\frac{\nu_{L}\sum_{j=1}^{n_{T}}\sigma_{L,\left(T_{m}+\tau+\left(b+1\right)n_{T}+j-1\right)h}\left(\mathbb{B}_{\left(b+1\right)n_{T}+j}-\mathbb{B}_{\left(b+1\right)n_{T}+j-1}\right)}{\sqrt{\Sigma_{h,b}^{*}}}+O_{\mathrm{a.s.}}\left(\frac{\Delta_{T_{n}}}{\sqrt{\Sigma_{h,b}^{*}}}\right)\right].
\end{align}
A similar argument can be used for the second term while in addition
for the denominator we use the fact that $\Sigma_{h,b-1}^{*}$ is
Lipschitz continuous and therefore $\Sigma_{h,b}^{*}-\Sigma_{h,b-1}^{*}=O_{\mathbb{P}}\left(n_{T}h\right)$,
which then gives $\sqrt{1+d_{h,b}}\sqrt{1+O_{\mathbb{P}}\left(n_{T}h\right)}=1+O_{\mathbb{P}}\left(d_{h,b}\right)$.
Let $\left\{ \mathbb{B}_{t}\right\} _{t\geq0}$ be a standard Wiener
process. We can then deduce that
\begin{align*}
\zeta_{b,h}^{*} & =\frac{\nu_{L}\sum_{j=1}^{n_{T}}\sigma_{L,\left(T_{m}+\tau+\left(b+1\right)n_{T}+j-1\right)h}\left(\mathbb{B}_{\left(b+1\right)n_{T}+j}-\mathbb{B}_{\left(b+1\right)n_{T}+j-1}\right)}{\sqrt{\Sigma_{h,b}^{*}}}\\
 & \quad+\frac{\nu_{L}\sum_{j=1}^{n_{T}}\sigma_{L,\left(T_{m}+\tau+bn_{T}+j-1\right)h}\left(\mathbb{B}_{bn_{T}+j}-\mathbb{B}_{bn_{T}+j-1}\right)}{\sqrt{\Sigma_{h,b-1}^{*}}}+\left(1+d_{h,b}\right)O_{\mathrm{a.s.}}\left(\frac{\Delta_{T_{n}}}{\sqrt{\Sigma_{h,b-1}^{*}}}\right).
\end{align*}
The stochastic order term in the last equation is, for some small
$\epsilon>0$, $O_{\mathbb{P}}\left(\log\left(T_{n}\right)/\left(T_{n}^{\epsilon/2}\right)\right)\rightarrow0$
where we have used $\Delta_{T_{n}}=T_{n}^{1/4}\log\left(T_{n}\right)$,
Condition \ref{Cond The-auxiliary-sequence LR VAR} and $\Sigma_{h,b}^{*}=O_{\mathbb{P}}\left(n_{T}\right)$.
Using the properties of the Wiener process, we have 
\begin{align*}
\left(\zeta_{b,h}^{*}\right)^{2} & =\frac{\nu_{L}^{2}\sum_{j=1}^{n_{T}}\sigma_{L,\left(T_{m}+\tau+\left(b+1\right)n_{T}+j-1\right)h}^{2}}{\sqrt{\Sigma_{b}^{*}}}\\
 & \quad+\frac{\nu_{L}^{2}\sum_{j=1}^{n_{T}}\sigma_{L,\left(T_{m}+\tau+bn_{T}+j-1\right)h}^{2}}{\sqrt{\Sigma_{b-1}^{*}}}+O_{\mathbb{P}}\left(\log\left(T_{n}\right)/T_{n}^{\epsilon/2}\right)\\
 & =2\nu_{L}^{2}+O_{\mathbb{P}}\left(\left(\log\left(T_{n}\right)\right)^{2}/T_{n}^{\epsilon}\right),
\end{align*}
 and therefore, $\widehat{\nu}_{L}^{2}=\nu_{L}^{2}+O_{\mathbb{P}}\left(\left(\log\left(T_{n}\right)\right)^{2}/T_{n}^{\epsilon}\right)$.
$\square$

\subsection{\label{subSection Proofs-of-Section Ito Vol}Proofs of Section \ref{Section Ito Vol}}

\subsubsection{Proof of Theorem \ref{Theoem Asymptotic H0 Distrbution Bmax and Qmax Ito Vol}}

See \citet{casini_FFb}.

\subsubsection{Proof of Theorem \ref{Theorem Local Asym Power}}

Let $\Delta_{h}\widetilde{e}_{k}^{\textdegree}\triangleq\Delta_{h}Y_{k}-\beta_{k}'\Delta_{h}X_{k-\tau}$,
where $\beta_{k}=\beta^{*}+\mu_{\beta,kh}/\left(\log\left(T_{n}\right)n_{T}\right)^{1/4}$.
Let 
\begin{align*}
\mathrm{\widetilde{MQ}}_{\mathrm{max,}h}^{*}\left(T_{n},\,\tau\right) & \triangleq\nu_{L}^{-1}\max_{i=n_{T},\ldots,\,T_{n}-n_{T}}\left|n_{T}^{-1}\sum_{j=i+1}^{i+n_{T}}\left(SL_{\psi,T_{m}+\tau+j-1}\left(\beta^{*}\right)-\zeta_{\mu,j,+}\right)\right.\\
 & -\left.n_{T}^{-1}\sum_{j=i-n_{T}+1}^{i}\left(SL_{\psi,T_{m}+\tau+j-1}\left(\beta^{*}\right)-\zeta_{\mu,j,-}\right)\right|.
\end{align*}
Our final goal is to show that $\left(\log\left(T_{n}\right)n_{T}\right)^{1/2}\left(\mathrm{V}_{\max,h}\left(T_{n},\,\tau\right)-\mathrm{MQ}_{\mathrm{max},h}\left(T_{n},\,\tau\right)\right)\overset{\mathbb{P}}{\rightarrow}0$,
where 
\begin{align*}
\mathrm{V}_{\max,h}\left(T_{n},\,\tau\right) & =\nu_{L}^{-1}\max_{i=n_{T},\ldots,\,T_{n}-n_{T}}\left|n_{T}^{-1}\sum_{j=i+1}^{i+n_{T}}\left(\Delta_{h}\widetilde{e}_{T_{m}+\tau+j-1}^{\textdegree}\right)^{2}-n_{T}^{-1}\sum_{j=i-n_{T}+1}^{i}\left(\Delta_{h}\widetilde{e}_{T_{m}+\tau+j-1}^{\textdegree}\right)^{2}\right|.
\end{align*}
The result of the theorem then follows from Corollary \ref{Corollary CR Null Distrb General}.
\begin{lem}
\label{Lemma LP 1}$\left(\log\left(T_{n}\right)n_{T}\right)^{1/2}\left(\mathrm{\widetilde{MQ}}{}_{\mathrm{max},h}^{*}\left(T_{n},\,\tau\right)-\mathrm{\widetilde{MQ}}_{\mathrm{max},h}\left(T_{n},\,\tau\right)\right)\overset{\mathbb{P}}{\rightarrow}0$. 
\end{lem}
\noindent\textit{Proof.} We have 
\begin{align}
\mathrm{\widetilde{MQ}}{}_{\mathrm{max},h}^{*} & \left(T_{n},\,\tau\right)-\mathrm{\widetilde{MQ}}_{\mathrm{max},h}\left(T_{n},\,\tau\right)\nonumber \\
 & \leq\nu_{L}^{-1}\max_{i=n_{T},\ldots,\,T_{n}-n_{T}}\left|n_{T}^{-1}\sum_{j=i+1}^{i+n_{T}}\left(SL_{\psi,T_{m}+\tau+j-1}\left(\beta^{*}\right)-SL_{\psi,T_{m}+\tau+j-1}\left(\widehat{\beta}_{T_{m}+j-1}\right)-\zeta_{\mu,j,+}\right)\right|\nonumber \\
 & \quad+\nu_{L}^{-1}\max_{i=n_{T},\ldots,\,T_{n}-n_{T}}\left|n_{T}^{-1}\sum_{j=i-n_{T}+1}^{i}\left(SL_{\psi,T_{m}+\tau+j-1}\left(\beta^{*}\right)-SL_{\psi,T_{m}+\tau+j-1}\left(\widehat{\beta}_{T_{m}+j-1}\right)-\zeta_{\mu,j,-}\right)\right|.\label{eq (LP 1)}
\end{align}
 Note that for any $j=i+1,\ldots,\,i+n_{T}$, 
\begin{align*}
SL_{\psi,T_{m}+\tau+j-1}\left(\beta^{*}\right) & -SL_{\psi,T_{m}+\tau+j-1}\left(\widehat{\beta}_{T_{m}+j-1}\right)\\
 & =L_{\psi,T_{m}+\tau+j-1}\left(\beta^{*}\right)-L_{\psi,T_{m}+\tau+j-1}\left(\widehat{\beta}_{T_{m}+j-1}\right)+o_{\mathbb{P}}\left(1/\left(\log\left(T_{n}\right)n_{T}\right)^{1/2}\right),
\end{align*}
where the $o_{\mathbb{P}}\left(1/\left(\log\left(T_{n}\right)n_{T}\right)^{1/2}\right)$
term arises from Lemma \ref{Lemma LP3} below. Focusing on the
first two terms we have
\begin{align}
L_{\psi,T_{m}+\tau+j-1}\left(\beta^{*}\right) & -L_{\psi,T_{m}+\tau+j-1}\left(\widehat{\beta}_{T_{m}+j-1}\right)-\zeta_{\mu,j,+}\nonumber \\
 & =\left(\beta^{*}-\widehat{\beta}_{T_{m}+j-1}\right)'\Delta_{h}\widetilde{X}{}_{T_{m}+j-1}\Delta_{h}\widetilde{X}'{}_{T_{m}+j-1}\left(\beta^{*}-\widehat{\beta}_{T_{m}+j-1}\right)-\zeta_{\mu,j,+}\nonumber \\
 & \quad+2\sigma_{e,\left(T_{m}+\tau+i\right)h}\left(h^{-1/2}\Delta_{h}W_{e,T_{m}+\tau+j-1}\right)\left(\beta^{*}-\widehat{\beta}_{T_{m}+j-1}\right)'\Delta_{h}\widetilde{X}{}_{T_{m}+j-1}.\label{eq (A2)-2-1}
\end{align}
We deal explicitly with the first term on the right-hand side above
in \eqref{eq (2) XX- SigmaX} below and show that it is $o_{\mathbb{P}}\left(\left(\log\left(T_{n}\right)n_{T}\right)^{-1/2}h^{1/4}\right)$.
Moving to the second term, by Theorem 13.3.7 in \citet{jacod/protter:12}
we have $n_{T}^{-1/2}\sum_{j=i+1}^{i+n_{T}}h^{-1/2}\Delta_{h}W_{e,T_{m}+\tau+j-1}\Delta_{h}\widetilde{X}{}_{T_{m}+j-1}=O_{\mathbb{P}}\left(1\right)$.
Using Assumption \ref{Ass Local Power}, $\widehat{\beta}_{T_{m}+j-1}-\beta^{*}=O_{\mathbb{P}}\left(1/\left(\log\left(T_{n}\right)n_{T}\right)^{-1/4}\right)$
uniformly in $j$ and thus, for any $\varepsilon>0$, 
\begin{align*}
\mathbb{P} & \left(\max_{i=n_{T},\ldots,\,T_{n}-n_{T}}\nu_{L}^{-1}\left(\log\left(T_{n}\right)n_{T}\right)^{1/2}\left|n_{T}^{-1}\sum_{j=i+1}^{i+n_{T}}\left(SL_{T_{m}+\tau+j-1}\left(\beta^{*}\right)-SL_{T_{m}+\tau+j-1}\left(\widehat{\beta}_{T_{m}+j-1}\right)\right)\right|>\varepsilon\right)\\
 & =\left(\frac{1}{\nu_{L}\varepsilon}\right)^{r}\left(\log\left(T_{n}\right)n_{T}\right)^{r/2}\sum_{i=n_{T}}^{T_{n}-n_{T}}\mathbb{E}\left[\left|n_{T}^{-1}\sum_{j=i+1}^{i+n_{T}}\left(SL_{T_{m}+\tau+j-1}\left(\beta^{*}\right)-SL_{T_{m}+\tau+j-1}\left(\widehat{\beta}_{T_{m}+j-1}\right)\right)\right|^{r}\right]\\
 & \leq\left(\frac{C}{\nu_{L}\varepsilon}\right)^{r}\left(\log\left(T_{n}\right)n_{T}\right)^{r/4}O_{\mathbb{P}}\left(n_{T}^{-r/2}\right)\rightarrow0,
\end{align*}
for $r>0$ sufficiently large. The same bound applies to the term
in \eqref{eq (LP 1)} and this proves the claim of the lemma. $\square$
\begin{lem}
\label{Lemma LP3}As $h\downarrow0$,
\begin{align*}
\nu_{L}^{-1}\max_{i=n_{T},\ldots,\,T_{n}-n_{T}}\left(\log\left(T_{n}\right)n_{T}\right)^{1/2}\left|n_{T}^{-1}\sum_{j=i+1}^{i+n_{T}}\overline{L}_{\psi,T_{m}+j-1}\left(\beta^{*}\right)-\sum_{j=i-n_{T}+1}^{i}\overline{L}_{\psi,T_{m}+j-1}\left(\beta^{*}\right)\right| & \overset{\mathbb{P}}{\rightarrow}0,
\end{align*}
and
\begin{align*}
\nu_{L}^{-1}\max_{i=n_{T},\ldots,\,T_{n}-n_{T}}\left(\log\left(T_{n}\right)n_{T}\right)^{1/2}\left|n_{T}^{-1}\sum_{j=i+1}^{i+n_{T}}\left(\overline{L}_{\psi,T_{m}+j-1}\left(\beta^{*}\right)-\overline{L}_{\psi,T_{m}+j-1}\left(\widehat{\beta}_{T_{m}+j-1}\right)\right)\right| & \overset{\mathbb{P}}{\rightarrow}0.
\end{align*}
\end{lem}
\noindent\textit{Proof.} By definition, 
\begin{align*}
\left|n_{T}^{-1}\sum_{j=i+1}^{i+n_{T}}\right. & \left.\overline{L}_{\psi,T_{m}+j-1}\left(\beta^{*}\right)-\sum_{j=i-n_{T}+1}^{i}\overline{L}_{\psi,T_{m}+j-1}\left(\beta^{*}\right)\right|\\
 & \leq\left|n_{T}^{-1}\sum_{j=i+1}^{i+n_{T}}\sum_{l=1}^{T_{m}+j-1-\tau}\frac{\left(\Delta_{h}\tilde{e}_{l+\tau}^{\textdegree}\right)^{2}}{T_{m}+j-1}-n_{T}^{-1}\sum_{j=i-n_{T}+1}^{i}\sum_{l=1}^{T_{m}+j-1-\tau}\frac{\left(\Delta_{h}\tilde{e}_{l+\tau}^{\textdegree}\right)^{2}}{T_{m}+j-1}\right|\\
 & \quad+\left|n_{T}^{-1}\sum_{j=i+1}^{i+n_{T}}\sum_{l=1}^{T_{m}+j-1-\tau}\frac{\mu'_{\beta,\left(l+\tau\right)h}\Delta_{h}\widetilde{X}{}_{l}\Delta_{h}\widetilde{X}'_{l}\mu{}_{\beta,\left(l+\tau\right)h}}{\left(T_{m}+j-1\right)\left(\log T_{n}n_{T}\right)^{1/2}}\right.\\
 & \quad\quad-\left.n_{T}^{-1}\sum_{j=i-n_{T}+1}^{i}\sum_{l=1}^{T_{m}+j-1-\tau}\frac{\mu'_{\beta,\left(l+\tau\right)h}\Delta_{h}\widetilde{X}{}_{l}\Delta_{h}\widetilde{X}_{l}'\mu{}_{\beta,\left(l+\tau\right)h}}{\left(T_{m}+j-1\right)\left(\log T_{n}n_{T}\right)^{1/2}}\right|\\
 & \quad+2\left|n_{T}^{-1}\sum_{j=i+1}^{i+n_{T}}\sum_{l=1}^{T_{m}+j-1-\tau}\frac{\Delta_{h}\tilde{e}_{l+\tau}^{\textdegree}\mu'_{\beta,\left(l+\tau\right)h}\Delta_{h}\widetilde{X}{}_{T_{m}+j-1}}{\left(T_{m}+j-1\right)\left(\log T_{n}n_{T}\right)^{1/4}}\right.\\
 & \quad\quad-\left.n_{T}^{-1}\sum_{j=i-n_{T}+1}^{i}\sum_{l=1}^{T_{m}+j-1-\tau}\frac{\Delta_{h}\tilde{e}_{l+\tau}^{\textdegree}\mu'_{\beta,\left(l+\tau\right)h}\Delta_{h}\widetilde{X}{}_{T_{m}+j-1}}{\left(T_{m}+j-1\right)\left(\log T_{n}n_{T}\right)^{1/4}}\right|\\
 & \triangleq A_{1,h}+A_{2,h}+A_{3,h}.
\end{align*}
Observe that the leading term is $A_{1,h}$ and thus it is sufficient
to establish a bound for it. By proceeding as in the proof of Lemma
\ref{Lemma QA2} we shall use $\left(T_{m}+i-1\right)^{-1}\sum_{l=1}^{T_{m}+i-1-\tau}\left(\Delta_{h}\tilde{e}_{l+\tau}^{\textdegree}\right)^{2}=O_{\mathbb{P}}\left(1\right)$.
Then, $A_{1,h}$ is less than 
\begin{align*}
\left|n_{T}^{-1}\right. & \sum_{j=i+1}^{i+n_{T}}\sum_{l=T_{m}+j-n_{T}-1-\tau}^{T_{m}+j-1-\tau}\frac{\left(\Delta_{h}\widetilde{e}_{l+\tau}^{\circ}\right)^{2}}{T_{m}+j-1}\\
 & -\left.n_{T}^{-1}\sum_{j=i-n_{T}+1}^{i}\sum_{l=1}^{T_{m}+j-1-\tau}\left(\Delta_{h}\tilde{e}_{l+\tau}^{\textdegree}\right)^{2}\left(\frac{1}{T_{m}+j+n_{T}-1}-\frac{1}{T_{m}+j-1}\right)\right|\\
 & =\frac{2n_{T}}{T_{m}}O_{\mathbb{P}}\left(1\right)+\frac{n_{T}}{T_{m}}O_{\mathbb{P}}\left(1\right).
\end{align*}
This leads to 
\begin{align*}
\left|n_{T}^{-1}\sum_{j=i+1}^{i+n_{T}}\overline{L}_{\psi,T_{m}+j-1}\left(\beta^{*}\right)-n_{T}^{-1}\sum_{j=i-n_{T}+1}^{i}\overline{L}_{\psi,T_{m}+j-1}\left(\beta^{*}\right)\right| & \leq CO_{\mathbb{P}}\left(\frac{n_{T}}{T_{m}}\right).
\end{align*}
Thus, for any $\varepsilon>0$, 
\begin{align}
\mathbb{P} & \left(\nu_{L}^{-1}\max_{i=n_{T},\ldots,\,T_{n}-n_{T}}\left(\log\left(T_{n}\right)n_{T}\right)^{1/2}\left|n_{T}^{-1}\sum_{j=i+1}^{i+n_{T}}\overline{L}_{\psi,T_{m}+j-1}\left(\beta^{*}\right)-n_{T}^{-1}\sum_{j=i-n_{T}+1}^{i}\overline{L}_{\psi,T_{m}+j-1}\left(\beta^{*}\right)\right|>\varepsilon\right)\nonumber \\
 & \leq\varepsilon^{-r}\sum_{i=n_{T}}^{T_{n}-n_{T}}\mathbb{E}\left[\left(\log\left(T_{n}\right)n_{T}\right)^{r/2}\left|n_{T}^{-1}\sum_{j=i+1}^{i+n_{T}}\overline{L}_{\psi,T_{m}+j-1}\left(\beta^{*}\right)-n_{T}^{-1}\sum_{j=i-n_{T}+1}^{i}\overline{L}_{\psi,T_{m}+j-1}\left(\beta^{*}\right)\right|^{r}\right]\nonumber \\
 & \leq\varepsilon^{-r}C\left(\log\left(T_{n}\right)n_{T}\right)^{r/2}O_{\mathbb{P}}\left(n_{T}^{r}T_{n}^{1-r}\right)\rightarrow0,\label{eq (A1a0)-1}
\end{align}
for $r>0$ sufficiently large in view of Condition \ref{Cond The-auxiliary-sequence}
and that $T_{n}=O\left(T_{m}\right)$. For the second claim of the
lemma, note that
\begin{align}
\overline{L}_{\psi,T_{m}+j-1}\left(\widehat{\beta}\right) & -\overline{L}_{\psi,T_{m}+j-1}\left(\beta^{*}\right)\nonumber \\
 & =\sum_{l=1}^{T_{m}+j-1-\tau}\frac{\left(\Delta_{h}\widetilde{e}_{l+\tau}\right)^{2}}{T_{m}+j-1-\tau}-\sum_{l=1}^{T_{m}+j-1-\tau}\frac{\left(\Delta_{h}\tilde{e}_{l+\tau}^{\textdegree}\right)^{2}}{T_{m}+j-1-\tau}\nonumber \\
 & =\frac{1}{T_{m}+j-1-\tau}\sum_{l=1}^{T_{m}+j-1-\tau}\left(\widehat{\beta}_{T_{m}+j-1}-\beta^{*}\right)'\Delta_{h}\widetilde{X}{}_{l}\Delta_{h}\widetilde{X}'_{l}\left(\widehat{\beta}_{T_{m}+j-1}-\beta^{*}\right)\label{eq (A1a)-1}\\
 & \quad-\frac{2}{T_{m}+j-1-\tau}\sum_{l=1}^{T_{m}+j-1-\tau}\left(\widehat{\beta}_{T_{m}+j-1}-\beta^{*}\right)'\Delta_{h}\widetilde{X}{}_{l}\tilde{e}_{l+\tau}^{\textdegree}.\label{eq (A1b)-1}
\end{align}
By Lemma \ref{Lemma Prelim A1}, $\left(T_{m}+j-1\right)^{-1}\sum_{l=1}^{T_{m}+j-1-\tau}\Delta_{h}\widetilde{X}{}_{l}\Delta_{h}\widetilde{X}'_{l}$
is $O_{\mathbb{P}}\left(1\right)$ while by Assumption \ref{Ass Local Power},
$\widehat{\beta}_{k}-\beta^{*}=O_{\mathbb{P}}\left(1/\left(\log\left(T_{n}\right)n_{T}\right)^{1/4}\right)$
uniformly in $k$. It follows that the term in equation \eqref{eq (A1a)-1}
is $O_{\mathbb{P}}\left(\left(\log\left(T_{n}\right)n_{T}\right)^{1/2}\right)$
whereas the term in \eqref{eq (A1b)-1} is such that
\begin{align*}
\frac{2}{T_{m}+j-1} & \sum_{l=1}^{T_{m}+j-1-\tau}\Delta_{h}\tilde{e}_{l+\tau}^{\textdegree}\left(\widehat{\beta}_{T_{m}+j-1}-\beta^{*}\right)'\Delta_{h}\widetilde{X}{}_{l}\\
 & \leq2C\sup_{j}\left\Vert \widehat{\beta}_{T_{m}+j-1}-\beta^{*}\right\Vert \frac{1}{T_{m}+j-1}\sum_{l=1}^{T_{m}+j-1-\tau}\Delta_{h}\tilde{e}_{l+\tau}^{\textdegree}\iota'\Delta_{h}\widetilde{X}_{l}\\
 & =o_{\mathbb{P}}\left(T_{m}^{-1/2}\left(\log\left(T_{n}\right)n_{T}\right)^{1/4}\right),
\end{align*}
where $\iota$ is a $q\times1$ unit vector and we have used the central
limit theorem in Lemma \ref{Lemma Prelim A1}-(iii). Therefore, upon
using the same argument that led to \eqref{eq (A1a0)-1} and Condition
\ref{Cond The-auxiliary-sequence}\textbf{ }we have the last claim
of the lemma. $\square$
\begin{lem}
\label{Lemma LP 2}$\left(\log\left(T_{n}\right)n_{T}\right)^{1/2}\left(\mathrm{V}_{\max,h}\left(T_{n},\,\tau\right)-\mathrm{\widetilde{MQ}}^{*}{}_{\mathrm{max},h}-\left(T_{n},\,\tau\right)\right)\overset{\mathbb{P}}{\rightarrow}0$. 
\end{lem}
\noindent\textit{Proof.} Note that $SL_{T_{m}+\tau+j-1}\left(\beta^{*}\right)$
can be expanded as follows: 
\begin{align}
SL_{T_{m}+\tau+j-1}\left(\widehat{\beta}\right) & =L_{\psi,T_{m}+\tau+j-1}\left(\widehat{\beta}\right)-\overline{L}_{\psi,T_{m}+\tau+j-1}\left(\widehat{\beta}\right)\label{eq (a) Lemma LP 2}\\
 & =\left(\Delta_{h}\widetilde{e}_{T_{m}+\tau+j-1}^{\textdegree}\right)^{2}+\left(\beta^{*}-\widehat{\beta}_{T_{m}+j-1}\right)'\Delta_{h}\widetilde{X}{}_{T_{m}+j-1}\Delta_{h}\widetilde{X}'_{T_{m}+j-1}\left(\beta^{*}-\widehat{\beta}_{T_{m}+j-1}\right)\nonumber \\
 & \quad-2\Delta_{h}\widetilde{e}_{T_{m}+\tau+j-1}^{\textdegree}\left(\beta^{*}-\widehat{\beta}_{T_{m}+j-1}\right)'\Delta_{h}\widetilde{X}{}_{T_{m}+j-1}-\overline{L}_{T_{m}+\tau+j-1}\left(\widehat{\beta}\right).\nonumber 
\end{align}
Then we can write (omitting the index from $\widehat{\beta}$), 
\begin{align}
\left(\log\left(T_{n}\right)n_{T}\right)^{1/2} & \left(\mathrm{V}_{\max,h}\left(T_{n},\,\tau\right)-\mathrm{\widetilde{MQ}}^{*}{}_{\mathrm{max},h}-\left(T_{n},\,\tau\right)\right)\label{eq (2) Lemma LP 2}\\
 & \leq\max_{i=n_{T},\ldots,\,T_{n}-n_{T}}\left(\log\left(T_{n}\right)n_{T}\right)^{1/2}\nu_{L}^{-1}\nonumber \\
 & \quad\times\left|n_{T}^{-1}\sum_{j=i+1}^{i+n_{T}}\left(\left(\beta^{*}-\widehat{\beta}\right)'\Delta_{h}\widetilde{X}{}_{T_{m}+j-1}\Delta_{h}\widetilde{X}'_{T_{m}+j-1}\left(\beta^{*}-\widehat{\beta}\right)-\zeta_{\mu,j,+}\right)\right.\\
 & \quad\quad\left.-n_{T}^{-1}\sum_{j=i-n_{T}+1}^{i}\left(\left(\beta^{*}-\widehat{\beta}\right)'\Delta_{h}\widetilde{X}{}_{T_{m}+j-1}\Delta_{h}\widetilde{X}'_{T_{m}+j-1}\left(\beta^{*}-\widehat{\beta}\right)-\zeta_{\mu,j,-}\right)\right|\nonumber \\
 & \quad+\max_{i=n_{T},\ldots,\,T_{n}-n_{T}}2\left(\log\left(T_{n}\right)n_{T}\right)^{1/2}\nu_{L}^{-1}\left|n_{T}^{-1}\sum_{j=i+1}^{i+n_{T}}\Delta_{h}\widetilde{e}_{T_{m}+\tau+j-1}^{\textdegree}\left(\beta^{*}-\widehat{\beta}\right)'\Delta_{h}\widetilde{X}{}_{T_{m}+j-1}\right.\nonumber \\
 & \quad\quad\left.-n_{T}^{-1}\sum_{j=i-n_{T}+1}^{i}\Delta_{h}\widetilde{e}_{T_{m}+\tau+j-1}^{\textdegree}\left(\beta^{*}-\widehat{\beta}\right)'\Delta_{h}\widetilde{X}{}_{T_{m}+j-1}\right|\nonumber \\
 & \quad+\max_{i=n_{T},\ldots,\,T_{n}-n_{T}}\left(\log\left(T_{n}\right)n_{T}\right)^{1/2}\nu_{L}^{-1}\nonumber \\
 & \quad\quad\times\left|n_{T}^{-1}\sum_{j=i+1}^{i+n_{T}}\overline{L}_{\psi,T_{m}+\tau+j-1}\left(\beta^{*}\right)-n_{T}^{-1}\sum_{j=i-n_{T}+1}^{i}\overline{L}_{\psi,T_{m}+\tau+j-1}\left(\beta^{*}\right)\right|\\
 & \triangleq A_{1,h}+A_{2,h}+A_{3,h}.\nonumber 
\end{align}
Our goal is to show that $A_{l,h}\overset{\mathbb{P}}{\rightarrow}0$
for $l=1,\,2,\,3$. By Lemma \ref{Lemma LP3} we know that $A_{3,h}\overset{\mathbb{P}}{\rightarrow}0$.
Let us focus on $A_{1,h}$. Note that 
\begin{align}
A_{1,h} & \leq\max_{i=n_{T},\ldots,\,T_{n}-n_{T}}\left(\log\left(T_{n}\right)n_{T}\right)^{1/2}\nu_{L}^{-1}\label{eq (1) LP3}\\
 & \quad\times\left|n_{T}^{-1}\sum_{j=i+1}^{i+n_{T}}\left(\left(\beta^{*}-\widehat{\beta}_{T_{m}+j-1}\right)\Delta_{h}\widetilde{X}{}_{T_{m}+j-1}\Delta_{h}\widetilde{X}'{}_{T_{m}+j-1}\left(\beta^{*}-\widehat{\beta}_{T_{m}+j-1}\right)-\zeta_{\mu,j,+}\right)\right|\nonumber \\
 & \quad+\max_{i=n_{T},\ldots,\,T_{n}-n_{T}}\left(\log\left(T_{n}\right)n_{T}\right)^{1/2}\nu_{L}^{-1}\nonumber \\
 & \quad\times\left|-n_{T}^{-1}\sum_{j=i-n_{T}+1}^{i}\left(\left(\beta^{*}-\widehat{\beta}_{T_{m}+j-1}\right)'\Delta_{h}\widetilde{X}{}_{T_{m}+j-1}\Delta_{h}\widetilde{X}'{}_{T_{m}+j-1}\left(\beta^{*}-\widehat{\beta}_{T_{m}+j-1}\right)-\zeta_{\mu,j,-}\right)\right|.\nonumber 
\end{align}
We have $\beta^{*}-\widehat{\beta}_{T_{m}+j-1}=\mu_{\beta,\left(T_{m}+j-1\right)h}/\left(\log\left(T_{n}\right)n_{T}\right)^{1/4}$
and 
\begin{align*}
h^{-1/4}\left(\sum_{j=i+1}^{i+n_{T}}\Delta_{h}\widetilde{X}{}_{T_{m}+j-1}\Delta_{h}\widetilde{X}'_{T_{m}+j-1}-\Sigma_{X,\left(T_{m}+i\right)h}\right) & =O_{\mathbb{P}}\left(1\right),
\end{align*}
by Theorem 13.3.7 in \citet{jacod/protter:12}. Upon using the property
of the trace operator we have 
\begin{align}
\left(\log\left(T_{n}\right)n_{T}\right)^{-1/2}n_{T}^{-1} & \sum_{j=i+1}^{i+n_{T}}\mu'_{\beta,\left(T_{m}+\tau+j-1\right)h}\left(\Delta_{h}\widetilde{X}{}_{T_{m}+j-1}\Delta_{h}\widetilde{X}'{}_{T_{m}+j-1}-\Sigma_{X,\left(T_{m}+i\right)h}\right)\mu_{\beta,\left(T_{m}+\tau+j-1\right)h}\label{eq (2) XX- SigmaX}\\
 & =O_{\mathbb{P}}\left(\left(\log\left(T_{n}\right)n_{T}\right)^{-1/2}h^{1/4}\right).\nonumber 
\end{align}
Thus, the first term on the right-hand side of \eqref{eq (1) LP3}
is less than
\begin{alignat*}{1}
C_{r}\left(\frac{1}{\nu_{L}\varepsilon}\right)^{r}\sum_{i=n_{T}}^{T_{n}-n_{T}}\mathbb{E}\left[\left|n_{T}^{-1}\sum_{j=i+1}^{i+n_{T}}\iota'\left(\Delta_{h}\widetilde{X}{}_{T_{m}+j-1}\Delta_{h}\widetilde{X}'{}_{T_{m}+j-1}-\zeta_{\mu,j,+}\right)\iota\right|^{r}\right] & \leq C_{r}\left(\frac{1}{\nu_{L}\varepsilon}\right)^{r}T_{n}h^{r/4}\rightarrow0,
\end{alignat*}
 for $r>0$ sufficiently large given that $h=O\left(T^{-1}\right)=O\left(T_{n}^{-1}\right)$
and $\varepsilon>0$. The same argument can be applied to the second
term of \eqref{eq (1) LP3} which then yields $A_{1,h}=o_{\mathbb{P}}\left(1\right)$.
It remains to consider $A_{2,h}$. It is sufficient to show that 
\begin{align}
\max_{i=n_{T},\ldots,\,T_{n}-n_{T}}\left(\log\left(T_{n}\right)n_{T}\right)^{1/2}\nu_{L}^{-1}\left|2n_{T}^{-1}\sum_{j=i+1}^{i+n_{T}}\Delta_{h}\widetilde{e}_{T_{m}+\tau+j-1}^{\textdegree}\left(\beta^{*}-\widehat{\beta}_{T_{m}+j-1}\right)'\Delta_{h}\widetilde{X}{}_{T_{m}+j-1}\right| & \overset{\mathbb{P}}{\rightarrow}0.\label{eq (2) Lemma LP3}
\end{align}
 By Theorem 13.3.7 in \citet{jacod/protter:12} we now have $n_{T}^{-1/2}\sum_{j=i+1}^{i+n_{T}}\Delta_{h}\widetilde{e}_{T_{m}+\tau+j-1}^{\textdegree}\Delta_{h}\widetilde{X}{}_{T_{m}+j-1}<\infty$.
By Marlov's inequality, for any $\varepsilon>0$ we have
\begin{align*}
\mathbb{P} & \left(\max_{i=n_{T},\ldots,\,T_{n}-n_{T}}\left(\log\left(T_{n}\right)n_{T}\right)^{1/2}\nu_{L}^{-1}\left|2n_{T}^{-1}\sum_{j=i+1}^{i+n_{T}}\Delta_{h}\widetilde{e}_{T_{m}+\tau+j-1}^{\textdegree}\left(\beta^{*}-\widehat{\beta}_{T_{m}+j-1}\right)'\Delta_{h}\widetilde{X}{}_{T_{m}+j-1}\right|>\varepsilon\right)\\
 & \leq C_{r}\left(\frac{2}{\nu_{L}\varepsilon}\right)^{r}\left(\log\left(T_{n}\right)n_{T}\right)^{r/2-r/4}\sum_{i=n_{T}}^{T_{n}-n_{T}}\mathbb{E}\left[\left|n_{T}^{-1}\sum_{j=i+1}^{i+n_{T}}\Delta_{h}\widetilde{e}_{T_{m}+\tau+j-1}^{\textdegree}\Delta_{h}\widetilde{X}{}_{T_{m}+j-1}\right|^{r}\right]\\
 & \leq C_{r}\left(\frac{2}{\nu_{L}\varepsilon}\right)^{r}\left(\log\left(T_{n}\right)n_{T}\right)^{r/2-r/4}O_{\mathbb{P}}\left(T_{n}n_{T}^{-r/2}\right)\rightarrow0
\end{align*}
$r>0$ sufficiently large. This gives \eqref{eq (2) Lemma LP3} and
thus $A_{2,h}\overset{\mathbb{P}}{\rightarrow}0$ which in turn concludes
the proof. $\square$

\medskip{}

\noindent\textit{Proof of Theorem \ref{Theorem Local Asym Power}.}
From Lemma \ref{Lemma LP 1}-\ref{Lemma LP 2} $\left(\log\left(T_{n}\right)n_{T}\right)^{1/2}\left(\mathrm{V}_{\max,h}\left(T_{n},\,\tau\right)-\mathrm{\widetilde{MQ}}{}_{\mathrm{max},h}-\left(T_{n},\,\tau\right)\right)\overset{\mathbb{P}}{\rightarrow}0$.
The result then follows from Corollary \ref{Corollary CR Null Distrb Quadratic}.
$\square$

\subsubsection{Proof of Corollary \ref{Corollary Local Power} }

\noindent\textit{Proof}. Since the statistic $\mathrm{\widetilde{MQ}}{}_{\mathrm{max},h}\left(T_{n},\,\tau\right)$
admits a limit theorem by Theorem \ref{Theorem Local Asym Power},
it is sufficient to show that, conditional on $\left\{ \sigma_{X,t}\right\} _{t\geq0}$,
for all $i=n_{T},\ldots,\,T_{n}-n_{T}$, 
\begin{align*}
\left(\log\left(T_{n}\right)n_{T}\right)^{1/2}c\left|n_{T}^{-1}\sum_{j=i+1}^{i+n_{T}}\zeta_{\mu,j,+}-n_{T}^{-1}\sum_{j=i-n_{T}+1}^{i}\zeta_{\mu,j,-}\right| & \rightarrow\infty,
\end{align*}
 or that
\begin{align}
n_{T}^{-1} & \sum_{j=i+1}^{i+n_{T}}\mu'_{\beta,\left(T_{m}+\tau+j-1\right)h}\sigma_{X,\left(T_{m}+\tau+i-1\right)h}\mu_{\beta,\left(T_{m}+\tau+j-1\right)h}\label{eq (1) Cor LP}\\
 & =n_{T}^{-1}\sum_{j=i-n_{T}+1}^{i}\mu'_{\beta,\left(T_{m}+\tau+j-1\right)h}\sigma_{X,\left(T_{m}+\tau+i-n_{T}-1\right)h}\mu_{\beta,\left(T_{m}+\tau+j-1\right)h}\nonumber 
\end{align}
for all $i=n_{T},\ldots,\,T_{n}-n_{T}$ does not hold. Suppose by
contradiction that \eqref{eq (1) Cor LP} holds. Due to the block-wise
structure of the statistic, we know that $\sigma_{X,\left(T_{m}+\tau+j-1\right)h}=\sigma_{X,\left(T_{m}+\tau+i-1\right)h}$
for all $j=i+1,\ldots,\,i+n_{T}$ and $\sigma_{X,\left(T_{m}+\tau+j-1\right)h}=\sigma_{X,\left(T_{m}+\tau+i-n_{T}-1\right)h}$
for all $j=i-n_{T},\ldots,\,i$. Then, \eqref{eq (1) Cor LP} implies
\begin{align*}
\mu'_{\beta,\left(T_{m}+\tau+i-1\right)h}\sigma_{X,\left(T_{m}+\tau+i-1\right)h}\mu_{\beta,\left(T_{m}+\tau+i-1\right)h} & =\mu'_{\beta,\left(T_{m}+\tau+i-n_{T}-1\right)h}=\sigma_{X,\left(T_{m}+\tau+i-n_{T}-1\right)h}\mu_{\beta,\left(T_{m}+\tau+i-n_{T}-1\right)h}
\end{align*}
 for all $i$. This holds if and only if the process $\left\{ z_{i}\right\} _{i=n_{T}}^{T_{n}-n_{T}}$
defined by
\begin{align*}
z_{i} & \triangleq\mu'_{\beta,\left(T_{m}+\tau+i-1\right)h}\sigma_{X,\left(T_{m}+\tau+i-1\right)h}\mu_{\beta,\left(T_{m}+\tau+i-1\right)h}
\end{align*}
is constant. But this is a contradiction because it is non-smooth
by assumption (if only $\mu_{\beta,\left(T_{m}+\tau+i-1\right)h}$
is non-smooth then $z_{i}$ is still non-smooth because $\sigma_{X,\left(T_{m}+\tau+i-1\right)h}>0$
$\mathbb{P}$-a.s. by assumption.) $\square$

\clearpage{}

\end{singlespace}

\newpage{}

%\appendixpagenumbering

\renewcommand{\thesection}{S.\Alph{section}}   

\section{\label{Section Additional Figures Supp}Additional Figures Related
to Section \ref{Section Simulation Study}}

\normalsize
\indent

%\appendixpagenumbering

\setcounter{figure}{0}
\renewcommand{\thefigure}{S-\arabic{figure}}
\renewcommand{\thepage}{S-\arabic{page}}   %\renewcommand{\thesection}{S.\arabic{section}}
\renewcommand{\thesection}{S.\Alph{section}}   

\begin{singlespace}
\noindent 
\begin{figure}[H]
\includegraphics[width=18cm,height=12cm]{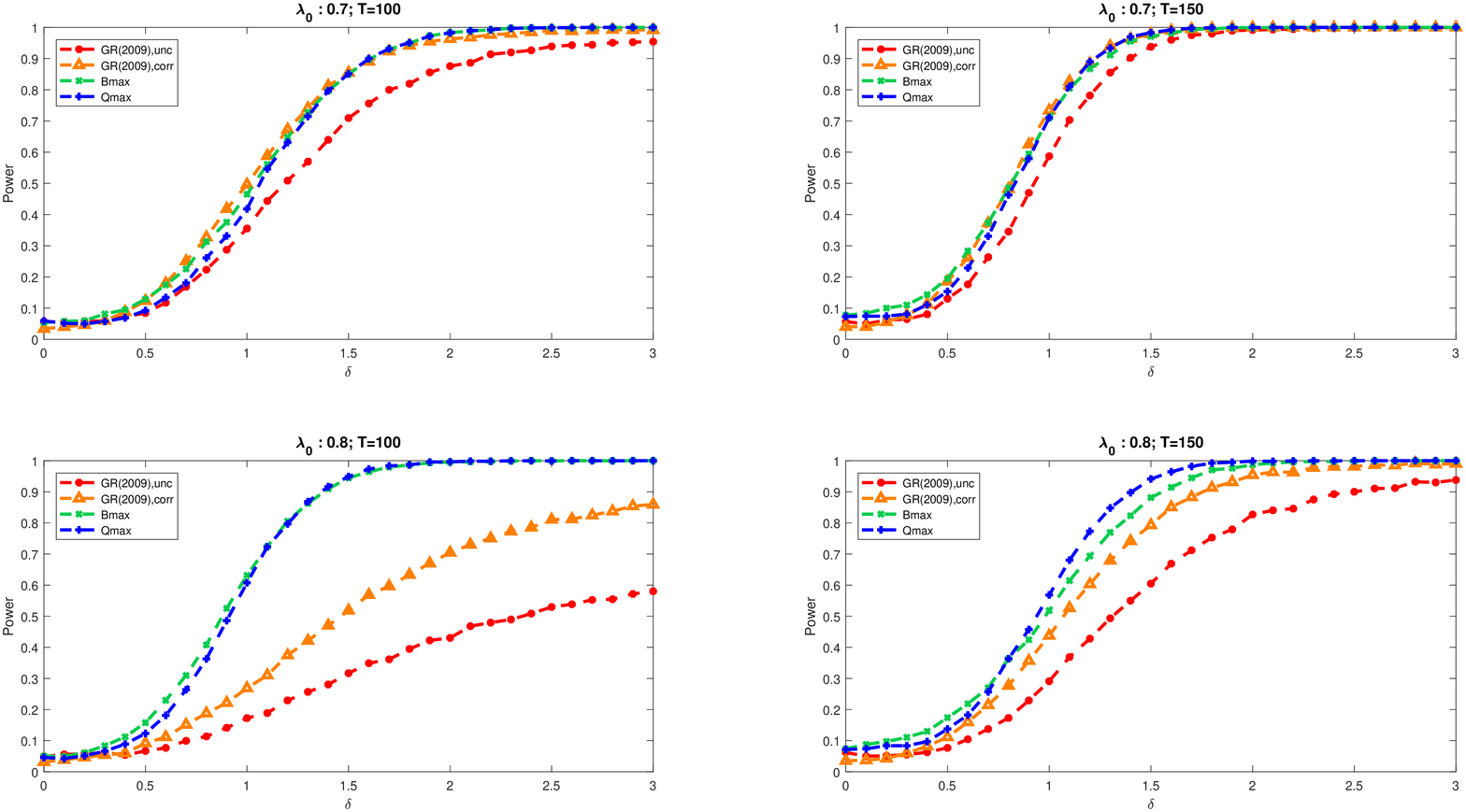}

{\footnotesize{}\caption{{\footnotesize{}\label{Fig_P1b_150}Small sample power functions for
model P1b: $Y_{t}=2.73-0.44X_{t-1}+\delta X_{t-1}\mathbf{1}\left\{ t>T_{b}^{0}\right\} +e_{t}$},{\footnotesize{}
where $X_{t-1}\sim\mathrm{i.i.d.}\mathscr{N}\left(1,\,1\right)$,
$e_{t}\sim\mathrm{i.i.d.}\mathscr{N}\left(0,\,1\right)$, and $T_{b}^{0}=T\lambda_{0}$.
The sample size is $T=100$ (left panels) and $T=150$ (right panels).
The fractional break date is $\lambda_{0}=0.7$ (top panels) and $\lambda_{0}=0.8$
(bottom panels). In-sample size is $T_{m}=0.4T$ while out-of-sample
size is $T_{n}=0.6T$. The green and blue broken lines correspond
to $\mathrm{B}_{\mathrm{max},h}$ and $\mathrm{Q}_{\mathrm{max},h}$,
respectively. The red and orange broken lines correspond to the $t^{\mathrm{stat}}$
of \citet{giacomini/rossi:09}, respectively, the uncorrected and
corrected version.}}
}{\footnotesize \par}
\end{figure}

\end{singlespace}

\begin{singlespace}
\noindent 
\begin{figure}[H]
\includegraphics[width=18cm,height=12cm]{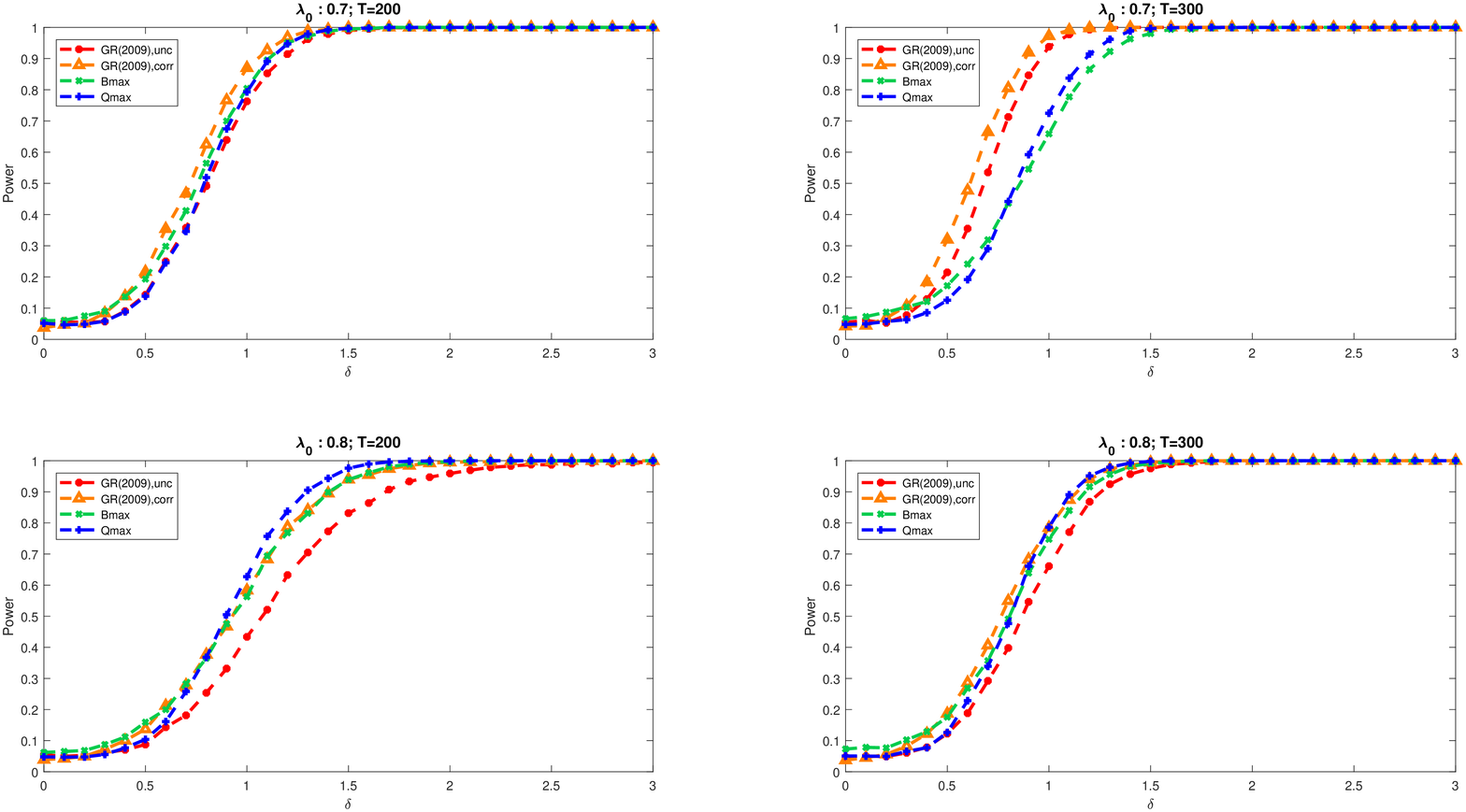}

{\footnotesize{}\caption{{\footnotesize{}\label{Fig_P1b_2300}mall sample power functions for
model P1b. The sample size is $T=200$ (left panels) and $T=300$
(right panels). The notes of Figure \ref{Fig_P1b_150} apply. }}
}{\footnotesize \par}
\end{figure}

\end{singlespace}

\begin{singlespace}
\noindent 
\begin{figure}[H]
\includegraphics[width=18cm,height=12cm]{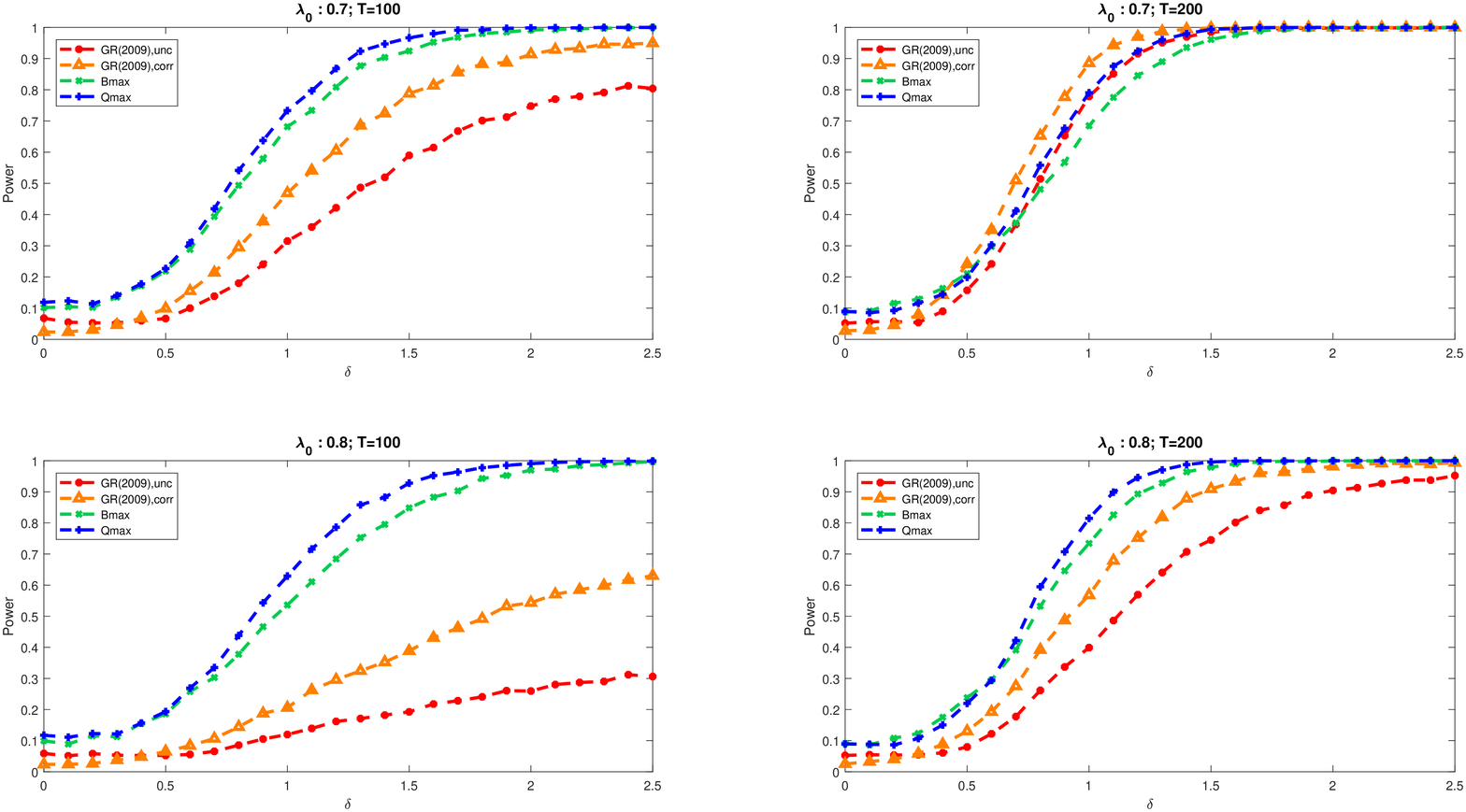}

{\footnotesize{}\caption{{\footnotesize{}\label{Fig_P2_120_078}Small sample power functions
for model P2: $Y_{t}=X_{t-1}+\delta X_{t-1}\mathbf{1}\left\{ t>T_{b}^{0}\right\} +e_{t}$
where $X_{t}$ is a Gaussian AR(1) with autoregressive coefficient
0.4 and unit variance, and $e_{t}\sim\mathrm{i.i.d.\,}\mathscr{N}\left(0,\,0.49\right)$.
The sample size is $T=100$ (left panel) and $T=200$ (right panel).
The fractional break date is $\lambda_{0}=0.7$ (top panel) and $\lambda_{0}=0.8$
(bottom panel). In-sample size is $T_{m}=0.4T$ while out-of-sample
size is $T_{n}=0.6T$. The green and blue broken lines correspond
to $\mathrm{B}_{\mathrm{max},h}$ and $\mathrm{Q}_{\mathrm{max},h}$,
respectively. The red and orange broken lines correspond to the $t^{\mathrm{stat}}$
of \citet{giacomini/rossi:09}, respectively, the uncorrected and
corrected version.}}
}{\footnotesize \par}
\end{figure}

\end{singlespace}

\begin{singlespace}
\noindent 
\begin{figure}[H]
\includegraphics[width=18cm,height=12cm]{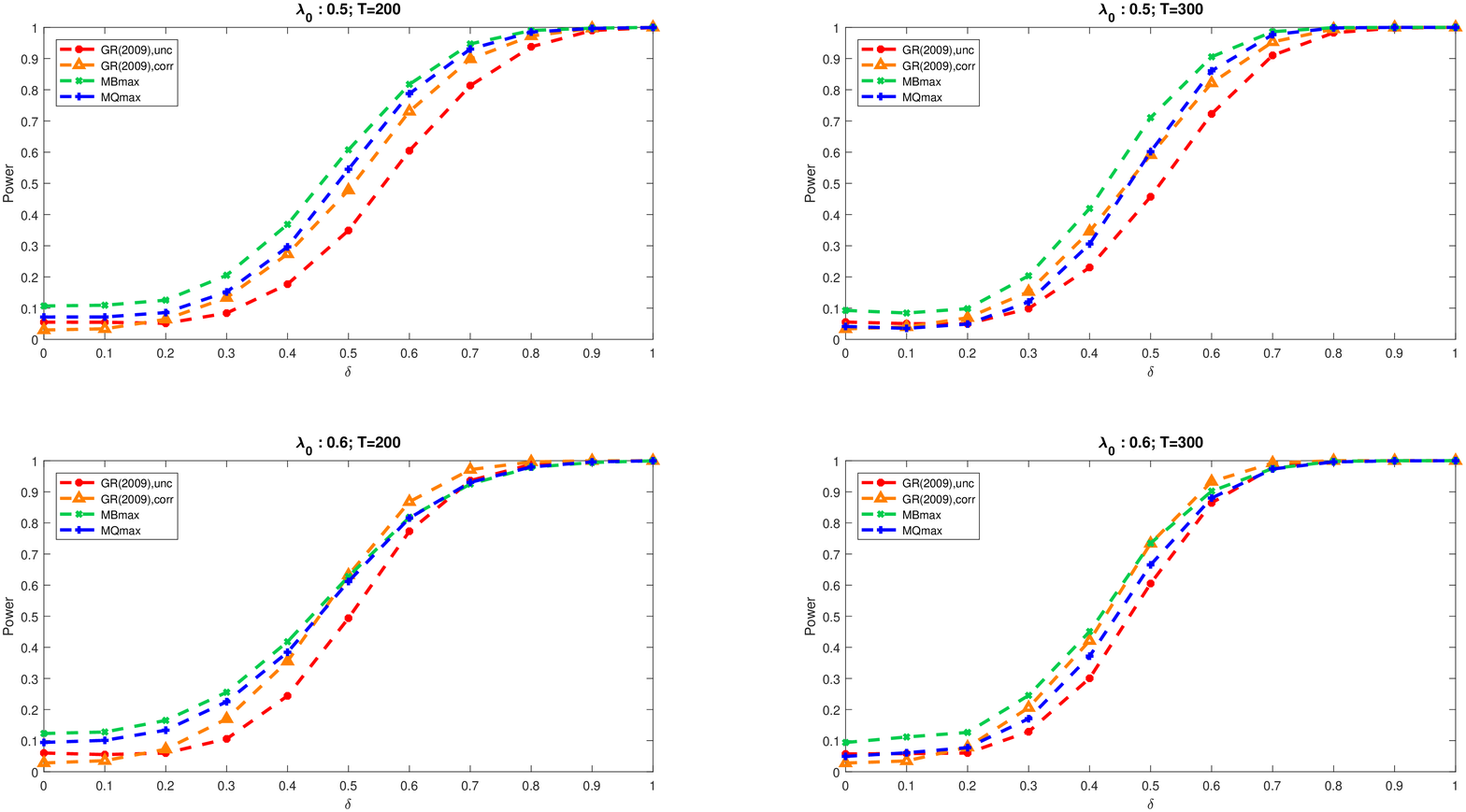}

{\footnotesize{}\caption{{\footnotesize{}\label{Fig_P3_230_056}Small sample power functions
for model P4 (recurrent break in mean): $Y_{t}=\beta_{t}+e_{t}$,
where $\beta_{t}$ switches between $\delta$ and $0$ every $p$
periods and $e_{t}\sim\mathrm{i.i.d.}\mathscr{N}\left(0,\,0.64\right)$.
We set $\left(T,\,p\right)=\left\{ \left(200,\,30\right),\,\left(300,\,40\right)\right\} $.
The fractional break date is $\lambda_{0}=0.5$ (top panels) and $\lambda_{0}=0.6$
(bottom panels). In-sample size is $T_{m}=T\lambda_{0}$ while out-of-sample
size is $T_{n}=T\left(1-\lambda_{0}\right)$. The green and blue broken
lines correspond to $\mathrm{B}_{\mathrm{max},h}$ and $\mathrm{Q}_{\mathrm{max},h}$,
respectively. The red and orange broken lines correspond to the $t^{\mathrm{stat}}$
of \citet{giacomini/rossi:09}, respectively, the uncorrected and
corrected version. }}
}{\footnotesize \par}
\end{figure}

\end{singlespace}

\begin{singlespace}
\noindent 
\begin{figure}[H]
\includegraphics[width=18cm,height=12cm]{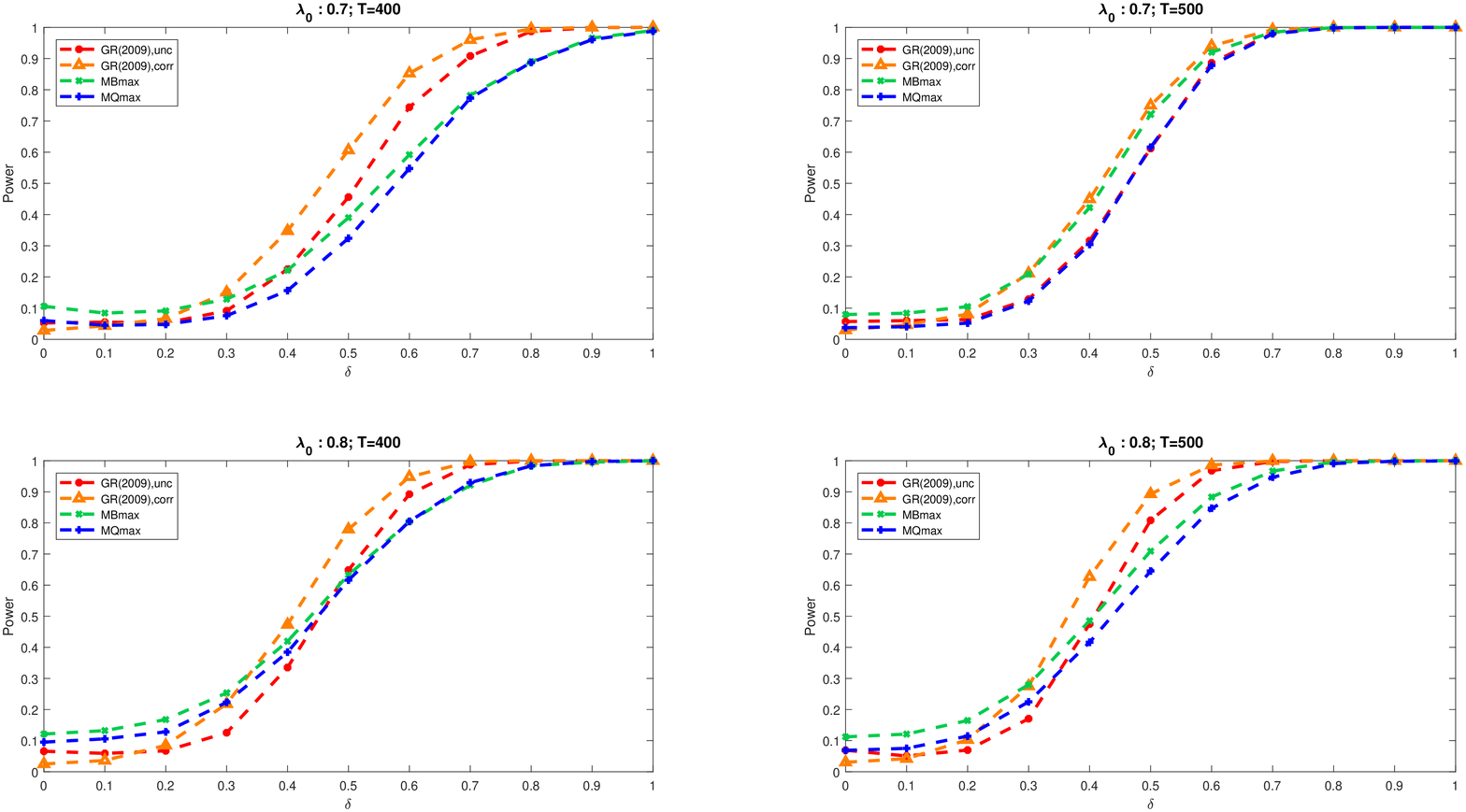}

{\footnotesize{}\caption{{\footnotesize{}\label{Fig_P3_450_078}Small sample power functions
for model P3. We set $\left(T,\,p\right)=\left\{ \left(400,\,30\right),\,\left(500,\,40\right)\right\} $.
The fractional break date is $\lambda_{0}=0.7$ (top panels) and $\lambda_{0}=0.8$
(bottom panels). In-sample size is $T_{m}=T\lambda_{0}$ while out-of-sample
size is $T_{n}=T\left(1-\lambda_{0}\right)$. The notes of Figure
\ref{Fig_P3_230_056} apply. }}
}{\footnotesize \par}
\end{figure}

\end{singlespace}

\noindent 
\begin{figure}[H]
\includegraphics[width=18cm,height=12cm]{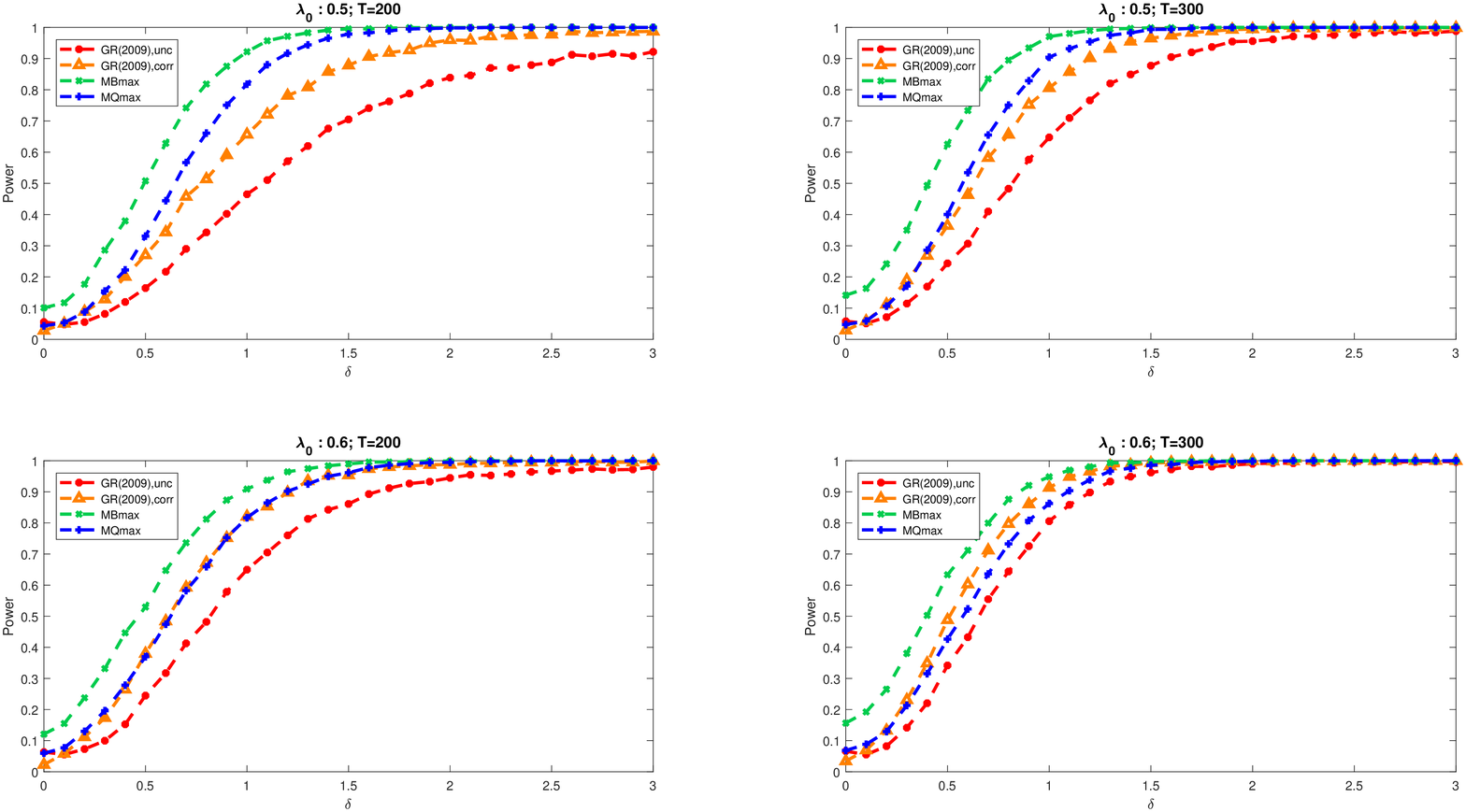}

{\footnotesize{}\caption{{\footnotesize{}\label{Fig_P5_230_056}Small sample power functions
for model P6 (recursive break in variance): $Y_{t}=\mu+\left(1+\beta_{t}\right)e_{t}$,
where $\beta_{t}$ switches between $\delta$ and $0$ every $p$
periods and $e_{t}\sim\mathrm{i.i.d.}\mathscr{N}\left(0,\,0.49\right)$.
We set $\left(T,\,p\right)=\left\{ \left(200,\,30\right),\,\left(300,\,40\right)\right\} $.
The fractional break date is $\lambda_{0}=0.5$ (top panels) and $\lambda_{0}=0.6$
(bottom panels). In-sample size is $T_{m}=T\lambda_{0}$ while out-of-sample
size is $T_{n}=T\lambda_{0}$. The green and blue broken lines correspond
to $\mathrm{B}_{\mathrm{max},h}$ and $\mathrm{Q}_{\mathrm{max},h}$,
respectively. The red and orange broken lines correspond to the $t^{\mathrm{stat}}$
of \citet{giacomini/rossi:09}, respectively, the uncorrected and
corrected version.}}
}{\footnotesize \par}
\end{figure}

\begin{figure}[H]
\includegraphics[width=18cm,height=12cm]{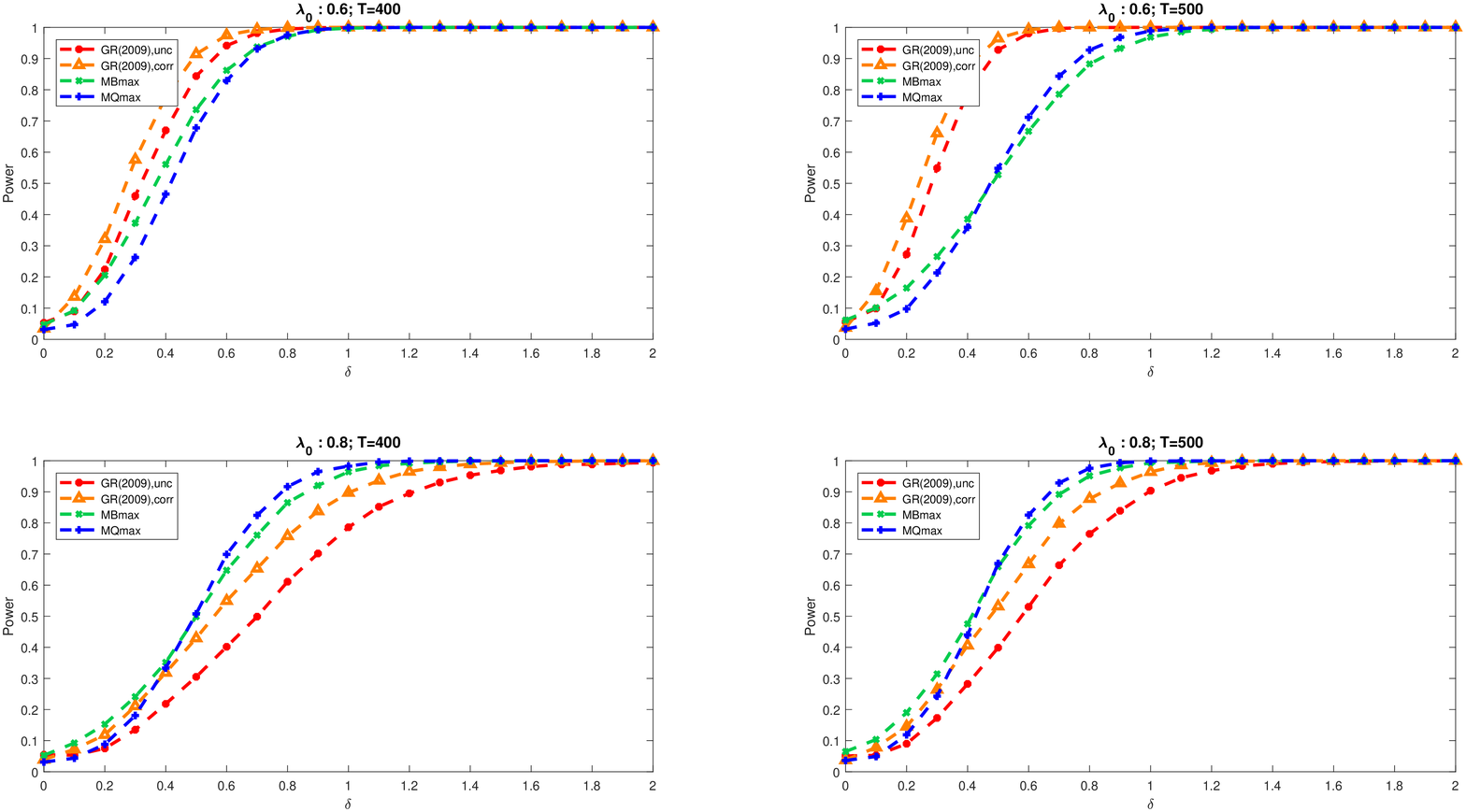}

{\footnotesize{}\caption{{\footnotesize{}\label{Fig_P4_450_068}Small sample power functions
for model P4 (single break in variance). The sample size is $T=400$
(left panels) and $T=500$ (right panels). The notes of Figure \ref{Fig_P4_230_068}
apply.}}
}{\footnotesize \par}
\end{figure}

\begin{singlespace}
\noindent 
\begin{figure}[H]
\includegraphics[width=18cm,height=12cm]{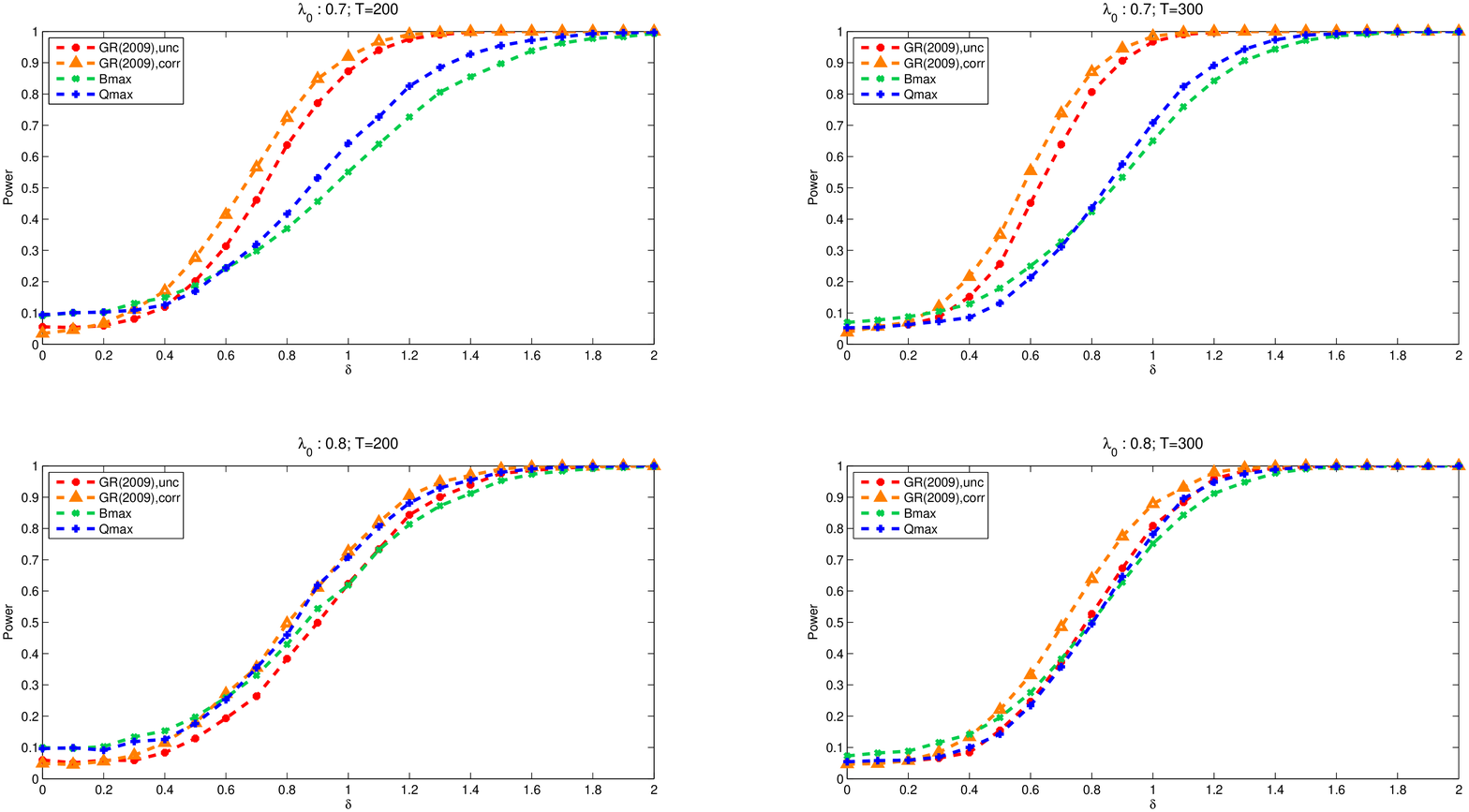}

{\footnotesize{}\caption{{\footnotesize{}\label{Fig_P5_2300_078}Small sample power functions
for model P5 (recursive break in variance). We set $\left(T,\,p\right)=\left\{ \left(200,\,30\right),\,\left(300,\,40\right)\right\} $.
The fractional break date is $\lambda_{0}=0.7$ (top panels) and $\lambda_{0}=0.8$
(bottom panels). The notes of Figure \ref{Fig_P5_230_056} apply.}}
}{\footnotesize \par}
\end{figure}

\end{singlespace}

\begin{singlespace}
\noindent 
\begin{figure}[H]
\includegraphics[width=18cm,height=12cm]{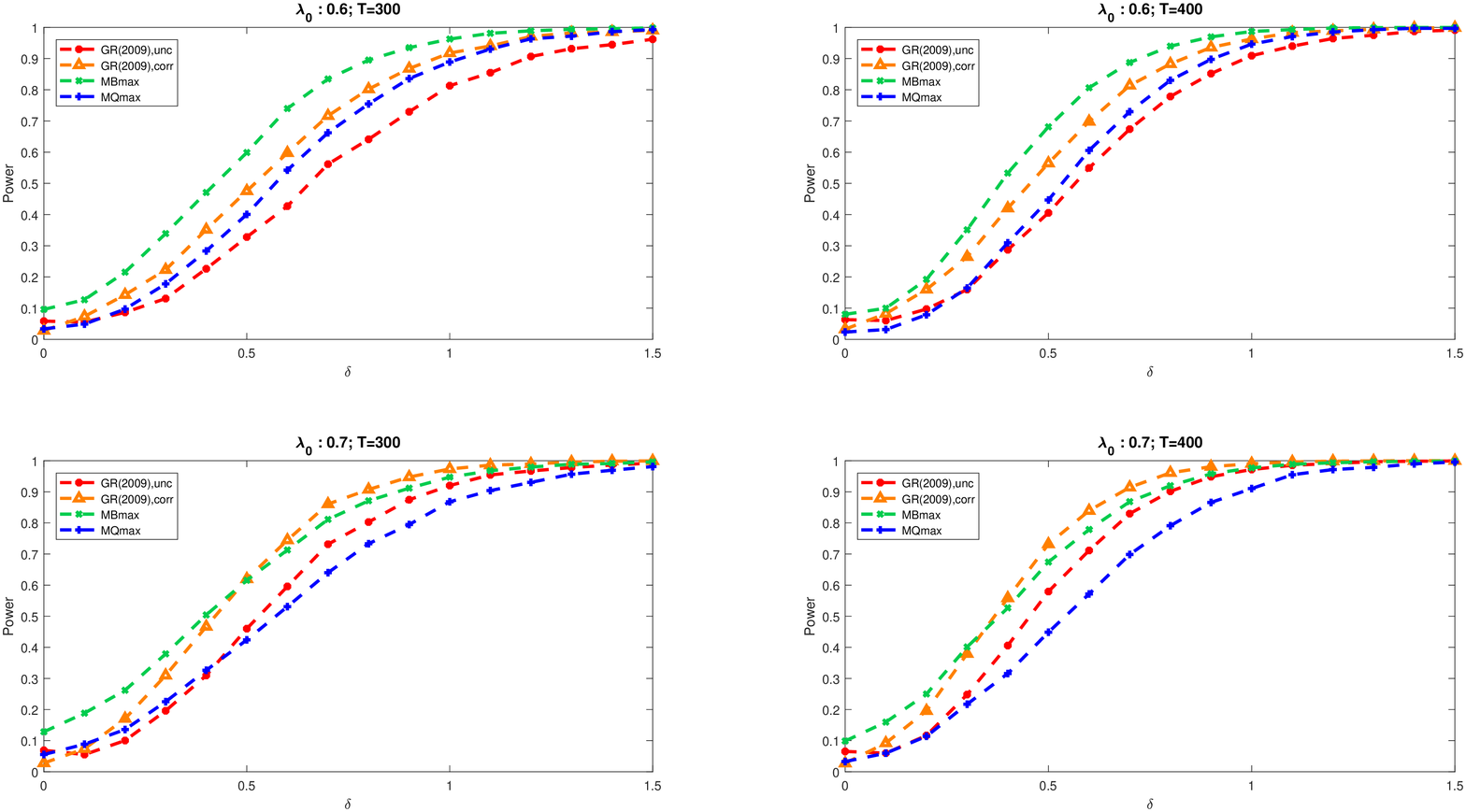}

{\footnotesize{}\caption{{\footnotesize{}\label{Fig_P5_3400_067-1}Small sample power functions
for model P5 (recurrent break in variance). The sample size is $T=300$
(left panels) and $T=400$ (right panels). The fractional break date
is $\lambda_{0}=0.6$ (top panels) and $\lambda_{0}=0.7$ (bottom
panels). The notes of Figure \ref{Fig_P5_230_056} apply.}}
}{\footnotesize \par}
\end{figure}

\end{singlespace}

\begin{singlespace}
\noindent 
\begin{figure}[H]
\includegraphics[width=18cm,height=12cm]{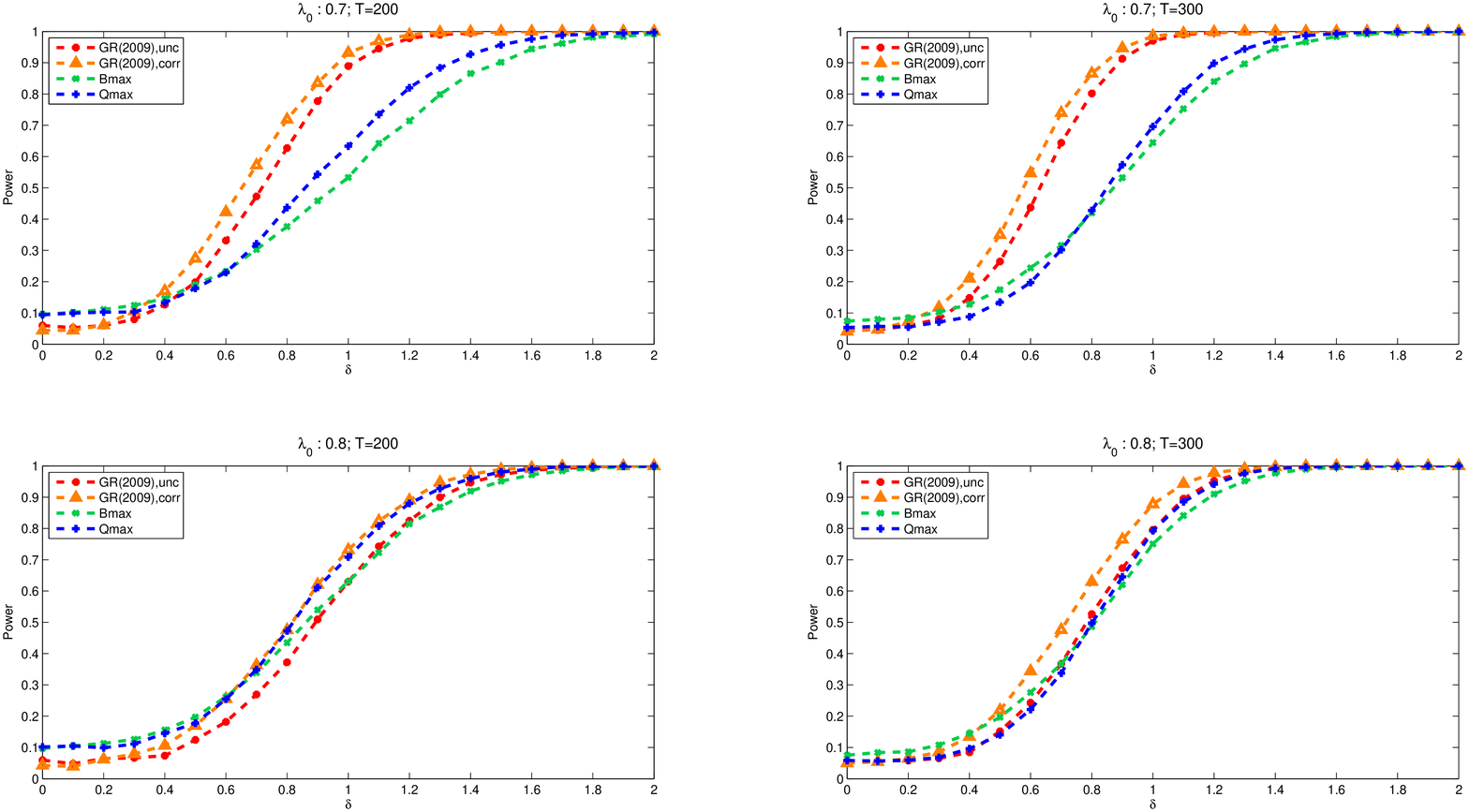}

{\footnotesize{}\caption{{\footnotesize{}\label{Fig_P6_2300_078}Small sample power functions
for model P6 (lagged dependent variables): $Y_{t}=\delta\mathbf{1}\left\{ t>T_{b}^{0}\right\} +0.3Y_{t-1}+e_{t}$,
$e_{t}\sim\mathrm{i.i.d.}\mathscr{N}\left(0,\,0.49\right)$. The sample
size is $T=200$ (left panels) and $T=300$ (right panels). The fractional
break date is $\lambda_{0}=0.7$ (top panels) and $\lambda_{0}=0.8$
(bottom panels). In-sample size is $T_{m}=0.4T$ while out-of-sample
size is $T_{n}=0.6T$. The green and blue broken lines correspond
to $\mathrm{B}_{\mathrm{max},h}$ and $\mathrm{Q}_{\mathrm{max},h}$,
respectively. The red and orange broken lines correspond to the $t^{\mathrm{stat}}$
of \citet{giacomini/rossi:09}, respectively, the uncorrected and
corrected version. }}
}{\footnotesize \par}
\end{figure}

\end{singlespace}

\begin{singlespace}
\noindent 
\begin{figure}[H]
\includegraphics[width=18cm,height=12cm]{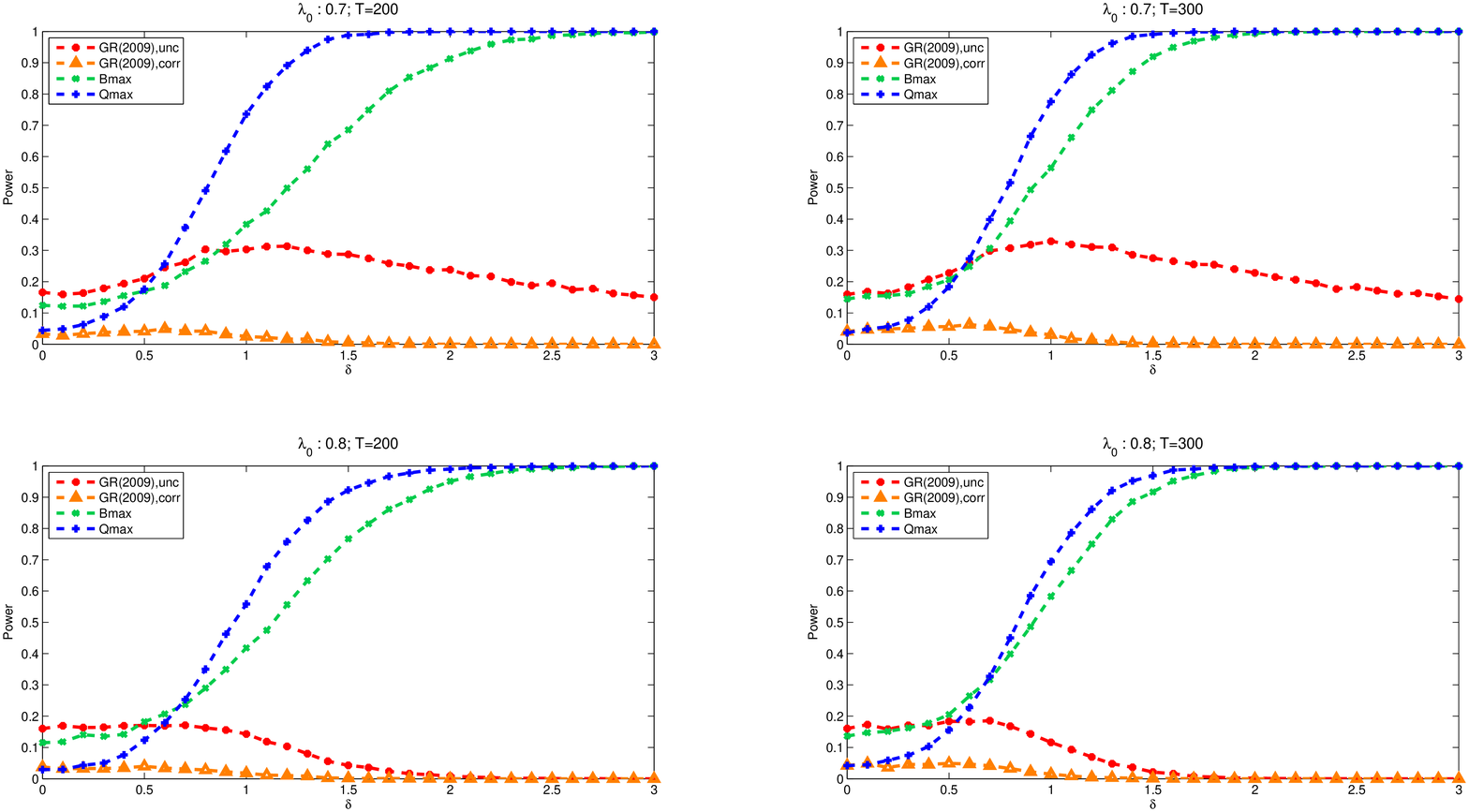}

{\footnotesize{}\caption{{\footnotesize{}\label{Fig_P8_2300_078}Small sample power functions
for model P8 (autocorrelated errors): $Y_{t}=1+X_{t-1}+\delta X_{t-1}\mathbf{1}\left\{ t>T_{b}^{0}\right\} +e_{t}$,
where $X_{t-1}\sim\mathrm{i.i.d.}\mathscr{N}\left(0,\,1.4\right)$
and $e_{t}=0.4u_{t-1}+u_{t}$, $u_{t}\sim\mathrm{i.i.d.}\mathscr{N}\left(0,\,1\right)$.
The sample size is $T=200$ (left panels) and $T=300$ (right panels).
The fractional break date is $\lambda_{0}=0.7$ (top panels) and $\lambda_{0}=0.8$
(bottom panels). In-sample size is $T_{m}=0.5T$ while out-of-sample
size is $T_{n}=0.5T$. The green and blue broken lines correspond
to $\mathrm{B}_{\mathrm{max},h}$ and $\mathrm{Q}_{\mathrm{max},h}$,
respectively. The red and orange broken lines correspond to the $t^{\mathrm{stat}}$
of \citet{giacomini/rossi:09}, respectively, the uncorrected and
corrected version.}}
}{\footnotesize \par}
\end{figure}

\end{singlespace}

\begin{singlespace}
\noindent 
\begin{figure}[H]
\includegraphics[width=18cm,height=12cm]{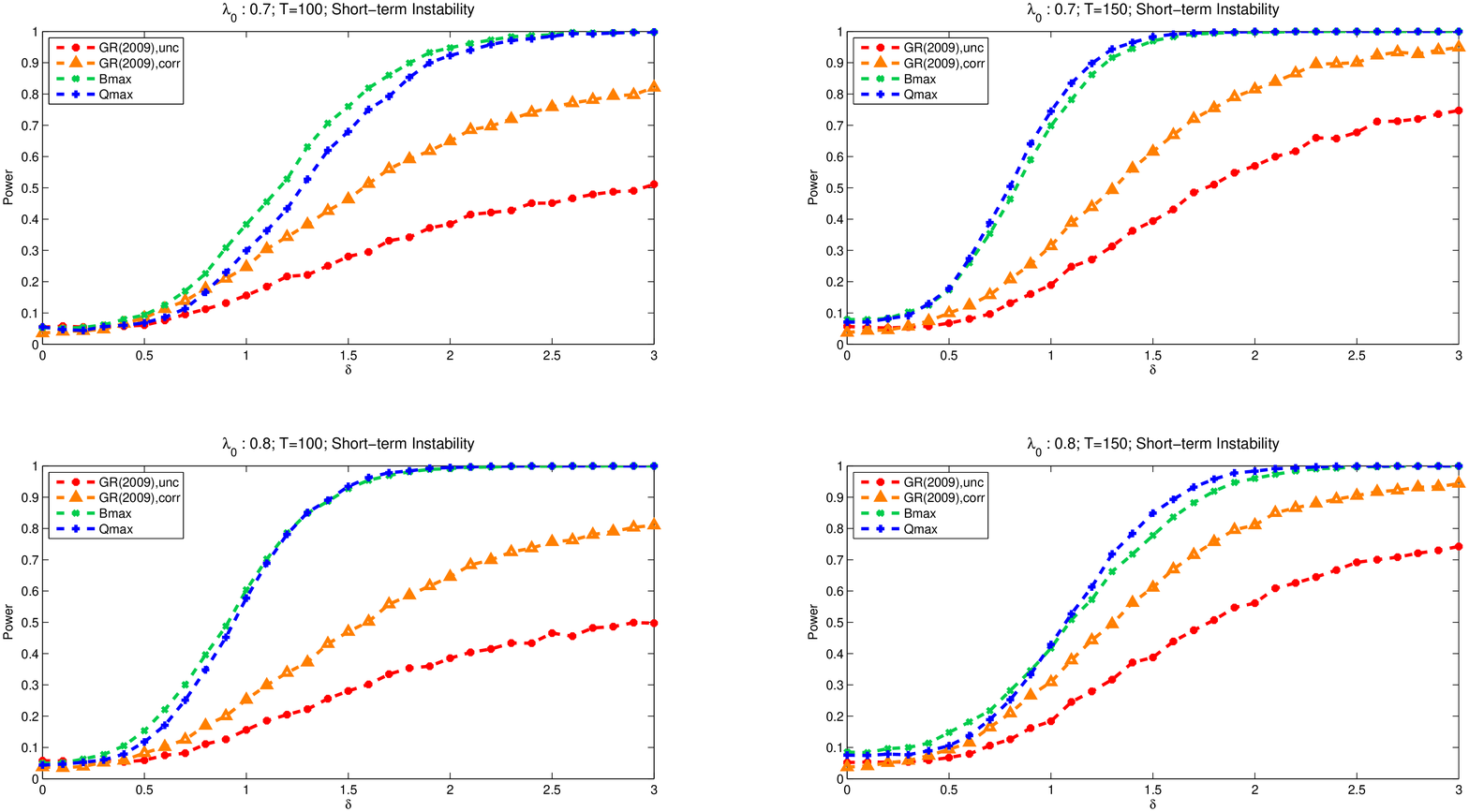}

{\footnotesize{}\caption{{\footnotesize{}\label{Fig_P1b_150_st}Small sample power functions
for model P1b with short-term instability: $Y_{t}=2.73-0.44X_{t-1}+\delta X_{t-1}\mathbf{1}\left\{ T_{b}^{0}<t\leq T_{b}^{0}+p\right\} +e_{t}$},{\footnotesize{}
where $X_{t-1}\sim\mathrm{i.i.d.}\mathscr{N}\left(1,\,1\right)$,
$e_{t}\sim\mathrm{i.i.d.}\mathscr{N}\left(0,\,1\right)$, and $T_{b}^{0}=T\lambda_{0}$.
We set $\left(T,\,p\right)=\left\{ \left(100,\,20\right),\,\left(150,\,25\right)\right\} $.
The fractional break date is $\lambda_{0}=0.7$ (top panels) and $\lambda_{0}=0.8$
(bottom panels). In-sample size is $T_{m}=0.4T$ while out-of-sample
size is $T_{n}=0.6T$. The green and blue broken lines correspond
to $\mathrm{B}_{\mathrm{max},h}$ and $\mathrm{Q}_{\mathrm{max},h}$,
respectively. The red and orange broken lines correspond to the $t^{\mathrm{stat}}$
of \citet{giacomini/rossi:09}, respectively, the uncorrected and
corrected version.}}
}{\footnotesize \par}
\end{figure}

\end{singlespace}

\begin{singlespace}
\noindent 
\begin{figure}[H]
\includegraphics[width=18cm,height=12cm]{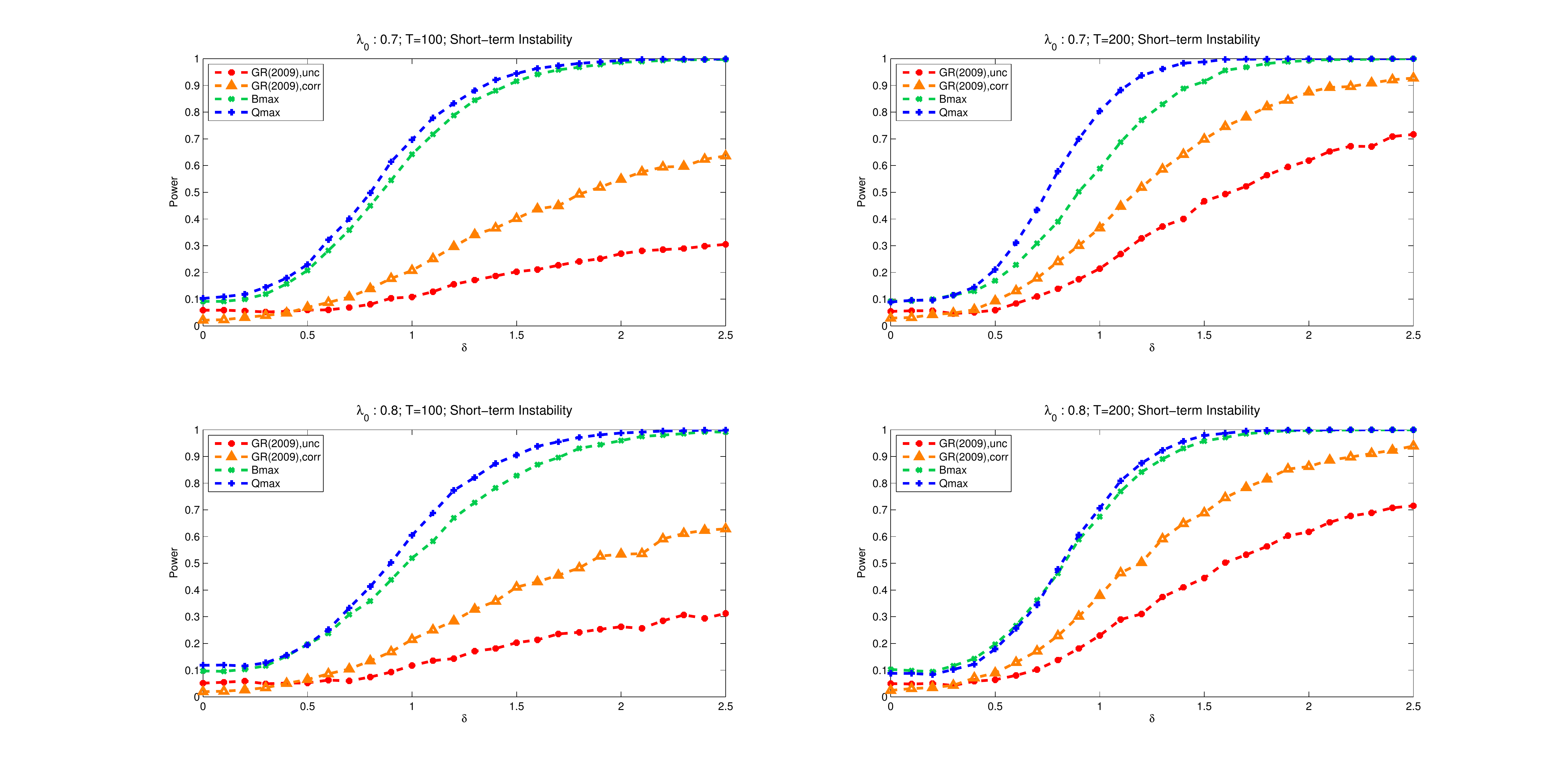}

{\footnotesize{}\caption{{\footnotesize{}\label{Fig_P2_1200_078_st}Small sample power functions
for model P2 with short-term instability: $Y_{t}=X_{t-1}+\delta X_{t-1}\mathbf{1}\left\{ T_{b}^{0}<t\leq T_{b}^{0}+p\right\} +e_{t}$,
where $X_{t-1}$ is a Gaussina AR(1) with autoregressive coefficient
0.4 and unit variance, and $e_{t}\sim\mathrm{i.i.d.}\mathscr{N}\left(0,\,0.49\right)$,
and $T_{b}^{0}=T\lambda_{0}$. We set $\left(T,\,p\right)=\left\{ \left(100,\,20\right),\,\left(200,\,30\right)\right\} $.
The fractional break date is $\lambda_{0}=0.7$ (top panels) and $\lambda_{0}=0.8$
(bottom panels). In-sample size is $T_{m}=0.4T$ while out-of-sample
size is $T_{n}=0.6T$. The green and blue broken lines correspond
to $\mathrm{B}_{\mathrm{max},h}$ and $\mathrm{Q}_{\mathrm{max},h}$,
respectively. The red and orange broken lines correspond to the $t^{\mathrm{stat}}$
of \citet{giacomini/rossi:09}, respectively, the uncorrected and
corrected version.}}
}{\footnotesize \par}
\end{figure}

\end{singlespace}

\newpage{}

%\appendixpagenumbering

\section{\label{Sec Additional-Monte-Carlo}Additional Monte Carlo Studies}

This section extends the small-sample analysis of Section \ref{Section Simulation Study}
to alternative forecasting schemes (recursive and rolling scheme)
and to the Linex loss function. Table \ref{Table Size S1 Recursive}-\ref{Table Size S1 Rolling}
report the empirical sizes at significance level $\alpha=0.05$ for
model S1 for all the tests considered. Let us first consider the case
regarding the recursive scheme. The statistic $\mathrm{Q}_{\mathrm{max},h}$
appears to be well-sized, even though slightly liberal when the in-sample
window is either too small or too large. The statistics $\mathrm{B}_{\mathrm{max},h}$
and $\mathrm{MQ}_{\mathrm{max},h}$ seem to display reasonable size
control, though they tend to overeject quite a bit whereas $\mathrm{MB}_{\mathrm{max},h}$
is excessively oversized and therefore not comparable with the other
tests. Turning to the tests of \citet{giacomini/rossi:09}, we note
that the uncorrected version $t^{\mathrm{stat}}$ performs well while
the corrected version is undersized, though not excessively, for all
sample sizes. Overall, we can conclude that all the statistics, with
exception of $\mathrm{MB}_{\mathrm{max},h}$, are comparable in term
of size and consequently it is fair to compare their power properties. 

We present the empirical power of the tests in Table \ref{Table P1 Recursive}-\ref{Table Size P1 Rolling ST}.
We consider model P1 with either long-lasting and short-lasting instabilities.
Under recursive scheme, Table \ref{Table P1 Recursive} shows that
all the statistics possess good power against break in a regression
coefficient. Since the instability begins at about middle sample (i.e.,
$\lambda_{0}=0.6$) and lasts for about 40\% of the total sample the
tests of \citet{giacomini/rossi:09} tend to have higher power when
the sample size is large. However, when the instability only lasts
for few consecutive periods (cf. Table \ref{Table P1 Recursive ST})
the gains in statistical power of our tests are substantial. This
is equivalent to what observed in the main paper regarding the fixed
scheme and it extends to the recursive and rolling scheme (cf. Table
\ref{Table Size P1 Rolling ST}). Table \ref{Table Size P1 Rolling}
reports the power comparison for the rolling scheme. All tests display
good power when the magnitude of the break is large. When $\delta=0.5,\,1$
the statistic $\mathrm{B}_{\mathrm{max},h}$ dominates the statistics
of \citet{giacomini/rossi:09}, especially the magnitude of the break
is small (i.e., $\delta=0.5$)\textemdash which constitutes a highly
relevant case in practice.

Overall, we confirm the same observations relevant for the fixed scheme
considered in the main text. Our test statistics show reasonable size
control, the only exception is $\mathrm{MB}_{\mathrm{max},h}$ which
turns out to be oversized. In terms of power properties, all tests
display good power while there are substantial power gains relative
to existing methods especially when the instability (i) is short-lasting
and/or (ii) is located toward the tail of the out-of-sample.

\newpage{}

\clearpage{}\pagebreak{}

\section{Tables}

%\end{appendices}

%\end{singlespace}

\normalsize
\indent

%\appendixpagenumbering

\setcounter{table}{0}
\renewcommand{\thetable}{S-\arabic{table}}

\begin{table}[H]
\caption{\label{Table Size S3}Empirical small sample size of forecast instability
tests based on model S3}
\begin{centering}
\begin{tabular}{cccccccccccc}
\hline 
 &  &  & \multicolumn{2}{c}{GR (2009)} & $\mathrm{B}_{\mathrm{max},h}$ & \multicolumn{2}{c}{$\mathrm{Q}_{\mathrm{max},h}$} & \multicolumn{2}{c}{$\mathrm{MB}_{\mathrm{max},h}$} & \multicolumn{2}{c}{$\mathrm{MQ}_{\mathrm{max},h}$}\tabularnewline
 &  &  & $t^{\mathrm{stat}}$ (uncorrected) & $t^{\mathrm{stat,c}}$ (corrected) &  & \multicolumn{6}{c}{}\tabularnewline
\hline 
\multicolumn{3}{l}{$\alpha=0.05$ } & \multicolumn{2}{c}{} & \tabularnewline
 & $T_{m}$ & $T_{n}$ &  &  &  & \multicolumn{2}{c}{} & \multicolumn{2}{c}{} & \multicolumn{2}{c}{}\tabularnewline
$T=100$ & 25 & 75 & 0.049 & 0.019 & 0.086 & \multicolumn{2}{c}{0.098} & \multicolumn{2}{c}{0.090} & \multicolumn{2}{c}{0.089}\tabularnewline
\multirow{2}{*}{} & 50 & 50 & 0.069 & 0.024 & 0.058 & \multicolumn{2}{c}{0.072} & \multicolumn{2}{c}{0.083} & \multicolumn{2}{c}{0.067}\tabularnewline
 & 75 & 25 & 0.039 & 0.016 & 0.081 & \multicolumn{2}{c}{0.111} & \multicolumn{2}{c}{0.092} & \multicolumn{2}{c}{0.091}\tabularnewline
$T=200$ & 50 & 150 & 0.049 & 0.025 & 0.076 & \multicolumn{2}{c}{0.072} & \multicolumn{2}{c}{0.138} & \multicolumn{2}{c}{0.089}\tabularnewline
\multirow{2}{*}{} & 100 & 100 & 0.057 & 0.026 & 0.070 & \multicolumn{2}{c}{0.073} & \multicolumn{2}{c}{0.106} & \multicolumn{2}{c}{0.068}\tabularnewline
 & 150 & 50 & 0.075 & 0.020 & 0.055 & \multicolumn{2}{c}{0.070} & \multicolumn{2}{c}{0.082} & \multicolumn{2}{c}{0.070}\tabularnewline
$T=300$ & 75 & 225 & 0.050 & 0.029 & 0.058 & \multicolumn{2}{c}{0.036} & \multicolumn{2}{c}{0.102} & \multicolumn{2}{c}{0.044}\tabularnewline
\multirow{2}{*}{} & 150 & 150 & 0.059 & 0.032 & 0.077 & \multicolumn{2}{c}{0.072} & \multicolumn{2}{c}{0.144} & \multicolumn{2}{c}{0.086}\tabularnewline
 & 225 & 75 & 0.065 & 0.025 & 0.096 & \multicolumn{2}{c}{0.103} & \multicolumn{2}{c}{0.152} & \multicolumn{2}{c}{0.123}\tabularnewline
$T=400$ & 100 & 300 & 0.054 & 0.032 & 0.061 & \multicolumn{2}{c}{0.041} & \multicolumn{2}{c}{0.123} & \multicolumn{2}{c}{0.046}\tabularnewline
\multirow{2}{*}{} & 200 & 200 & 0.051 & 0.035 & 0.065 & \multicolumn{2}{c}{0.048} & \multicolumn{2}{c}{0.111} & \multicolumn{2}{c}{0.052}\tabularnewline
 & 300 & 100 & 0.068 & 0.031 & 0.067 & \multicolumn{2}{c}{0.063} & \multicolumn{2}{c}{0.115} & \multicolumn{2}{c}{0.074}\tabularnewline
 &  &  &  &  &  &  &  & \multicolumn{2}{c}{} &  & \tabularnewline
\multicolumn{3}{l}{$\alpha=0.10$ } &  &  &  &  &  &  &  &  & \tabularnewline
 & $T_{m}$ & $T_{n}$ & \multicolumn{2}{c}{} &  & \multicolumn{2}{c}{} & \multicolumn{2}{c}{} & \multicolumn{2}{c}{}\tabularnewline
$T=100$ & 25 & 75 & 0.109 & 0.069 & 0.136 & \multicolumn{2}{c}{0.152} & \multicolumn{2}{c}{0.191} & \multicolumn{2}{c}{0.165}\tabularnewline
\multirow{2}{*}{} & 50 & 50 & 0.107 & 0.069 & 0.095 & \multicolumn{2}{c}{0.118} & \multicolumn{2}{c}{0.118} & \multicolumn{2}{c}{0.095}\tabularnewline
 & 75 & 25 & 0.134 & 0.060 & 0.112 & \multicolumn{2}{c}{0.152} & \multicolumn{2}{c}{0.125} & \multicolumn{2}{c}{0.128}\tabularnewline
$T=200$ & 50 & 150 & 0.100 & 0.078 & 0.125 & \multicolumn{2}{c}{0.113} & \multicolumn{2}{c}{0.199} & \multicolumn{2}{c}{0.133}\tabularnewline
\multirow{2}{*}{} & 100 & 100 & 0.106 & 0.073 & 0.108 & \multicolumn{2}{c}{0.111} & \multicolumn{2}{c}{0.160} & \multicolumn{2}{c}{0.108}\tabularnewline
 & 150 & 50 & 0.101 & 0.078 & 0.101 & \multicolumn{2}{c}{0.105} & \multicolumn{2}{c}{0.112} & \multicolumn{2}{c}{0.091}\tabularnewline
$T=300$ & 75 & 225 & 0.102 & 0.081 & 0.103 & \multicolumn{2}{c}{0.071} & \multicolumn{2}{c}{0.159} & \multicolumn{2}{c}{0.077}\tabularnewline
\multirow{2}{*}{} & 150 & 150 & 0.111 & 0.079 & 0.119 & \multicolumn{2}{c}{0.112} & \multicolumn{2}{c}{0.189} & \multicolumn{2}{c}{0.129}\tabularnewline
 & 225 & 75 & 0.114 & 0.068 & 0.144 & \multicolumn{2}{c}{0.159} & \multicolumn{2}{c}{0.197} & \multicolumn{2}{c}{0.170}\tabularnewline
$T=400$ & 100 & 300 & 0.097 & 0.082 & 0.109 & \multicolumn{2}{c}{0.079} & \multicolumn{2}{c}{0.193} & \multicolumn{2}{c}{0.096}\tabularnewline
\multirow{2}{*}{} & 200 & 200 & 0.089 & 0.106 & 0.075 & \multicolumn{2}{c}{0.079} & \multicolumn{2}{c}{0.171} & \multicolumn{2}{c}{0.088}\tabularnewline
 & 300 & 100 & 0.112 & 0.079 & 0.104 & \multicolumn{2}{c}{0.110} & \multicolumn{2}{c}{0.164} & \multicolumn{2}{c}{0.104}\tabularnewline
\hline 
\end{tabular}
\par\end{centering}
\noindent\begin{minipage}[t]{1\columnwidth}%
{\small{}Model S3. The notes of Table \ref{Table S1} apply.}%
\end{minipage}
\end{table}

\begin{table}[H]
\caption{\label{Table S4}Empirical small sample size of forecast instability
tests based on model S4}
\begin{centering}
\begin{tabular}{cccccccccccc}
\hline 
 &  &  & \multicolumn{2}{c}{GR (2009)} & $\mathrm{B}_{\mathrm{max},h}$ & \multicolumn{2}{c}{$\mathrm{Q}_{\mathrm{max},h}$} & \multicolumn{2}{c}{$\mathrm{MB}_{\mathrm{max},h}$} & \multicolumn{2}{c}{$\mathrm{MQ}_{\mathrm{max},h}$}\tabularnewline
 &  &  & $t^{\mathrm{stat}}$ (uncorrected) & $t^{\mathrm{stat,c}}$ (corrected) &  & \multicolumn{6}{c}{}\tabularnewline
\hline 
\multicolumn{3}{l}{$\alpha=0.05$ } & \multicolumn{2}{c}{} & \tabularnewline
 & $T_{m}$ & $T_{n}$ &  &  &  & \multicolumn{2}{c}{} & \multicolumn{2}{c}{} & \multicolumn{2}{c}{}\tabularnewline
$T=100$ & 25 & 75 & 0.055 & 0.016 & 0.089 & \multicolumn{2}{c}{0.099} & \multicolumn{2}{c}{0.149} & \multicolumn{2}{c}{0.122}\tabularnewline
\multirow{2}{*}{} & 50 & 50 & 0.077 & 0.023 & 0.064 & \multicolumn{2}{c}{0.073} & \multicolumn{2}{c}{0.074} & \multicolumn{2}{c}{0.063}\tabularnewline
 & 75 & 25 & 0.066 & 0.027 & 0.042 & \multicolumn{2}{c}{0.044} & \multicolumn{2}{c}{0.086} & \multicolumn{2}{c}{0.092}\tabularnewline
$T=200$ & 50 & 150 & 0.055 & 0.028 & 0.112 & \multicolumn{2}{c}{0.121} & \multicolumn{2}{c}{0.141} & \multicolumn{2}{c}{0.089}\tabularnewline
\multirow{2}{*}{} & 100 & 100 & 0.060 & 0.029 & 0.069 & \multicolumn{2}{c}{0.063} & \multicolumn{2}{c}{0.110} & \multicolumn{2}{c}{0.069}\tabularnewline
 & 150 & 50 & 0.084 & 0.020 & 0.063 & \multicolumn{2}{c}{0.068} & \multicolumn{2}{c}{0.075} & \multicolumn{2}{c}{0.063}\tabularnewline
$T=300$ & 75 & 225 & 0.056 & 0.032 & 0.067 & \multicolumn{2}{c}{0.040} & \multicolumn{2}{c}{0.108} & \multicolumn{2}{c}{0.048}\tabularnewline
\multirow{2}{*}{} & 150 & 150 & 0.047 & 0.029 & 0.062 & \multicolumn{2}{c}{0.039} & \multicolumn{2}{c}{0.144} & \multicolumn{2}{c}{0.085}\tabularnewline
 & 225 & 75 & 0.059 & 0.034 & 0.045 & \multicolumn{2}{c}{0.020} & \multicolumn{2}{c}{0.147} & \multicolumn{2}{c}{0.120}\tabularnewline
$T=400$ & 100 & 300 & 0.051 & 0.036 & 0.076 & \multicolumn{2}{c}{0.038} & \multicolumn{2}{c}{0.132} & \multicolumn{2}{c}{0.055}\tabularnewline
\multirow{2}{*}{} & 200 & 200 & 0.052 & 0.034 & 0.052 & \multicolumn{2}{c}{0.027} & \multicolumn{2}{c}{0.121} & \multicolumn{2}{c}{0.054}\tabularnewline
 & 300 & 100 & 0.051 & 0.037 & 0.032 & \multicolumn{2}{c}{0.013} & \multicolumn{2}{c}{0.109} & \multicolumn{2}{c}{0.071}\tabularnewline
 &  &  &  &  &  &  &  & \multicolumn{2}{c}{} &  & \tabularnewline
\multicolumn{3}{l}{$\alpha=0.10$ } &  &  &  &  &  &  &  &  & \tabularnewline
 & $T_{m}$ & $T_{n}$ & \multicolumn{2}{c}{} &  & \multicolumn{2}{c}{} & \multicolumn{2}{c}{} & \multicolumn{2}{c}{}\tabularnewline
$T=100$ & 25 & 75 & 0.054 & 0.072 & 0.143 & \multicolumn{2}{c}{0.154} & \multicolumn{2}{c}{0.198} & \multicolumn{2}{c}{0.168}\tabularnewline
\multirow{2}{*}{} & 50 & 50 & 0.070 & 0.065 & 0.100 & \multicolumn{2}{c}{0.120} & \multicolumn{2}{c}{0.115} & \multicolumn{2}{c}{0.096}\tabularnewline
 & 75 & 25 & 0.062 & 0.069 & 0.069 & \multicolumn{2}{c}{0.073} & \multicolumn{2}{c}{0.124} & \multicolumn{2}{c}{0.125}\tabularnewline
$T=200$ & 50 & 150 & 0.053 & 0.073 & 0.120 & \multicolumn{2}{c}{0.118} & \multicolumn{2}{c}{0.202} & \multicolumn{2}{c}{0.137}\tabularnewline
\multirow{2}{*}{} & 100 & 100 & 0.064 & 0.078 & 0.107 & \multicolumn{2}{c}{0.104} & \multicolumn{2}{c}{0.152} & \multicolumn{2}{c}{0.106}\tabularnewline
 & 150 & 50 & 0.082 & 0.069 & 0.095 & \multicolumn{2}{c}{0.119} & \multicolumn{2}{c}{0.118} & \multicolumn{2}{c}{0.097}\tabularnewline
$T=300$ & 75 & 225 & 0.046 & 0.084 & 0.111 & \multicolumn{2}{c}{0.081} & \multicolumn{2}{c}{0.179} & \multicolumn{2}{c}{0.085}\tabularnewline
\multirow{2}{*}{} & 150 & 150 & 0.060 & 0.073 & 0.131 & \multicolumn{2}{c}{0.118} & \multicolumn{2}{c}{0.207} & \multicolumn{2}{c}{0.134}\tabularnewline
 & 225 & 75 & 0.068 & 0.078 & 0.129 & \multicolumn{2}{c}{0.142} & \multicolumn{2}{c}{0.184} & \multicolumn{2}{c}{0.160}\tabularnewline
$T=400$ & 100 & 300 & 0.045 & 0.085 & 0.122 & \multicolumn{2}{c}{0.084} & \multicolumn{2}{c}{0.192} & \multicolumn{2}{c}{0.091}\tabularnewline
\multirow{2}{*}{} & 200 & 200 & 0.056 & 0.088 & 0.118 & \multicolumn{2}{c}{0.084} & \multicolumn{2}{c}{0.178} & \multicolumn{2}{c}{0.089}\tabularnewline
 & 300 & 100 & 0.060 & 0.086 & 0.072 & \multicolumn{2}{c}{0.050} & \multicolumn{2}{c}{0.159} & \multicolumn{2}{c}{0.111}\tabularnewline
\hline 
\end{tabular}
\par\end{centering}
\noindent\begin{minipage}[t]{1\columnwidth}%
{\small{}Model S4. The notes of Table \ref{Table S1} apply.}%
\end{minipage}
\end{table}

\begin{table}[H]
\caption{\label{Table S6}Empirical small sample size of forecast instability
tests based on model S6}
\begin{centering}
\begin{tabular}{ccccccc||cc||cc||c}
\hline 
 &  &  & \multicolumn{2}{c}{GR (2009)} & $\mathrm{B}_{\mathrm{max},h}$ & \multicolumn{2}{c}{$\mathrm{Q}_{\mathrm{max},h}$} & \multicolumn{2}{c}{$\mathrm{MB}_{\mathrm{max},h}$} & \multicolumn{2}{c}{$\mathrm{MQ}_{\mathrm{max},h}$}\tabularnewline
 &  &  & $t^{\mathrm{stat}}$ (uncorrected) & $t^{\mathrm{stat,c}}$ (corrected) &  & \multicolumn{6}{c}{}\tabularnewline
\hline 
\multicolumn{3}{l}{$\alpha=0.05$ } & \multicolumn{2}{c}{} & \tabularnewline
 & $T_{m}$ & $T_{n}$ &  &  &  & \multicolumn{2}{c}{} & \multicolumn{2}{c}{} & \multicolumn{2}{c}{}\tabularnewline
$T=100$ & 25 & 75 & 0.171 & 0.005 & 0.146 & \multicolumn{2}{c}{0.081} & \multicolumn{2}{c}{0.237} & \multicolumn{2}{c}{0.040}\tabularnewline
\multirow{2}{*}{} & 50 & 50 & 0.163 & 0.005 & 0.142 & \multicolumn{2}{c}{0.076} & \multicolumn{2}{c}{0.134} & \multicolumn{2}{c}{0.020}\tabularnewline
 & 75 & 25 & 0.177 & 0.035 & 0.113 & \multicolumn{2}{c}{0.052} & \multicolumn{2}{c}{0.108} & \multicolumn{2}{c}{0.033}\tabularnewline
$T=200$ & 50 & 150 & 0.170 & 0.021 & 0.155 & \multicolumn{2}{c}{0.059} & \multicolumn{2}{c}{0.087} & \multicolumn{2}{c}{0.022}\tabularnewline
\multirow{2}{*}{} & 100 & 100 & 0.165 & 0.032 & 0.121 & \multicolumn{2}{c}{0.045} & \multicolumn{2}{c}{0.193} & \multicolumn{2}{c}{0.024}\tabularnewline
 & 150 & 50 & 0.182 & 0.042 & 0.092 & \multicolumn{2}{c}{0.044} & \multicolumn{2}{c}{0.127} & \multicolumn{2}{c}{0.018}\tabularnewline
$T=300$ & 75 & 225 & 0.172 & 0.028 & 0.129 & \multicolumn{2}{c}{0.044} & \multicolumn{2}{c}{0.216} & \multicolumn{2}{c}{0.025}\tabularnewline
\multirow{2}{*}{} & 150 & 150 & 0.164 & 0.046 & 0.148 & \multicolumn{2}{c}{0.040} & \multicolumn{2}{c}{0.251} & \multicolumn{2}{c}{0.021}\tabularnewline
 & 225 & 75 & 0.168 & 0.049 & 0.138 & \multicolumn{2}{c}{0.045} & \multicolumn{2}{c}{0.228} & \multicolumn{2}{c}{0.021}\tabularnewline
$T=400$ & 100 & 300 & 0.161 & 0.034 & 0.135 & \multicolumn{2}{c}{0.041} & \multicolumn{2}{c}{0.280} & \multicolumn{2}{c}{0.023}\tabularnewline
\multirow{2}{*}{} & 200 & 200 & 0.155 & 0.046 & 0.128 & \multicolumn{2}{c}{0.038} & \multicolumn{2}{c}{0.228} & \multicolumn{2}{c}{0.021}\tabularnewline
 & 300 & 100 & 0.185 & 0.061 & 0.126 & \multicolumn{2}{c}{0.036} & \multicolumn{2}{c}{0.184} & \multicolumn{2}{c}{0.018}\tabularnewline
 &  &  &  &  &  & \multicolumn{2}{c}{} & \multicolumn{2}{c}{} & \multicolumn{2}{c}{}\tabularnewline
\multicolumn{3}{l}{$\alpha=0.10$ } &  &  &  & \multicolumn{2}{c}{} & \multicolumn{2}{c}{} & \multicolumn{2}{c}{}\tabularnewline
 & $T_{m}$ & $T_{n}$ & \multicolumn{2}{c}{} &  & \multicolumn{2}{c}{} & \multicolumn{2}{c}{} & \multicolumn{2}{c}{}\tabularnewline
$T=100$ & 25 & 75 & 0.251 & 0.017 & 0.205 & \multicolumn{2}{c}{0.152} & \multicolumn{2}{c}{0.294} & \multicolumn{2}{c}{0.073}\tabularnewline
\multirow{2}{*}{} & 50 & 50 & 0.235 & 0.060 & 0.179 & \multicolumn{2}{c}{0.105} & \multicolumn{2}{c}{0.178} & \multicolumn{2}{c}{0.052}\tabularnewline
 & 75 & 25 & 0.252 & 0.063 & 0.139 & \multicolumn{2}{c}{0.105} & \multicolumn{2}{c}{0.189} & \multicolumn{2}{c}{0.047}\tabularnewline
$T=200$ & 50 & 150 & 0.249 & 0.041 & 0.219 & \multicolumn{2}{c}{0.127} & \multicolumn{2}{c}{0.332} & \multicolumn{2}{c}{0.063}\tabularnewline
\multirow{2}{*}{} & 100 & 100 & 0.226 & 0.060 & 0.179 & \multicolumn{2}{c}{0.106} & \multicolumn{2}{c}{0.254} & \multicolumn{2}{c}{0.049}\tabularnewline
 & 150 & 50 & 0.186 & 0.057 & 0.130 & \multicolumn{2}{c}{0.095} & \multicolumn{2}{c}{0.294} & \multicolumn{2}{c}{0.038}\tabularnewline
$T=300$ & 75 & 225 & 0.241 & 0.053 & 0.186 & \multicolumn{2}{c}{0.113} & \multicolumn{2}{c}{0.301} & \multicolumn{2}{c}{0.060}\tabularnewline
\multirow{2}{*}{} & 150 & 150 & 0.233 & 0.067 & 0.198 & \multicolumn{2}{c}{0.101} & \multicolumn{2}{c}{0.329} & \multicolumn{2}{c}{0.041}\tabularnewline
 & 225 & 75 & 0.240 & 0.092 & 0.188 & \multicolumn{2}{c}{0.093} & \multicolumn{2}{c}{0.288} & \multicolumn{2}{c}{0.037}\tabularnewline
$T=400$ & 100 & 300 & 0.227 & 0.061 & 0.222 & \multicolumn{2}{c}{0.098} & \multicolumn{2}{c}{0.359} & \multicolumn{2}{c}{0.056}\tabularnewline
\multirow{2}{*}{} & 200 & 200 & 0.229 & 0.091 & 0.194 & \multicolumn{2}{c}{0.090} & \multicolumn{2}{c}{0.302} & \multicolumn{2}{c}{0.046}\tabularnewline
 & 300 & 100 & 0.236 & 0.093 & 0.168 & \multicolumn{2}{c}{0.091} & \multicolumn{2}{c}{0.245} & \multicolumn{2}{c}{0.040}\tabularnewline
\hline 
\end{tabular}
\par\end{centering}
\noindent\begin{minipage}[t]{1\columnwidth}%
{\small{}Model S6. The notes of Table \ref{Table S1} apply.}%
\end{minipage}
\end{table}

\begin{table}[H]
\caption{\label{Table Size S1 Recursive}Empirical small sample size of forecast
instability tests based on model S1; Recursive scheme}
\begin{centering}
\begin{tabular}{ccccccc||cc||cc||c}
\hline 
 &  &  & \multicolumn{2}{c}{GR (2009)} & $\mathrm{B}_{\mathrm{max},h}$ & \multicolumn{2}{c}{$\mathrm{Q}_{\mathrm{max},h}$} & \multicolumn{2}{c}{$\mathrm{MB}_{\mathrm{max},h}$} & \multicolumn{2}{c}{$\mathrm{MQ}_{\mathrm{max},h}$}\tabularnewline
 &  &  & $t^{\mathrm{stat}}$ (uncorrected) & $t^{\mathrm{stat,c}}$ (corrected) &  & \multicolumn{6}{c}{}\tabularnewline
\hline 
\multicolumn{3}{l}{$\alpha=0.05$ } & \multicolumn{2}{c}{} & \tabularnewline
 & $T_{m}$ & $T_{n}$ &  &  &  & \multicolumn{2}{c}{} & \multicolumn{2}{c}{} & \multicolumn{2}{c}{}\tabularnewline
$T=100$ & 25 & 75 & 0.060 & 0.028 & 0.108 & \multicolumn{2}{c}{0.071} & \multicolumn{2}{c}{0.194} & \multicolumn{2}{c}{0.111}\tabularnewline
\multirow{2}{*}{} & 50 & 50 & 0.067 & 0.026 & 0.060 & \multicolumn{2}{c}{0.075} & \multicolumn{2}{c}{0.120} & \multicolumn{2}{c}{0.067}\tabularnewline
 & 75 & 25 & 0.079 & 0.021 & 0.109 & \multicolumn{2}{c}{0.060} & \multicolumn{2}{c}{0.159} & \multicolumn{2}{c}{0.069}\tabularnewline
$T=200$ & 50 & 150 & 0.053 & 0.033 & 0.108 & \multicolumn{2}{c}{0.071} & \multicolumn{2}{c}{0.182} & \multicolumn{2}{c}{0.096}\tabularnewline
\multirow{2}{*}{} & 100 & 100 & 0.052 & 0.028 & 0.089 & \multicolumn{2}{c}{0.064} & \multicolumn{2}{c}{0.146} & \multicolumn{2}{c}{0.076}\tabularnewline
 & 150 & 50 & 0.055 & 0.022 & 0.079 & \multicolumn{2}{c}{0.054} & \multicolumn{2}{c}{0.118} & \multicolumn{2}{c}{0.055}\tabularnewline
$T=300$ & 75 & 225 & 0.063 & 0.027 & 0.084 & \multicolumn{2}{c}{0.042} & \multicolumn{2}{c}{0.197} & \multicolumn{2}{c}{0.091}\tabularnewline
\multirow{2}{*}{} & 150 & 150 & 0.049 & 0.033 & 0.028 & \multicolumn{2}{c}{0.096} & \multicolumn{2}{c}{0.178} & \multicolumn{2}{c}{0.089}\tabularnewline
 & 225 & 75 & 0.063 & 0.026 & 0.108 & \multicolumn{2}{c}{0.072} & \multicolumn{2}{c}{0.197} & \multicolumn{2}{c}{0.091}\tabularnewline
$T=400$ & 100 & 300 & 0.059 & 0.029 & 0.078 & \multicolumn{2}{c}{0.050} & \multicolumn{2}{c}{0.156} & \multicolumn{2}{c}{0.080}\tabularnewline
\multirow{2}{*}{} & 200 & 200 & 0.055 & 0.048 & 0.064 & \multicolumn{2}{c}{0.048} & \multicolumn{2}{c}{0.143} & \multicolumn{2}{c}{0.065}\tabularnewline
 & 300 & 100 & 0.059 & 0.025 & 0.135 & \multicolumn{2}{c}{0.037} & \multicolumn{2}{c}{0.24} & \multicolumn{2}{c}{0.112}\tabularnewline
\hline 
\end{tabular}
\par\end{centering}
\noindent\begin{minipage}[t]{1\columnwidth}%
{\small{}The table reports the rejection probabilities of $5\%$-level
tests proposed in the paper and those proposed by \citet{giacomini/rossi:09}
{[}(abbreviated GR (2009){]} for model S1. For all methods we use
the recursive forecasting scheme. $T=T_{m}+T_{n}$, where $T$ is
the total sample size, $T_{m}$ is the size of the in-sample window
and $T_{n}$ is the size of the out-of-sample window. $m_{T}$ is
set equal to the smallest integer allowed by Condition \ref{Cond The-auxiliary-sequence}.
Based on 5,000 replications.}%
\end{minipage}
\end{table}

\begin{table}[H]
\caption{\label{Table Size S1 Rolling}Empirical small sample size of forecast
instability tests based on model S1; Rolling scheme}
\begin{centering}
\begin{tabular}{ccccccc||cc||cc||c}
\hline 
 &  &  & \multicolumn{2}{c}{GR (2009)} & $\mathrm{B}_{\mathrm{max},h}$ & \multicolumn{2}{c}{$\mathrm{Q}_{\mathrm{max},h}$} & \multicolumn{2}{c}{$\mathrm{MB}_{\mathrm{max},h}$} & \multicolumn{2}{c}{$\mathrm{MQ}_{\mathrm{max},h}$}\tabularnewline
 &  &  & $t^{\mathrm{stat}}$ (uncorrected) & $t^{\mathrm{stat,c}}$ (corrected) &  & \multicolumn{6}{c}{}\tabularnewline
\hline 
\multicolumn{3}{l}{$\alpha=0.05$ } & \multicolumn{2}{c}{} & \tabularnewline
 & $T_{m}$ & $T_{n}$ &  &  &  & \multicolumn{2}{c}{} & \multicolumn{2}{c}{} & \multicolumn{2}{c}{}\tabularnewline
$T=100$ & 25 & 75 & 0.514 & 0.018 & 0.126 & \multicolumn{2}{c}{0.128} & \multicolumn{2}{c}{0.185} & \multicolumn{2}{c}{0.128}\tabularnewline
\multirow{2}{*}{} & 50 & 50 & 0.064 & 0.008 & 0.090 & \multicolumn{2}{c}{0.089} & \multicolumn{2}{c}{0.133} & \multicolumn{2}{c}{0.075}\tabularnewline
 & 75 & 25 & 0.072 & 0.019 & 0.131 & \multicolumn{2}{c}{0.104} & \multicolumn{2}{c}{0.165} & \multicolumn{2}{c}{0.064}\tabularnewline
$T=200$ & 50 & 150 & 0.364 & 0.038 & 0.094 & \multicolumn{2}{c}{0.078} & \multicolumn{2}{c}{0.177} & \multicolumn{2}{c}{0.063}\tabularnewline
\multirow{2}{*}{} & 100 & 100 & 0.058 & 0.017 & 0.082 & \multicolumn{2}{c}{0.073} & \multicolumn{2}{c}{0.132} & \multicolumn{2}{c}{0.064}\tabularnewline
 & 150 & 50 & 0.078 & 0.023 & 0.087 & \multicolumn{2}{c}{0.071} & \multicolumn{2}{c}{0.112} & \multicolumn{2}{c}{0.038}\tabularnewline
$T=300$ & 75 & 225 & 0.230 & 0.010 & 0.074 & \multicolumn{2}{c}{0.055} & \multicolumn{2}{c}{0.131} & \multicolumn{2}{c}{0.095}\tabularnewline
\multirow{2}{*}{} & 150 & 150 & 0.054 & 0.036 & 0.082 & \multicolumn{2}{c}{0.061} & \multicolumn{2}{c}{0.180} & \multicolumn{2}{c}{0.084}\tabularnewline
 & 225 & 75 & 0.065 & 0.026 & 0.118 & \multicolumn{2}{c}{0.072} & \multicolumn{2}{c}{0.193} & \multicolumn{2}{c}{0.069}\tabularnewline
$T=400$ & 100 & 300 & 0.168 & 0.020 & 0.084 & \multicolumn{2}{c}{0.060} & \multicolumn{2}{c}{0.099} & \multicolumn{2}{c}{0.163}\tabularnewline
\multirow{2}{*}{} & 200 & 200 & 0.057 & 0.036 & 0.088 & \multicolumn{2}{c}{0.053} & \multicolumn{2}{c}{0.148} & \multicolumn{2}{c}{0.064}\tabularnewline
 & 300 & 100 & 0.060 & 0.023 & 0.088 & \multicolumn{2}{c}{0.083} & \multicolumn{2}{c}{0.059} & \multicolumn{2}{c}{0.057}\tabularnewline
\hline 
\end{tabular}
\par\end{centering}
\noindent\begin{minipage}[t]{1\columnwidth}%
{\small{}Model S1; rolling scheme. The notes of Table \ref{Table Size S1 Recursive}
apply.}%
\end{minipage}
\end{table}

\begin{table}[H]
\caption{\label{Table P1 Recursive}Empirical small sample power of forecast
instability tests based on model P1; Recursive scheme}
\begin{centering}
\begin{tabular}{ccccccc||cc||cc||c}
\hline 
 &  &  & \multicolumn{2}{c}{GR (2009)} & $\mathrm{B}_{\mathrm{max},h}$ & \multicolumn{2}{c}{$\mathrm{Q}_{\mathrm{max},h}$} & \multicolumn{2}{c}{$\mathrm{MB}_{\mathrm{max},h}$} & \multicolumn{2}{c}{$\mathrm{MQ}_{\mathrm{max},h}$}\tabularnewline
 &  &  & $t^{\mathrm{stat}}$ (uncorrected) & $t^{\mathrm{stat,c}}$ (corrected) &  & \multicolumn{6}{c}{}\tabularnewline
\hline 
\multicolumn{3}{l}{$\delta=0.5$ } & \multicolumn{2}{c}{} & \tabularnewline
 & $T_{m}$ & $T_{n}$ &  &  &  & \multicolumn{2}{c}{} & \multicolumn{2}{c}{} & \multicolumn{2}{c}{}\tabularnewline
$T=100$ & 50 & 50 & 0.071 & 0.056 & 0.083 & \multicolumn{2}{c}{0.063} & \multicolumn{2}{c}{0.103} & \multicolumn{2}{c}{0.064}\tabularnewline
$T=200$ & 100 & 100 & 0.080 & 0.094 & 0.101 & \multicolumn{2}{c}{0.069} & \multicolumn{2}{c}{0.161} & \multicolumn{2}{c}{0.099}\tabularnewline
$T=300$ & 150 & 150 & 0.098 & 0.127 & 0.107 & \multicolumn{2}{c}{0.086} & \multicolumn{2}{c}{0.208} & \multicolumn{2}{c}{0.136}\tabularnewline
$T=400$ & 200 & 200 & 0.095 & 0.125 & 0.093 & \multicolumn{2}{c}{0.062} & \multicolumn{2}{c}{0.188} & \multicolumn{2}{c}{0.106}\tabularnewline
 &  &  &  &  &  & \multicolumn{2}{c}{} & \multicolumn{2}{c}{} & \multicolumn{2}{c}{}\tabularnewline
\multicolumn{3}{l}{$\delta=1$ } &  &  &  & \multicolumn{2}{c}{} & \multicolumn{2}{c}{} & \multicolumn{2}{c}{}\tabularnewline
 & $T_{m}$ & $T_{n}$ & \multicolumn{2}{c}{} &  & \multicolumn{2}{c}{} & \multicolumn{2}{c}{} & \multicolumn{2}{c}{}\tabularnewline
$T=100$ & 50 & 50 & 0.244 & 0.253 & 0.206 & \multicolumn{2}{c}{0.151} & \multicolumn{2}{c}{0.167} & \multicolumn{2}{c}{0.117}\tabularnewline
$T=200$ & 100 & 100 & 0.445 & 0.401 & 0.260 & \multicolumn{2}{c}{0.223} & \multicolumn{2}{c}{0.403} & \multicolumn{2}{c}{0.298}\tabularnewline
$T=300$ & 150 & 150 & 0.615 & 0.679 & 0.218 & \multicolumn{2}{c}{0.249} & \multicolumn{2}{c}{0.508} & \multicolumn{2}{c}{0.400}\tabularnewline
$T=400$ & 200 & 200 & 0.745 & 0.814 & 0.240 & \multicolumn{2}{c}{0.268} & \multicolumn{2}{c}{0.588} & \multicolumn{2}{c}{0.432}\tabularnewline
 &  &  &  &  &  & \multicolumn{2}{c}{} & \multicolumn{2}{c}{} & \multicolumn{2}{c}{}\tabularnewline
\multicolumn{3}{l}{$\delta=1.5$ } &  &  &  & \multicolumn{2}{c}{} & \multicolumn{2}{c}{} & \multicolumn{2}{c}{}\tabularnewline
 & $T_{m}$ & $T_{n}$ &  &  &  & \multicolumn{2}{c}{} & \multicolumn{2}{c}{} & \multicolumn{2}{c}{}\tabularnewline
$T=100$ & 50 & 50 & 0.706 & 0.720 & 0.483 & \multicolumn{2}{c}{0.350} & \multicolumn{2}{c}{0.308} & \multicolumn{2}{c}{0.208}\tabularnewline
$T=200$ & 100 & 100 & 0.946 & 0.958 & 0.660 & \multicolumn{2}{c}{0.603} & \multicolumn{2}{c}{0.821} & \multicolumn{2}{c}{0.701}\tabularnewline
$T=300$ & 150 & 150 & 0.995 & 0.997 & 0.495 & \multicolumn{2}{c}{0.666} & \multicolumn{2}{c}{0.909} & \multicolumn{2}{c}{0.833}\tabularnewline
$T=400$ & 200 & 200 & 1 & 1 & 0.606 & \multicolumn{2}{c}{0.782} & \multicolumn{2}{c}{0.974} & \multicolumn{2}{c}{0.921}\tabularnewline
 &  &  &  &  &  & \multicolumn{2}{c}{} & \multicolumn{2}{c}{} & \multicolumn{2}{c}{}\tabularnewline
\multicolumn{3}{l}{$\delta=2$ } &  &  &  & \multicolumn{2}{c}{} & \multicolumn{2}{c}{} & \multicolumn{2}{c}{}\tabularnewline
 & $T_{m}$ & $T_{n}$ &  &  &  & \multicolumn{2}{c}{} & \multicolumn{2}{c}{} & \multicolumn{2}{c}{}\tabularnewline
$T=100$ & 50 & 50 & 0.968 & 0.970 & 0.780 & \multicolumn{2}{c}{0.566} & \multicolumn{2}{c}{0.504} & \multicolumn{2}{c}{0.301}\tabularnewline
$T=200$ & 100 & 100 & 1 & 1 & 0.943 & \multicolumn{2}{c}{0.906} & \multicolumn{2}{c}{0.987} & \multicolumn{2}{c}{0.940}\tabularnewline
$T=300$ & 150 & 150 & 1 & 1 & 0.846 & \multicolumn{2}{c}{0.948} & \multicolumn{2}{c}{0.997} & \multicolumn{2}{c}{0.989}\tabularnewline
$T=400$ & 200 & 200 & 1 & 1 & 0.924 & \multicolumn{2}{c}{0.990} & \multicolumn{2}{c}{1} & \multicolumn{2}{c}{1}\tabularnewline
\hline 
\end{tabular}
\par\end{centering}
\noindent\begin{minipage}[t]{1\columnwidth}%
{\small{}The table reports the rejection probabilities of $5\%$-level
tests proposed in the paper and those proposed by \citet{giacomini/rossi:09}
{[}(abbreviated GR (2009){]} for model P1 with short-term instability.
For all methods we use the rolling forecasting scheme. $T=T_{m}+T_{n}$,
where $T$ is the total sample size, $T_{m}$ is the size of the in-sample
window and $T_{n}$ is the size of the out-of-sample window. $\lambda_{0}=0.6$
and $m_{T}$ is set equal to the smallest integer allowed by Condition
\ref{Cond The-auxiliary-sequence}. Based on 5,000 replications.}%
\end{minipage}
\end{table}

\begin{table}[H]
\caption{\label{Table P1 Recursive ST}Empirical small sample power of forecast
instability tests based on model P1; Recursive scheme; Short-term
instability}
\begin{centering}
\begin{tabular}{ccccccc||cc||cc||c}
\hline 
 &  &  & \multicolumn{2}{c}{GR (2009)} & $\mathrm{B}_{\mathrm{max},h}$ & \multicolumn{2}{c}{$\mathrm{Q}_{\mathrm{max},h}$} & \multicolumn{2}{c}{$\mathrm{MB}_{\mathrm{max},h}$} & \multicolumn{2}{c}{$\mathrm{MQ}_{\mathrm{max},h}$}\tabularnewline
 &  &  & $t^{\mathrm{stat}}$ (uncorrected) & $t^{\mathrm{stat,c}}$ (corrected) &  & \multicolumn{6}{c}{}\tabularnewline
\hline 
\multicolumn{3}{l}{$\delta=0.5$ } & \multicolumn{2}{c}{} & \tabularnewline
 & $T_{m}$ & $T_{n}$ &  &  &  & \multicolumn{2}{c}{} & \multicolumn{2}{c}{} & \multicolumn{2}{c}{}\tabularnewline
$T=100$ & 50 & 50 & 0.060 & 0.031 & 0.064 & \multicolumn{2}{c}{0.053} & \multicolumn{2}{c}{0.071} & \multicolumn{2}{c}{0.039}\tabularnewline
$T=200$ & 100 & 100 & 0.053 & 0.044 & 0.091 & \multicolumn{2}{c}{0.052} & \multicolumn{2}{c}{0.133} & \multicolumn{2}{c}{0.071}\tabularnewline
$T=300$ & 150 & 150 & 0.058 & 0.053 & 0.093 & \multicolumn{2}{c}{0.051} & \multicolumn{2}{c}{0.184} & \multicolumn{2}{c}{0.106}\tabularnewline
$T=400$ & 200 & 200 & 0.049 & 0.050 & 0.075 & \multicolumn{2}{c}{0.049} & \multicolumn{2}{c}{0.154} & \multicolumn{2}{c}{0.097}\tabularnewline
 &  &  &  &  &  & \multicolumn{2}{c}{} & \multicolumn{2}{c}{} & \multicolumn{2}{c}{}\tabularnewline
\multicolumn{3}{l}{$\delta=1$ } &  &  &  & \multicolumn{2}{c}{} & \multicolumn{2}{c}{} & \multicolumn{2}{c}{}\tabularnewline
 & $T_{m}$ & $T_{n}$ & \multicolumn{2}{c}{} &  & \multicolumn{2}{c}{} & \multicolumn{2}{c}{} & \multicolumn{2}{c}{}\tabularnewline
$T=100$ & 50 & 50 & 0.058 & 0.049 & 0.097 & \multicolumn{2}{c}{0.059} & \multicolumn{2}{c}{0.057} & \multicolumn{2}{c}{0.0294}\tabularnewline
$T=200$ & 100 & 100 & 0.073 & 0.088 & 0.244 & \multicolumn{2}{c}{0.155} & \multicolumn{2}{c}{0.269} & \multicolumn{2}{c}{0.179}\tabularnewline
$T=300$ & 150 & 150 & 0.088 & 0.136 & 0.157 & \multicolumn{2}{c}{0.132} & \multicolumn{2}{c}{0.455} & \multicolumn{2}{c}{0.348}\tabularnewline
$T=400$ & 200 & 200 & 0.100 & 0.128 & 0.155 & \multicolumn{2}{c}{0.117} & \multicolumn{2}{c}{0.486} & \multicolumn{2}{c}{0.353}\tabularnewline
 &  &  &  &  &  & \multicolumn{2}{c}{} & \multicolumn{2}{c}{} & \multicolumn{2}{c}{}\tabularnewline
\multicolumn{3}{l}{$\delta=1.5$ } &  &  &  & \multicolumn{2}{c}{} & \multicolumn{2}{c}{} & \multicolumn{2}{c}{}\tabularnewline
 & $T_{m}$ & $T_{n}$ &  &  &  & \multicolumn{2}{c}{} & \multicolumn{2}{c}{} & \multicolumn{2}{c}{}\tabularnewline
$T=100$ & 50 & 50 & 0.086 & 0.090 & 0.193 & \multicolumn{2}{c}{0.112} & \multicolumn{2}{c}{0.072} & \multicolumn{2}{c}{0.040}\tabularnewline
$T=200$ & 100 & 100 & 0.156 & 0.203 & 0.629 & \multicolumn{2}{c}{0.487} & \multicolumn{2}{c}{0.641} & \multicolumn{2}{c}{0.493}\tabularnewline
$T=300$ & 150 & 150 & 0.254 & 0.351 & 0.398 & \multicolumn{2}{c}{0.371} & \multicolumn{2}{c}{0.865} & \multicolumn{2}{c}{0.758}\tabularnewline
$T=400$ & 200 & 200 & 0.329 & 0.415 & 0.432 & \multicolumn{2}{c}{0.434} & \multicolumn{2}{c}{0.942} & \multicolumn{2}{c}{0.870}\tabularnewline
 &  &  &  &  &  & \multicolumn{2}{c}{} & \multicolumn{2}{c}{} & \multicolumn{2}{c}{}\tabularnewline
\multicolumn{3}{l}{$\delta=2$ } &  &  &  & \multicolumn{2}{c}{} & \multicolumn{2}{c}{} & \multicolumn{2}{c}{}\tabularnewline
 & $T_{m}$ & $T_{n}$ &  &  &  & \multicolumn{2}{c}{} & \multicolumn{2}{c}{} & \multicolumn{2}{c}{}\tabularnewline
$T=100$ & 50 & 50 & 0.130 & 0.151 & 0.347 & \multicolumn{2}{c}{0.181} & \multicolumn{2}{c}{0.107} & \multicolumn{2}{c}{0.052}\tabularnewline
$T=200$ & 100 & 100 & 0.325 & 0.415 & 0.930 & \multicolumn{2}{c}{0.842} & \multicolumn{2}{c}{0.928} & \multicolumn{2}{c}{0.839}\tabularnewline
$T=300$ & 150 & 150 & 0.533 & 0.652 & 0.736 & \multicolumn{2}{c}{0.692} & \multicolumn{2}{c}{0.993} & \multicolumn{2}{c}{0.97}\tabularnewline
$T=400$ & 200 & 200 & 0.724 & 0.819 & 0.809 & \multicolumn{2}{c}{0.805} & \multicolumn{2}{c}{1} & \multicolumn{2}{c}{1}\tabularnewline
\hline 
\end{tabular}
\par\end{centering}
\noindent\begin{minipage}[t]{1\columnwidth}%
{\small{}Model P1; recursive scheme. The notes of Table \ref{Table P1 Recursive}
apply.}%
\end{minipage}
\end{table}

\begin{table}[H]
\caption{\label{Table Size P1 Rolling}Empirical small sample power of forecast
instability tests based on model P1; Rolling scheme}
\begin{centering}
\begin{tabular}{ccccccc||cc||cc||c}
\hline 
 &  &  & \multicolumn{2}{c}{GR (2009)} & $\mathrm{B}_{\mathrm{max},h}$ & \multicolumn{2}{c}{$\mathrm{Q}_{\mathrm{max},h}$} & \multicolumn{2}{c}{$\mathrm{MB}_{\mathrm{max},h}$} & \multicolumn{2}{c}{$\mathrm{MQ}_{\mathrm{max},h}$}\tabularnewline
 &  &  & $t^{\mathrm{stat}}$ (uncorrected) & $t^{\mathrm{stat,c}}$ (corrected) &  & \multicolumn{6}{c}{}\tabularnewline
\hline 
\multicolumn{3}{l}{$\delta=0.5$ } & \multicolumn{2}{c}{} & \tabularnewline
 & $T_{m}$ & $T_{n}$ &  &  &  & \multicolumn{2}{c}{} & \multicolumn{2}{c}{} & \multicolumn{2}{c}{}\tabularnewline
$T=100$ & 50 & 50 & 0.158 & 0.059 & 0.429 & \multicolumn{2}{c}{0.063} & \multicolumn{2}{c}{0.168} & \multicolumn{2}{c}{0.097}\tabularnewline
$T=200$ & 100 & 100 & 0.144 & 0.089 & 0.643 & \multicolumn{2}{c}{0.079} & \multicolumn{2}{c}{0.201} & \multicolumn{2}{c}{0.112}\tabularnewline
$T=300$ & 150 & 150 & 0.098 & 0.104 & 0.777 & \multicolumn{2}{c}{0.093} & \multicolumn{2}{c}{0.246} & \multicolumn{2}{c}{0.136}\tabularnewline
$T=400$ & 200 & 200 & 0.052 & 0.033 & 0.809 & \multicolumn{2}{c}{0.073} & \multicolumn{2}{c}{0.146} & \multicolumn{2}{c}{0.068}\tabularnewline
 &  &  &  &  &  & \multicolumn{2}{c}{} & \multicolumn{2}{c}{} & \multicolumn{2}{c}{}\tabularnewline
\multicolumn{3}{l}{$\delta=1$ } &  &  &  & \multicolumn{2}{c}{} & \multicolumn{2}{c}{} & \multicolumn{2}{c}{}\tabularnewline
 & $T_{m}$ & $T_{n}$ & \multicolumn{2}{c}{} &  & \multicolumn{2}{c}{} & \multicolumn{2}{c}{} & \multicolumn{2}{c}{}\tabularnewline
$T=100$ & 50 & 50 & 0.342 & 0.165 & 0.485 & \multicolumn{2}{c}{0.107} & \multicolumn{2}{c}{0.362} & \multicolumn{2}{c}{0.259}\tabularnewline
$T=200$ & 100 & 100 & 0.439 & 0.329 & 0.717 & \multicolumn{2}{c}{0.218} & \multicolumn{2}{c}{0.456} & \multicolumn{2}{c}{0.324}\tabularnewline
$T=300$ & 150 & 150 & 0.629 & 0.571 & 0.812 & \multicolumn{2}{c}{0027} & \multicolumn{2}{c}{0.662} & \multicolumn{2}{c}{0.541}\tabularnewline
$T=400$ & 200 & 200 & 0.878 & 0.876 & 0.855 & \multicolumn{2}{c}{0.299} & \multicolumn{2}{c}{0.827} & \multicolumn{2}{c}{0.700}\tabularnewline
 &  &  &  &  &  & \multicolumn{2}{c}{} & \multicolumn{2}{c}{} & \multicolumn{2}{c}{}\tabularnewline
\multicolumn{3}{l}{$\delta=1.5$ } &  &  &  & \multicolumn{2}{c}{} & \multicolumn{2}{c}{} & \multicolumn{2}{c}{}\tabularnewline
 & $T_{m}$ & $T_{n}$ &  &  &  & \multicolumn{2}{c}{} & \multicolumn{2}{c}{} & \multicolumn{2}{c}{}\tabularnewline
$T=100$ & 50 & 50 & 0.686 & 0.448 & 0.623 & \multicolumn{2}{c}{0.236} & \multicolumn{2}{c}{0.662} & \multicolumn{2}{c}{0.541}\tabularnewline
$T=200$ & 100 & 100 & 0.879 & 0.801 & 0.882 & \multicolumn{2}{c}{0.620} & \multicolumn{2}{c}{0.827} & \multicolumn{2}{c}{0.700}\tabularnewline
$T=300$ & 150 & 150 & 0.968 & 0.974 & 0.902 & \multicolumn{2}{c}{0.685} & \multicolumn{2}{c}{0.895} & \multicolumn{2}{c}{0.778}\tabularnewline
$T=400$ & 200 & 200 & 0.991 & 0.997 & 0.932 & \multicolumn{2}{c}{0.801} & \multicolumn{2}{c}{0.930} & \multicolumn{2}{c}{0.843}\tabularnewline
 &  &  &  &  &  & \multicolumn{2}{c}{} & \multicolumn{2}{c}{} & \multicolumn{2}{c}{}\tabularnewline
\multicolumn{3}{l}{$\delta=2$ } &  &  &  & \multicolumn{2}{c}{} & \multicolumn{2}{c}{} & \multicolumn{2}{c}{}\tabularnewline
 & $T_{m}$ & $T_{n}$ &  &  &  & \multicolumn{2}{c}{} & \multicolumn{2}{c}{} & \multicolumn{2}{c}{}\tabularnewline
$T=100$ & 50 & 50 & 0.929 & 0.800 & 0.768 & \multicolumn{2}{c}{0.398} & \multicolumn{2}{c}{0.761} & \multicolumn{2}{c}{0.398}\tabularnewline
$T=200$ & 100 & 100 & 0.997 & 0.993 & 0.977 & \multicolumn{2}{c}{0.924} & \multicolumn{2}{c}{0.977} & \multicolumn{2}{c}{0.925}\tabularnewline
$T=300$ & 150 & 150 & 1 & 0.999 & 0.996 & \multicolumn{2}{c}{0.958} & \multicolumn{2}{c}{0.969} & \multicolumn{2}{c}{0.9580}\tabularnewline
$T=400$ & 200 & 200 & 1 & 1 & 0.980 & \multicolumn{2}{c}{0.987} & \multicolumn{2}{c}{0.980} & \multicolumn{2}{c}{0.988}\tabularnewline
\hline 
\end{tabular}
\par\end{centering}
\noindent\begin{minipage}[t]{1\columnwidth}%
{\small{}The table reports the rejection probabilities of $95\%$-level
tests proposed in the paper and those proposed by \citet{giacomini/rossi:09}
{[}(abbreviated GR (2009){]} for model P1. For all methods we use
the rolling forecasting scheme. $T=T_{m}+T_{n}$, where $T$ is the
total sample size, $T_{m}$ is the size of the in-sample window and
$T_{n}$ is the size of the out-of-sample window. }$\lambda_{0}=0.6${\small{}
and $m_{T}$ is set equal to the smallest integer allowed by Condition
\ref{Cond The-auxiliary-sequence}. Based on 5,000 replications.}%
\end{minipage}
\end{table}

\begin{table}[H]
\caption{\label{Table Size P1 Rolling ST}Empirical small sample power of forecast
instability tests based on model P1; Rolling scheme; Short-term instability}
\begin{centering}
\begin{tabular}{ccccccc||cc||cc||c}
\hline 
 &  &  & \multicolumn{2}{c}{GR (2009)} & $\mathrm{B}_{\mathrm{max},h}$ & \multicolumn{2}{c}{$\mathrm{Q}_{\mathrm{max},h}$} & \multicolumn{2}{c}{$\mathrm{MB}_{\mathrm{max},h}$} & \multicolumn{2}{c}{$\mathrm{MQ}_{\mathrm{max},h}$}\tabularnewline
 &  &  & $t^{\mathrm{stat}}$ (uncorrected) & $t^{\mathrm{stat,c}}$ (corrected) &  & \multicolumn{6}{c}{}\tabularnewline
\hline 
\multicolumn{3}{l}{$\delta=0.5$ } & \multicolumn{2}{c}{} & \tabularnewline
 & $T_{m}$ & $T_{n}$ &  &  &  & \multicolumn{2}{c}{} & \multicolumn{2}{c}{} & \multicolumn{2}{c}{}\tabularnewline
$T=100$ & 50 & 50 & 0.159 & 0.059 & 0.429 & \multicolumn{2}{c}{0.063} & \multicolumn{2}{c}{0.208} & \multicolumn{2}{c}{0.102}\tabularnewline
$T=200$ & 100 & 100 & 0.114 & 0.062 & 0.631 & \multicolumn{2}{c}{0.067} & \multicolumn{2}{c}{0.132} & \multicolumn{2}{c}{0.072}\tabularnewline
$T=300$ & 150 & 150 & 0.106 & 0.070 & 0.778 & \multicolumn{2}{c}{0.097} & \multicolumn{2}{c}{0.182} & \multicolumn{2}{c}{0.095}\tabularnewline
$T=400$ & 200 & 200 & 0.055 & 0.048 &  & \multicolumn{2}{c}{} & \multicolumn{2}{c}{0.141} & \multicolumn{2}{c}{0.065}\tabularnewline
 &  &  &  &  &  & \multicolumn{2}{c}{} & \multicolumn{2}{c}{} & \multicolumn{2}{c}{}\tabularnewline
\multicolumn{3}{l}{$\delta=1$ } &  &  &  & \multicolumn{2}{c}{} & \multicolumn{2}{c}{} & \multicolumn{2}{c}{}\tabularnewline
 & $T_{m}$ & $T_{n}$ & \multicolumn{2}{c}{} &  & \multicolumn{2}{c}{} & \multicolumn{2}{c}{} & \multicolumn{2}{c}{}\tabularnewline
$T=100$ & 50 & 50 & 0.169 & 0.071 & 0.431 & \multicolumn{2}{c}{0.064} & \multicolumn{2}{c}{0.267} & \multicolumn{2}{c}{0.151}\tabularnewline
$T=200$ & 100 & 100 & 0.170 & 0.112 & 0.702 & \multicolumn{2}{c}{0.193} & \multicolumn{2}{c}{0.261} & \multicolumn{2}{c}{0.149}\tabularnewline
$T=300$ & 150 & 150 & 0.186 & 0.153 & 0.789 & \multicolumn{2}{c}{0.186} & \multicolumn{2}{c}{0.456} & \multicolumn{2}{c}{0.304}\tabularnewline
$T=400$ & 200 & 200 & 0.198 & 0.173 & 0.816 & \multicolumn{2}{c}{0.184} & \multicolumn{2}{c}{0.459} & \multicolumn{2}{c}{0.294}\tabularnewline
 &  &  &  &  &  & \multicolumn{2}{c}{} & \multicolumn{2}{c}{} & \multicolumn{2}{c}{}\tabularnewline
\multicolumn{3}{l}{$\delta=1.5$ } &  &  &  & \multicolumn{2}{c}{} & \multicolumn{2}{c}{} & \multicolumn{2}{c}{}\tabularnewline
 & $T_{m}$ & $T_{n}$ &  &  &  & \multicolumn{2}{c}{} & \multicolumn{2}{c}{} & \multicolumn{2}{c}{}\tabularnewline
$T=100$ & 50 & 50 & 0.229 & 0.117 & 0.469 & \multicolumn{2}{c}{0.093} & \multicolumn{2}{c}{0.462} & \multicolumn{2}{c}{0.320}\tabularnewline
$T=200$ & 100 & 100 & 0.297 & 0.231 & 0.875 & \multicolumn{2}{c}{0.601} & \multicolumn{2}{c}{0.621} & \multicolumn{2}{c}{0.408}\tabularnewline
$T=300$ & 150 & 150 & 0.402 & 0.355 & 0.849 & \multicolumn{2}{c}{0.521} & \multicolumn{2}{c}{0.858} & \multicolumn{2}{c}{0.727}\tabularnewline
$T=400$ & 200 & 200 & 0.454 & 0.439 & 0.885 & \multicolumn{2}{c}{0.625} & \multicolumn{2}{c}{0.919} & \multicolumn{2}{c}{0.781}\tabularnewline
 &  &  &  &  &  & \multicolumn{2}{c}{} & \multicolumn{2}{c}{} & \multicolumn{2}{c}{}\tabularnewline
\multicolumn{3}{l}{$\delta=2$ } &  &  &  & \multicolumn{2}{c}{} & \multicolumn{2}{c}{} & \multicolumn{2}{c}{}\tabularnewline
 & $T_{m}$ & $T_{n}$ &  &  &  & \multicolumn{2}{c}{} & \multicolumn{2}{c}{} & \multicolumn{2}{c}{}\tabularnewline
$T=100$ & 50 & 50 & 0.328 & 0.1842 & 0.518 & \multicolumn{2}{c}{0.149} & \multicolumn{2}{c}{0.854} & \multicolumn{2}{c}{0.748}\tabularnewline
$T=200$ & 100 & 100 & 0.508 & 0.428 & 0.977 & \multicolumn{2}{c}{0.925} & \multicolumn{2}{c}{0.965} & \multicolumn{2}{c}{0.904}\tabularnewline
$T=300$ & 150 & 150 & 0.687 & 0.645 & 0.940 & \multicolumn{2}{c}{0.880} & \multicolumn{2}{c}{0.996} & \multicolumn{2}{c}{0.976}\tabularnewline
$T=400$ & 200 & 200 & 0.991 & 0.994 & 0.966 & \multicolumn{2}{c}{0.952} & \multicolumn{2}{c}{0.999} & \multicolumn{2}{c}{0.990}\tabularnewline
\hline 
\end{tabular}
\par\end{centering}
\noindent\begin{minipage}[t]{1\columnwidth}%
{\small{}Model P1; rolling scheme. The notes of Table \ref{Table Size P1 Rolling}
apply.}%
\end{minipage}
\end{table}

\begin{table}[H]
\caption{\label{Table Size S1 Linex}Empirical small sample size of forecast
instability tests based on model S1; Linex Loss}
\begin{centering}
\begin{tabular}{ccccccc||cc||cc||c}
\hline 
 &  &  & \multicolumn{2}{c}{GR (2009)} & $\mathrm{B}_{\mathrm{max},h}$ & \multicolumn{2}{c}{$\mathrm{Q}_{\mathrm{max},h}$} & \multicolumn{2}{c}{$\mathrm{MB}_{\mathrm{max},h}$} & \multicolumn{2}{c}{$\mathrm{MQ}_{\mathrm{max},h}$}\tabularnewline
 &  &  & $t^{\mathrm{stat}}$ (uncorrected) & $t^{\mathrm{stat,c}}$ (corrected) &  & \multicolumn{6}{c}{}\tabularnewline
\hline 
\multicolumn{3}{l}{$\alpha=0.05$ } & \multicolumn{2}{c}{} & \tabularnewline
 & $T_{m}$ & $T_{n}$ &  &  &  & \multicolumn{2}{c}{} & \multicolumn{2}{c}{} & \multicolumn{2}{c}{}\tabularnewline
$T=100$ & 25 & 75 & 0.051 & 0.010 & 0.263 & \multicolumn{2}{c}{0.166} & \multicolumn{2}{c}{0.266} & \multicolumn{2}{c}{0.170}\tabularnewline
\multirow{2}{*}{} & 50 & 50 & 0.083 & 0.015 & 0.119 & \multicolumn{2}{c}{0.069} & \multicolumn{2}{c}{0.152} & \multicolumn{2}{c}{0.103}\tabularnewline
 & 75 & 25 & 0.116 & 0.011 & 0.143 & \multicolumn{2}{c}{0.098} & \multicolumn{2}{c}{0.164} & \multicolumn{2}{c}{0.128}\tabularnewline
$T=200$ & 50 & 150 & 0.058 & 0.013 & 0.185 & \multicolumn{2}{c}{0.091} & \multicolumn{2}{c}{0.309} & \multicolumn{2}{c}{0.158}\tabularnewline
\multirow{2}{*}{} & 100 & 100 & 0.074 & 0.016 & 0.165 & \multicolumn{2}{c}{0.088} & \multicolumn{2}{c}{0.236} & \multicolumn{2}{c}{0.132}\tabularnewline
 & 150 & 50 & 0.102 & 0.011 & 0.122 & \multicolumn{2}{c}{0.074} & \multicolumn{2}{c}{0.157} & \multicolumn{2}{c}{0.097}\tabularnewline
$T=300$ & 75 & 225 & 0.057 & 0.023 & 0.168 & \multicolumn{2}{c}{0.070} & \multicolumn{2}{c}{0.293} & \multicolumn{2}{c}{0.123}\tabularnewline
\multirow{2}{*}{} & 150 & 150 & 0.069 & 0.019 & 0.192 & \multicolumn{2}{c}{0.094} & \multicolumn{2}{c}{0.322} & \multicolumn{2}{c}{0.164}\tabularnewline
 & 225 & 75 & 0.102 & 0.015 & 0.195 & \multicolumn{2}{c}{0.112} & \multicolumn{2}{c}{0.261} & \multicolumn{2}{c}{0.172}\tabularnewline
$T=400$ & 100 & 300 & 0.058 & 0.026 & 0.210 & \multicolumn{2}{c}{0.080} & \multicolumn{2}{c}{0.354} & \multicolumn{2}{c}{0.129}\tabularnewline
\multirow{2}{*}{} & 200 & 200 & 0.068 & 0.024 & 0.183 & \multicolumn{2}{c}{0.083} & \multicolumn{2}{c}{0.297} & \multicolumn{2}{c}{0.137}\tabularnewline
 & 300 & 100 & 0.089 & 0.016 & 0.152 & \multicolumn{2}{c}{0.080} & \multicolumn{2}{c}{0.237} & \multicolumn{2}{c}{0.133}\tabularnewline
 &  &  &  &  &  & \multicolumn{2}{c}{} & \multicolumn{2}{c}{} & \multicolumn{2}{c}{}\tabularnewline
\multicolumn{3}{l}{$\alpha=0.10$ } &  &  &  & \multicolumn{2}{c}{} & \multicolumn{2}{c}{} & \multicolumn{2}{c}{}\tabularnewline
 & $T_{m}$ & $T_{n}$ & \multicolumn{2}{c}{} &  & \multicolumn{2}{c}{} & \multicolumn{2}{c}{} & \multicolumn{2}{c}{}\tabularnewline
$T=100$ & 25 & 75 & 0.096 & 0.054 & 0.218 & \multicolumn{2}{c}{0.117} & \multicolumn{2}{c}{0.317} & \multicolumn{2}{c}{0.213}\tabularnewline
\multirow{2}{*}{} & 50 & 50 & 0.114 & 0.051 & 0.150 & \multicolumn{2}{c}{0.084} & \multicolumn{2}{c}{0.190} & \multicolumn{2}{c}{0.128}\tabularnewline
 & 75 & 25 & 0.166 & 0.044 & 0.178 & \multicolumn{2}{c}{0.126} & \multicolumn{2}{c}{0.213} & \multicolumn{2}{c}{0.164}\tabularnewline
$T=200$ & 50 & 150 & 0.152 & 0.051 & 0.171 & \multicolumn{2}{c}{0.110} & \multicolumn{2}{c}{0.397} & \multicolumn{2}{c}{0.227}\tabularnewline
\multirow{2}{*}{} & 100 & 100 & 0.128 & 0.058 & 0.213 & \multicolumn{2}{c}{0.122} & \multicolumn{2}{c}{0.315} & \multicolumn{2}{c}{0.181}\tabularnewline
 & 150 & 50 & 0.154 & 0.047 & 0.179 & \multicolumn{2}{c}{0.117} & \multicolumn{2}{c}{0.206} & \multicolumn{2}{c}{0.134}\tabularnewline
$T=300$ & 75 & 225 & 0.100 & 0.060 & 0.240 & \multicolumn{2}{c}{0.110} & \multicolumn{2}{c}{0.373} & \multicolumn{2}{c}{0.166}\tabularnewline
\multirow{2}{*}{} & 150 & 150 & 0.121 & 0.063 & 0.259 & \multicolumn{2}{c}{0.146} & \multicolumn{2}{c}{0.394} & \multicolumn{2}{c}{0.219}\tabularnewline
 & 225 & 75 & 0.145 & 0.054 & 0.238 & \multicolumn{2}{c}{0.143} & \multicolumn{2}{c}{0.333} & \multicolumn{2}{c}{0.218}\tabularnewline
$T=400$ & 100 & 300 & 0.110 & 0.078 & 0.270 & \multicolumn{2}{c}{0.120} & \multicolumn{2}{c}{0.463} & \multicolumn{2}{c}{0.212}\tabularnewline
\multirow{2}{*}{} & 200 & 200 & 0.110 & 0.071 & 0.257 & \multicolumn{2}{c}{0.121} & \multicolumn{2}{c}{0.380} & \multicolumn{2}{c}{0.135}\tabularnewline
 & 300 & 100 & 0.136 & 0.057 & 0.212 & \multicolumn{2}{c}{0.127} & \multicolumn{2}{c}{0.300} & \multicolumn{2}{c}{0.176}\tabularnewline
\hline 
\end{tabular}
\par\end{centering}
\noindent\begin{minipage}[t]{1\columnwidth}%
{\small{}The table reports the rejection probabilities of $5\%$-level
tests proposed in the paper and those proposed by \citet{giacomini/rossi:09}
{[}(abbreviated GR (2009){]} for model S1. For all methods we use
the recursive forecasting scheme. $T=T_{m}+T_{n}$, where $T$ is
the total sample size, $T_{m}$ is the size of the in-sample window
and $T_{n}$ is the size of the out-of-sample window. $m_{T}$ is
set equal to the smallest integer allowed by Condition \ref{Cond The-auxiliary-sequence}.
Based on 5,000 replications.}%
\end{minipage}
\end{table}

\end{document}